\documentclass{ut-thesis}

\usepackage{amsmath,amsthm,amsfonts,mathrsfs,caption,subcaption,graphicx,mathtools,xfrac,dsfont}
\usepackage{amssymb}
\usepackage{hyperref,xcolor}
\theoremstyle{plain}

  \theoremstyle{plain}
  
  \theoremstyle{remark}

\usepackage{cleveref}
\crefname{equation}{equation}{equations}
\Crefname{equation}{Equation}{Equations}
\crefrangelabelformat{equation}{(#3#1#4--#5#2#6)}

\crefmultiformat{equation}{(#2#1#3}{, #2#1#3)}{#2#1#3}{#2#1#3}
\Crefmultiformat{equation}{(#2#1#3}{, #2#1#3)}{#2#1#3}{#2#1#3}

\degree{Doctor of Philosophy}
\department{Physics}
\gradyear{2019}
\author{Jesse Cole Cresswell}
\title{Quantum Information Approaches to Quantum Gravity}

\newcommand{\n}{\noindent}
\newcommand{\tr}{\mathrm{tr}}
\newcommand{\parder}[2]{\frac{\partial {#1}}{\partial {#2}}}
\newcommand{\pa}{\partial}
\newcommand{\ot}{\otimes}
\newcommand{\id}{\mathbb{I}}
\newcommand{\op}{\mathcal{O}}
\newcommand{\h}{\mathcal{H}}
\newcommand{\ve}{\text{vec}\, }
\newcommand{\veT}{\text{vec}^T }
\newcommand{\hess}{\mathscr{H}}
\newcommand{\eq}[1]{\begin{equation}\begin{aligned}#1\end{aligned}\end{equation}}
\newcommand{\had}{\hat{a}^\dagger\vphantom{n}}
\newcommand{\ha}{\hat{a}\vphantom{n}}
\newcommand{\iu}{\text{i}}
\newcommand{\eu}{\text{e}}
\newcommand{\ket}[1]{\left|#1\right\rangle}
\newcommand{\bra}[1]{\left\langle#1\right|}

\newtheorem*{theorem*}{Theorem}

\newcommand{\al}{\alpha}\newcommand{\be}{\beta}\newcommand{\ga}{\gamma}\newcommand{\de}{\delta}\newcommand{\ep}{\epsilon}\newcommand{\te}{\theta}\newcommand{\la}{\lambda}\newcommand{\si}{\sigma}\newcommand{\om}{\omega}

\newcommand{\Ga}{\Gamma}\newcommand{\De}{\Delta}\newcommand{\Te}{\Theta}\newcommand{\La}{\Lambda}\newcommand{\Si}{\Sigma}\newcommand{\Om}{\Omega}

\begin{document}

\begin{preliminary}

\maketitle


\begin{abstract}


In this thesis we apply techniques from quantum information theory to study quantum gravity within the framework of the anti-de Sitter / conformal field theory correspondence (AdS/CFT). A great deal of interest has arisen around how quantum information ideas in CFT translate to geometric features of the quantum gravitational theory in AdS. Through AdS/CFT, progress has been made in understanding the structure of entanglement in quantum field theories, and in how gravitational physics can emerge from these structures. However, this understanding is far from complete and will require the development of new tools to quantify correlations in CFT.

This thesis presents refinements of a duality between operator product expansion (OPE) blocks in the CFT, giving the contribution of a conformal family to the OPE, and geodesic integrated fields in AdS which are diffeomorphism invariant quantities. This duality was originally discovered in the maximally symmetric setting of pure AdS dual to the CFT ground state. In less symmetric states the duality must be modified. Working with excited states within AdS$_3$/CFT$_2$, this thesis shows how the OPE block decomposes into more fine-grained CFT observables that are dual to AdS fields integrated over non-minimal geodesics. These constructions are presented for several classes of asymptotically AdS spacetimes.

Additionally, this thesis contains results on the dynamics of entanglement measures for general quantum systems, not necessarily confined to quantum gravity. The quantification of quantum correlations is the main objective of quantum information theory, and it is crucial to understand how they are generated dynamically. Results are presented for the family of quantum R\'enyi entropies and entanglement negativity. R\'enyi entropies are studied for general dynamics by imposing special initial conditions. Around pure, separable initial states, all R\'enyi entropies grow with the same timescale at leading, and next-to-leading order. For negativity, mathematical tools are developed for the differentiation of non-analytic matrix functions with respect to constrained arguments. These tools are used to construct analytic expressions for derivatives of negativity. We establish bounds on the rate of change of state distinguishability under arbitrary dynamics, and the rate of entanglement growth for closed systems.

\end{abstract}



\begin{dedication}
To my parents Kimberley and Larry

\end{dedication}


\vspace{.5in}  


\begin{acknowledgements}
The work in this thesis could not have been completed without the constant support of my entire family, and especially Chantelle.

I am grateful to my supervisor A.W. Peet for the countless discussions, and numerous research opportunities they provided over the years.

I would like to thank my coauthors Ian Jardine, Aaron Goldberg, and Ilan Tzitrin for their collaborative efforts, as well as present and past group members Thomas de Beer, Daniel O'Keeffe, Callum Quigley, and Zaq Carson. Additionally, thanks to my colleagues and officemates Hudson Pimenta, Kevin Marshall, Nichol\'as Quesada, Jaspreet Sahota for the many discussions we shared. 

I also appreciate the guidance of my committee members Erich Poppitz and Harald Pfeiffer throughout the program.
\end{acknowledgements}

\tableofcontents


\listoffigures


\end{preliminary}


\chapter{Introduction}\label{ch:intro}

It is well known that the Einstein-Hilbert action for gravity, plus higher curvature corrections, represents a non-renormalizable theory if the spacetime metric is treated as a dynamical quantum field, since the coupling constant $(16\pi G_N)^{-1}$ has positive mass dimension. Although there is no obstruction to applying quantum field theory techniques to general relativity at low energies where there will be an effective field theory description, we do not currently understand the theory's ultraviolet (UV) completion. The Anti-de Sitter / conformal field theory correspondence (AdS/CFT) \cite{Maldacena1999} is one of the most common tools used in our pursuit of a true theory of quantum gravity, as it allows us to reframe questions of quantum gravity in terms of non-gravitational quantum field theories (QFT), in particular, ones with manifest scaling symmetry giving a straightforward UV completion. 

While most work on AdS/CFT has been geared towards understanding quantum gravity in terms of CFTs, the correspondence also works in the other direction. Conformal field theories arise in numerous models of physical systems, especially in condensed matter physics where they often describe the physics of quantum critical systems since generically the endpoints of renormalization group flows are CFTs. In some cases strongly coupled critical systems can be described more simply by a dual gravitational theory, since AdS/CFT is a strong-weak duality \cite{Hartnoll2009,McGreevy2009}. When the CFT is strongly coupled, quantum effects in the gravity theory are suppressed and we recover a semiclassical theory. A similar application arises in quantum chromodynamics (QCD) for describing the quark-gluon plasma produced in relativistic heavy ion collisions \cite{Liu2014b}, while AdS/CFT has also been used to describe effective theories of relativistic hydrodynamics \cite{Bhattacharyya2008,Haehl2016}.

This thesis is primarily concerned with the application of techniques from quantum information theory to quantum gravity through the AdS/CFT correspondence. As CFTs are ordinary quantum theories, states typically exhibit quantum correlations which can be quantified using ideas from quantum information theory. Interest in the overlap of these disciplines blossomed when it was realized quantum information becomes encoded in the geometry of the dual quantum gravity theories, and even leads to gravitational dynamics governed by the Einstein equations. The influx of quantum information ideas to quantum gravity has had a reciprocal effect, with the development of new techniques for measuring quantum correlations in general, and new insights into the properties of well established quantities. In this thesis we will provide an overview of the interchange of ideas between these fields. The novel work forming the body of this thesis is split into two parts. The first half concerns the refinement of a specific duality within AdS/CFT which stems from measures of quantum information that were designed with quantum gravity in mind. In the remainder, we study the dynamical properties of well-established entanglement measures in general quantum systems.

In this introduction we will make the case that quantum information tools have diverse application in AdS/CFT, and play a large role in our understanding of how gravitational phenomena such as black holes and dynamical spacetimes can emerge from non-gravitational field theories. After a review of the most important aspects of conformal field theories, we study the holotype for measures of quantum information in QFT and CFT, entanglement entropy. We discuss the important physical implications of its universal behaviour, but also the shortcomings of its traditional definition. A deeper dive into algebraic aspects of QFT allows us to construct mathematically consistent measures of quantum information while simultaneously exposing the foundational entanglement properties of quantum field theories. We then build off these notions to explore how entanglement in CFT can be understood geometrically in quantum gravity via the AdS/CFT correspondence, and the implications entanglement has for the emergence of spacetime.

\section{Conformal field theory}

Compared to ordinary relativistic quantum field theories, conformal field theories have a significantly enlarged symmetry group which allows for greater theoretical control. An enormous literature has been developed on the unique properties of this class of theories along with special mathematical techniques that take advantage of the extra symmetry. This section will only include a brief introduction to the principal advantages of CFTs that have allowed the study of AdS/CFT to flourish over the past two decades. A more comprehensive introduction can be found in the standard textbook \cite{Francesco1997}.

Relativistic quantum field theories including the Standard Model obey the Poincar\'e symmetry of spacetime transformations that leaves the Minkowski metric invariant, along with possibly some internal symmetry groups. Poincar\'e transformations include rotations, boosts, and translations comprising the $\mathbb{R}^{1,d-1}\rtimes SO(1,d-1)$ group, where $d$ is the total number of spacetime dimensions. In addition to this, conformal theories are invariant under scaling operations, or dilatations, as well as so-called special conformal transformations, which can be viewed as the composite operation of an inversion, translation, and inversion. In total, these operations form the $SO(2,d)$ conformal group ($SO(1,d+1)$ in Euclidean signature), and have the overall property of preserving the metric up to a scale factor $g_{\mu\nu}'(x')=\La(x)g_{\mu\nu}(x)$.

The generators of each type of transformation can be expressed as
\begin{equation}
\begin{alignedat}{2}
&P_\mu&&=-i\pa_\mu,\\
&L_{\mu\nu}&&=i(x_\mu\pa_\nu-x_\nu\pa_\mu),\\
&D &&=-ix^\mu \pa_\mu,\\
&K_{\mu}&&=-i(2x_\mu x^\nu \pa_\nu-x^\nu x_\nu \pa_\mu).
\end{alignedat}
\end{equation}
These generate translations, rotations and boosts, dilatations, and special conformal transformations respectively, while satisfying the $so(2,d)$ conformal algebra
\begin{equation}
\begin{alignedat}{2}
&[D,P_\mu]&&=iP_\mu,\\
&[D,K_\mu]&&=-iK_\mu,\\
&[K_\mu,P_\nu]&&=2i(\eta_{\mu\nu} D-L_{\mu\nu}),\\
&[K_\si,L_{\mu\nu}]&&=i(\eta_{\si\mu}K_\nu-\eta_{\si\nu}K_\mu),\\
&[P_\si,L_{\mu\nu}] &&=i(\eta_{\si\mu}P_\nu-\eta_{\si\nu}P_{\mu}),\\
&[L_{\mu\nu},L_{\si\rho}]&&=i(\eta_{\nu\si}l_{\mu\rho} +\eta_{\mu\rho}L_{\nu\si}-\eta_{\mu\si}L_{\nu\rho}-\eta_{\nu\rho}L_{\mu\si}).
\end{alignedat}
\end{equation}

An important property of conformal fields is their scaling dimension $\De$ in response to a dilatation
\begin{equation}
\phi(\la x)=\la^{-\De}\phi(x).
\end{equation}
In more generality, the transformation of a spinless field under a general conformal transformation is
\begin{equation}\label{eq:quasiprimary}
\phi'(x')=\left| \frac{\pa x'}{\pa x}\right|^{-\De/d}\phi(x).
\end{equation}
Any field that transforms in this way, with $\left| \frac{\pa x'}{\pa x}\right|$ the Jacobian of the transformation, is called quasi-primary. Fields of this type play a major role in AdS/CFT due to their highly constrained properties.

One of the most powerful consequences of conformal symmetry is that the 2-point and 3-point correlation functions of  quasi-primary operators are almost entirely fixed. For instance, the 2-point function of spinless fields contains only a single arbitrary constant $C_{ij}$ which can be determined by the normalization of the fields,
\begin{equation}\label{eq:2-ptfunction}
\langle \phi_i(x_i)\phi_j(x_j)\rangle=\frac{C_{ij}}{x_{ij}^{2\Delta}},
\end{equation}
where $x_{ij}\equiv |x_i-x_j|$, and the fields $\phi_i$ and $\phi_j$ must have the same scaling dimension $\De$ for a non-zero result. Similarly, the structure of 3-point functions is mostly fixed,
\begin{equation}\label{eq:3-ptfunction}
\langle \phi_i(x_i)\phi_j(x_j)\phi_k(x_k)\rangle=\frac{C_{ijk}}{x_{ij}^{\Delta_i+\Delta_j-\Delta_k}x_{jk}^{\Delta_j+\Delta_k-\Delta_i} x_{ik}^{\Delta_k+\Delta_i-\Delta_j}},
\end{equation}
which involves the set of theory-dependent 3-point coefficients $C_{ijk}$. Correlation functions of higher numbers of fields are not fixed in the same way. When there are at least 4 distinct positions involved we can form conformally invariant combinations of the points called cross ratios. For instance, 4-point functions can involve an arbitrary function $g(u,v)$ of the combinations
\begin{equation}
u=\frac{x_{12}^2x_{34}^2}{x_{13}^2x_{24}^2},\quad v=\frac{x_{14}^2x_{23}^2}{x_{13}^2x_{24}^2}.
\end{equation}
A particularly useful way to write the 4-point function of scalars for later purposes is
\begin{equation}\label{eq:4pointfunction}
\langle \phi_1(x_1)\phi_2(x_2)\phi_3(x_3)\phi_4(x_4)\rangle =\left(\frac{x_{24}^2}{x_{14}^2}\right)^{\tfrac{1}{2} \De_{12}}\left(\frac{x_{14}^2}{ x_{13}^2}\right)^{\tfrac{1}{2} \De_{34}}{g(u,v)\over (x_{12}^2)^{\tfrac{1}{2} (\De_1+\De_2)}(x_{34}^2)^{\tfrac{1}{2}(\De_3+\De_4)}},
\end{equation}
where $\De_{ij}=\De_i-\De_j$.

Despite the freedom in $n$-point functions, there are powerful techniques to relate them to sums of $(n-1)$-point functions, which can be applied iteratively eventually reaching the fixed 2 and 3-point structures. The fundamental tool enabling this is the operator product expansion (OPE) which expresses a product of quasi-primary operators in terms of a local basis of other operators in the theory \cite{Belavin1984}. For two scalar operators $\op_{i}(x_i)$ and $\op_{j}(x_j)$ with scaling dimensions $\De_i$ and $\De_j$ respectively, the OPE in the limit $x\to0$ takes the form
\begin{equation}\label{eq:ope1}
\op_{i}(x)\op_{j}(0) =  \sum_{k}C_{ijk}\left|x\right|^{\Delta_{k}-\Delta_{i}-\Delta_{j}}\big(1+b_{1}\,x^{\mu}\partial_{\mu}+b_{2}\,x^{\mu}x^{\nu}\partial_{\mu}\partial_{\nu}+\ldots\big)\op_{k}(0).
\end{equation}
In this expression $\op_k(x_k)$ represents another quasi-primary operator in the theory with dimension $\De_k$, and the sum is over all such operators. The derivative terms act on $\op_k(x_k)$ to produce descendant operators which are not quasiprimary, but are in the same representation of the conformal group as $\op_k(x_k)$. Additionally, the constants $b_i$ are completely fixed by conformal symmetry. This can be seen by taking the OPE of two operators inside a 3-point function, resulting in a sum of derivatives of the fixed 2-point structure \eqref{eq:2-ptfunction},
\begin{equation}
\left\langle \phi(y)\op_{i}(x_i)\op_{j}(x_j) \right\rangle = \sum_{k}C_{ijk}\left|x_i-x_j\right|^{\Delta_{k}-\Delta_{i}-\Delta_{j}}\big(1+b_{1}\,x_j^{\mu}\partial^{x_j}_{\mu}+b_{2}\,x_j^{\mu}x_j^{\nu}\partial^{x_j}_{\mu}\partial^{x_j}_{\nu}+\ldots\big)\left\langle\phi(y)\op_{k}(x_j)\right\rangle.
\end{equation}
 Consistency between the derivatives of 2-point functions and the 3-point structure itself \eqref{eq:3-ptfunction} determines the $b_i$ coefficients \cite{Belavin1984}. Hence, the OPE can be used recursively to write $n$-point functions in terms of the CFT data consisting of the spectrum of quasiprimary operators $\op_k(x_k)$ with their associated dimensions $\De_k$, and the 3-point coefficients $C_{ijk}$. While the OPE can be used in any QFT, its special form in CFT becomes even more useful as the expansion is not only valid in the $x_i\to x_j$ limit, but is absolutely convergent at finite separations, as long as no other operators are within a radius of $|x_i-x_j|$ \cite{Pappadopulo2012,Rychkov2016}.

It is also important to note that in the special case of $d=2$ CFTs, the conformal group becomes much larger than the expected $SO(2,2)$ or $SO(1,3)$. This is most easily expressed in Euclidean signature, related to the Lorentzian case by Wick rotation. In this case we can utilize complex coordinates $z,\bar z$ on the plane, in which case any (anti-)holomorphic function $f(z)$ ($\bar f( \bar z)$) gives a valid conformal map. Such transformations can be generated by $l_n=-z^{n+1} \pa_z$ and $\bar l_n=-\bar z^{n+1} \pa_{\bar z}$ for all $n\in\mathbb{Z}$, satisfying the Witt algebras
\begin{align}\begin{aligned}
[l_n,l_m]&=(n-m)l_{n+m},\\
[\bar{l}_n,\bar l_m]&=(n-m)\bar l_{n+m},\\
[l_n,\bar l_m]&=0.
\end{aligned}\end{align}
Notice that the subgroup generated by elements $l_n$ over $\mathbb{C}$ with $n=\{-1,0,1\}$ is $SL(2,\mathbb{C})$, isomorphic to $SO(1,3)$. This important subgroup is often called the global conformal subgroup. In the two-dimensional context, we reserve the terminology ``primary" for fields which transform like \eqref{eq:quasiprimary} under all conformal transformations, while quasiprimary fields may transform like \eqref{eq:quasiprimary} under only the global subgroup. In the quantum theory, due to the trace anomaly of the stress tensor, the Witt algebras are replaced by their unique central extension, the Virasoro algebra
\begin{equation}\label{eq:Virasoroalg}
[L_n,L_m]=(n-m)L_{n+m}+\frac{c}{12}(m^3-m)\de_{m+n,0},
\end{equation}
which incorporates the constant central charge $c$. This constant plays a significant role in 2D CFTs, including in the transformation properties of the stress tensor, a non-quasiprimary field, and appears in physical quantities like the entanglement entropy which we turn to next.

\section{Entanglement entropy in quantum field theory}\label{sec:EEQFT}

In the context of QFT, entanglement entropy is an important quantity for expressing one of the key features of typical low energy states such as the vacuum; they are highly entangled. This fact underpins many of the interesting recent developments in quantum gravity, such as the connection between spatial entanglement and the emergence of spacetime \cite{VanRaamsdonk2010}. It is perhaps best expressed by the Reeh-Schlieder theorem of algebraic quantum field theory \cite{Reeh1961} which states that starting with the vacuum of a QFT in Minkowski spacetime, the states generated by smeared operators supported in an arbitrarily small region of spacetime are dense in the entire vacuum sector of the Hilbert space of the theory. In essence, the vacuum state is spatially entangled between any two local regions of the spacetime, such that operations in any region can affect any other \cite{Witten2018}. To understand the consequences of this statement, it is of great interest to quantify the amount of entanglement in a quantum system. The most elementary tool for this is the entanglement entropy.

The origin of entanglement entropy is in quantum information theory where it is defined in terms of the reduced density matrices of a state $\left | \psi\right\rangle$ with respect to a bipartition of the Hilbert space. If the Hilbert space factorizes as $\mathcal{H}=\mathcal{H}_A\ot\mathcal{H}_B$, then the subsystem $A$ described by $\rho_A=\tr_B \left | \psi\right\rangle\left\langle \psi \right |$ has entanglement entropy
\begin{equation}\label{eq:EEdefinition}
S(\rho_A)=-\tr_A(\rho_A\log\rho_A).
\end{equation}
This quantity will be zero for all pure states with no entanglement between subsystems $A$ and $B$, so called separable states of the form $\left | \psi\right\rangle= \left| \psi\right\rangle_A\ot\left| \psi \right\rangle_B$, and is non-zero for all non-separable states. The value of $S(\rho_A)$ is directly tied to the ability of performing operational tasks that rely on using entanglement as a resource, at least in pure states \cite{Plenio2005}.

Despite this simple characterization of entanglement for pure states in quantum mechanics, when dealing with quantum fields things are significantly more complicated. First, the dimensionality of the Hilbert space of a quantum field theory is typically infinite, so that a definition like \eqref{eq:EEdefinition} would involve tracing out an infinite number of degrees of freedom. While the dimensionality alone is not an insurmountable difficulty, after all quantum harmonic oscillators have infinite dimensional Hilbert spaces but still can have sensible entanglement entropies, it does already suggest that \eqref{eq:EEdefinition} may be divergent in QFT. In the algebraic approach to QFT local algebras associated to spatial subsystems are of a type where a normalized trace cannot be defined so that the existence of the trace used in \eqref{eq:EEdefinition} cannot be taken for granted. Second, the Hilbert space of a QFT does not necessarily factorize across spatial bipartitions, the typical counterexample being gauge theories \cite{Casini2014,Ghosh2015,Lin2018}. To identify gauge invariant states, the physical states of the theory, one must look at the system overall, and not simply at subregions. In other words, gauge constraints relate degrees of freedom at different spatial locations meaning that they do not factorize along spatial lines. As a result, it becomes difficult to uniquely define what is meant by the reduced density matrix. At the very least, choices must be made as to which degrees of freedom are or are not traced out near the boundary of a particular partition. Some QFTs escape this caveat, such as the lattice regularized free boson theory which does factorize over lattice sites, but even this is plagued by the third problem of UV divergences. In the continuum limit there are modes of the field at arbitrarily small scales, and in typical states these UV modes will be entangled across any partition. This again suggests that \eqref{eq:EEdefinition}, when it is even possible to define the reduced density matrix, will be UV divergent. Theories can be regulated with a UV cutoff, for example by working on a lattice with a minimum scale, which can in some cases assuage the three concerns mentioned here, but it is a delicate matter to remove the cutoff while maintaining physically sensible results. 

\begin{figure}
    \centering
    \begin{subfigure}[b]{0.40\textwidth}
        \includegraphics[width=0.9\textwidth]{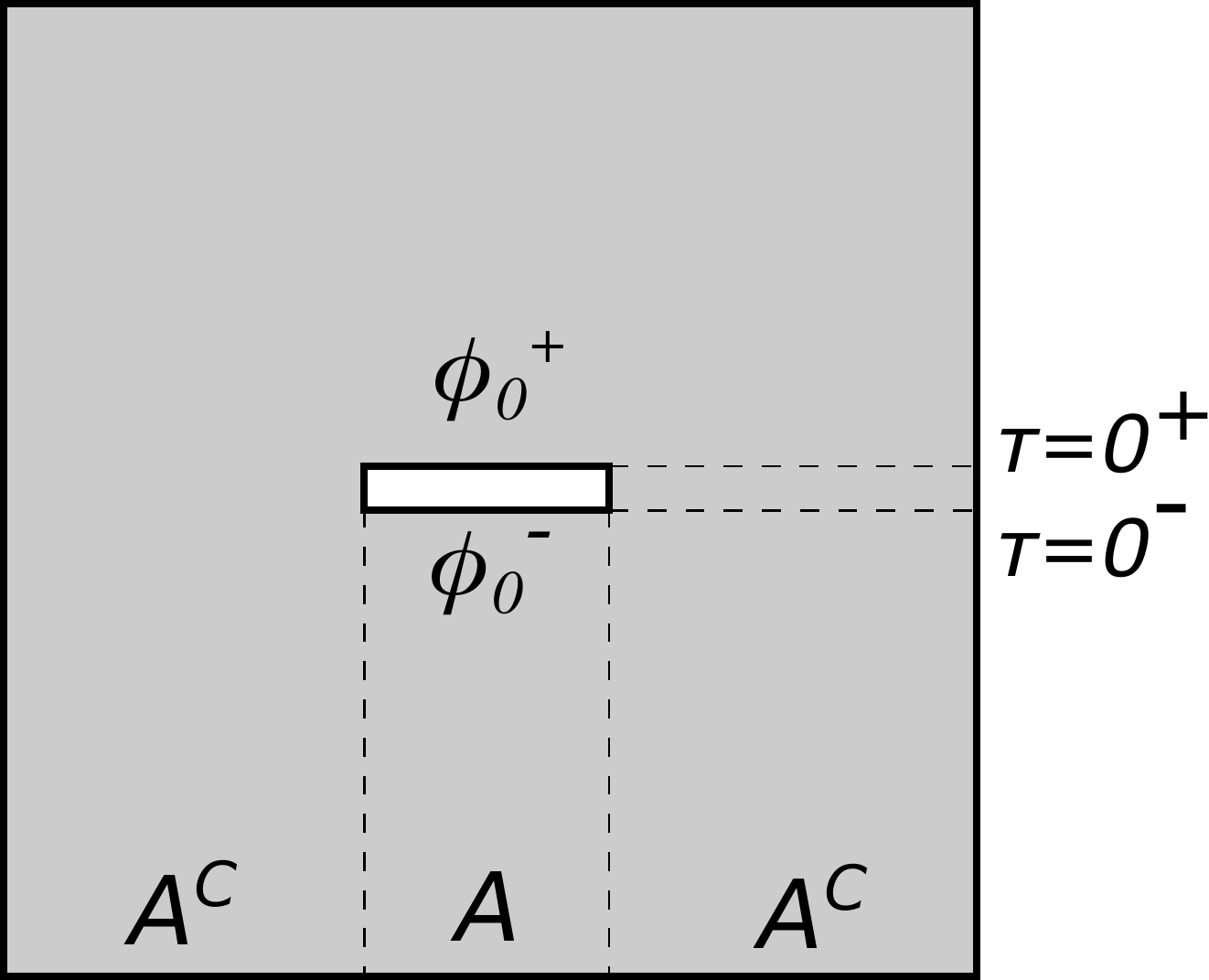}
        \caption{}
        \label{fig:pathintegral1}
    \end{subfigure}
     \ \qquad
    \begin{subfigure}[b]{0.31\textwidth}
        \includegraphics[width=0.9\textwidth]{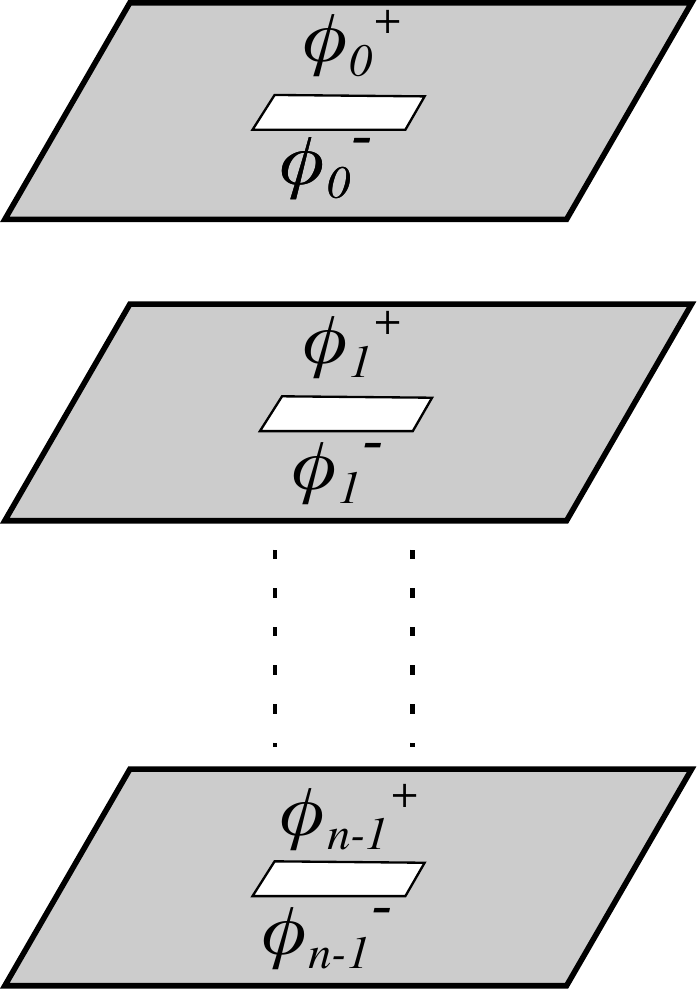}
        \caption{}
        \label{fig:pathintegral2}
    \end{subfigure}
    \caption[Euclidean path integrals]{(a) Euclidean path integral preparing the reduced density matrix $\rho_A$. The ground state wavefunction boundary conditions are inserted on either side of the cut along interval $A$. (b) $\tr_A \rho_A^n$ is prepared by taking $n$ copies of the path integral preparing $\rho_A$, and identifying their boundary conditions cyclically, such that $\phi_j^+=\phi_{j+1}^-$, and additionally $\phi_{n-1}^+=\phi_0^-$, forming an $n$-sheeted Riemann surface of integration. }\label{fig:pathintegral}
\end{figure}

The groundbreaking replica trick technique \cite{Calabrese2004,Calabrese2009} allowed the first systematic calculations of entanglement entropy for lattice regularized QFTs in 2 dimensions, putting previous arguments about its expected behaviour on solid footing \cite{Srednicki1993,Holzhey1994}. In the replica trick, one starts with a generalization of entanglement entropy to the R\'enyi entropies
\begin{equation}\label{eq:Renyidefinition}
S_n(\rho_A)=\frac{1}{1-n}\log \tr_A\rho_A^n,
\end{equation}
from which the entanglement entropy is recovered in the $n\to 1 $ limit since $S(\rho_A)=-[\frac{\pa}{\pa n} \tr_A \rho_A^n]_{n=1}$. This definition is simpler to handle, since it involves the logarithm of the trace (a number) rather than the logarithm of the density matrix itself. Strictly speaking, the powers of the reduced density matrix are only guaranteed to be defined for positive integers $n$, yet the limit $n\to1$ requires us to extend the definition to the reals. This can be a subtle matter as there is often not a unique analytic continuation given the data of $S_n(\rho_A)$ at the positive integers. Nevertheless, an individual copy of $\rho_A$ can be prepared as followed. The ground state wavefunctional for the field $\phi(\tau, x)$ corresponds to a Euclidean path integral with boundary conditions inserted at $\tau=0$,
\begin{equation}
\Psi(\phi_0(x))=\int_{\tau=-\infty}^{\phi(\tau=0,x)=\phi_0^-(x)} \mathcal{D}\phi\, e^{-S_E(\phi)}.
\end{equation}
The integration from $\tau=-\infty$ to $0$ of the exponential of the Euclidean action $S_E(\phi)$ damps out any possible excitations, ensuring we prepare the ground state. A similar construction integrating from $\infty$ to $0$ with boundary conditions $\phi_{0}(x)^+$ gives the complex conjugate $\bar\Psi$, and together this constructs the density matrix $\rho=\Psi(\phi_0^-(x))\bar\Psi(\phi_{0}^+(x))$. Tracing over the complement of the region $A$ has the effect of setting $\phi_0^+(x)=\phi_{0}^-(x)$ when $x\in A^c$. Hence, the reduced density matrix of the ground state for an interval $A$ along $\tau=0$ is given by the path integral over the Euclidean space with a cut along $A$, and boundary conditions inserted on either side of the cut, as depicted in Fig. \ref{fig:pathintegral1}.

We move from $\rho_A$ to $\tr_A\rho_A^n$ by preparing $n$ copies of the path integral just described, and stitching together the copies sequentially along $A$ by matching the boundary conditions as $\phi_0^+(x)=\phi_1^-(x)$, $\phi_1^+(x)=\phi_2^-(x)$, ..., $\phi_{n-1}^+(x)=\phi_0^-(x)$. In total, $\tr_A\rho_A^n$ is computed by the Euclidean path integral over the $n$-sheeted Riemann surface given by the stitching procedure as shown in Fig. \ref{fig:pathintegral2}. Then, taking the derivative and $n\to 1$ limit produces the ground state entanglement entropy of the single region $A$. This is easier said than done, since actually performing the path integral can be difficult. For 2d CFTs in particular the path integral can actually be evaluated, and the complete result is \cite{Calabrese2004,Calabrese2009}
\begin{equation}\label{eq:2dCFTintervalEE}
S(\rho_A)=\frac{c}{3} \log \frac{l}{a},
\end{equation}
with $l$ the length of interval $A$, and $a$ the lattice spacing. Furthermore, $c$ is the central charge of the CFT appearing in the Virasoro algebra \eqref{eq:Virasoroalg}, but this is the only theory-dependent detail of the CFT that enters this universal result. 

A crucial feature of this expression is the UV divergence as we take the continuum limit $a\to0$. In fact this divergence does not depend on our choice of the ground state, as it also appears in the entanglement entropy calculated when other operators are inserted into the path integral \cite{Alcaraz2011,Berganza2012}. Changing the state while leaving the entangling region the same only changes the entanglement entropy by a finite amount. In other words, every state looks like the vacuum at short enough distances, and will therefore have the same leading UV divergence. From general arguments we always expect that $S(\rho_A)$ for a region in QFT will have a divergence proportional to the area of the boundary of $A$ \cite{Bombelli1986,Srednicki1993,Holzhey1994},
\begin{equation}\label{eq:arealaw1}
S(\rho_A)=c_0 \frac{\text{Area}(\pa A)}{a^{d-2}}+...
\end{equation}
 with $c_0$ a theory dependent constant, and $d$ the total dimensionality of spacetime. This area law expresses the idea that it is UV scale entanglement across the codimension-2 entangling surface $\pa A$ that leads to divergences. We note that the previous result \eqref{eq:2dCFTintervalEE} is slightly different because with one spatial dimension, the entangling surface is just a set of points, and the power law divergence $a^{-(d-2)}$ is tempered to a logarithmic divergence. Following these milestone results on calculating entanglement entropy in QFT, many variations and extensions have been explored, including working with higher dimensional theories, more complicated entangling regions, using quantum fields with different spins and statistics, and many forms of interactions.
 
 Instead of entanglement entropy it can be more useful to study UV finite quantities like the relative entropy, defined for factorizable Hilbert spaces as
 \begin{equation}\label{eq:relativeentropy1}
S(\rho_A||\si_A)=\tr_A(\rho_A\log\rho_A)-\tr_A(\rho_A\log\si_A),
\end{equation}
or the mutual information
\begin{equation}\label{eq:mutualinformation1}
I(A:B)=S(\rho_A)+S(\rho_B)-S(\rho_{AB}).
\end{equation}
Relative entropy is a measure of distinguishability of states, since it is zero if and only if $\rho_A=\si_A$, otherwise being positive, and acts as a distance measure between states in Hilbert space (though it is not a metric as it is not symmetric). In the relative entropy, two states $\rho$ and $\si$ are compared on the region $A$. Since both exhibit the same UV divergence, the difference of the two entropy-like terms can be finite\footnote{The second term in \eqref{eq:relativeentropy1} can be divergent for other reasons, namely when $\rho_A$ has support on the kernel of $\si_A$. In this case the states are perfectly distinguishable.}. The mutual information measures the total correlations, both quantum and classical, shared by regions $A$ and $B$ in the potentially mixed state $\rho_{AB}$. It is also finite since it can be written as the relative entropy between $\rho_{AB}$, and the uncorrelated product of reduced density matrices $\rho_A$ and $\rho_B$,
\begin{equation}\label{eq:mutualinformation2}
I(A:B)=S(\rho_{AB}||\rho_A\ot\rho_B).
\end{equation}
 
 The definitions given so far for entanglement entropies, reduced density matrices, and relative entropies are borrowed from quantum information theory and rely on assumptions which cannot honestly hold in QFT. Eq.\eqref{eq:EEdefinition} uses the factorization of the Hilbert space in order to separate degrees of freedom in $A$ from those in $A^c$, but this and the calculation leading to \eqref{eq:2dCFTintervalEE} make use of a lattice regularization. Furthermore it is not apparent that a trace over $A^c$ can be normalized appropriately when removing a non-countably infinite number of degrees of freedom associated with $A^c$ in the continuum theory. It is therefore desirable to work directly with the continuum theory and avoid these assumptions. However this requires a great overhaul to the definitions already made. 
 
 Starting with a QFT in Minkowski spacetime, such as a Hermitian scalar field $\phi(x)$, the vacuum state $\left | \Om\right\rangle$ can be used to build the vacuum sector Hilbert space $\mathcal{H}_0$ by acting with smeared operators $\phi_f=\int d^{d}x f(x)\phi(x)$ for smooth functions $f(x)$, 
 \begin{equation}\label{eq:buildpsi}
\left | \Psi_{\vec f}\right\rangle=\phi_{f_1}\phi_{f_2}...\phi_{f_n}\left | \Om\right\rangle.
\end{equation}
 Allowing the functions $f_i$ to have support on the entire spacetime ensures that all states in the vacuum sector $\mathcal{H}_0$ are generated. This is overkill though, since we expect that data limited to an initial value hypersurface, or Cauchy slice $\Si$, should be sufficient to generate $\mathcal{H}_0$. Therefore we can restrict to functions $ f_i$ supported in an open neighbourhood $\mathcal{U}$ of $\Si$. These statements apply directly to the continuum theory and make no assumptions about possible factorizations of $\mathcal{H}_0$. 
 
 The primary result indicative of the major role entanglement plays in QFT pushes the previous restriction to the extreme: the Reeh-Schlieder theorem allows us to restrict $f_i$ to have support in a neighbourhood $\mathcal{U}_{\mathcal{V}}$ of an arbitrarily small open set $\mathcal{V}\subset\Si$, and still generate $\mathcal{H}_0$ with states of the form \eqref{eq:buildpsi} \cite{Reeh1961,Haag1996,Witten2018}. In more detail, one can show that any proposed state $\left | \chi\right\rangle$ which is orthogonal to all states of the form \eqref{eq:buildpsi} for $f_i$ supported in $\mathcal{U}_{\mathcal{V}}$ must also be orthogonal to all states created without the restriction to $\mathcal{U}_{\mathcal{V}}$. This argument relies on the Hamiltonian satisfying $H \left | \Om\right\rangle=0$ and this being the lower bound of the operator. But then $\left | \chi\right\rangle$ is orthogonal to all states in $\mathcal{H}_0$, and must be zero. Considering the local algebra of operators supported in $\mathcal{U}_{\mathcal{V}}$ as $\mathcal{A}_\mathcal{U}$, we say that a state is cyclic with respect to $\mathcal{A}_\mathcal{U}$ when the states $a\left | \Psi\right\rangle$ for $a\in\mathcal{A}_\mathcal{U}$ are dense in $\mathcal{H}_0$. From a physical point of view, the cyclicity of the vacuum is indicative of the nonlocal vacuum fluctuations we expect to be present in a QFT. An experimenter working in a local region can perform measurements that exploit these fluctuations to produce any state in the vacuum sector over the entire spacetime.

  When $\mathcal{V}$ is a small open subset of $\Si$ we can consider the complement of its closure to be $\mathcal{V}'$, another open set spacelike separated from $\mathcal{V}$. Then any operator supported in the neighbourhood $\mathcal{U}_{\mathcal{V}}$ will commute with operators supported in $\mathcal{U}_{\mathcal{V'}}$, so long as the neighbourhoods are also taken to be spacelike separated. Consider one such operator $a$ in $\mathcal{U}_{\mathcal{V}}$, and suppose that it annihilates the vacuum, $a\left | \Om\right\rangle=0$. It would then follow that
 \begin{equation}
a\phi_{f(x_1)}\phi_{f(x_2)}...\phi_{f(x_n)}\left | \Om\right\rangle=0,\quad x_i\in \mathcal{U}_{\mathcal{V'}},
\end{equation}
by commuting $a$ onto $\left | \Om\right\rangle$. But the Reeh-Schlieder theorem applies to $\mathcal{U}_{\mathcal{V'}}$ as well, and implies that the states $\phi_{f(x_1)}\phi_{f(x_2)}...\phi_{f(x_n)}\left | \Om\right\rangle$ are dense in $\mathcal{H}_0$, so that $a$ annihilates all states in $\mathcal{H}_0$ and must be the zero operator, $a=0$. Clearly a similar construction would apply to any region $\mathcal{U}_{\mathcal{V}}$ of the spacetime which is spacelike separated from some open region to which Reeh-Schlieder can be applied. We say that a state $\left | \Psi\right\rangle$ is separating with respect to $\mathcal{A}_\mathcal{U}$ when $a\left | \Psi\right\rangle=0$ implies $a=0$ for any $a\in\mathcal{A}_\mathcal{U}$. The implication is that a state that is cyclic for $\mathcal{A}_{\mathcal{U}}$ will be separating for any other algebra of local operators which all commute with $\mathcal{A}_\mathcal{U}$. From a physical point of view, the separating property implies that there are no localized operators which annihilate the vacuum. Conversely, true particle excitations represented by states orthogonal to the vacuum cannot be produced by localized operators \cite{Knight1961}. Despite this, it is a mistake to think of the vacuum as empty. The separating property also implies that the vacuum has non-zero overlap with all states created by operators in $\mathcal{A}_\mathcal{U}$ for open regions $\mathcal{U}$.

In this language, the Reeh-Schlieder theorem says that the vacuum of our QFT is cyclic and separating for any local algebra $\mathcal{A}_\mathcal{U}$ constructed as above. In fact the theorem can be extended to states in the theory which have bounded energy, meaning that cyclic separating states are commonplace in QFT. Let us now unpack these properties and understand the role of entanglement in its implications. If $b$ is any operator supported in the region $\mathcal{U}_{\mathcal{V'}}$, then there must exist another operator $a$ supported in $\mathcal{U}_{\mathcal{V}}$ such that
\begin{equation}\label{eq:statecompare}
a \left| \Om\right\rangle= b\left| \Om\right\rangle,
\end{equation}
since either local algebra generates the same vacuum sector Hilbert space\footnote{This type of relationship, an operational symmetry of the state, motivates a method of measuring entanglement which we have developed in \cite{Cresswell2019a}.}. We are free to choose $\mathcal{U}_{\mathcal{V}}$ arbitrarily small, and to be located at the other end of the universe from $\mathcal{U}_{\mathcal{V'}}$, yet the approximation between the states in \eqref{eq:statecompare} can be made arbitrarily good. At face value this may seem to violate the sacred principles of causality and locality, however it is important to note two things. First, the operator $a$ will typically not be simple, nor unitary. We note that since $a$ and $b$ commute, $\left\langle a\Om|b|a\Om\right\rangle=\left\langle \Om|b a^\dag a|\Om\right\rangle$, but this need not be equal to $\left\langle \Om|b|\Om\right\rangle$. $b$ could represent some operator with very small vacuum expectation value, but a large expectation value in the mimicking state $a \left| \Om\right\rangle$, and still there is no contradiction. The Reeh-Schlieder theorem does not guarantee that an $a$ exists that is also unitary. Hence, it is also not guaranteed that we will be able to implement $a$ as a physical operation, i.e. as $e^{i Ht}$ for some Hamiltonian over which we have experimental control. The more relevant consequence is what this implies about the entanglement structure of the vacuum. The possibility of recreating the action of a local operator with other, very distant, local operators is saying that the vacuum of a QFT contains quantum correlations between any two regions of the spacetime. These intrinsic correlations are what allow local, non-unitary operations on a state to affect correlation functions at distant, spacelike separated points. There is no violation of causality in the same sense that Bell pairs do not transmit information superluminally when measurements are performed on one qubit in the pair, despite the non-classical correlations between measurements attributed to the entanglement of the pair. 

It is also important that the Hilbert space of our QFT does not, in truth, factorize into a (infinite) product of Hilbert spaces associated to small regions $\mathcal{U}_{\mathcal{V}}$. If there were some factorization into Hilbert spaces representing subsets of $\Si$, roughly $\mathcal{H}_{\mathcal{V}}\ot\mathcal{H}_{\mathcal{V'}}$, then there would exist in the theory separable states like $\left|\Psi_\mathcal{V}\right\rangle\ot\left| \Psi_\mathcal{V'}\right\rangle$. This type of state would share none of the interesting physics displayed by the entangled vacuum. Apart from the evidence given in the area law \eqref{eq:arealaw1} that all low energy QFT states share the UV divergent entanglement entropy of the vacuum, ruling out the separable state, we also expect that on small enough scales any state should behave like vacuum, and emphatically $\left|\Psi_\mathcal{V}\right\rangle\ot\left| \Psi_\mathcal{V'}\right\rangle$ does not. We are forced to conclude that there is no factorization structure of the Hilbert space according to spatial regions of the spacetime. But without this there is also no notion of the reduced density matrix $\rho_\mathcal{V}$, and definitions like \eqref{eq:EEdefinition} or \eqref{eq:relativeentropy1} cannot be used. Luckily, there are additional tools in algebraic QFT that allow the construction of a relative entropy function with the same properties as \eqref{eq:relativeentropy1}, and that reduces to \eqref{eq:relativeentropy1} when a factorization structure is put in place (e.g. when a lattice discretization is used).

In order to construct a relative entropy function directly in the continuum of a QFT without reference to reduced density matrices, it is necessary to borrow a result from the theory of von Neumann algebras. The major result of Tomita-Takesaki theory \cite{Haag1996} is that there exists an operator $S_\Psi$, sometimes called the Tomita operator, associated to any state $\left|\Psi\right\rangle$ which is cyclic separating for $\mathcal{A}_{\mathcal{U}}$ with the property
\begin{equation}
S_\Psi a\left|\Psi\right\rangle=a^\dag \left|\Psi\right\rangle, \quad \forall a\in \mathcal{A}_{\mathcal{U}}.
\end{equation}
There are several immediate facts one can derive about the Tomita operator, but as it is only an intermediate step towards our goal, we only mention the most important one for us which is its invertibility. This property is clear since $S_\Psi^2=1$. Hence, it has a polar decomposition which we write
\begin{equation}
S_\Psi=J_\Psi \De_\Psi^{1/2},
\end{equation}
in terms of the modular conjugation operator $J_\Psi$ and modular operator $\De_\Psi$. The latter is really our object of study. It is Hermitian and positive definite, since it can be shown that $\De_\Psi=S_\Psi^\dag S_\Psi$, and furthermore from $S_\Psi \left|\Psi\right\rangle=S_\Psi^\dag \left|\Psi\right\rangle=\left|\Psi\right\rangle$ we obtain that for any function of $\De_\Psi$, $f(\De_\Psi)\left|\Psi\right\rangle=f(1)\left|\Psi\right\rangle$.

The relative entropy must compare two states, so mirroring the construction of $\De_\Psi$ the relative Tomita operator can be introduced as
\begin{equation}
S_{\Psi|\Phi} a\left|\Psi\right\rangle=a^\dag \left|\Phi\right\rangle, \quad \forall a\in \mathcal{A}_{\mathcal{U}}.
\end{equation}
Once again $\left|\Psi\right\rangle$ must be cyclic separating for $\mathcal{A}_{\mathcal{U}}$ assuring that any $a^\dag \left|\Phi\right\rangle\in \mathcal{H}_0$ can be produced by acting on $\left|\Psi\right\rangle$. While there is no strict constraint on $\left|\Phi\right\rangle$, it is easiest to assume that it is cyclic separating for $\mathcal{A}_{\mathcal{U}}$ as well, in which case there would also exist another relative Tomita operator 
\begin{equation}
S_{\Phi|\Psi} a\left|\Phi\right\rangle=a^\dag \left|\Psi\right\rangle.
\end{equation}
Obviously $S_{\Phi|\Psi}S_{\Psi|\Phi}=1$ so again $S_{\Psi|\Phi}$ is invertible. Now the relative modular operator is defined in the same way as before; from the polar decomposition $S_{\Psi|\Phi}=J_{\Psi|\Phi}\De_{\Psi|\Phi}^{1/2}$ it can be shown that,
\begin{equation}\label{eq:relativemodularoperator}
\De_{\Psi|\Phi}=S^\dag_{\Psi|\Phi}S_{\Psi|\Phi},
\end{equation}
where the relative modular operator $\De_{\Psi|\Phi}$ is Hermitian and positive definite.

From these definitions Araki constructed the relative entropy between states $\left|\Psi\right\rangle$ and $\left|\Phi\right\rangle$, compared in the region $\mathcal{U}_{\mathcal{V}}$ as \cite{Araki1976},
\begin{equation}\label{eq:aqftrelent}
S(\Psi||\Phi;\mathcal{U}_{\mathcal{V}})=-\left\langle \Psi|\log \De_{\Psi|\Phi} |\Psi\right\rangle.
\end{equation}
To see why this could be a sensible definition for a measure of distinguishability between $\left|\Psi\right\rangle$ and $\left|\Phi\right\rangle$, consider when the states are related by $\left|\Phi\right\rangle=u' \left|\Psi\right\rangle$ for a unitary element $u'\in\mathcal{A}_{\mathcal{U'}}$, the commuting algebra of $\mathcal{A}_{\mathcal{U}}$. Then the measurement of any operator $a\in\mathcal{A}_{\mathcal{U}}$ in the state $\left|\Psi\right\rangle$ would produce
\begin{equation}
\left\langle \Psi|a|\Psi\right\rangle=\left\langle \Psi\right|a ({u'}^\dag u')\left|\Psi\right\rangle=\left\langle \Phi\right|a\left|\Phi\right\rangle.
\end{equation}
In other words, the states $\left|\Psi\right\rangle$ and $\left|\Phi\right\rangle$ give the same results for all measurements of operators in $\mathcal{A}_{\mathcal{U}}$, and thus are indistinguishable in $\mathcal{U}_\mathcal{V}$, as the local unitary outside of $\mathcal{U}_{\mathcal{V}}$ does not affect the physics in region $\mathcal{V}$. Indeed we find for these two states that the relative entropy $S(\Psi||\Phi;\mathcal{U}_{\mathcal{V}})$ is zero\footnote{For $\left|\Psi\right\rangle$ and $\left|\Phi\right\rangle=u'\left|\Psi\right\rangle$, the relative Tomita operator behaves as $S_{\Psi|u'\Psi} a\left| \Psi\right\rangle=a^\dag\left| u' \Psi\right\rangle=u'a^\dag \left|\Psi\right\rangle=u' S_\Psi \left| \Psi\right\rangle$, so we identify $S_{\Psi|u'\Psi}=u'S_{\Psi}$. Hence, by definition \eqref{eq:relativemodularoperator} the relative modular operator also reduces to the ordinary modular operator for $\left| \Psi \right\rangle$, $\De_{\Psi|u'\Psi}=S_{\Psi}{u'}^\dag u'S_{\Psi}=\De_\Psi$. Then we find that $\log(\De_\Psi)\left|\Psi\right\rangle =\log(1)\left|\Psi\right\rangle=0$, and the relative entropy between the states vanishes.}, $\left|\Psi\right\rangle$ and $\left|\Phi\right\rangle$ are indistinguishable in $\mathcal{U}_{\mathcal{V}}$. Furthermore, this definition obeys the same properties as the quantum information definition \eqref{eq:relativeentropy1}; it is non-negative for any two states, and monotonically increasing as we enlarge the region in which measurements can take place,
\begin{equation}
S(\Psi||\Phi;\mathcal{U}_2)\geq S(\Psi||\Phi;\mathcal{U}_1) \quad \text{for} \quad \mathcal{U}_1\subset \mathcal{U}_2.
\end{equation}
Intuitively, with access to additional measurements an experimenter has better prospects for distinguishing the states, so the relative entropy must increase for larger regions. These are significant properties in the quantum information context. Positivity implies positivity for mutual information \eqref{eq:mutualinformation2}, equivalently subadditivity of the entanglement entropy
\begin{equation}\label{eq:subadditivity1}
S(\rho_{A})+S(\rho_{B})\geq S(\rho_{AB}),
\end{equation}
 through \eqref{eq:mutualinformation1}. Monotonicity implies the highly non-trivial strong subadditivity result for entropies, 
\begin{equation}\label{eq:SSA1}
S(\rho_{AB})+S(\rho_{BC})\geq S(\rho_B)+S(\rho_{ABC}),
\end{equation}
as this can be rewritten $I(A:BC)\geq I(A:B)$ or indeed $S(\rho_{ABC}||\rho_{A}\ot\rho_{BC})\geq S(\rho_{AB}||\rho_{A}\ot\rho_{B})$ which is monotonicity. In the quantum information context, these are the only inequalities needed to completely characterize the structure of allowed relative entropy values \cite{Ibinson2007}. 

To summarize, in this section we have explored one of the crucial facts about low energy states in QFTs: they are highly entangled between any two local regions. This can be seen explicitly in calculations of the entanglement entropy for lattice regularized systems, where a universal UV divergence is found, best expressed by the area law \eqref{eq:arealaw1}. Alternatively, it can be seen as a corollary of the Reeh-Schlieder theorem from algebraic QFT which shows that the vacuum of a QFT is cyclic separating for any local subalgebra, and means that any local operation on the vacuum can be reproduced by another local operation \eqref{eq:statecompare}, potentially in a spacelike separated region. We have also shown that relative entropy can be defined directly in the continuum for QFTs \eqref{eq:aqftrelent} where it is free from UV divergences and does not assume anything about the factorization structure of the Hilbert space. Our next focus will be on the far-reaching consequences of entanglement in holography, and what it tells us about the emergence of spacetime.

\section{AdS/CFT}
The Anti-de Sitter/conformal field theory correspondence has been continuously revolutionizing our understanding of quantum gravity since its discovery \cite{Maldacena1999,Witten1998,Gubser1998}. AdS/CFT posits that there is an exact mathematical duality between certain CFTs in $d$ dimensions, with theories of quantum gravity in $d+1$ dimensional asymptotically AdS spacetimes. Due to the difference in dimensionality of the theories, the correspondence is said to be holographic, and is often referred to simply as ``holography" \cite{Hooft1993,Susskind1995}. Originally understood in the context of string theory constructions, where extended objects could be described equivalently by a supersymmetric CFT or a string theory on a product of AdS with other compact manifolds, the correspondence has more recently been found to apply under quite general conditions \cite{Heemskerk2009}. Specifically, whenever a boundary CFT admits a large $N$ expansion, and has a sparse spectrum of light operators, it should have a bulk AdS description, where the AdS radius is large compared to the Planck length.

These conjectures have exceedingly important implications for our understanding of quantum gravity. While at present there is no direct construction of a well understood, experimentally verified theory of quantum gravity, AdS/CFT gives us an alternate approach to the problem. Since the CFTs are non-gravitational theories on a much firmer theoretical footing, they can be used to define what we mean by quantum gravity in AdS, and explore its properties. Still, we do not have a full understanding of how CFTs can manifest all aspects of quantum gravity, such as the emergence of an extra holographic bulk dimension \cite{deHaro2001}, locality of bulk observables \cite{Hamilton2006}, their gauge invariance under diffeomorphisms \cite{Heemskerk2012}, or the apparent loss of information through black hole evolution \cite{Almheiri2013,Harlow2016}. Evidence for the correspondence consists of a dictionary that translates observables and other quantities between the two sides of the duality. For instance, every CFT has a spin 2 stress tensor $T$ as the conserved current for translation symmetry, and this is dual to the metric tensor $g_{\mu\nu}$ of the gravitational theory. The matching of other dual fields, like scalar CFT operators to scalar bulk fields, is accomplished by calculating their correlation functions. Using the AdS propagator one finds the 2-point function of a free scalar field with mass $m^2=\De(\De-d)$ to be
\begin{equation}
\left\langle\phi(x_1)\phi(x_2)\right\rangle\sim |x_1-x_2|^{-2\De}
\end{equation}
in the limit where $x_1$ and $x_2$ are points close to the asymptotic boundary. We see that the 2-point function matches the form \eqref{eq:2-ptfunction} for a scalar CFT operator with dimension $\De$, which was dictated entirely by conformal symmetry. 

Most of the early checks on AdS/CFT involved the matching of correlations functions, as the GKPW conjecture \cite{Gubser1998,Witten1998} provided a concrete formulation of the correspondence in these terms:
\begin{equation}
Z_{\text{grav}}[\phi_0^i(x);\pa M]=\left\langle \exp \left(-\sum_i\int d^d x  \ \phi_0^i(x)\op^i(x)\right)\right\rangle_{\text{CFT\ on\ }\pa M}.
\end{equation}
This states that the gravitational partition function in an asymptotically AdS spacetime $M$, with light bulk fields $\phi^i(x)$, can be computed by the generating functional of correlation functions in the dual CFT on the spacetime $\pa M$, the conformal boundary of $M$. It includes all light fields in the bulk effective theory, with their sources at the boundary $\phi_0^i(x)$, and dual low-dimension CFT operators $\op^i(x)$. Remarkably, since CFTs are scale invariant and therefore UV complete, the GKPW dictionary has the potential to non-perturbatively define a UV complete theory of quantum gravity.

\begin{figure}
    \centering
        \includegraphics[width=0.4\textwidth]{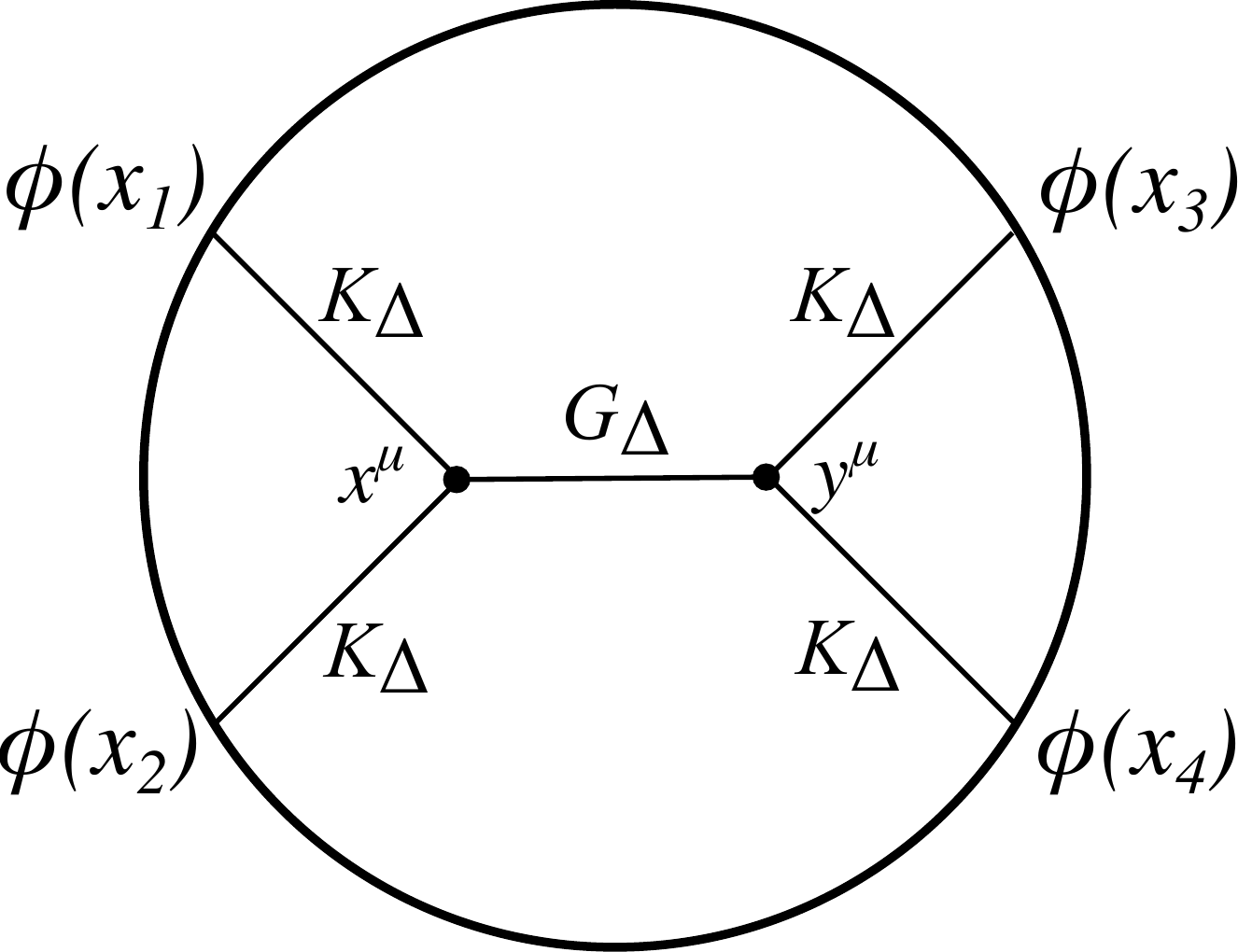}
        \caption[Scalar exchange Witten diagram]{A tree level Witten diagram showing a scalar exchange contribution to the CFT four point function $\left \langle \op(x_1)\op(x_2)\op(x_3)\op(x_4)\right\rangle$. The bulk interaction vertices $x^\mu,y^\mu$ should be integrated over the entire spacetime. Other diagrams for other exchange channels, as well as exchange of other fields such as gravitons will also contribute.}
        \label{fig:WittenDiagram}
\end{figure}

Higher-point CFT correlation functions, such as a four-point function of scalar quasi-primaries\\ $\left \langle \op(x_1)\op(x_2)\op(x_3)\op(x_4)\right\rangle$, can be computed holographically by Witten diagrams, a diagrammatic expansion in bulk field couplings. This construction operates on the observation that the normalizable modes of a bulk field $\phi(r,x)$ extrapolated to the boundary $r\to\infty$, with an appropriate normalization factor to account for the field's falloff, produces a quantity that behaves exactly like a CFT quasi-primary \cite{Banks1998,Balasubramanian1999c,Harlow2011},
\begin{equation}
\lim_{r\to\infty} r^\De \phi(r,x)=\op(x).
\end{equation}
The mass of the field determines the rate of falloff $r^{-\De}$ near the boundary via $m^2=\De(\De-d)$, but also dictates the scaling dimension of the CFT object. Witten diagrams connect the boundary fields $\phi(r=\infty, x)$ to bulk interaction vertices $y^\mu=(r,y)$ with a bulk-to-boundary propagator, the Green's function which solves 
\begin{equation}
(\square_g+m^2)K_\De(r,x;y)=r^{\De-d}\de^{d}(x,y).
\end{equation}
Bulk vertices are connected to each other via the bulk-to-bulk propagator of bulk fields involved in the interaction, which could include gravitons, or scalar exchange as in Fig. \ref{fig:WittenDiagram}. In the latter example, the propagator would solve
\begin{equation}
(\square_g+m^2)G_\De(x^\mu,y^\mu)=\de^{d+1}(x^\mu,y^\mu),
\end{equation}
in the fixed AdS background. The Witten diagram then shows us the elements needed for this contribution to the 4-point function; the diagram in Fig. \ref{fig:WittenDiagram} represents the contribution
\begin{equation}
\int g\, d^{d+1}x\, d^{d+1}y \, K_\De(x^\mu,x_1)K_\De(x^\mu,x_2)G_\De(x^\mu,y^\mu)K_\De(y^\mu,x_3)K_\De(y^\mu,x_4).
\end{equation}
An enormous amount of work has gone into calculating correlation functions in this way and comparing them to the structures dictated by CFT \cite{DHoker2002,Skenderis2002}.

\begin{figure}
    \centering
    \begin{subfigure}[b]{0.35\textwidth}
        \includegraphics[width=0.9\textwidth]{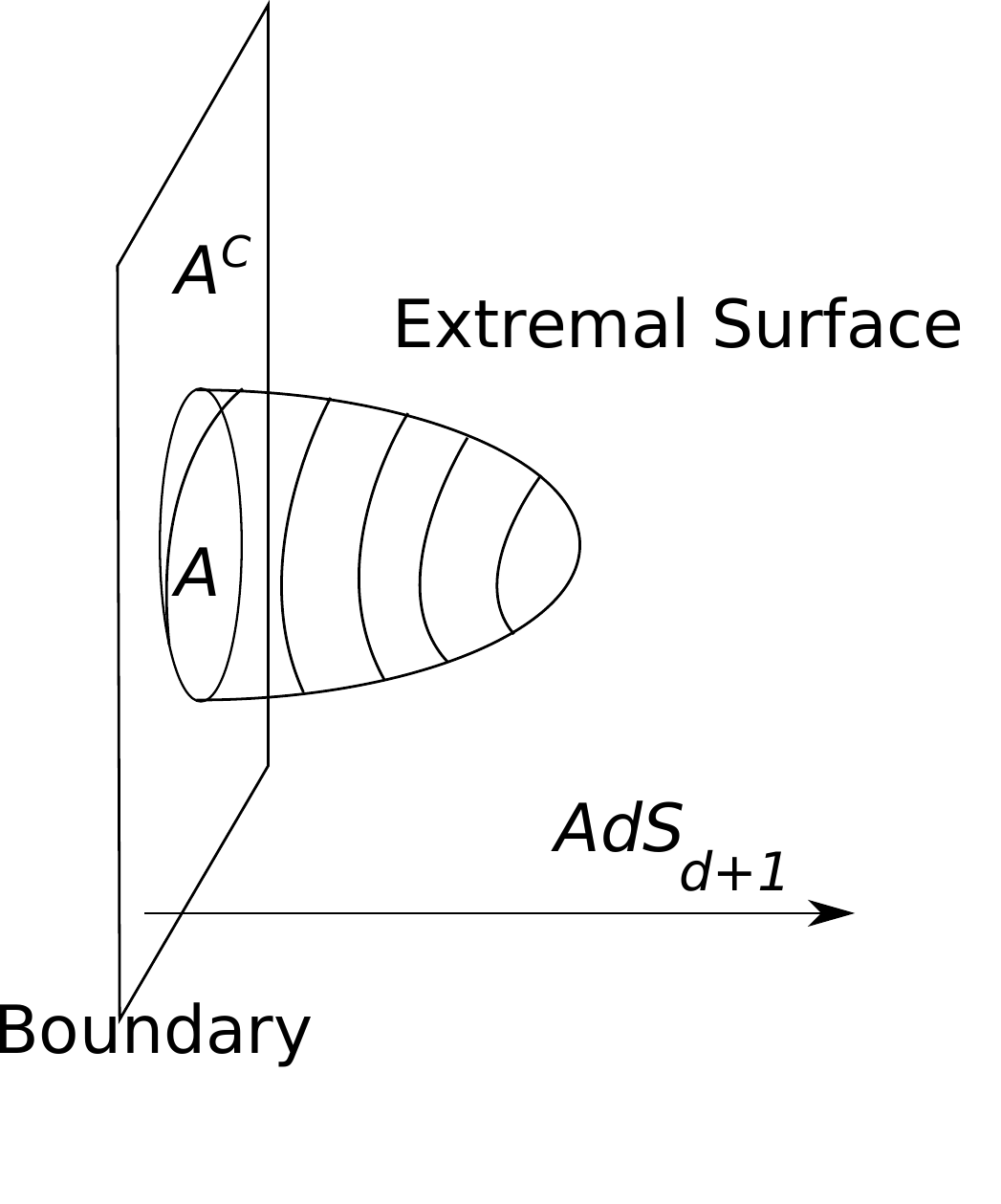}
        \caption{}
        \label{fig:RTsurface}
    \end{subfigure}
     \ \quad 
    \begin{subfigure}[b]{0.45\textwidth}
        \includegraphics[width=0.9\textwidth,trim={0 1.3cm 0 0},clip]{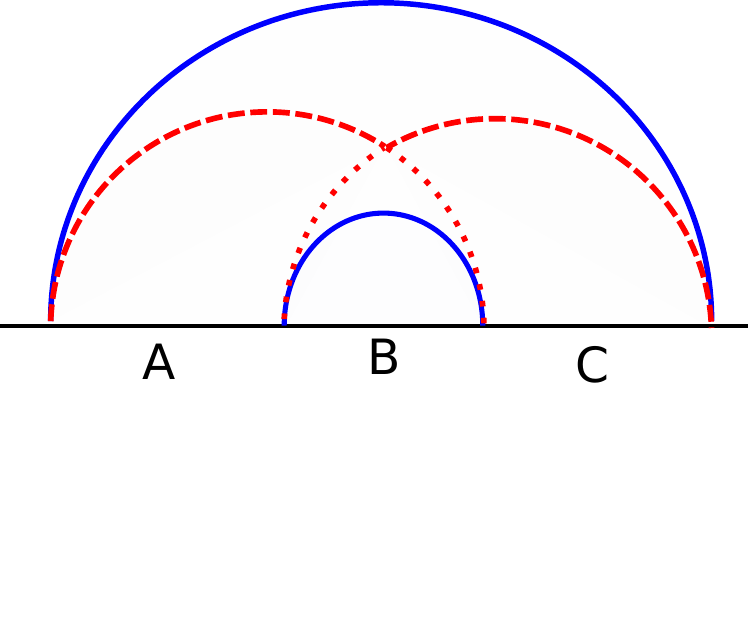}
        \caption{}
        \label{fig:SSA}
    \end{subfigure}
    \caption[Ryu-Takayanagi extremal surfaces]{(a) The entanglement entropy of the boundary CFT between $A$ and $A^c$ is computed holographically by the area of an extremal surface homologous to $A$ that extends into the bulk. (b) Example configuration showing strong subadditivity \eqref{eq:SSA1} of holographic entanglement entropy in AdS$_3$. The solid blue geodesics compute $S(\rho_B)$ and $S(\rho_{ABC})$, while the red geodesics computing $S(\rho_{AB})$ and $S(\rho_{BC})$ have been divided and recombined into a dashed line homologous to $ABC$ and a dotted line homologous to $B$. The latter two are not geodesics, so have greater length than their counterparts, showing $S(\rho_{AB})+S(\rho_{BC})\geq S(\rho_B)+S(\rho_{ABC})$. }\label{fig:RTandSSA}
\end{figure}

The aspects of holography that will be of most interest in this thesis are some that go beyond the matching of correlation functions. In recent years, a number of concepts from quantum information theory have become highly applicable in holography due to their geometric realizations in the bulk. The primary example is the entanglement entropy of a subregion, which we saw takes a universal form for a single interval in a 2d CFT vacuum state \eqref{eq:2dCFTintervalEE}. Holographically, entanglement entropy is dual to the area of an extremal surface $\ga_A$ attached to the boundary such that it is homologous to the CFT subregion $A$ \cite{Ryu2006,Hubeny2007,Lewkowycz2013,Dong2016a}, see Fig. \ref{fig:RTsurface},
\begin{equation}\label{eq:RyuTakayanagi}
S_A=\min_{\substack{\ga_A\\ \pa\ga_A= \pa A}}\frac{\text{Area}(\ga_A)}{4 G_N}.
\end{equation}
Using this Ryu-Takayanagi prescription in AdS$_3$, the minimal surface is just a geodesic attached to the boundary at the ends of a CFT interval (or set of intervals). A simple geometrical calculation for the AdS$_3$ geodesic length subtending a boundary interval of length $l$ produces
\begin{equation}
S_A=\frac{R_{\text{AdS}}}{2G_N}\log\frac{l}{\ep_{UV}}.
\end{equation}
Since distances near the boundary become large in AdS, the length of the geodesic is divergent, and we regulate by placing a cutoff surface at a radial distance $\ep_{UV}$ from the boundary. Furthermore, the Brown-Henneaux formula \cite{Brown1986},
\begin{equation}
c=\frac{3 R_{\text{AdS}}}{2 G_N},
\end{equation}
provides a relationship between the central charge $c$ of the dual CFT$_2$, and both the AdS radius $R_{\text{AdS}}$ and gravitational Newton constant $G_N$, from the conformal algebra of asymptotic symmetries in AdS$_3$. Together, these results exactly reproduce the CFT entanglement entropy \eqref{eq:2dCFTintervalEE} with the expected UV divergence. The matching between \eqref{eq:RyuTakayanagi} in asymptotically AdS geometries and \eqref{eq:EEdefinition} in dual CFT states has been tested in a wide variety of scenarios \cite{Nishioka2009,Headrick2010,Headrick2014,Freedman2017}. 

Notably, the geometric interpretation allows for very simple proofs of some important properties of entanglement entropy, such as subadditivity \eqref{eq:subadditivity1} and strong subadditivity \eqref{eq:SSA1} \cite{Headrick2007}. In quantum information theory the proof of strong subadditivity relies on the highly non-trivial result that relative entropy is monotonic. Yet, in holography its proof can be summed up digrammatically, see Fig. \ref{fig:SSA}. Interestingly, entanglement entropy is more constrained in holography as compared with general quantum systems, since it satisfies a number of additional inequalities. The first new inequality found is called monogamy of mutual information and constrains the entanglement shared between three regions \cite{Hayden2013},
\begin{equation}\label{eq:MMI1}
S(\rho_{AB})+S(\rho_{AC})+ S(\rho_{BC})\geq S(\rho_{A})+S(\rho_{B})+S(\rho_{C})+S(\rho_{ABC}).
\end{equation}
 Additional inequalities for higher numbers of regions have been reported, but a complete characterization of the extra conditions obeyed by \eqref{eq:RyuTakayanagi} is still a topic of active research \cite{Bao2015,Hubeny2018,Hubeny2018a}. These results have direct bearing on the structure of CFT states which can give rise to holographic geometries \cite{Cui2018}.

Holographic entanglement entropy was the first example of how entanglement and other information theoretic properties of a CFT can be geometrized in the bulk. A number of widely used ideas from quantum information theory have been realized in similar ways, including entanglement negativity \cite{Rangamani2014,Chaturvedietal2018}, quantum Fisher information \cite{Lashkari2016}, quantum error correcting codes \cite{Almheiri2015,Pastawski2015,Mintun2015,Harlow2017}, and modular flow \cite{Jafferis2016,Faulkner2016,Faulkner2017a,Faulkner2018}. As examples of recent progress we will look over two other cases in more detail: entanglement of purification and complexity.

One recently discovered duality that is fairly similar to entanglement entropy is the entanglement of purification \cite{Umemoto2018,Nguyen2018}, defined by quantum information theorists in the following way \cite{Terhal2002}. Given a mixed state $\rho_{AB}$ we can create a pure state $\left|\psi\right\rangle$ on a larger space by introducing ancilla systems $A'B'$ and entangling $AB$ with $A'B'$. This is done in such a way that tracing out $A'B'$ reproduces $\rho_{AB}$. Then one can evaluate the von Neumann entropy \eqref{eq:EEdefinition} between $AA'$ and $BB'$. However, many purifications are possible for different ancilla systems, so the entanglement of purification is defined as the least entropy over all possible purifications,
\begin{equation}\label{eq:EoP}
E_P(\rho_{AB})=\min_{\rho_{AB}=\tr_{A'B'}\left|\psi\right\rangle\left\langle\psi\right|}S(\rho_{AA'}).
\end{equation}
In fact, for pure states $\rho_{AB}=\left|\psi\right\rangle\left\langle\psi\right|$ there is no need to introduce $A'B'$ for purification, so this construction reduces to the entanglement entropy $S(\rho_A)=S(\rho_B)$ exactly. For mixed states the definitions are not equivalent, and in holography we find a different interpretation of $E_P(\rho_A)$. 

\begin{figure}
    \centering
        \includegraphics[width=0.3\textwidth]{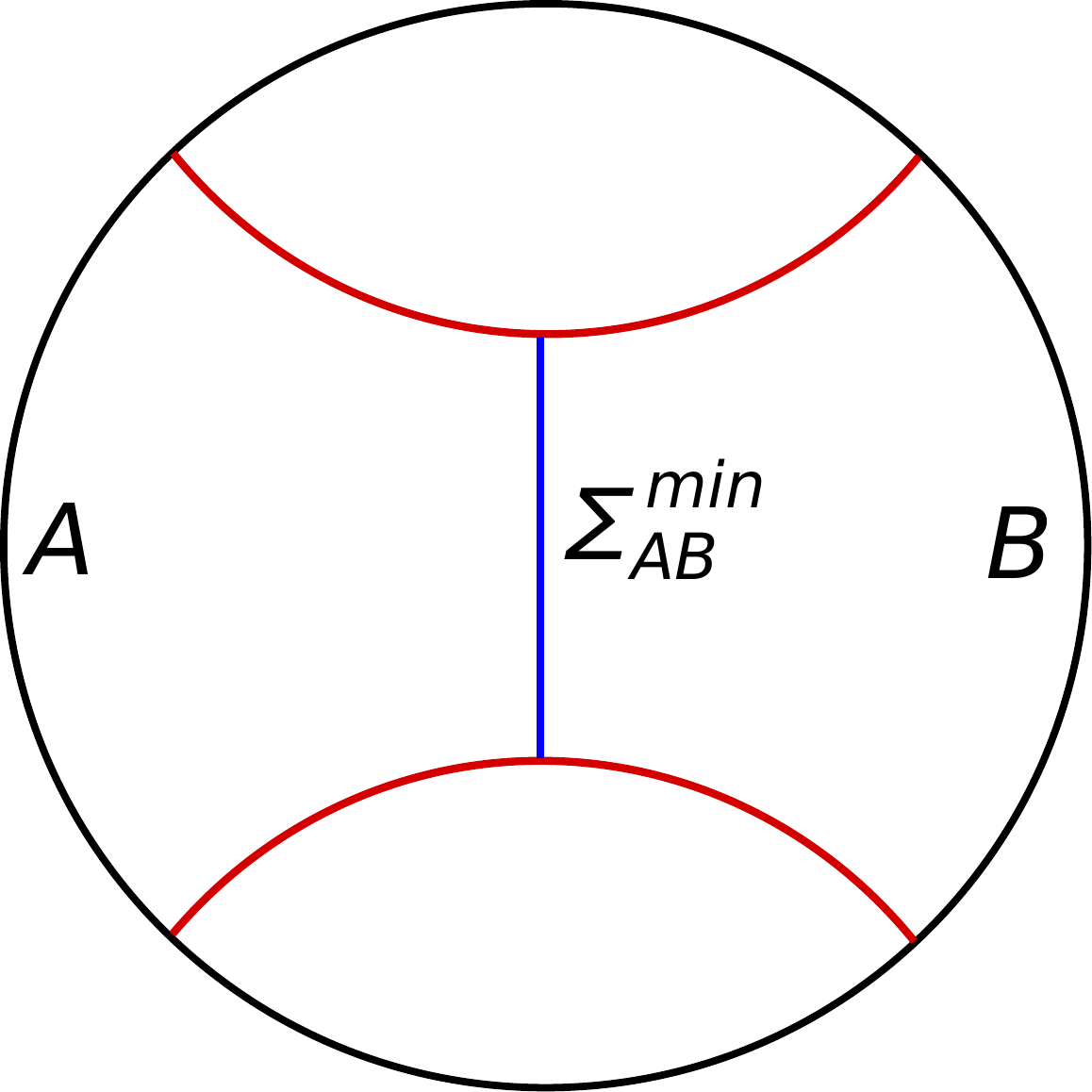}
        \caption[Entanglement wedge cross section]{The entanglement of purification can be calculated on a timeslice of AdS as the area of the minimal surface $\Si_{AB}^{min}$ (or length of geodesic segment in $d+1=3$ as pictured) that separates $A$ from $B$ in the entanglement wedge of $AB$ (interior of red geodesics). }
        \label{fig:EoPsurface}
\end{figure}

Let $A$ and $B$ represent two boundary subregions. The entanglement wedge of $AB$ is the region surrounded by the Ryu-Takayanagi extremal surface homologous to $AB$, and we break it up into two disjoint components separated by a surface $\Si_{AB}$. There are many bulk surfaces which can separate $A$ from $B$ within the entanglement wedge, but we select the one with minimal area $\Si_{AB}^{min}$, see Fig. \ref{fig:EoPsurface}, and define the entanglement wedge cross section in a similar manner to the Ryu-Takayanagi entanglement entropy,
\begin{equation}\label{eq:EWCS}
E_W(\rho_{AB})=\min_{\Si_{Ab}}{ \left[ \frac{A(\Si_{AB})}{4 G_N} \right]}.
\end{equation}
There is strong evidence supporting the conjectured duality that the CFT entanglement of purification for the region $AB$ equals the entanglement wedge cross section,
\begin{equation}
E_P(\rho_{AB})=E_W(\rho_{AB}),
\end{equation}
which has been accumulated in a number of papers exploring aspects and extensions of the conjecture \cite{Bao2018,Bao2019,Caputa2019,Dutta2019}. The significance of this duality is that it can be used in conjunction with entanglement entropy to isolate quantum correlations. Entanglement entropy and entanglement of purification both have the undesirable feature that they are sensitive to the classical correlations in mixed states. But the two measures account for classical correlations in different ways so that by comparing $S(\rho_A)$ and $E_P(\rho_{AB})$ it is possible to isolate the contributions of genuine quantum correlations \cite{Bhattacharyya2019}.

Another conjectured duality for which a great deal of evidence has arisen has to do with the quantum information theory concept of complexity \cite{Nielsen2006}. The circuit complexity of a quantum state is defined as the number of simple quantum logic gates needed to construct the state from a fixed reference state. Clearly there are some ambiguities in this definition which we should address. For quantum circuits acting on qubit systems, it is well known that it suffices to choose a small number of different logic gates to reproduce the action of any possible circuit. Such a gate set is called a universal set. When defining the complexity of a state it certainly matters which gates we are permitted to utilize. If we allow any gates whatsoever, then any state should have a complexity of one (or zero), since there exists some operation to prepare the state directly from the reference. By limiting ourselves to a fixed universal gate set we can meaningfully compare the number of gates needed to construct different states. Still, there will be many different choices of universal sets, and these choices will potentially provide different complexities for the same state. Furthermore, the choice of reference state will also greatly affect the definition of complexity, and there is no \emph{a priori} reason to prioritize some particular state as a reference over any other. Hence, complexity is only defined relative to some reference state and some universal gate set; it is not an absolute concept.

Complexity can be considered as a distance measure on Hilbert space, but is quite different in character than the relative entropy which we mentioned earlier. The relative entropy \eqref{eq:relativeentropy1} between two orthogonal pure state diverges, since orthogonal states are perfectly distinguishable with the appropriate measurement protocol, e.g. a projective measurement onto one of the states. However, in some sense this can produce unintuitive distances. If we have a pure state of a large system comprised of thousands of qubits each in the computational $\left|0\right\rangle$ state, $\left|\psi\right\rangle=\left|0_1\right\rangle...\left|0_{3000}\right\rangle$, then when we flip the state of the $n$th qubit, $\left|\psi'\right\rangle=\left|0_1\right\rangle...\left|1_n\right\rangle...\left|0_{3000}\right\rangle$, we obtain an orthogonal state with infinite relative entropy. But overall we may consider a single bit flip to be a small change. Using complexity as a distance measure, the two states $\left|\psi\right\rangle$, $\left|\psi'\right\rangle$ could have a distance of 1, since a single bit flip gate enacts the change. A state like $\left|\psi''\right\rangle=\left|1_1\right\rangle...\left|1_{3000}\right\rangle$ has the same infinite relative entropy with $\left|\psi\right\rangle$, but much higher relative complexity. Despite the freedoms in the definition of complexity, it is possible to identify some definite features.

 For general interacting quantum systems with random local dynamics complexity is known to grow approximately linearly for very long times. Starting from a reference state, a discrete evolution where individual gates are chosen randomly and applied to the state almost always takes us to more complex states rather than less complex states. This is simply due to the size of Hilbert space; at almost every step in the evolution there are many more branches outward to new parts of Hilbert space than branches taking us back towards previously seen, or less complex states. This is very similar to the microscopic explanation of the second law of thermodynamics, and can be stated in qubit systems as a \emph{second law of complexity}: if complexity is less than its maximum, exponential in the number of qubits, then it will increase with overwhelming likelihood into the future and the past \cite{Brown2018}. Since complexity will almost always grow by one unit at each step in the evolution, given that we only apply one gate per step, then the rate of growth is approximately linear in time for exponentially long times. Although CFTs are not simply qubit systems, the notion of complexity can still be applied to them. We can think of individual logic gates as small changes to, for example, the global phase, the position or momentum of the wavefunction, the entanglement between two field modes, or the scale of a mode \cite{Jefferson2017}. The second law still applies for random applications of these gates to parts of the system.

The real motivation for introducing complexity in the context of AdS/CFT was the observed discrepancy between timescales of evolution in black holes and their dual thermal CFT states. An eternal black hole in AdS is characterized by a wormhole, or Einsten-Rosen bridge (ERB), which connects two asymptotic AdS boundaries. This system has a boundary dual comprised of two CFTs in an entangled state, such that the reduced state of either CFT is thermal, with temperature equal to the black hole temperature \cite{Maldacena2001}. This is the so-called thermofield double state,
\begin{equation}
\left|\Psi_{\text{TFD}}\right\rangle=\frac{1}{\sqrt{Z(\beta)}}\sum_n e^{-\be E_n /2}\left|E_n\right\rangle_l\ot\left|E_n\right\rangle_r,
\end{equation}
where $\left|E_n\right\rangle_{l,r}$ are energy eigenstates of the left and right CFTs, $T=\be^{-1}$ is the temperature, and  the partition function $Z(\be)$ normalizes the state.
 In the CFT state, small perturbations thermalize within a short scrambling time, proportional to $T^{-1}\log S$, where $S$ is the entropy, similar to the rapid scrambling of information thrown into a black hole \cite{Sekino2008}. After this time entropy is maximized, and there is no apparent evolution of typical observables like correlation functions in the CFT. In contrast, the volume of the ERB continues to grow linearly long after the scrambling time \cite{Susskind2014,Stanford2014}. It was conjectured that the circuit complexity of the CFT state should be dual to this long lasting growth, since even after thermalization, fluctuations to nearby thermal states cause complexity to increase continuously. 
 
 These ideas were formalized in a number of distinct conjectures for the gravitational quantity dual to CFT complexity \cite{Alishahiha2015,Brown2016a,Brown2016,Carmi2017}. It should not be too surprising that multiple candidates for a dual quantity have been proposed given that the circuit complexity is inherently ambiguous due to the choice of reference state and gate set. One proposal suggests that the complexity is dual to the volume of the maximal codimension-1 surface $\si$ through the bulk that reaches the boundary at the timeslice $\Si$ where the CFT state is defined,
\begin{equation}
C_V=\max_{\Si=\pa \si} \frac{\text{vol}(\si)}{G_N L}.
\end{equation}
To achieve the correct units, an extra length scale $L$ is added, which may be the AdS radius or black hole horizon length, but again this ambiguity hearkens back to the inherent difficulties of defining complexity. A second proposal suggests that the complexity should be given by the gravitational action of the Wheeler-DeWitt patch, the causal development of the same surface $\si$ mentioned before,
\begin{equation}
C_A= \frac{I_{\text{WDW}}}{\pi \hbar}.
\end{equation}
Both of these quantities exhibit linear growth for exponentially long times, well beyond the scrambling time for thermalization.

There is currently no consensus on which definition is ``correct", and probably there can be no unique dual for CFT complexity for the reasons discussed. It is entirely possibly that there exists a large class of gravitational quantities that exhibit late time linear growth, which would roughly correspond to different choices of gate set and reference state for defining complexity. Yet, the study of complexity is a worthwhile endeavour as a novel and very distinct quantity in dynamical systems with applications to the black hole information paradox \cite{Susskind2016}.

Beyond simply matching quantities on the two sides of the duality, the information theoretic dualities we have mentioned suggest deeper connections between the nature of spacetime in quantum gravity and entanglement. It has been suggested that CFT entanglement is responsible for stitching together the bulk spacetime, based on the observation that two unentangled CFTs are dual to two disconnected geometries whereas two entangled CFTs can be dual to a single connected spacetime where a wormhole joins the two asymptotic regions through the bulk. As entanglement is removed from the CFT state, the bulk spacetime pinches off and becomes disconnected \cite{Maldacena2001,VanRaamsdonk2010}. This idea was expanded into the conjecture that wormholes and entangled particles are fundamentally the same under the moniker ``ER=EPR" \cite{Maldacena2013}. 

Holographic entanglement also gives us a handle on the structure of spacetime \cite{Bianchi2014}. If we suppose a bulk geometry exists and is such that extremal surfaces attached to the boundary reproduce the patterns of entanglement in the dual CFT state, then in principle the bulk metric should be reconstructable from entanglement entropy data in the regions that can be reached by the extremal surfaces. In practice this is a difficult approach because the problem is highly overconstrained. Typically the metric of the spacetime is specified by a number of functions of a few coordinates and parameters, whereas the boundary entanglement entropies give some function on the space of subregions of the boundary. It will then only be for very special CFT states that a geometry exists, but this is exactly what happens in known examples of the duality. In more practical terms, individual points in the bulk can be located through extraneous singularities of boundary correlators \cite{Maldacena2017}, and the metric at these points reconstructed up to a conformal factor through boundary data \cite{Engelhardt2016}. Alternatively, the metric can be reconstructed up to a conformal factor through the Lie algebra generated by modular Hamiltonians of all spherical CFT subregions \cite{Kabat2018}.

An additional obstacle to reconstruction of the bulk metric through entanglement entropy data is that in many geometries boundary anchored extremal surfaces do not reach all parts of the bulk. Regions not reached are in the ``entanglement shadow" \cite{Balasubramanian2014,Myers2014,Headrick2014a,Freivogel2014}. Based on the statement of the AdS/CFT correspondence, we expect that the entire bulk should be encoded in the CFT state, including these shadow regions, so it behooves us to develop better probes which can reach such regions. One clue in this direction comes from gauged systems, where there can be internal degrees of freedom that are not spatially organized, but which contribute to entanglement. This type of entanglement in a CFT state is not captured by Ryu-Takayanagi surfaces which are only sensitive to spatially organized entanglement at leading order in $\tfrac{1}{G_N}$\footnote{The leading order behaviour of holographic entanglement entropy is purely geometrical and hence spatially organized, but corrections to the Ryu-Takayanagi formula may exhibit different behaviour \cite{Faulkner2013,Dong2014}.}. These ideas led to the introduction of \emph{entwinement}, computed in the CFT by first lifting the gauge constraints, computing entanglement entropy, and then enforcing gauge invariance. This construction is dual in AdS$_3$ to the length of non-minimal boundary anchored geodesics which reach into the entanglement shadow \cite{Balasubramanian2015,Balasubramanian2016}. The need to develop more fine grained probes of entanglement for understanding holographic spacetimes is a major motivation for this thesis.

The idea of bulk reconstruction through boundary data has been approached from many directions, including those just mentioned, but initially was accomplished by assuming the existence of a bulk AdS spacetime and using equations of motion to reconstruct local bulk fields in terms of smeared boundary operators \cite{Hamilton2006}. In order to isolate a bulk point with these methods, one must involve a sufficiently large boundary region such that its causal extension into the bulk contains that point. It is, however, undesirable to assume the very features we are hoping to reconstruct. Using more modern techniques involving modular flow, local bulk fields can be reconstructed using CFT considerations only \cite{Kabat2017}. Interestingly, it has also become apparent that sometimes local bulk fields can be reconstructed even if they lie outside the causal wedge of a boundary region, as long as they are contained within the Ryu-Takayanagi extremal surface homologous to that region \cite{Wall2014,Headrick2014b,Dong2016}. Generally, the idea of subregion-subregion duality is a refinement of the global AdS/CFT proposal that suggests that there should be a holographic dictionary mapping all bulk data within a bulk subregion to boundary data in a corresponding subregion \cite{Czech2012}.

Not only can the structure of spacetime potentially be reproduced via CFT entanglement data, but gravitational dynamics can also be found through laws of entanglement. Spacetime geometries which are consistent with CFT entanglement entropies according to the Ryu-Takayanagi relation \eqref{eq:RyuTakayanagi} satisfy the Einstein equations perturbatively  \cite{Faulkner2014,Lashkari2014a}. In fact, this can be shown independently of AdS/CFT, that is, without assuming the existence of a quantum gravitational bulk dual, and hence it implies a direct emergence of gravitation from CFT entanglement \cite{Faulkner2017}. Along these lines, entropic inequalities also imply gravitational energy conditions \cite{Lashkari2015,Lashkari2016a}, an avenue of study which has lead to numerous developments in AdS/CFT, QFT, and even classical gravity on the general validity of energy conditions \cite{Bousso2016,Bousso2015,Faulkner2016,Hartman2017,Wall2017a,Balakrishnan2017}.

The survey of topics presented here hopefully conveys the diversity of applications of quantum information theory to AdS/CFT, and its contributions to our understanding of quantum gravity in general.

\section{Outline}
This thesis is quantized as follows. In Chapters \ref{ch:KSCD} and \ref{ch:OPE} we detail the construction of new fine-grained observables in the AdS$_3$/CFT$_2$ correspondence associated to the non-minimal boundary anchored geodesics which appear for non-pure AdS spacetimes. Chapter \ref{ch:KSCD} begins by recounting the construction of a third spacetime in the correspondence, kinematic space, which allows us to connect some aspects of the bulk and boundary theories more easily. This intermediary spacetime reveals a duality between OPE blocks in the CFT, which give the contribution to the OPE from an entire conformal family of operators in the theory, and geodesic integrated bulk fields, a conveniently diffeomorphism invariant set of observables in the bulk. However, the original arguments establishing this duality relied heavily on the symmetries of pure AdS, and were therefore significantly limited in scope. We will show how the duality must be modified when the bulk is not pure AdS, using the example of conical defect spacetimes \cite{Cresswell2017}. In the bulk, the presence of non-minimal boundary anchored geodesics corresponds to a decomposition of OPE blocks in the CFT. We will establish a duality between the quantities in this decomposition, which are individually valid observables in the CFT, and geodesic integrated bulk fields, where the geodesics can be minimal or non-minimal.

In Chapter \ref{ch:OPE} we explore the new duality further in an important class of spacetimes obtained as quotients of AdS$_3$ with respect to elements of its isometry group. Instead of relying on symmetry arguments, we will construct explicit coordinate maps between pure AdS$_3$ and the quotient spacetimes \cite{Cresswell2018}. Then, we will use these maps to demonstrate how non-minimal geodesics arise due to the non-analyticities in the maps. The same maps, in the boundary limit, will be used to transform OPE blocks in the CFT. Again, the non-analyticities in the maps alter the structure of the OPE blocks in a way that mirrors the appearance of non-minimal geodesics. The transformed OPE blocks in all cases considered admit a decomposition into more fine-grained CFT observables. We argue that the new observables are dual to geodesic integrated bulk fields, where the geodesics can be minimal, wind around the singularity in the bulk, or even cross through horizons. We will conclude these sections with further connections to recent ideas in the literature.

In the latter parts of this thesis we will consider entanglement phenomena in general quantum systems, not specifically focused on AdS/CFT, but potentially applicable there. In Chapter \ref{ch:interlude}, we will briefly present a quantum information theoretic introduction to entanglement measures and their properties to reorient the reader. In Chapter \ref{ch:Renyi} the focus will be on entanglement dynamics of general systems, as measured by the family of R\'enyi entropies \cite{Cresswell2017a}. Instead of choosing a specific physical system, we retain general applicability by focussing on common initial conditions. We show that, starting from initially pure, unentangled states, the leading order growth of entanglement is characterized by a timescale which has the same form for all R\'enyi entropies. Since the R\'enyi entropies as a family completely characterize the bipartite entanglement properties of a pure state, the timescale can be considered universal.

The universal growth of R\'enyi entropies raises questions around the behaviour of distinct entanglement measures. In particular, one of the most commonly used entanglement measures, called negativity, does not conform to the growth behaviour described above. In Chapter \ref{ch:neg} we conduct an independent perturbative analysis of negativity and find that additional mathematical tools are required \cite{Cresswell2019}. These new tools are extended from an underdeveloped branch of mathematics known as patterned matrix calculus. After constructing the perturbative expansion of negativity, we compare its dynamical behaviour to R\'enyi entropies, and investigate the structural differences between these measures. The result is summarized in a theorem describing a class of functions for which patterned matrix derivatives are equivalent to ordinary matrix derivatives, and hence the additional complications of patterned matrix calculus can safely be glossed over. R\'enyi entropies belong to this class, while negativity does not, which neatly explains why a different approach is necessary for the perturbative expansion of negativity. Finally, we will discuss several other quantities which are commonly used in quantum information theory, and which do not belong to the class in the theorem. These examples will show how relevant patterned matrix calculus can be within this branch of physics as a whole.

\chapter{Kinematic space for conical defects}\label{ch:KSCD}

\emph{This chapter is based on the paper \cite{Cresswell2017} published in JHEP.}
\section{Introduction}

\begin{figure}
    \centering
        \includegraphics[width=0.4\textwidth]{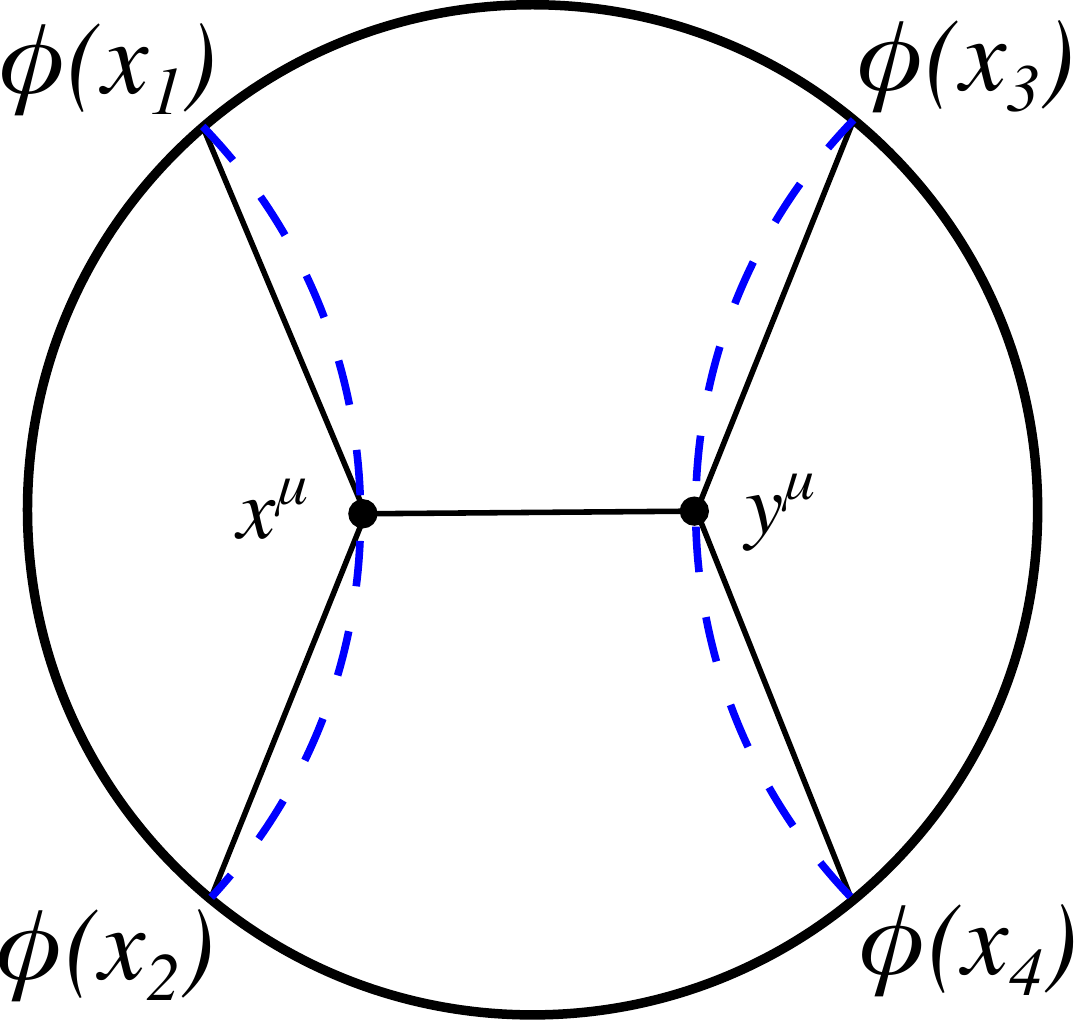}
        \caption[Scalar exchange geodesic Witten diagram]{A geodesic Witten diagram computes a conformal partial wave, the contribution to a CFT four point function from the conformal family of a single quasi-primary operator. The bulk interaction vertices $x^\mu,y^\mu$ are only integrated over geodesics connecting the boundary insertion points.}
        \label{fig:GeodesicWittenDiagram}
\end{figure}

Even before the AdS/CFT correspondence \cite{Maldacena1999,Witten1998,Gubser1998} provided a physical duality between conformal field theories and theories of quantum gravity in Anti-de Sitter spacetimes, CFT quantities had been mathematically represented in terms of bulk fields \cite{Ferrara1971,Ferrara1972}. These ideas relating contributions to conformal blocks and integrals of bulk fields over geodesics have reemerged recently in the context of geodesic Witten diagrams \cite{Hijano2015,Hijano2016}. Whereas a four-point Witten diagram with bulk vertices integrated over the entire bulk calculates a contribution to a full CFT four-point function, integrating the vertices only over geodesics connecting boundary insertions as in Fig. \ref{fig:GeodesicWittenDiagram} computes a conformal partial wave. The conformal partial wave represents the contribution of a primary operator and its descendants to the four-point function, and somehow knows about the geodesic structure of AdS.

In more detail, a 4-point function of identical scalars like $\langle \op(x_1)\op(x_2)\op(x_3)\op(x_4)\rangle$, which has the structure \eqref{eq:4pointfunction}, can be reduced using the OPE \eqref{eq:ope1} twice, taking $x_1\to x_2$ and $x_3\to x_4$
\begin{equation}
\langle \op(x_1)\op(x_2)\op(x_3)\op(x_4)\rangle=\sum_{k,k'}C_{\op\op k}C_{\op\op k'}D(x_{12},\pa_{x_2},k,\ell)D(x_{34},\pa_{x_4},k',\ell')\left\langle \op_{k,\ell}(x_2)\op_{k',\ell'}(x_4)\right\rangle.
\end{equation}
 Here $D(x_{12},\pa_{x_2})$ stands in for the fixed differential operator appearing in the OPE according to the quasi-primary $\op_{k,\ell}$ of dimension $\De_k$ and spin $\ell$, with tensor indices implied. Since the 2-point function is only non-zero when the two operators have the same dimension, we take them to be identical for simplicity, and define conformal partial waves $W_{\De_k,\ell}(u,v)$ from
\begin{equation}\label{eq:confpartialwaves}
\langle \op(x_1)\op(x_2)\op(x_3)\op(x_4)\rangle=\sum_{k}C_{\op\op k}^2W_{\De_k,\ell}(x_i).
\end{equation}
We see that this represents the contribution to the 4-point function from the exchange of $\op_{k,\ell}$ and its descendants in the channel $(12)(34)$. If we imagine a projector onto the conformal family of $\op_{k,\ell}$, schematically
\begin{equation}
P_{k,\ell}=\sum_n \left| P^n \op_{k,\ell}\right\rangle\left\langle P^n \op_{k,\ell}\right |,
\end{equation}
with $P^n$ the operator that generates the $n$th descendant, then the conformal partial wave would be
\begin{equation}
W_{\De_k,\ell}(x_i)\sim\langle \op(x_1)\op(x_2)P_{k,\ell}\op(x_3)\op(x_4)\rangle.
\end{equation}

We also can define conformal blocks which only depend on the conformally invariant cross ratios,
\begin{equation}
G_{\De_k,\ell}(u,v)=x_{12}^{2\De_\op}x_{34}^{2\De_\op}W_{\De_k,\ell}(x_i).
\end{equation}
Hence conformal blocks and conformal partial waves contain the same physical content, but the former is manifestly conformally invariant. Comparing \eqref{eq:confpartialwaves} and \eqref{eq:4pointfunction}, the conformal blocks are related to the arbitrary function $g(u,v)$ appearing in the 4-point function by
\begin{equation}
g(u,v)=\sum_k C_{\op\op k}^2 G_{\De_k,\ell}(u,v). 
\end{equation}
The fact that conformal blocks (or partial waves) can be computed by geodesic Witten diagrams \cite{Hijano2015,Hijano2016} is not only a technical boon for calculations, but provides a new understanding of how AdS 4-point amplitudes can be contained within the rigid structure of CFT correlation functions, and has led to many new developments along these lines \cite{Fukuda2018,Kobayashi2018,DAS2019397,Prudenziati2019}. As we will explain, a very similar bulk to boundary correspondence holds when considering the bare OPE itself.

 A new approach to the AdS/CFT correspondence has shed more light on the connection between composite operators in the OPE, and integrated bulk fields. The authors of \cite{Czech2015,Czech2016} proposed the use of an auxiliary space that interpolates between the bulk and boundary theories, similar to the space used in \cite{Boer2015}. The auxiliary space, called kinematic space, functions as a way of organizing the non-local degrees of freedom which lead to diffeomorphism invariant quantities in the bulk gravity theory. Whereas local bulk fields fail to satisfy diffeomorphism invariance, a field integrated over a boundary anchored geodesic or otherwise attached to the boundary with a geodesic dressing can be invariant \cite{Donnelly2015,Donnelly2016}. Boundary anchored geodesics in asymptotically AdS spacetimes meet the boundary at pairs of spacelike or null separated points suggesting a relation to bi-local CFT operators. Such composite operators are easily described in terms of the OPE. Both a geodesic integrated field and the basis of non-local operators forming the OPE can be viewed as fields on kinematic space leading to a diffeomorphism invariant entry into the AdS/CFT dictionary.

Several proposals have been made as to how kinematic space should be defined from the bulk and boundary. Kinematic space was originally presented as the space of bulk geodesics with a measure derived from their lengths in terms of the Crofton form used in integral geometry (as opposed to differential geometry) \cite{Czech2015}. Since the length of a minimal geodesic is holographically related to entanglement entropy in AdS$_3$/CFT$_2$ \cite{Ryu2006}, a boundary description of kinematic space was given as the space of boundary intervals with the metric defined in terms of the differential entropy of those intervals \cite{Balasubramanian2014}.\footnote{This approach was recently inverted to derive the universal parts of the entanglement entropy in a CFT with a boundary from knowledge of the kinematic space \cite{Bhowmick2017}.} In order to generalize the kinematic space approach to higher dimensional systems, later approaches defined points in kinematic space as oriented bulk geodesics, and simultaneously as ordered pairs of boundary points \cite{Czech2016}. 

In the case of a pure AdS$_3$ geometry, these approaches are consistent since there is a unique geodesic connecting each pair of spacelike separated boundary points. Other well known locally AdS$_3$ geometries can have several geodesics connecting each pair of boundary points, namely conical defects and BTZ black holes \cite{Banados1992,Banados1993,Deser1984}. There are two diverging ways to modify the definition of what constitutes a kinematic space point in such cases. Any spacelike separated pair of boundary points will be connected by a unique minimal geodesic, so the bulk definition can exclude non-minimal geodesics from kinematic space with no need to change the boundary definition. Alternatively, non-minimal geodesics can be considered as points with the same standing as minimal ones, in which case ordered pairs of boundary points alone will not fill out kinematic space. Excluding non-minimal geodesics is not desirable due to the generic fact that minimal geodesics do not reach all depths of the bulk. The region probed by non-minimal geodesics is known as the entanglement shadow \cite{Balasubramanian2015,Freivogel2014,Balasubramanian2016}. A full description of the bulk in terms of kinematic space can only succeed when non-minimal geodesics are included. This forces a change to the definition of kinematic space from the boundary point of view.

In this chapter, we take up the issue of non-minimal geodesics in kinematic space, and the matter of an equivalent boundary definition of points in the simplest geometry exhibiting this feature, the static conical defects in three bulk dimensions. In Section \ref{secbulk} the geometry of the conical defect kinematic space is derived in two ways. The first is a simple application of the differential entropy definition applied to geodesics of all lengths. The second follows \cite{Zhang2017} in noting that the conical defects can be obtained as a quotient of pure AdS$_3$. Under this quotient classes of geodesics are identified, producing a quotient on kinematic space, and leading to a result equivalent to the first approach. In Section \ref{secboundary} the metric of kinematic space is extracted from OPE blocks in the CFT. By mapping to a convenient covering CFT system we find that conventional OPE blocks can be broken down further than done before using the method of images. Individual image contributions to the OPE blocks contain information about subregions of kinematic space that, when combined, reproduce the same space identified from the bulk. Intuition from previous uses of the method of images to calculate correlation functions holographically suggests an association between partial OPE blocks in the CFT and geodesics of a fixed winding number in the bulk. Kinematic space provides a realm where the connection between these objects can be made precise, as is shown in Section \ref{secdis}. We conclude by isolating the contribution to the full OPE block from individual bulk geodesics, minimal or non-minimal, which connect the boundary insertion points. This extends the holographic dictionary established in \cite{Czech2016} between OPE blocks and geodesic integrated operators, and provides more fine-grained information about the holographic contributions to the blocks.

\section{Kinematic space from the bulk}\label{secbulk}

In this section we focus on static conical defect spacetimes and consider the kinematic space for a constant time slice. We show that the differential entropy approach \cite{Czech2015}, and the quotient approach \cite{Zhang2017} produce different fundamental regions of the same kinematic space, but are entirely equivalent. 

\subsection{Review of geometries}

 In global coordinates, the universal cover of AdS$_3$ has the metric
\begin{equation}\label{eqgads}
ds^2=R^2_{\mathrm{AdS}}(-\cosh^2\rho\ dt^2+d\rho^2+\sinh^2\rho\ d\phi^2),
\end{equation}
with $t\in\mathbb{R},~\rho\in \mathbb{R^+}$, and $\phi\in[0,2\pi]$ with the identification $\phi=\phi+2\pi$. Throughout this chapter the ``unwrapped" time coordinate $t$ of the universal cover will be used. The AdS$_3$ geometry can be understood as a surface embedded in the higher dimensional flat space $\mathbb{R}^{(2,2)}$ with metric
\begin{equation}
ds^2=-dU^2-dV^2+dX^2+dY^2.
\end{equation}
The AdS$_3$ metric is induced by restricting to a hyperbolic surface
\begin{equation}
-U^2-V^2+X^2+Y^2=-R^2_{\mathrm{AdS}}.
\end{equation}
The parameter $R_{\mathrm{AdS}}$ is the AdS length scale which will be set to unity throughout the remainder of this chapter. The metric in global coordinates is obtained from the embedding equations
\begin{equation}
U=\cosh\rho\cos{t},~~V=\cosh{\rho}\sin{t},~~X=\sinh\rho\cos\phi,~~Y=\sinh\rho\sin\phi.
\end{equation}

\begin{figure}
    \centering
    \begin{subfigure}[b]{0.33\textwidth}
        \includegraphics[width=0.9\textwidth]{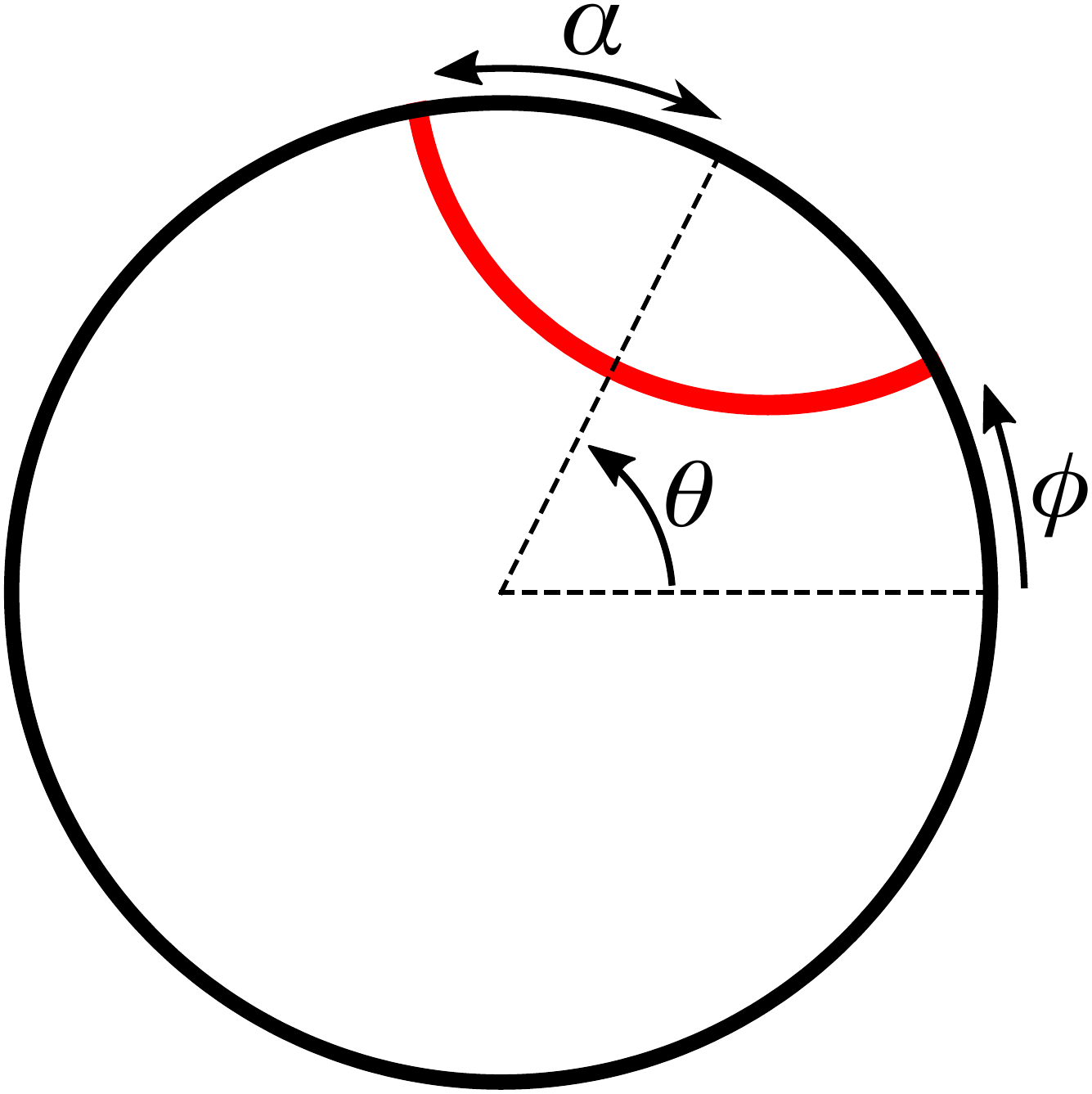}
        \caption{}
        \label{fig:H2coords}
    \end{subfigure}
     \ \ 
    \begin{subfigure}[b]{0.31\textwidth}
        \includegraphics[width=0.9\textwidth]{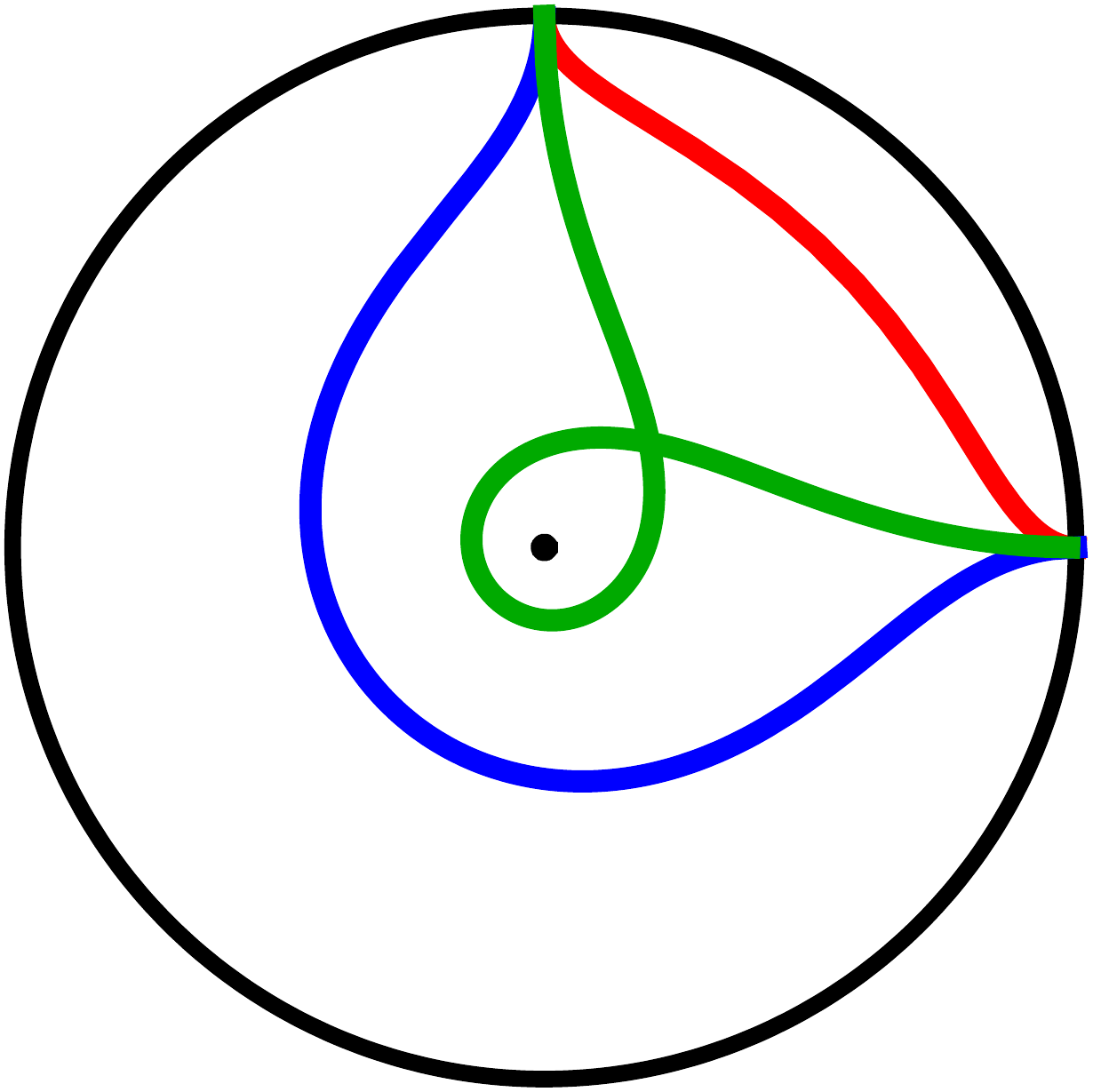}
        \caption{}
        \label{fig:CDPoincareDiskGeodesics2}
    \end{subfigure}
     \ \ 
    \begin{subfigure}[b]{0.31\textwidth}
        \includegraphics[width=0.9\textwidth]{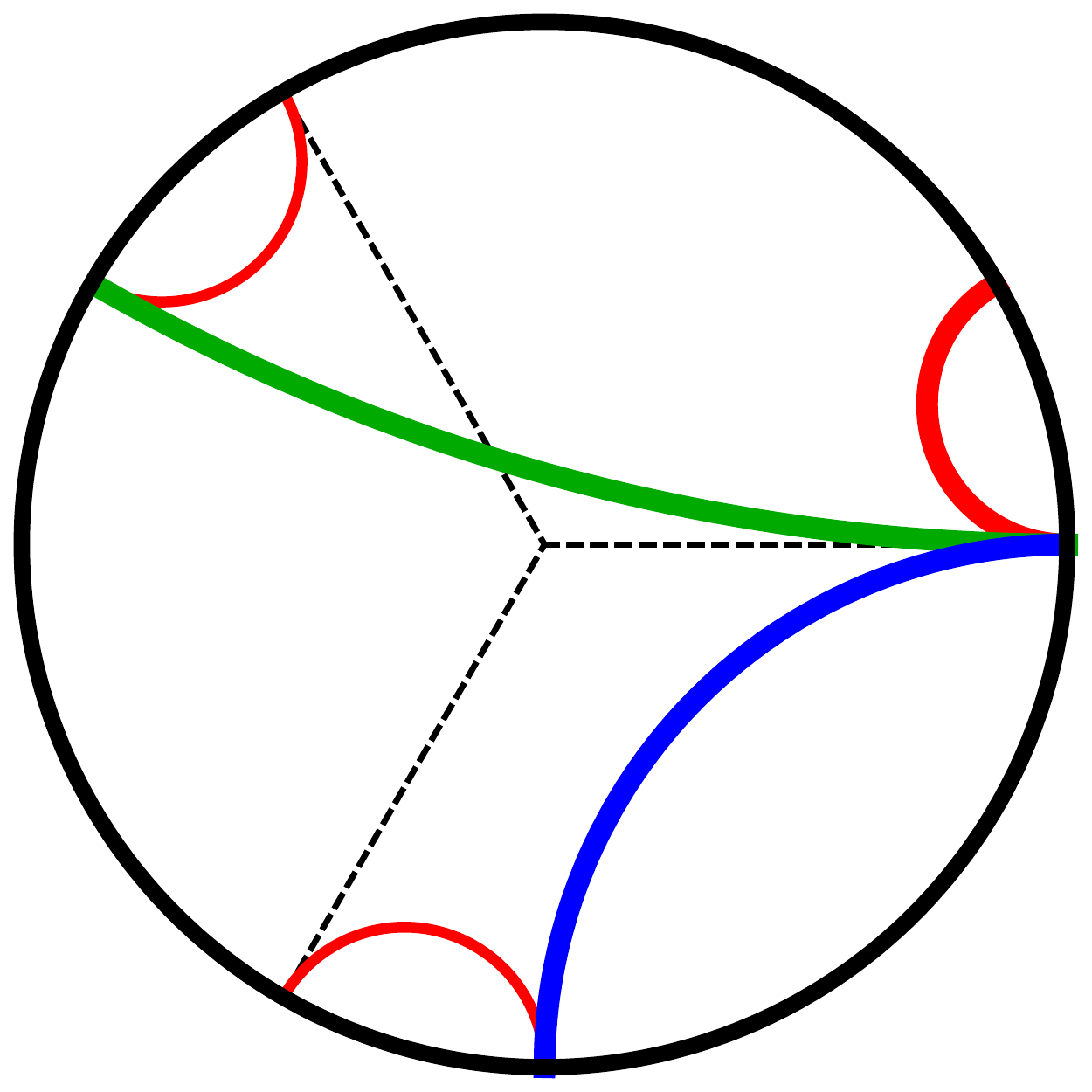}
        \caption{}
        \label{fig:PoincareDiskGeodesics}
    \end{subfigure}
    \caption[Geodesics in the Poincar\'e disk]{(a) The Poincare disk showing a geodesic and its kinematic coordinates. (b) A spatial slice of a conical defect, with $N=3$ for illustration, showing three geodesics subtending the same boundary interval with winding numbers $n=0,1,2$ respectively in order of increasing length. (c)~The covering space of the conical defect showing identified wedges, and preimages of the corresponding geodesics in Figure (b). Also shown are two equivalent images of the minimal geodesic (red).}\label{fig:1}
\end{figure}

For visual representations it will be useful to consider the Poincar\'e disk. By taking a constant time slice $t=0$, equivalently $V=0$, the metric induced from $\mathbb{R}^{(1,2)}$ is that of the hyperbolic plane $\mathbb{H}_2$,
\begin{equation}
ds^2=d\rho^2+\sinh^2\rho \ d\phi^2.
\end{equation}
This describes a two sheeted hyperboloid in $\mathbb{R}^{(1,2)}$ with disconnected parts above and below the $U=0$ plane. The tips of the sheets are located at $(-1,0,0)$ and $(1,0,0)$ in the $(U,X,Y)$ embedding coordinates. The Poincar\'e disk can be obtained by projecting the $U>1$ sheet onto the $U=0$ plane through the point $(-1,0,0)$. In the disk, boundary anchored geodesics are described by the particularly simple equation
\begin{equation}
\tanh{\rho} \ \cos{(\phi-\te)}=\cos \al.
\end{equation}
Here $\te$ denotes the angular coordinate of the center of the geodesic, and $\al\in[0,\pi]$ is the half-opening angle. Pictorially, geodesics in the Poincar\'e disk are arcs of circles that meet the boundary at right angles as in Figure \ref{fig:H2coords}. 

Conical defect spacetimes can be obtained as a quotient of AdS$_3$ by identifying surfaces of constant $\phi$ leaving an angular coordinate with a smaller period. In global coordinates the metric is simply
\begin{equation}\label{eqCDmetric}
ds^2=-\cosh^2\rho\ dt^2+d\rho^2+\sinh^2\rho\ d\tilde \phi^2,
\end{equation}
where now $\tilde \phi=\tilde \phi+\frac{2\pi}{N}$. The parameter $N\in(1,\infty)$ gives the strength of the defect. This metric is no longer a solution of the vacuum Einstein equations everywhere but requires a pointlike source at the origin. The defect can be viewed as a static particle of mass $M$ where $4G_{N} M=1-{1}/{N}$. The mass must stay below the black hole limit $M={1}/{4G_{N}}$, which corresponds to $N\to\infty$. For the special cases where $N$ is an integer, the spacetime is a cyclic orbifold AdS$_3/\mathbb{Z}_N$. Some example geodesics in the $t=0$ slice of the conical defect are shown in Figure \ref{fig:CDPoincareDiskGeodesics2}, and the corresponding geodesics of AdS$_3$ in Figure \ref{fig:PoincareDiskGeodesics}.

The kinematic space corresponding to the Poincar\'e disk was investigated in \cite{Czech2015} and found to be a two dimensional de Sitter geometry. For ease of comparison the dS$_2$ spacetime can be embedded in the same $\mathbb{R}^{(1,2)}$ where it is a one-sheeted hyperboloid given by
\begin{equation}
-U^2+X^2+Y^2=1.
\end{equation}
The embedding equations 
\begin{equation}
U=\sinh{t},~~X=\cosh t\cos \te,~~Y=\cosh t\sin\te,
\end{equation}
lead to the dS$_2$ metric in global coordinates,
\begin{equation}\label{dS}
ds^2=-dt^2+\cosh^2t \ d\te^2.
\end{equation}
Conformal or ``kinematic" coordinates $(\al,\te)$ will be used often in this chapter as they naturally fit with the description of kinematic space as the space of geodesics in AdS$_3$. The transformation $\cosh t=1/\sin\al$, where now $\al\in[0,\pi]$, leads to the dS$_2$ metric
\begin{equation}\label{kinematic}
ds^2=\frac{-d\alpha^2+d\theta^2}{\sin^2\alpha}. 
\end{equation}
With these conventions laid out, the remainder of this section briefly recounts the derivation of the kinematic space geometry for pure AdS$_3$, then details two methods of obtaining the kinematic space for conical defects from the bulk.

\subsection{Kinematic space from differential entropy}\label{secdifferentialentropy}

In \cite{Czech2015} a definition of kinematic space for constant time slices of AdS$_3$ in terms of differential entropy was derived from integral geometry. Each interval of the boundary, denoted by an ordered pair of points $(u,v)$, corresponds to a point in kinematic space covered by null coordinates $(u,v)$. The kinematic space metric in these coordinates was found to be
\begin{equation}\label{eqCrofton}
ds^2=\frac{\pa^2 S(u,v)}{\pa u\pa v} \ du dv,
\end{equation}
where $S(u,v)$ was the length of the shortest oriented geodesic connecting the ends of the interval $(u,v)$ through the bulk. Since the length of a minimal geodesic is holographically interpreted as the entanglement entropy of the interval it subtends, the quantity $\pa^2 S/\pa u\pa v$ was dubbed differential entropy \cite{Balasubramanian2014,Myers2014,Espindola2017}. However, many interesting spacetimes including the conical defects and BTZ black holes have multiple geodesics connecting pairs of spacelike separated boundary points. Non-minimal geodesics do not correspond to entanglement between spatial regions, but have been conjectured to describe correlations between internal degrees of freedom \cite{Balasubramanian2015}. Because of this potential interest, and their importance in the geodesic approximation for correlation functions \cite{Balasubramanian1999,Asplund2014}, in this chapter the differential entropy definition will be expanded to include non-minimal geodesics.

For the constant time slice of AdS$_3$, there is a unique oriented geodesic connecting each ordered pair of boundary points so the issue of non-minimal geodesics in eq. \eqref{eqCrofton} does not arise. Geodesics can be labelled by their half-opening angle $\al$ and centre angle $\te$, and have length
\begin{equation}\label{eqAdSgdlength}
S(\al)=\frac{1}{2G_N}\log \frac{2 \sin\al}{\mu},
\end{equation}
where $\mu$ serves as a gravitational infrared cutoff \cite{Czech2014}. By transforming between kinematic coordinates and null coordinates using $u=\te-\al$, and $v=\te+\al$, eq. \eqref{eqCrofton} can be applied to find
\begin{align}\label{eqds2KS}
\begin{aligned}
ds^2&=\frac{1}{8 G_N}\frac{1}{\sin^2[(v-u)/{2}]} \ du dv\\
&=\frac{1}{8 G_N}\frac{-d\alpha^2+d\theta^2}{\sin^2\alpha}.
\end{aligned}
\end{align}
\begin{figure}
	\centering
        \includegraphics[width=0.48\textwidth]{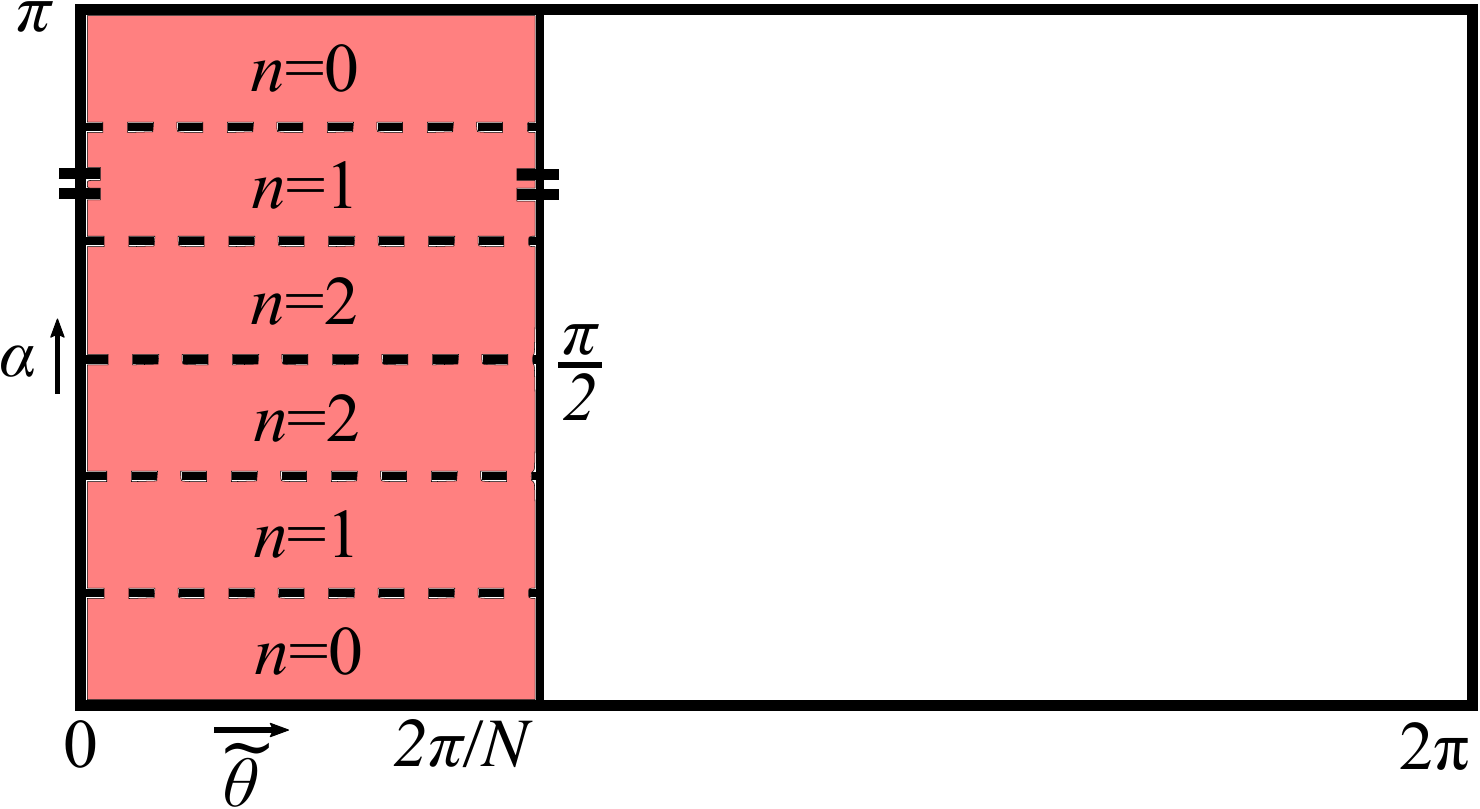}
    \caption[Kinematic space with $\mathbb{Z}_N$ identification]{The Penrose diagram for the dS$_2$ kinematic space for pure AdS$_3$ is shown as the full rectangular region, with $\te=\te+2\pi$. For the conical defect case the differential entropy definition of kinematic space produces a vertical strip subregion. The same angular identification which gives the conical defect from AdS$_3$ also gives the kinematic space. $N=3$ is shown for illustration throughout most of this chapter.}\label{fig:KSCDVertical}
\end{figure}

Thus, the kinematic space of a constant time slice of AdS$_3$ is dS$_2$ according to the differential entropy definition, as shown in Figure \ref{fig:KSCDVertical}. The $\al<\pi/2$ and $\al>\pi/2$ halves are mapped into one another under orientation reversal which acts as $\al\to \pi-\al$ and $\te\to\te+\pi$. The geodesics with $\al=\pi/2$ cut straight across the Poincar\'e disk and have maximal length.

Now consider a constant time slice of the conical defect geometry eq. \eqref{eqCDmetric}. Since the total angle around the boundary is $2\pi/N$, the centre angle of a geodesic will now be denoted $\tilde \te\in[0,2\pi/N]$. Once again, for any pair of boundary points there is a unique minimal geodesic connecting them through the bulk. Minimal geodesics have half-opening angles in the domain $\al\in[0,\pi/2N]$, and by reversing orientations with $\al\to \pi-\al$ and $\tilde \te\to\tilde \te+\pi/N$, also the domain $\al\in[(2N{-}1)\pi/2N,\pi]$. Minimal geodesics cover the top and bottom regions of kinematic space in Figure \ref{fig:KSCDVertical}.

In contrast to AdS$_3$, there can be non-minimal geodesics connecting pairs of boundary points. It will be useful to label geodesics and their corresponding regions in kinematic space by the number of times they wind around the defect, $n$. The cases of integer and non-integer $N$ will be treated separately for clarity.

For integer $N$ there are $N-1$ non-minimal geodesics connecting each pair of boundary points, with winding numbers $1\leq n\leq N-1$. Geodesics with winding number $n$ fill in the regions of kinematic space 
\begin{equation}\label{eqalpharanges}
\al\in \left(\frac{n\pi}{2N},\frac{(n+1)\pi}{2N}\right],\quad  \al\in\left[\frac{(2N-n-1)\pi}{2N}, \frac{(2N-n)\pi}{2N}\right),
\end{equation}
where these domains are related by orientation reversal. The upper and lower halves of kinematic space are divided by geodesics with $\al=\pi/2$ which touch the conical defect. On the covering AdS$_3$ space, these are the straight lines through the origin of the Poincar\'e disk. In total there are $2N$ equally sized regions on kinematic space in the $(\al,\tilde \te)$ coordinates.

For non-integer $N$, the maximally winding geodesics have $n=\left\lfloor N\right\rfloor$ and live near the centre line $\al=\pi/2$. There are fewer maximally winding geodesics than other classes, filling out a truncated region 
\begin{equation}
\al\in\left(\frac{(\left\lfloor N\right\rfloor-1)\pi}{2N},\frac{(\left\lfloor N\right\rfloor+1)\pi}{2N}\right).
\end{equation}
Other winding numbers follow eq. \eqref{eqalpharanges}. Each pair of boundary points is connected by $\left\lfloor N\right\rfloor$ or $\left\lfloor N\right\rfloor-1$ geodesics, depending on their angular separation.

The differential entropy definition eq. \eqref{eqCrofton} can be applied to show that the geometry on kinematic space remains locally dS$_2$ for any $N$. The key fact is that minimal and non-minimal geodesics still have lengths given by eq. \eqref{eqAdSgdlength} \cite{Czech2014}. Treating the types on equal footings from the point of view of kinematic space and using $u=\tilde \te-\al$, $v=\tilde \te+\al$ once again gives\footnote{A previous paper \cite{Asplund2016} describing the kinematic spaces for several locally AdS$_3$ geometries, including conical defects, chose to consider only minimal geodesics, and hence found different kinematic space geometries.}
\begin{equation}\label{eqdiffentKS}
ds^2=\frac{1}{8 G_N}\frac{-d\alpha^2+d{\tilde\theta}^2}{\sin^2\alpha}.
\end{equation}
The kinematic space for a constant time slice of a conical defect has the same dS$_2$ metric as the AdS$_3$ case, but with the angular coordinate identified as $\tilde \te\sim\tilde \te+{2\pi}/{N}$. This was expected since the static conical defects are locally AdS$_3$, only differing by the global identification along the angular coordinate. The identification does not affect the lengths of the remaining geodesics. From the differential entropy perspective, the conical defect kinematic space is found by taking an angular quotient of the AdS$_3$ kinematic space; the same quotient that produces the conical defect from pure AdS$_3$ itself. In the next section we show how the quotient acts on geodesics in the covering space, displaying the inherent ambiguities involved in defining kinematic space.
%

\subsection{Kinematic space from boundary anchored geodesics}\label{secbulkKS}

The bulk calculation of the kinematic space for conical defects is more enlightening when the defects are viewed from the perspective of the covering space, AdS$_3$. In particular, it provides motivation for treating minimal and non-minimal geodesics on equal footing in the definition of kinematic space, since there is no real distinction between the types when viewed in the cover. All spacelike geodesics of the conical defect descend from the covering space; the quotient that produces the conical defect divides geodesics into equivalence classes.

\begin{figure}
    \centering
    \begin{subfigure}[b]{0.3\textwidth}
        \includegraphics[width=0.9\textwidth]{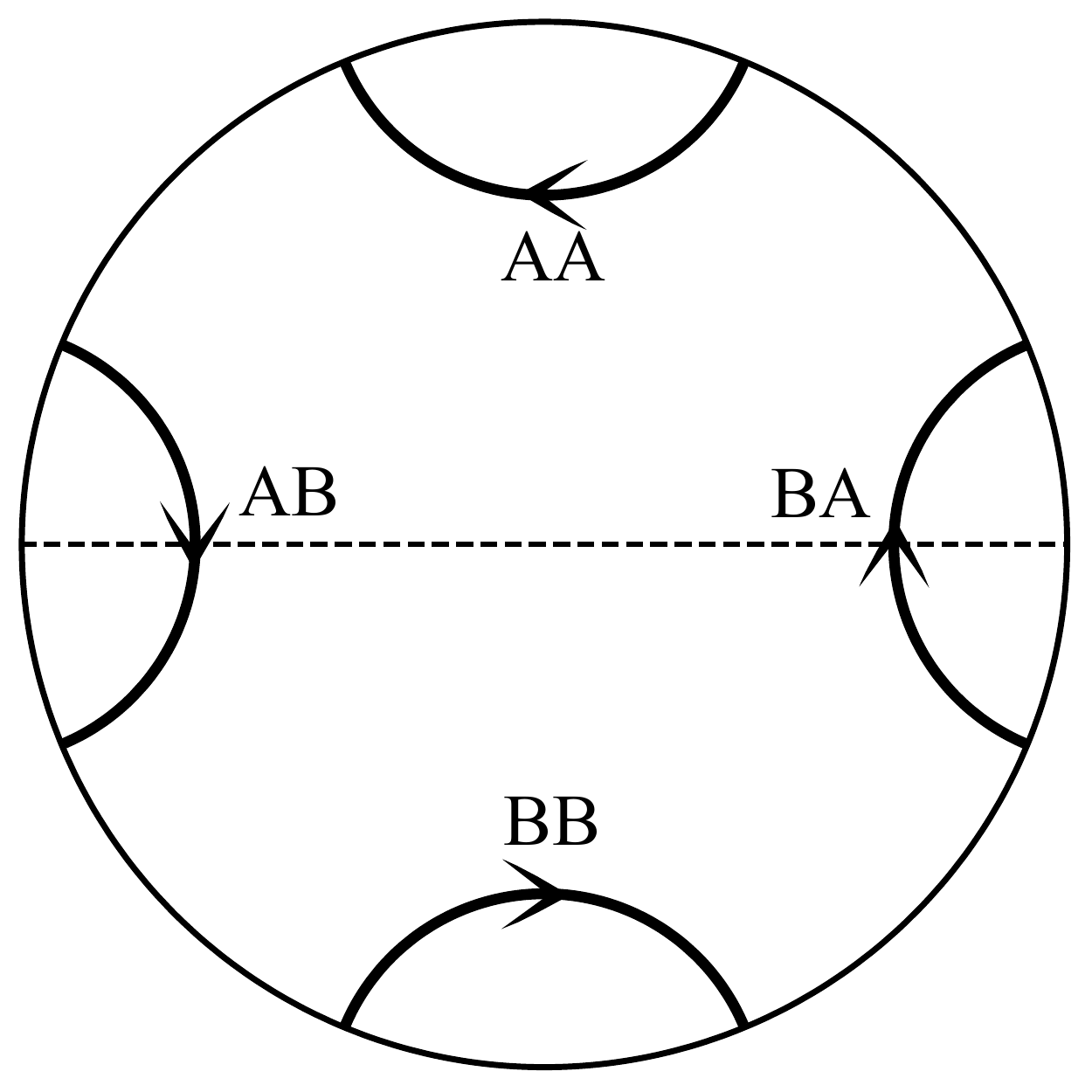}
        \caption{}
        \label{fig:4geodesics}
    \end{subfigure}
     \qquad \qquad
    \begin{subfigure}[b]{0.48\textwidth}
        \includegraphics[width=0.9\textwidth]{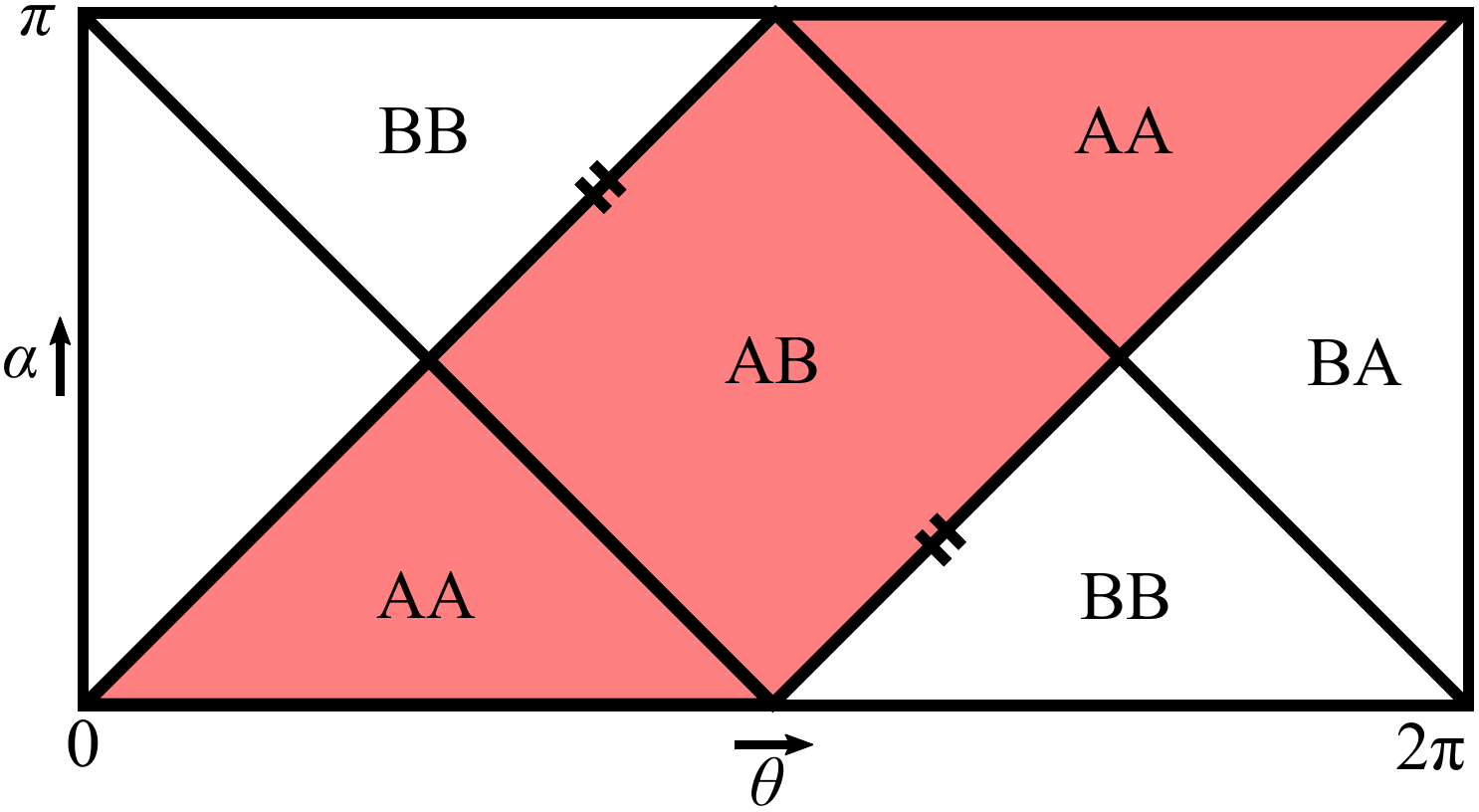}
        \caption{}
        \label{fig:Z2Stripe}
    \end{subfigure}
    \caption[Conical defect kinematic space with geodesic classes labeled]{(a) Oriented geodesics in the Poincar\'e disk labelled by their endpoint locations. The $N=2$ wedges are shown with two identified boundaries A and B. (b) Regions of kinematic space labelled by the boundaries each geodesic ends on. For the conical defect with $N=2$, the shaded diagonal strip is a fundamental domain equivalent to the vertical strip.}\label{fig:4}
\end{figure}

As an explicit example, consider the case of $N=2$. The covering space of $\mathbb{H}_2/\mathbb{Z}_2$ is shown in Figure \ref{fig:4geodesics}. The covering space can be split into two regions with boundaries labelled $A$ and $B$, which are identified under the quotient. Boundary anchored geodesics on this slice can be grouped into four classes $\{AA,BB,AB,BA\}$ depending on the boundary region their endpoints lie on. The locations of the classes on kinematic space are shown in Figure \ref{fig:Z2Stripe}.

Under the $\mathbb{Z}_2$ quotient, $BB$ geodesics are mapped into $AA$ geodesics. Similarly, $BA$ geodesics are mapped into $AB$ geodesics. Therefore, all geodesics in $\mathbb{H}_2/\mathbb{Z}_2$ can be generated by the classes $\{AA,AB\}$, and the $\mathbb{Z}_2$ action. The number of unique geodesics in the conical defect slice is greatly reduced, and similarly for points on kinematic space. As is shown in Figure \ref{fig:Z2Stripe}, the kinematic space for the $N=2$ conical defect slice is a diagonal strip of width $\te=\pi$, with the identification $\te= \te+\pi$. However, there are many equivalent ways to choose the fundamental region under the quotient action. If, for example, the classes $\{AA,BA\}$ had been chosen as fundamental, the diagonal strip would point in the opposite direction. Similarly, the entire strip can be shifted by any amount in the $\te$ direction. There is nothing to distinguish these choices, so as in \cite{Zhang2017} a conventional choice has been made.
 
  The quotient only changes the global identification of points in the spacetime, and the geodesics within it. The relationship between nearby geodesics in the kinematic space metric are locally unchanged. While the origin of AdS$_3$ is a fixed point of the quotient, there are no oriented geodesics which are left invariant. From the perspective of kinematic space, the quotient is freely acting, so the metric is expected to be locally unchanged, and the topology to be invariant. This is in contrast to the kinematic space of the BTZ black hole found in \cite{Zhang2017}. The quotient of AdS$_3$ which produces a BTZ black hole has no fixed points so there are no curvature singularities in the BTZ spacetime, but there are geodesics which are fixed under the quotient which changes the topology of kinematic space from a single cylinder to two.
 
  For the more general case of a $\mathbb{Z}_N$ quotient, there are $N^2$ distinct classes of oriented geodesics from the number of ways we can choose two ordered endpoints. The number of distinct regions in kinematic space is $N(N+1)$, one for each of the $N^2$ classes, and one extra for each of the $N$ boundaries. The fundamental region is a diagonal strip with width given by $2\pi/N$ since $\te=\te+2\pi/N$ is identified. This also describes the fundamental region for arbitrary $N$.

 The two approaches presented here, using the differential entropy definition eq. \eqref{eqCrofton}, and studying how the quotient identifies geodesics on the covering space both produce a locally dS$_2$ spacetime but naturally pick out different regions of the kinematic space. The differential entropy definition picks out a vertical strip, while the classification of endpoints on the covering space produces a diagonal strip. However, it is clear from the latter approach that there are many equivalent choices of fundamental region, each with its own merits. The diagonal choice contains some geodesics which have boundary position $\te>2\pi/N$ on the cover. The vertical choice only contains geodesics which are centred at boundary coordinates $\te<2\pi/N$. Since it is easiest to label geodesics with kinematic coordinates $\al\in(0,\pi)$ and $\tilde\te\in[0,2\pi/N]$, the vertical strip will be used in the rest of this chapter. 
 
\section{Kinematic space from the boundary}\label{secboundary}

\subsection{Kinematic space metric from conformal symmetry}

In \cite{Czech2016} a definition of kinematic space from the boundary theory was given: each point in kinematic space corresponds to an ordered pair of CFT points.\footnote{In \cite{Czech2016} and \cite{Boer2016} it was shown that an equivalent definition can be made in terms of boundary causal diamonds.} For pure AdS$_3$/CFT$_2$ restricted to a time slice, each ordered pair of CFT points singles out a unique spacelike boundary anchored geodesic so this definition is entirely natural. In the full time dependent geometry, conformal symmetry alone fixes the metric on kinematic space to be
\begin{equation}\label{eqdSdSmetric}
ds^2=4\frac{I_{\mu\nu}(x_1-x_2)}{|x_1-x_2|^2} \ dx_1^\mu dx_2^\nu,
\end{equation}
where 
\begin{equation}
I_{\mu\nu}(x_1-x_2)=\eta_{\mu\nu}-2\frac{(x_1-x_2)_\mu(x_1-x_2)_\nu}{(x_1-x_2)^2},
\end{equation}
is the inversion tensor. The numerical prefactor in the metric is chosen by convention. The two CFT points $x_1^\mu$ and $x_2^\mu$ form a pair of lightlike coordinates on kinematic space with the strange signature $(2,2)$. Since kinematic space is not to be viewed as a physical space, but only as a useful auxiliary space for translating between the bulk and boundary, this is not a concern.

In order to get back the dS$_2$ metric found from the bulk, it is easiest to perform a coordinate transformation from the planar set $x_1^\mu=\{t_1,x_1\}$ to kinematic coordinates on the cylinder. The two pairs of kinematic coordinates are defined through
\begin{align}\label{eq4kinecoords}
\begin{aligned}
\tan \al&=\frac{1}{2}\left(t_1-t_2+x_1-x_2\right),\quad \te=\frac{1}{2}\left(t_1+t_2+(x_1+x_2)\right),\\
\tan \bar\al&=\frac{1}{2}\left(t_1-t_2-(x_1-x_2)\right),\quad \bar\te=\frac{1}{2}\left(t_1+t_2-(x_1+x_2)\right).
\end{aligned}
\end{align}
In terms of these coordinates the kinematic space metric is two copies of the dS$_2$ metric in eq. \eqref{kinematic},
\begin{equation}\label{eqds2ds2}
ds^2=\frac{-d\alpha^2+d\theta^2}{2\sin^2\alpha}+\frac{-d\bar\alpha^2+d\bar \theta^2}{2\sin^2\bar\alpha}.
\end{equation}
Thus the kinematic space for global AdS$_3$, and the dual vacuum state of a CFT$_2$ is dS$_2\times$dS$_2$. When we restrict to a constant time slice by setting $t_1=t_2=0$ we see from eq. \eqref{eq4kinecoords} that $\bar \al$ and $\bar \te$ become redundant coordinates fixed in terms of $\{\al,\te\}$, and that eq. \eqref{eqds2ds2} becomes eq. \eqref{eqdiffentKS}, up to the arbitrarily chosen prefactor.

When the bulk spacetime has non-minimal geodesics, there is no longer a one-to-one correspondence between pairs of CFT points and bulk geodesics. In such a case the argument above cannot be applied. In order to reproduce the quotient structure of kinematic space for conical defects seen in section \ref{secbulk}, another approach must be taken. We take the point of view espoused in \cite{Karch2017}; OPE blocks in the CFT should be viewed as free fields on kinematic space, and their equation of motion reflects the geometry of kinematic space.

\subsection{OPE blocks}\label{secOPEblocks}

In \cite{Czech2016}, the operator product expansion (OPE) of two scalar CFT operators was broken into OPE blocks, and these blocks were identified as fields on kinematic space. Two scalar operators $\op_i(x_1)$ and $\op_j(0)$ in a planar CFT with conformal weights $\De_i$ and $\De_j$ respectively can be expanded in terms of a local basis of operators at the origin,
\begin{equation}\
\op_{i}(x)\op_{j}(0) =  \sum_{k}C_{ijk}\left|x\right|^{\Delta_{k}-\Delta_{i}-\Delta_{j}}\big(1+b_{1}\,x^{\mu}\partial_{\mu}+b_{2}\,x^{\mu}x^{\nu}\partial_{\mu}\partial_{\nu}+\ldots\big)\op_{k}(0).
\end{equation}
This is the OPE, where the quasi-primaries $\op_k(0)$, and their descendants given by the derivative terms, form the basis of operators at the origin. Notably, the $b_n$ coefficients are completely fixed by conformal symmetry, while the $C_{ijk}$ are simply constants, but are theory-dependent. Each term in the sum has a characteristic scaling dimension $\De_k$, the dimension of the quasi-primary $\op_k$, and represents the contribution to the OPE of the entire conformal family of $\op_k$. Each of these terms can be packaged into a new operator $\mathcal{B}^{ij}_k(x_1,x_2)$ called an OPE block, and the OPE can be written as
\begin{equation}\label{eqOPEblock}
\op_{i}\left(x_1\right)\op_{j}\left(x_2\right) = \left|x_1-x_2\right|^{-\Delta_{i}-\Delta_{j}}\sum_{k}C_{ijk}\mathcal{B}_{k}^{ij}\left(x_1,x_2\right).
\end{equation}

Since the OPE blocks depend on a pair of CFT points, the two points where operators in the OPE are inserted, it is natural to view them as fields on kinematic space. A major insight of \cite{Czech2016} was that the Casimir eigenvalue equation satisfied in the CFT by the OPE blocks can be interpreted as a wave equation. The differential representation of the CFT Casimir operator appropriate for OPE blocks is the Laplacian in the kinematic space metric eq. \eqref{eqds2ds2}. This gives yet another prescription for determining the kinematic space for a CFT state which is applicable when arguments from conformal symmetry alone are not sufficient, as advocated for recently in \cite{Karch2017}. In the following section, we will show how this prescription can be modified and used to obtain the kinematic space for excited CFT states dual to conical defects, in agreement with the results of section \ref{secbulk}. First, we review how the bulk metric of AdS$_3$ can be determined from a quadratic CFT$_2$ Casimir in a differential representation appropriate for scalar fields, and how the bilocal scalar representation of OPE blocks gives the metric on kinematic space. These initial cases have been summarized in \cite{Czech2016,Boer2016}.

In a 2d CFT, the global conformal group $SO(2,2)$ forms a subgroup of the larger Virasoro symmetry group. The global subgroup corresponds holographically to the isometries of pure AdS$_3$ with appropriate boundary conditions, while the other generators of the Virasoro group are associated to transformations which preserve the asymptotic boundary \cite{Brown1986}. The global conformal generators $L_{0,\pm1}, \ \bar L_{0,\pm 1}$ in the standard basis satisfy two copies of the Witt algebra
\begin{equation}\label{Witt}
[L_n,L_m]=(n-m)L_{n+m},\quad [\bar L_n,\bar L_m]=(n-m)\bar L_{n+m}, \quad [L_n,\bar L_m]=0.
\end{equation}
When acting on conformal operators, the algebra is represented by some differential operators $\mathcal{L}_{n}$ as
\begin{equation}\label{eqdiffrep}
[L_n,\op_k(x)]=\mathcal{L}_{n} \op_k(x),
\end{equation}
which depend on the $SO(2,2)$ representation of $\op_k$. 

The quadratic Casimir operator
\begin{equation}\label{eqc2}
\mathcal{C}_2=-\frac{1}{2}L^{AB}L_{AB}=-2L_0^2+(L_1 L_{-1}+L_{-1}L_1)\ + \ (L\to\bar L),
\end{equation}
commutes with all the global conformal generators.\footnote{Our conventions are as in \cite{Boer2016}.} Here, $L_{AB}$ is written as an $SO(2,2)$ Lorentz operator in the embedding space formalism \cite{Dolan2004}. Quasi-primary operators $\op_k(x)$ are eigenoperators of this Casimir obeying
\begin{equation}\label{eqcas}
[\mathcal{C}_2,\op_k(x)]=-\frac{1}{2}\mathcal{L}^{AB}\mathcal{L}_{AB}\op_k(x)=C_{k} \op_k(x),
\end{equation}
where for a quasi-primary with scaling dimension $\De_k$ and spin $l_k$ the eigenvalue is
\begin{equation}\label{eqcaseival}
C_k=\De_k(\De_k - d)-l_k(l_k+d-2).
\end{equation}
The same eigenvalue applies to the conformal Casimir in higher dimensional CFTs although we only consider $d=2$ here. Since descendants of $\op_k(x)$ are obtained through the action of conformal generators which commute with $\mathcal{C}_2$, descendants obey the same Casimir eigenvalue equation. Thus, Casimir eigenvalues classify irreducible representations of the global conformal group.

The holographic interpretation of the Casimir equation \eqref{eqcas} depends on the representation used for the conformal generators. As an example, consider a scalar quasi-primary operator $\op_k$ with dimension $\De_k$, dual to a massive bulk scalar field $\varphi$. In terms of right and left moving planar CFT coordinates $\xi=x+t$, $\bar \xi=x-t$, the appropriate differential representation of the global conformal generators is
\begin{equation}
\mathcal{L}_{-1}=\pa_\xi,\quad \mathcal{L}_0=-\xi\pa_\xi-\frac{1}{2}\De_k,\quad \mathcal{L}_1=\xi^2\pa_\xi+\xi\De_k,
\end{equation}
and similarly for barred generators with $\xi\to\bar\xi$. An explicit calculation of eq. \eqref{eqcas} using eq. \eqref{eqc2} verifies that $[\mathcal{C}_2,\op_k(x)]=\De_k(\De_k - 2)\op_k(x)$. 

Holographically, the global conformal generators correspond with AdS$_3$ isometries. Scale/radius duality prescribes that the scaling dimension $\De_k$ be replaced by the radial scale operator $z\pa_z$. Then the conformal generators become
\begin{equation}
\eta_{-1}=\pa_\xi,\quad \eta_0=-\xi\pa_\xi-\frac{1}{2}z\pa_z,\quad \eta_1=\xi^2\pa_\xi+\xi z\pa_z,
\end{equation}
with a barred sector given by $\xi\to\bar\xi$. These operators still satisfy the algebra eq. \eqref{Witt} under the Lie bracket. However, this algebra now admits a non-trivial extension 
\begin{equation}
\eta_1\to\xi^2\pa_\xi+\xi z\pa_z-z^2\pa_{\bar\xi},\quad \bar\eta_1\to\bar\xi^2\pa_{\bar\xi}+\bar\xi z\pa_z-z^2\pa_{\xi},
\end{equation}
which leaves the Lie brackets between all elements unchanged, and which vanishes in the boundary limit $z\to0$. Using the extended algebra, and replacing $\op_k$ by its dual field, the Casimir equation \eqref{eqcas} becomes
\begin{equation}\label{eqcasadslaplacian}
\left(z\pa_z-z^2\pa_z^2-4z^2\pa_\xi \pa_{\bar\xi}\right)\varphi=-\square_{\mathrm{AdS}}\varphi=-m^2\varphi,
\end{equation}
which is the Klein-Gordon equation for a massive scalar field in Poincar\'e AdS$_3$, with $m^2=-\De_k(\De_k-2)$ \cite{Maldacena1998}. In the $\De_k$ scalar representation, the global conformal Casimir can be identified as the AdS$_3$ Laplacian, $\mathcal{C}_2=-\square_{\mathrm{AdS}}$. 

In a similar manner, the Laplacian for the kinematic space of the CFT$_2$ vacuum state can be derived from the Casimir in an appropriate representation. The authors of \cite{Czech2016} identified this representation from the transformation properties of OPE blocks, the  natural candidates for fields on kinematic space. Under a conformal transformation, a spin-zero local operator with scaling dimension $\De_i$ transforms as
\begin{equation}\label{eq:CFTtransformation}
\op_i\left(x\right)\to \Om\left(x'\right)^{\De_i} \op_i\left(x'\right),\quad \Om\left(x'\right)=\det\left(\frac{\pa x'^\mu}{\pa x^\nu}\right),
\end{equation}
while
\begin{equation}
\left|x_1-x_2\right|\to\left(\Om\left(x_1'\right)\Om\left(x_2'\right)\right)^{-1/2}\left|x_1'-x_2'\right|.
\end{equation}
From eq. \eqref{eqOPEblock}, these transformation laws imply that OPE blocks obey
\begin{equation}
\mathcal{B}_{k}^{ij}\left(x_1,x_2\right)\to\left(\frac{\Om(x_1')}{\Om(x_2')} \right)^{(\De_i-\De_j)/2}\mathcal{B}_{k}^{ij}\left(x_1',x_2'\right).
\end{equation}
Restricting to the case of $\De_i=\De_j$ shows that the equal-weight OPE block transforms in a spinless, $\De=0$ representation in each of its coordinates. This is the same transformation law as a pair of dimensionless scalar operators $\varphi_1(x_1)\varphi_2(x_2)$. The action of the conformal generators on this pair is, from eq. \eqref{eqdiffrep},
\begin{align}
\begin{aligned}
{[L_n,\varphi_1(x_1)\varphi_2(x_2)]}&=[L_n,\varphi_1(x_1)]\varphi_2(x_2)+\varphi_1(x_1)[L_n,\phi_2(x_2)]\\
&=(\mathcal{L}_{n,1}+\mathcal{L}_{n,2})\varphi_1(x_1)\varphi_2(x_2),
\end{aligned}
\end{align}
where $\mathcal{L}_{n,k}$ is the $\De=0,l=0$ differential representation of $L_n$ acting only on the $x_k$ coordinates. The OPE block is a linear combination of a single quasi-primary and its descendants, so it satisfies a Casimir eigenvalue equation with the same eigenvalue \eqref{eqcaseival} as the quasi-primary,
\begin{equation}\label{eqbilocalcas}
[\mathcal{C}_2,\mathcal{B}_{k}\left(x_1,x_2\right)]=-\frac{1}{2}(\mathcal{L}^{AB}_1+\mathcal{L}^{AB}_2)(\mathcal{L}_{AB,1}+\mathcal{L}_{AB,2})\mathcal{B}_{k}\left(x_1,x_2\right)=C_k\mathcal{B}_{k}\left(x_1,x_2\right).
\end{equation}
Employing an explicit representation for the conformal generators will produce a differential equation for the OPE blocks which can be interpreted as a Klein-Gordon equation on kinematic space.

From the global AdS$_3$ Killing vectors
\begin{equation}\label{eqKV}
\small
\begin{split}
\xi_{-1}&=\frac{1}{2}e^{-i(t+\phi)}(\tanh(\rho)\pa_t +i\pa_\rho+\coth(\rho)\pa_\phi),\\
\xi_0&=\frac{1}{2}(\pa_t+\pa_\phi),\\
 \xi_1&=\frac{1}{2}e^{i(t+\phi)}(\tanh(\rho)\pa_t -i\pa_\rho+\coth(\rho)\pa_\phi),\\
\end{split}
\quad \quad 
\begin{split}
\bar\xi_{-1}&=\frac{1}{2}e^{-i(t-\phi)}(\tanh(\rho)\pa_t +i\pa_\rho-\coth(\rho)\pa_\phi),\\
\bar\xi_0&=\frac{1}{2}(\pa_t-\pa_\phi),\\
 \bar \xi_1&=\frac{1}{2}e^{i(t-\phi)}(\tanh(\rho)\pa_t -i\pa_\rho-\coth(\rho)\pa_\phi),
\end{split}
 \normalsize
\end{equation}
we can obtain a differential representation of the conformal generators on the cylinder by taking the $\rho\to\infty$ boundary limit \cite{Brown1986}, 
\begin{equation}\label{eqcylgen}
\begin{split}
    \mathcal{L}_{-1}&=\frac{1}{2}e^{-i(t+\phi)}(\pa_t +\pa_\phi),\\
\mathcal{L}_0&=\frac{1}{2}(\pa_t+\pa_\phi),\\
 \mathcal{L}_1&=\frac{1}{2}e^{i(t+\phi)}(\pa_t +\pa_\phi),\\
  \end{split}
  \quad \quad 
  \begin{split}
    \bar{\mathcal{L}}_{-1}&=\frac{1}{2}e^{-i(t-\phi)}(\pa_t -\pa_\phi),\\
\bar{\mathcal{L}}_0&=\frac{1}{2}(\pa_t-\pa_\phi),\\
 \bar{\mathcal{L}}_1&=\frac{1}{2}e^{i(t-\phi)}(\pa_t -\pa_\phi).
  \end{split}
\end{equation}
Using this representation to calculate the Casimir in its bilocal scalar representation \eqref{eqbilocalcas} requires computing
\begin{equation}\label{eqbilocalcas2}
-\frac{1}{2}\mathcal{L}_{AB,1}\mathcal{L}^{AB}_1-\frac{1}{2}\mathcal{L}_{AB,2}\mathcal{L}^{AB}_2+\mathcal{L}_{AB,1}\mathcal{L}^{AB}_2,
\end{equation}
as in eq. \eqref{eqc2}. This task is simplified since the two terms which act on only a single coordinate do not contribute. This can be verified directly from the representation \eqref{eqcylgen}, or by noting that $L_{AB,i}L^{AB}_i$ acting on $\mathcal{B}_{k}\left(x_1,x_2\right)$ produces the eigenvalue \eqref{eqcaseival}, which vanishes for the $\De=0,\ l=0$ representation appropriate for the equal-weight OPE blocks in $d=2$. 

The term with mixed derivatives does not vanish. It is
\begin{equation}
\mathcal{L}_{AB,1}\mathcal{L}^{AB}_2=-4\left(\bar{\mathcal{L}}_{0,1}\bar{\mathcal{L}}_{0,2}+\mathcal{L}_{0,1} \mathcal{L}_{0,2}\right)+2\left[\bar{\mathcal{L}}_{-1,1}\bar {\mathcal{L}}_{1,2}+\mathcal{L}_{1,1}\mathcal{L}_{-1,2}+\bar{\mathcal{L}}_{1,1}\bar{\mathcal{L}}_{-1,2}+\mathcal{L}_{-1,1}\mathcal{L}_{1,2}\right],
\end{equation}
where the second index indicates which point in the pair $(x_1,x_2)$ the operator acts on. Using eq. \eqref{eqcylgen} leads to 
\begin{align}
\begin{aligned}
\mathcal{L}_{AB,1}\mathcal{L}^{AB}_2=-2\left(\pa_{t_1}\pa_{t_2}+\pa_{\phi_1}\pa_{\phi_2}\right)&+\cos\left(t_1-t_2+\phi_1-\phi_2\right)\left(\pa_{t_1} +\pa_{\phi_1}\right)\left(\pa_{t_2} +\pa_{\phi_2}\right)\\
&+\cos\left(t_1-t_2-\left(\phi_1-\phi_2\right)\right)\left(\pa_{t_1} -\pa_{\phi_1}\right)\left(\pa_{t_2} -\pa_{\phi_2}\right).
\end{aligned}\end{align}
This operator simplifies greatly if we introduce coordinates analogous to the kinematic coordinates used in eq. \eqref{eq4kinecoords},\footnote{There is no longer a $\tan$ because this transformation is between sets of coordinates on the cylinder.}
\begin{equation}\label{eqcombinedtrans}
\begin{split}
    \al&=\frac{1}{2}\left(t_1-t_2+(\phi_1-\phi_2)\right),\\
\bar \al&=\frac{1}{2}\left(t_1-t_2-(\phi_1-\phi_2)\right),\\
  \end{split}
  \quad \quad 
  \begin{split}
    \te&=\frac{1}{2}\left(t_1+t_2+\phi_1+\phi_2\right),\\
\bar \te&=\frac{1}{2}\left(t_1+t_2-(\phi_1+\phi_2)\right),\\
  \end{split}
\end{equation}
which leads to
\begin{align}
\begin{aligned}
\mathcal{L}_{AB,1}\mathcal{L}^{AB}_2=-2\sin^2 \al\left(-\pa_\al^2+\pa_\te^2\right)-2\sin^2\bar \al\left(-\pa_{\bar \al}^2+\pa_{\bar \te}^2\right).
\end{aligned}
\end{align}
The Casimir equation for the OPE block is then
\begin{align}\label{eqopecas}
\begin{aligned}
\small
{\left[\mathcal{C}_2,\mathcal{B}_{k}\left(x_1,x_2\right)\right]}&=\left[-2\sin^2 \al \left(-{\pa_\al^2}{+}\pa_\te^2\right)-2\sin^2\bar \al\left(-{\pa_{\bar \al}^2}{+}\pa_{\bar \te}^2\right)\right]\mathcal{B}_{k}\left(x_1,x_2\right)=\De_k({\De_k}{-}2)\mathcal{B}_{k}.
\normalsize
\end{aligned}
\end{align}
It is easy to check that this operator is the scalar Laplacian in the dS$_2\times$dS$_2$ metric \eqref{eqds2ds2} found from conformal symmetry arguments. This motivates the interpretation of an OPE block as a negative mass scalar field propagating freely on kinematic space \cite{Czech2016},
\begin{equation}\label{eqOPEblockeom}
\left(\square_{dS}+\bar{\square}_{dS}\right)\mathcal{B}_{k}\left(x_1,x_2\right)=m^2\mathcal{B}_{k},
\end{equation}
with the mass term $m^2=-\De_k(\De_k-2)$ given by the Casimir eigenvalue \eqref{eqcaseival} for the quasi-primary of the block. Again, kinematic space is meant to be a useful auxiliary space, not a physical one, so the appearance of negative mass fields is not a concern.

In the following section the equal-time OPE will be considered for CFTs dual to conical defects. Setting $t_1=t_2=0$ in eq. \eqref{eqcombinedtrans} and eliminating two redundant coordinates in eq. \eqref{eqopecas} leads to the Laplacian for a single dS$_2$ spacetime,
\begin{align}\label{eqequaltimelaplacian}
\begin{aligned}
{[\mathcal{C}_2,\mathcal{B}_{k}\left(t=0,\al,\te\right)]}&=-4\sin^2 \al\left(-\frac{\pa^2}{\pa\al^2}+\frac{\pa}{\pa\te^2}\right)\mathcal{B}_{k}\left(t=0,\al,\te\right).
\end{aligned}
\end{align}

\subsection{CFT dual to conical defects}\label{sec:CFTCD}

Conical defect spacetimes can be created by adding a particle to pure AdS and are dual to certain excited states of the boundary theory \cite{Balasubramanian1999,Lunin2003}. The dual CFT is discretely gauged and lives on a cylinder with an angular identification inherited from the bulk. For the conical defects with integer $N$ it is often useful to consider a covering CFT living on the boundary of pure AdS$_3$ that ungauges the discrete $\mathbb{Z}_N$ symmetry \cite{Balasubramanian2015}.\footnote{The covering CFT only inherits a Virasoro symmetry group when $N$ is an integer \cite{Boer2011}.} Physical, gauge invariant quantities in the base CFT can be computed from appropriately symmetrized quantities on the cover. This method of images on the cover is a common way to calculate correlation functions of operators in the base CFT \cite{Balasubramanian1999,Balasubramanian2003,Arefeva2016,Arefeva2016a,Arefeva2017}. It is important to note that the covering CFT is not identical to the base CFT, as there are many non-symmetrized quantities on the cover that do not correspond to physical, gauge invariant quantities on the base. In addition, the two theories do not share the same central charge. In line with section \ref{secbulk}, quantities on the base where $\tilde\phi\in[0,2\pi/N]$ will be marked with a tilde to distinguish them from quantities on the cover where $\phi\in[0,2\pi]$.

Restricting to integer $N$, a base operator $\tilde\op(t,\tilde\phi)$ of dimension $\De$ can be represented on the cover by a symmetrized operator
\begin{equation}\label{eqsym}
\tilde\op\left(t,\tilde\phi\right)=\frac{1}{N}\sum_{m=0}^{N-1} \exp{\left(i\frac{2\pi m}{N}\frac{\pa}{\pa \phi}\right)}\op\left(t,\phi\right),
\end{equation}
where $\op(t,\phi)$ is an operator on the cover of the same dimension $\De$, with $ \phi\in[0,2\pi]$, and the first copy $(m=0)$ is inserted at $\phi=\tilde\phi$ by convention.\footnote{The equality of scaling dimensions here is a consequence of unitarity in a 1+1d CFT and may not be guaranteed in higher dimensions.} This convention is somewhat arbitrary. It reflects the freedom to choose a fundamental domain on the kinematic space, as will become clear. The timelike coordinates of the two theories are simply identified, and we work on a fixed time slice in both cases. The generators of rotation
\begin{equation}
\exp{\left(i\frac{2\pi m}{N}\frac{\pa}{\pa\phi}\right)},
\end{equation}
are conformal generators that have the effect of permuting through copies of $\op(t,\phi)$ equally spaced around the circle. An equivalent expression to eq. \eqref{eqsym} is
\begin{equation}
\tilde\op\left(t,\tilde\phi\right)=\frac{1}{N}\sum_{m=0}^{N-1} \op\left(t,\phi+\frac{2\pi m}{N}\right).
\end{equation}

\subsubsection{Partial OPE block decomposition}

The main goal of this section will be to obtain a symmetrized expression for the equal-time base OPE in terms of cover OPE blocks. That expression can then be used to determine the appropriate Casimir eigenvalue equation for the blocks, and in turn the kinematic space geometry can be inferred. The base OPE of equal-time operators inserted at locations $(t,\tilde\phi_1)$ and $(t,\tilde\phi_2)$ with $\tilde\phi_1>\tilde\phi_2$ is of the form

 \begin{equation}\label{equsualOPE}
\tilde\op_{i}\left(t,\tilde\phi_1\right)\tilde\op_{j}\left(t,\tilde\phi_2\right) = \left[2-2\cos(\tilde \phi_1-\tilde\phi_2)\right]^{-\Delta}\sum_{k}\tilde C_{ijk}\tilde{\mathcal{B}}_{k}\left(t,\tilde\phi_1,\tilde\phi_2\right),
\end{equation}
where $\tilde{\mathcal{B}}_{k}\left(t,\tilde\phi_1,\tilde\phi_2\right)$ are the equal-time base OPE blocks.\footnote{Here, operators on the cylinder have been rescaled relative to the planar operators used in section \ref{secOPEblocks}, see \cite{Pappadopulo2012} for example. In the OPE limit $\tilde\phi_2\to\tilde\phi_1$ where the curvature of the cylinder becomes unimportant, one recovers the form of eq. \eqref{eqOPEblock} for a planar CFT.}   Again, $\De_i=\De_j=\De$ so the indices $i,j$ on the OPE blocks are dropped for brevity. The base OPE can be rewritten using eq. \eqref{eqsym}, after which the OPE between cover operators can be broken into OPE blocks to get
\begin{align}
\begin{aligned}\label{eqdoublesumexpansion}
\tilde{\mathcal{O}}_i(t,  \tilde\phi_1) \tilde{\mathcal{O}}_j(t,  \tilde\phi_2)  =& \frac{1}{N^2}\sum_{a=0}^{N-1} \sum_{b=0}^{N-1}  \exp{\left(i\frac{2\pi a}{N}\frac{\pa}{\pa \phi_1}\right)}  \exp{\left(i\frac{2\pi b}{N}\frac{\pa}{\pa  \phi_2}\right)}  {\mathcal{O}}_i(t, \phi_1) {\mathcal{O}}_j(t, \phi_2) \\
=& \frac{1}{N^2}\sum_{a=0}^{N-1} \sum_{b=0}^{N-1}  \exp{\left(i\frac{2\pi a}{N}\frac{\pa}{\pa \phi_1}\right)}  \exp{\left(i\frac{2\pi b}{N}\frac{\pa}{\pa \phi_2}\right)} \\
&\cdot \left[ \left|2-2\cos(\phi_1-\phi_2)\right|^{-\De}\sum_{k} C_{ijk}{\mathcal{B}}_{k}\left(t, \phi_1,\phi_2\right)\right].
\end{aligned}
\end{align}
The structure constants and OPE blocks may be different on the cover compared to the base, and are differentiated by a tilde. 

 Now, in the covering space we introduce kinematic coordinates of the form (cf. \eqref{eqcombinedtrans})
\begin{equation}\label{eq:kinematiccoordinates}
\al=\frac{1}{2}(\phi_1-\phi_2),\quad \te=\frac{1}{2}(\phi_1+\phi_2).
\end{equation}
The permutation generators can be rewritten 
\begin{equation}
\exp{\left(i\frac{2\pi a}{N}\frac{\pa}{\pa  \phi_1}\right)}\exp{\left(i\frac{2\pi b}{N}\frac{\pa}{\pa  \phi_2}\right)}=\exp{\left(i\frac{2\pi (a-b)}{N}\frac{\pa}{\pa  \phi_1}\right)}\exp{\left(i\frac{2\pi b}{N}\frac{\pa}{\pa  \te}\right)}.
\end{equation}
The $N^2$ terms in the double sum \eqref{eqdoublesumexpansion} can be reorganized into a more appealing form 
\begin{align}\label{eqphiteexpansion}
\begin{aligned}
\tilde{\mathcal{O}}_i(t, \tilde\phi_1) \tilde{\mathcal{O}}_j(t, \tilde\phi_2&)  =\\ \frac{1}{N^2}\sum_{k} C_{ijk} &\sum_{m=0}^{N-1} \exp{\left(i\frac{2\pi m}{N}\frac{\pa}{\pa \phi_1}\right)}  \left[ \left[2-2\cos( 2\al)\right]^{-\De}\sum_{b=0}^{N-1} \exp{\left(i\frac{2\pi b}{N}\frac{\pa}{\pa  \te}\right)}\mathcal{B}_{k}\left(t,\al,\te\right)\right].
\end{aligned}
\end{align}
The interior sum over $b$ accounts for the $N$ terms where both points $\phi_1$ and $\phi_2$ are shifted by the same amount, that is $a=b$. In this case $\al$ is fixed; the $\pa/\pa{\te}$ generator permutes between images of the pair of operators on the cover. From the bulk viewpoint, $\pa/\pa{\te}$ permutes through the $N$ images of a geodesic that are identified under $\mathbb{Z}_N$. In the exterior sum, the $\pa/\pa{\phi_1}$ generators increase the angular distance between the insertion points. In bulk terms, $\pa/\pa\phi_1$ changes the winding number of geodesics connecting the boundary points. 

\begin{figure}
      \centering
        \includegraphics[width=.3\textwidth]{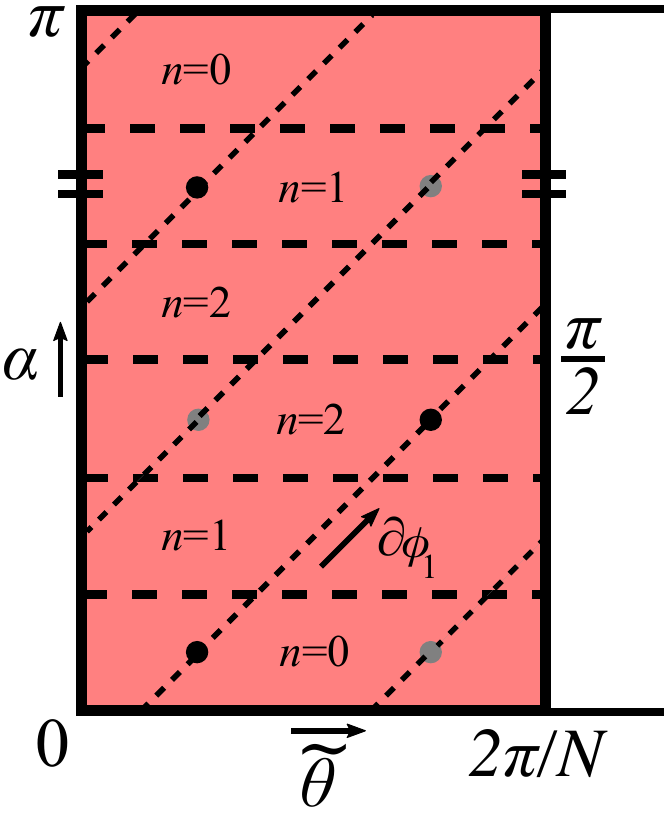}
        \caption[Orbits of null coordinate on kinematic space]{On kinematic space, $\phi_1$ is a null coordinate. In terms of cover operators, acting with $\exp{(i\frac{2\pi}{N}\frac{\pa}{\pa \phi_1})}$ increases the angular separation $\al$. In terms of conical defect geodesics, acting once with $\exp{(i\frac{2\pi}{N}\frac{\pa}{\pa \phi_1})}$ increases the winding number while leaving the endpoints fixed. All winding numbers are reached by acting with $\exp{(i\frac{2\pi}{N}\frac{\pa}{\pa \phi_1})}$ generators, in contrast to $\exp{(i\frac{2\pi}{N}\frac{\pa}{\pa \al})}$ generators. }
        \label{fig:KSCDdiagonalpath}
\end{figure}

It may seem more natural to use $\pa/\pa{ \al}$ generators along with the $\pa/\pa \te$ generators. However, when $N$ is an even integer, acting with $\pa/\pa\al$ alone does not reach images on the cover of all separations $\al$. In bulk terms, not all winding numbers for geodesics with a given orientation can be reached with $\pa/\pa\al$ generators alone. In order to reach all images for all integer $N$, a combination of $\pa/\pa\te$ and one of $\pa/\pa\phi_1$ or $\pa/\pa\phi_2$ is needed, as illustrated in Figure \ref{fig:KSCDdiagonalpath}.

\begin{figure}
    \centering
    \begin{subfigure}[b]{0.3\textwidth}
        \includegraphics[width=\textwidth]{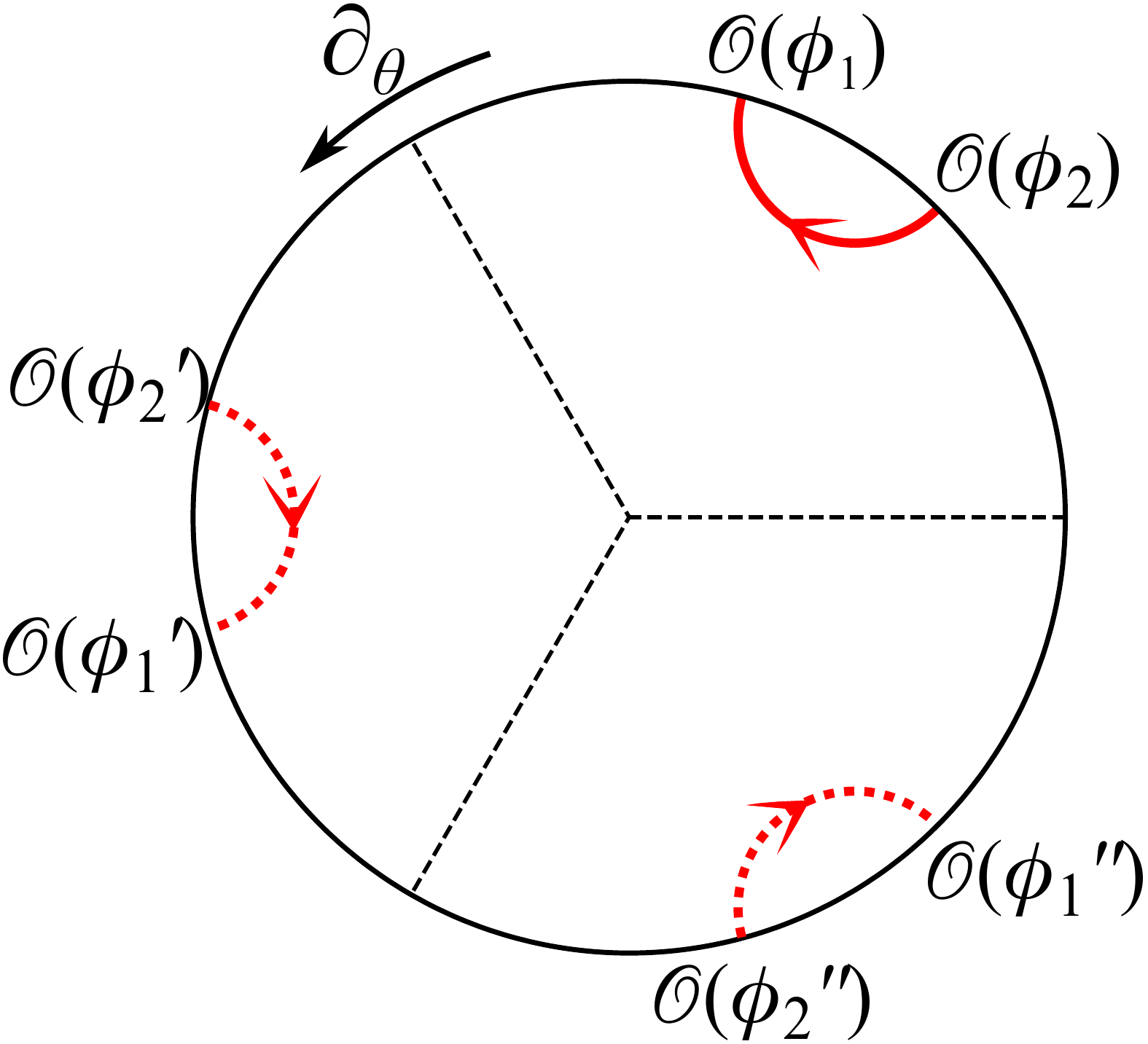}
        \caption{$\mathcal{B}_{k,m=0}$, $n=0$ geodesics.}
        \label{fig:3RedGDs}
    \end{subfigure}
    \quad
    \begin{subfigure}[b]{0.3\textwidth}
        \includegraphics[width=\textwidth]{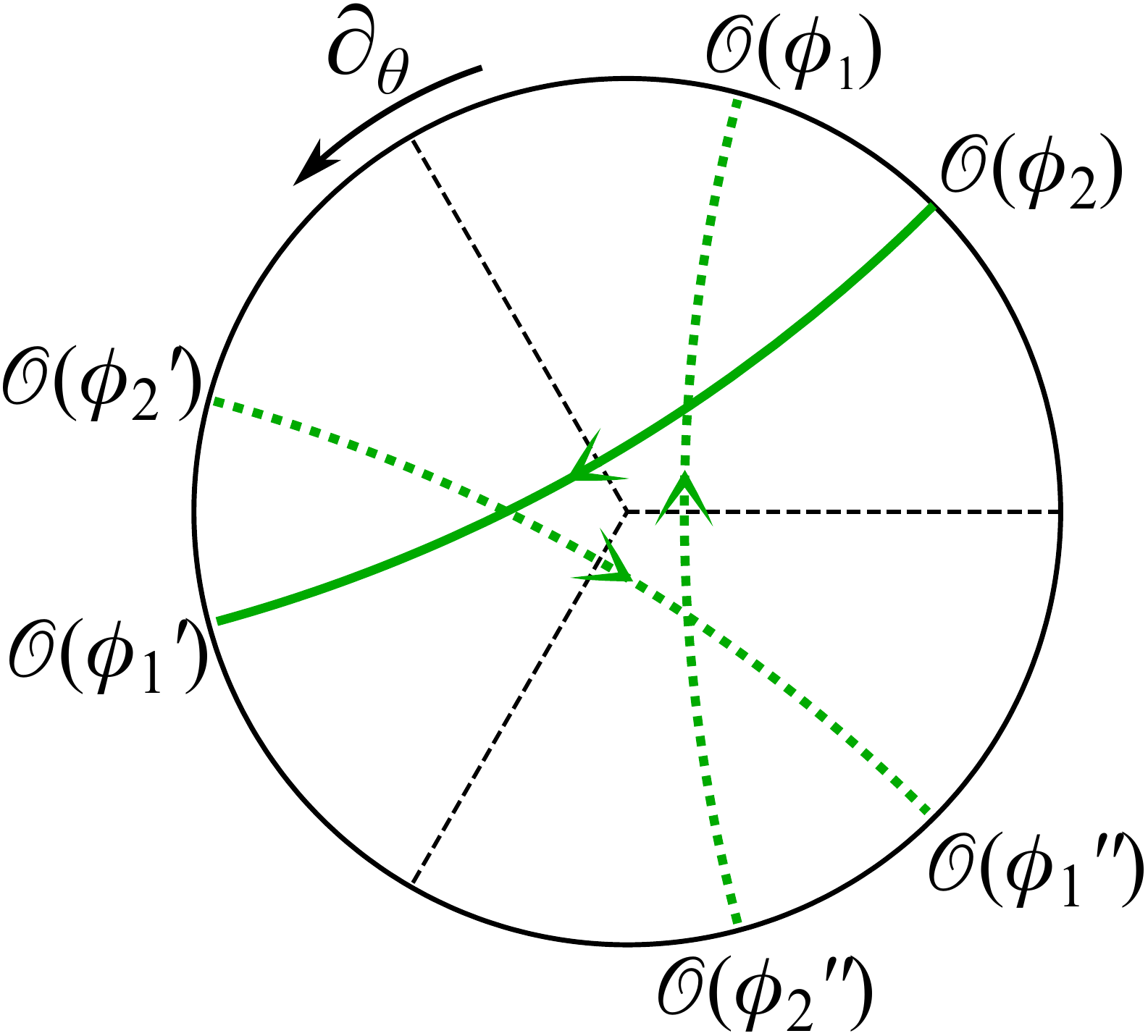}
        \caption{$\mathcal{B}_{k,m=1},~n=2$~geodesics}
        \label{fig:3RGreenGDs}
    \end{subfigure}
     \quad
    \begin{subfigure}[b]{0.3\textwidth}
        \includegraphics[width=\textwidth]{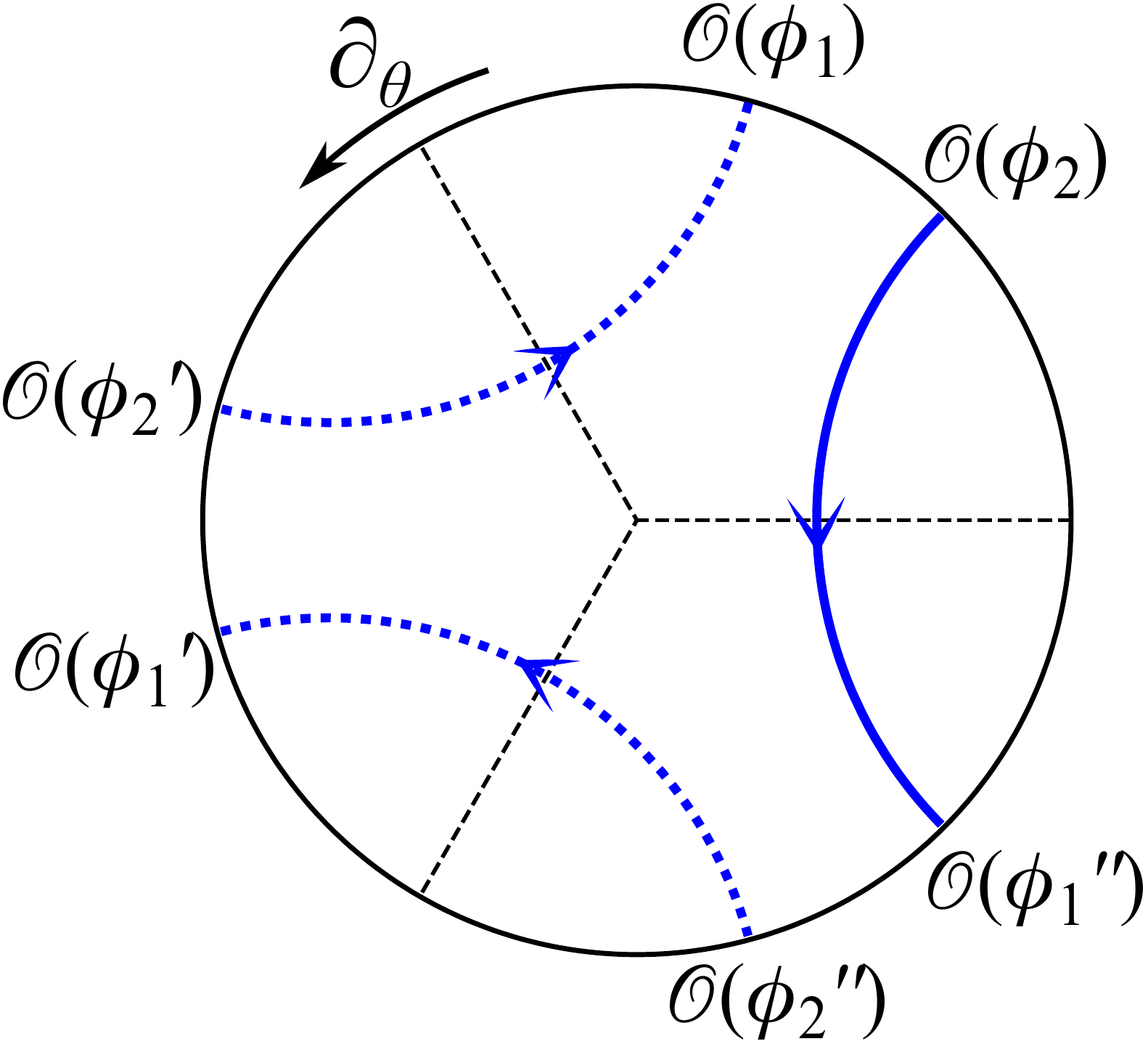}
        \caption{$\mathcal{B}_{k,m=2}$, $n=1$ geodesics}
        \label{fig:3BlueGDs}
    \end{subfigure}
          \caption[Cover OPE image points with geodesic interpretation]{(a)-(c) The contributions to the base OPE from symmetrized pairs of operators at fixed angular separation in the covering CFT are encapsulated in the $\mathcal{B}_{k,m}$ blocks. The corresponding oriented bulk geodesics are displayed to show the pairings. Note that only two operators are inserted on the boundary at a time, but all image locations are displayed here for comparison.}\label{fig:3GDs}
\end{figure}

The form of eq. \eqref{eqphiteexpansion} suggests the definition of a more fine-grained OPE block which is symmetrized on the cover,
\begin{equation}\label{eqbkn}
{\mathcal{B}}_{k,m}\left(t,\al,\te\right)=\frac{1}{N}\left[2-2\cos(2\al)\right]^{-\De}\sum_{b=0}^{N-1} \exp{\left(i\frac{2\pi b}{N}\frac{\pa}{\pa  \te}\right)}{\mathcal{B}}_{k}\left(t, \al,\te\right),
\end{equation}
where $\al$ takes on a fixed value $\al_m$ within each term of this block. We emphasize that since this ``partial" OPE block is $\mathbb{Z}_N$ symmetrized it is a valid observable on the base theory. The OPE of the base theory is then put in the suggestive form (cf. \eqref{eqsym})
\begin{align}\label{eqbk}
\begin{aligned}
\tilde{\mathcal{O}}_i(t, \tilde\phi_1) \tilde{\mathcal{O}}_j(t, \tilde\phi_2)  =& \sum_{k} C_{ijk} \frac{1}{N}\sum_{m=0}^{N-1} \exp{\left(i\frac{2\pi m}{N}\frac{\pa}{\pa \phi_1}\right)}  \mathcal{B}_{k,m}(t,\al_m,\te).
\end{aligned}
\end{align}
The partial OPE blocks ${\mathcal{B}}_{k,m}\left(t, \al_m,\te\right)$ encapsulate the contribution to the base OPE from ordered pairs of cover operators at a common distance $\al_m=\al+m\pi/N$, and $\phi_1>\phi_2$ as in Figure \ref{fig:3GDs}.

The base OPE blocks $\tilde {\mathcal{B}}_{k}$ in the decomposition \eqref{equsualOPE} group the contributions to the OPE from the conformal family of the primary $\tilde \op_k$. In rearranging the sums to get \eqref{eqphiteexpansion} we lose this interpretation for the partial OPE blocks  $\mathcal{B}_{k,m}(t,\al_m,\te)$. It is not immediately clear what CFT operator contributions these blocks group together. However, we will find that the partial OPE blocks have a clear interpretation in the bulk; they organize the contributions to the base OPE from bulk geodesics of fixed winding numbers.

For each block $ {\mathcal{B}}_{k,m}(t, \al_m, \te)$, the coordinate $ \te$ is in the domain $[0,2\pi/N]$ since $\te=\tilde\te$ was set by convention. We can always choose $ \te$ in this fundamental domain, even though its full domain on the covering space is $[0,2\pi]$, because the symmetry generators in eq. \eqref{eqbkn} permute through all the images of $ \te$ symmetrically. The choice of fundamental domain for this coordinate is the same as the choice for a fundamental domain of kinematic space made in  eq. \eqref{eqdiffentKS} and section \ref{secbulkKS}. 

Importantly, in a single ${\mathcal{B}}_{k,m}(t, \al_m, \te)$ block the coordinate $\al_m$ is restricted to a domain of size $\pi/2N$. To see this, consider the $m=0$ block where the image points have the smallest separation $\al$ and are connected by a geodesic of winding number $n=0$ through the bulk. Fix $\phi_2$ and allow $\phi_1$ to take on different values. Keeping $n=0$ and $\phi_1>\phi_2$ requires $\phi_1$ to stay in the domain $(\phi_2,\phi_2+\pi/N)$. Over this domain $\al_{(m=0)}\in(0,\pi/2N)$ so the $ {\mathcal{B}}_{k,0} $ block corresponds to $\al$ in this range. Increasing $m\to1$ moves the $\phi_1$ insertion to its next image at $\phi_1+2\pi/N$, so the $ {\mathcal{B}}_{k,1} $ block has $\al_{(m=1)}\in(\pi/N,3\pi/2N)$ and corresponds to geodesics of winding number $n=2$. The relationship between $m$ and $n$ is piecewise linear, and differs for even or odd integer $N$. For odd $N$, 
\begin{equation}\label{phi1n-odd}
N \ {\rm{odd:}}\quad 
\begin{array}{| c || c | c | c | c | c | c | c | c |c | c | c |}\hline
m & 0 & 1 & 2 & \ldots & \left\lfloor N/2 \right\rfloor{-}1 &
\left\lfloor N/2 \right\rfloor & 
\left\lceil N/2 \right\rceil & \left\lceil N/2 \right\rceil{+}1 &
\ldots & N{-}2 & N{-}1  \cr\hline
n & 0 & 2 & 4 & \ldots & N{-}3 & N{-1} & N{-}2 & N{-}4 & \ldots & 3 & 1  \cr\hline
\end{array} \,,
\end{equation}
while for even $N$,
\begin{equation}\label{phi1n-even}
N\ {\rm{even:}}\quad
\begin{array}{| c || c | c | c | c | c | c | c | c | c | c |}\hline
m & 0 & 1 & 2 & \ldots & N/2{-}1 & N/2 & 
N/2+1 & \ldots & N{-}2 & N{-}1 \cr\hline
n & 0 & 2 & 4 & \ldots & N{-}2 & N{-}1 & N{-}3 & \ldots & 3 & 1 \cr\hline
\end{array} \,.
\end{equation} 
All values of the winding number $n$ are reached by the $N$ applications of the $\pa/\pa \phi_1$ generator for both odd and even $N$.
In summary, with our conventions each partial OPE block ${\mathcal{B}}_{k,m}(t,\al_m,\te)$ lives in a restricted domain $ \te\in(0,2\pi/N)$ and $\al_m\in (m\pi/N,m\pi/N+\pi/2N)$.

\subsubsection{Partial OPE block Casimir equations}

It was noted in eq. \eqref{eqbilocalcas} that an OPE block satisfies a Casimir equation with the same eigenvalue as the quasi-primary $ \op_k$ it is built from. Since the Casimir operator commutes with all elements of the global conformal group, the $ {\mathcal{B}}_{k,m}$ blocks satisfy the same Casimir equation as the $ {\mathcal{B}}_{k}$ blocks from which they are built \eqref{eqbkn}, with the same eigenvalue, 
\begin{align}
\begin{aligned}
{[\mathcal{C}_2,{\mathcal{B}}_k\left(t,\al, \te\right)]}&= C_k{ \mathcal{B}}_k,\\
\implies [\mathcal{C}_2,{\mathcal{B}}_{k,m}]&=\frac{1}{N}\left[2-2\cos(2\al)\right]^{-\De}\sum_{b=0}^{N-1} \exp{\left(i\frac{2\pi b}{N}\frac{\pa}{\pa  \te}\right)}[\mathcal{C}_2,{\mathcal{B}}_{k}]=C_k {\mathcal{B}}_{k,m}.
\end{aligned}
\end{align}
The differential representation of $\mathcal{C}_2$ must be adapted for the ${\mathcal{B}}_{k,m}$ blocks compared to the $\tilde{\mathcal{B}}_k$ blocks because the conformal generators of the base and cover theory are not the same. 

\begin{figure}
      \centering
        \includegraphics[width=.475\textwidth]{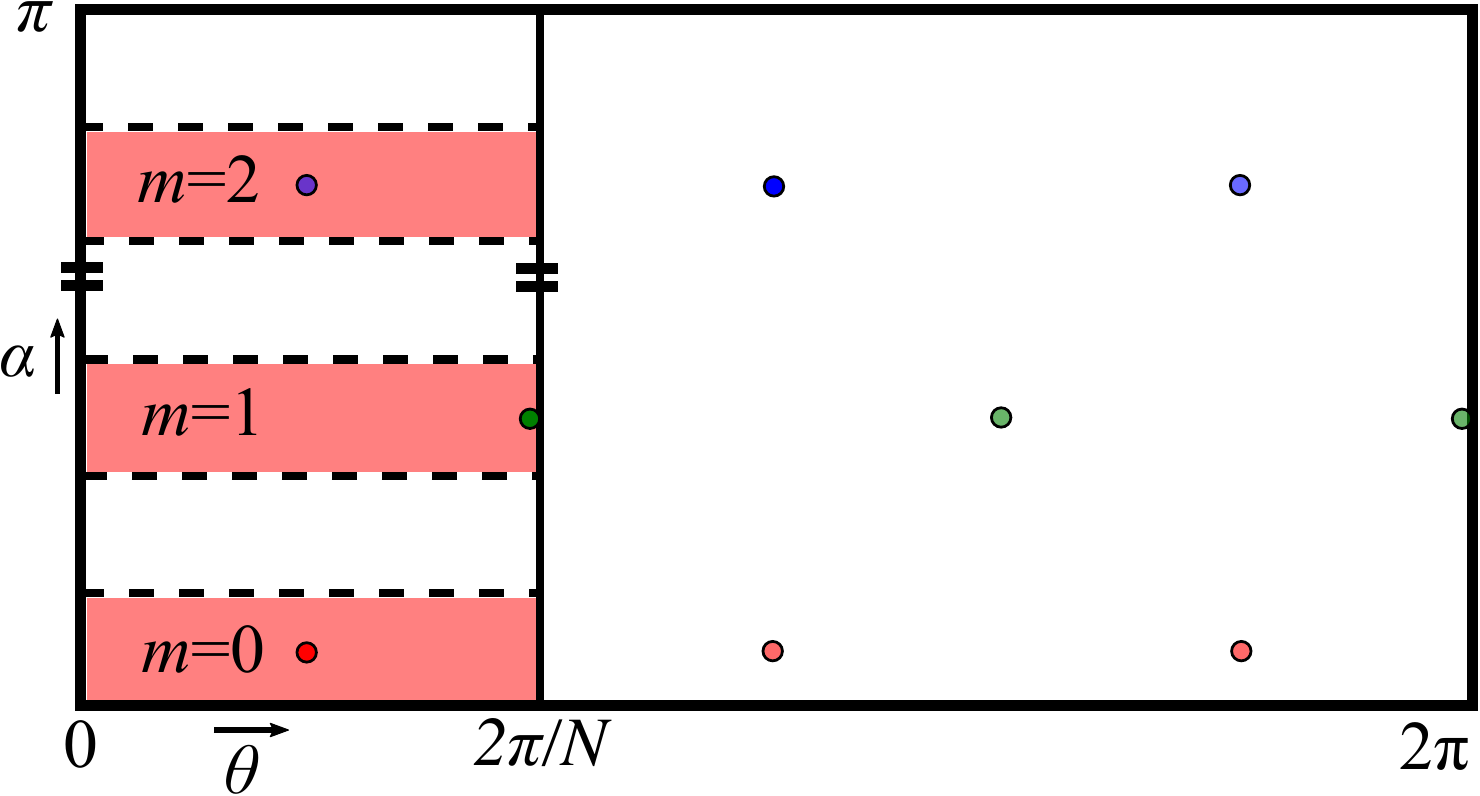}
        \caption[Conical defect kinematic space regions from individual partial OPE blocks]{Individual $\mathcal{B}_{k,m}$ blocks give rise to one of the shaded regions of kinematic space. The corresponding geodesics from figure \ref{fig:3GDs} are shown as points. The gaps are filled out by including contributions from the orientation reversed blocks with $\phi_1<\phi_2$. These correspond to the orientation reversed versions of the geodesics in figure \ref{fig:3GDs}.}
        \label{fig:KSCDverticalblocks}
\end{figure}

While the conical defect is dual to an excited state of the base CFT, the covering CFT is in its ground state \cite{Balasubramanian2015}. For this reason, the differential form of the Casimir operator acting on the ${\mathcal{B}}_{k,m}$ blocks is given by eq. \eqref{eqbilocalcas2} using a representation such as in eq. \eqref{eqcylgen}. The only difference that appears in the calculation leading to the Laplacian on kinematic space, eq. \eqref{eqequaltimelaplacian}, is the restricted coordinate domain of ${\mathcal{B}}_{k,m}(\al_m,\te)$: $ \te\in(0,2\pi/N)$ and $\al_m\in (m\pi/N,m\pi/N+\pi/2N)$. Thus the Casimir equation for the ${\mathcal{B}}_{k,m}$ blocks is
\begin{equation}\label{eqcasequationtildebkn}
{[\mathcal{C}_2,{\mathcal{B}}_{k,m}\left(t,\al_m, \te\right)]}=-4\sin^2(\al_m)\left(-\frac{\pa^2}{\pa \al_m^2}+\frac{\pa^2}{\pa\te^2}\right){\mathcal{B}}_{k,m}\left(t,\al_m,\te\right)=C_k{\mathcal{B}}_{k,m}\left(t, \al_m, \te\right),
\end{equation}
which suggests the metric for the kinematic space of the single ${\mathcal{B}}_{k,m}$ block is
\begin{equation}\label{eqksmetricn}
ds_m^2=\frac{1}{\sin^2\al_m}(-d{\al}_m^2+d\te^2).
\end{equation}
This is a subregion of dS$_2$ with the restricted coordinate range as indicated above. Each of the $N$ ${\mathcal{B}}_{k,m}$ blocks gives rise to a region of kinematic space in the same vertical strip of width $\te\in[0,2\pi/N]$ but with differing ranges of $\al$, as depicted in Figure \ref{fig:KSCDverticalblocks}. The union of these $N$ regions cover half of the vertical strip, but are not all connected because of how the winding number jumps as one insertion point is permuted through its images, recall tables \eqref{phi1n-odd} and \eqref{phi1n-even}. The indicated half of the vertical strip was obtained by taking $\phi_1>\phi_2$ for the $m=0$ block and acting with $\pa/\pa\phi_1$ generators. By starting with $\phi_1<\phi_2$ for the $m=0$ block and following the same construction with $\pa/\pa\phi_2$ in the place of $ \pa/\pa\phi_1$, one fills out the remaining regions of kinematic space. This is made more clear with a view of the bulk picture in Figure \ref{fig:3GDs} where interchanging the roles of $\phi_1$ and $\phi_2$ reverses the orientation of the connecting geodesics.

The base OPE in eq. \eqref{eqbk} receives contributions from each ${\mathcal{B}}_{k,m}$ with both $\phi_1>\phi_2$ and $\phi_1<\phi_2$. Taking the union of the regions identified from each ${\mathcal{B}}_{k,m}$ shows that the kinematic space for the excited states dual to a timeslice of AdS$_3/\mathbb{Z}_N$ can be identified as de Sitter with an identified angular coordinate $ \te=\te+2\pi/N$. In other words, the kinematic space of a static conical defect is a quotient of the kinematic space for pure AdS$_3$, as anticipated in \cite{Czech2015,Czech2016,Asplund2016}. This is the same kinematic space geometry, up to the choice of fundamental region, that was determined from the differential entropy prescription of eq. \eqref{eqdiffentKS}, and the analysis of boundary anchored geodesics under the $\mathbb{Z}_N$ quotient in section \ref{secbulkKS}. 

Just as kinematic space from the bulk point of view can be divided into regions by the winding number of geodesics as in Figure \ref{fig:KSCDVertical}, from the CFT perspective kinematic space is built up from the contributions to the OPE by images of fixed separation $\al_m$. This suggests that there should be a connection between the partial $\mathcal{B}_{k,m}$ OPE blocks and geodesics of a fixed winding number associated to $m$. In the following section and chapter we will further explore the properties of the new observables, partial OPE blocks, and will clarify the connection to bulk geodesics and their images under quotients.
\section{Discussion}\label{secdis}

In this chapter we have shown that the kinematic space for a constant time slice of a static conical defect spacetime is a quotient of the kinematic space for time slices of pure AdS$_3$. This fact was anticipated in \cite{Czech2015c,Czech2016,Asplund2016} since all locally AdS$_3$ spacetimes can be obtained as a quotient of AdS$_3$ itself, with geodesics of AdS$_3$ descending to geodesics of the quotient space. From the bulk our results were derived from the original differential entropy prescription, and by studying how the quotient acts on geodesics. The two approaches led to different subregions of the full dS$_2$ kinematic space for pure AdS$_3$, but it was argued that the subregions were equivalent fundamental domains under the identifications. 

From the CFT point of view kinematic space had previously been defined as the space of ordered pairs of points. For a CFT dual to pure AdS$_3$ there is a one-to-one correspondence between ordered pairs of points and bulk geodesics, making it consistent with the bulk definition. Then, conformal symmetry can be used to derive a unique metric on the space of pairs of points, matching the bulk results. However, the one-to-one correspondence is not a typical feature of locally AdS$_3$ spacetimes. While the possibility of including non-minimal geodesics in the description of kinematic space has been considered previously from the bulk \cite{Czech2015c,Asplund2016,Zhang2017}, there has been no clear generalization of the boundary point of view. In this chapter we showed that the metric of the kinematic space for conical defects can be inferred from the Casimir equation of partial OPE blocks. Excited states in a discretely gauged CFT dual to conical defects can be related to the ground state of a covering CFT, and gauge invariant operators in the base descend from symmetrized operators in the cover. This allows the base OPE blocks to be broken up into distinct contributions from pairs of image operators on the cover at each possible angular separation. These contributions are encapsulated in partial OPE blocks which were shown to satisfy a wave equation. The Laplacian appearing in the wave equation is that of a subregion of dS$_2$, which allows us to infer the metric of patches of kinematic space. The base OPE is a sum of partial OPE blocks, while the union of patches matches the kinematic space identified by bulk arguments. 

The method of images provides the solution to the lack of a one-to-one correspondence between pairs of points and geodesics in this case. When both the bulk and boundary are lifted to their covering spaces, non-minimal geodesics become minimal geodesics connecting distinct image points. The fact that each partial OPE block corresponds to a specific range of $\al$ on the CFT covering space is very similar to how the $\al$ coordinate on kinematic space arranges geodesics by their winding number. This suggests a holographic interpretation for the partial OPE blocks: the block ${\mathcal{B}}_{k,m}$ represents the contribution to the base OPE from a single class of bulk geodesics with fixed winding number $n$ related to $m$ by tables \eqref{phi1n-odd} or \eqref{phi1n-even}. Thus the partial OPE blocks allow for a more fine-grained understanding of the holographic contributions to the OPE. To confirm this suspicion we now consider the holographic dictionary entry relating OPE blocks and bulk fields integrated over geodesics that was established in \cite{Czech2016}, and find that partial OPE blocks are dual to bulk fields integrated over individual minimal or non-minimal geodesics. 

\subsection{Duality between OPE blocks and geodesic integrals of bulk fields}

In \cite{Czech2016} it was noted that a bulk scalar field integrated over a geodesic of AdS$_3$ satisfies the same differential equation on kinematic space as a scalar OPE block.\footnote{See also \cite{Cunha2016,Guica2016} for an independent development of the connection between geodesic operators and OPE blocks.} By verifying that the two quantities also obeyed the same initial conditions a holographic dictionary entry was established for pure AdS$_3$: \emph{OPE blocks are dual to integrals of bulk local fields along geodesics}. The derivation of this dictionary entry relies heavily on the fact that both pure AdS$_3$ and its kinematic space dS$_2\times$dS$_2$ are homogeneous spaces with the same isometry group. This allowed the authors to derive a kinematic space equation of motion for the integrated field by relating the action of the isometries on the field and on the geodesics. In contrast, the conical defect spacetimes are not homogeneous spaces. The defect traces out a worldline that is not invariant under boosts. Nevertheless, progress can be made on extending the dictionary entry to the conical defect case by working on the covering space. For continuity, we will review the essential points of the derivation of the dictionary entry in pure AdS$_3$. Full details can be found in \cite{Czech2016}.

Consider a massive scalar field $\varphi_{AdS}(x)$ on AdS$_3$ integrated over a boundary anchored geodesic $\Ga$ in a constant time slice of the geometry,
\begin{equation}\label{eqxray}
R[\varphi_{AdS}](\Ga)=\int_\Ga ds \ \varphi_{AdS}(x).
\end{equation}
This ``X-ray" transform of $\varphi_{AdS}(x)$ is naturally viewed as a field on kinematic space because it is a function of geodesics, i.e. points in kinematic space.\footnote{When the integration is performed over an extremal surface in a higher dimensional theory this is known as a Radon transform, used first in a holographic context in \cite{Lin2015}.} 

Let $g$ be an isometry of AdS$_3$. The scalar field is invariant under the isometry but its argument is shifted, $\varphi'_{AdS}(x)=\varphi_{AdS}(g^{-1}\cdot x)$. Integrating the shifted field over a geodesic $\Ga$ is equivalent to integrating the original field over a shifted geodesic $g\cdot \Ga$, noting that all isometries of AdS$_3$ map geodesics into geodesics. In terms of the X-ray transform this is expressed as
\begin{equation}\label{eqshifts}
R[\varphi'_{AdS}](\Ga)=\int_\Ga  ds\ \varphi_{AdS}(g^{-1}\cdot x) =\int_{g\cdot\Ga} ds\ \varphi_{AdS}(x) =R[\varphi_{AdS}](g\cdot \Ga).
\end{equation}
A shift in the argument of $\varphi_{AdS}(x)$ can be compensated by a shift in the argument of $R(\Ga)$.

When $g$ is an element of the isometry group near the identity, the action of $g$ on the field is described by the group generators
\begin{equation}\label{eqisomads}
\varphi'_{AdS}(x)=(1-\om^{AB}L_{AB}^{x})\varphi_{AdS}(x),
\end{equation}
where $L_{AB}^{x}$ is an isometry generator of AdS written with embedding space indices, and $\om^{AB}$ is the antisymmetric matrix parameterizing the isometry. In a similar way, the action of $g$ on the X-ray transform is
\begin{equation}\label{eqisomks}
R[\varphi_{AdS}](g\cdot  \Ga)=(1+\om^{AB}L_{AB}^{\Ga})R[\varphi_{AdS}](\Ga),
\end{equation}
where $L_{AB}^{\Ga}$ is an isometry generator on the kinematic space of geodesics. Applying eqs. \eqref{eqisomads} and \eqref{eqisomks} to eq. \eqref{eqshifts} produces the remarkable intertwining relation of isometry generators
\begin{equation}\label{eqintertwining}
L_{AB}^\Ga R[\varphi_{AdS}](\Ga)=-R[L_{AB}^x \varphi_{AdS}](\Ga).
\end{equation}
Applying the same relation twice produces quadratic Casimirs \eqref{eqc2}, in their respective representations of the isometry group;
\begin{equation}\label{eqintertwinedcasimirs}
\mathcal{C}^{\Ga}_2 R[ \varphi_{AdS}](\Ga)=R[\mathcal{C}^{x}_2\varphi_{AdS}(x)](\Ga).
\end{equation}

The subsequent step of the derivation relies crucially on the properties of homogeneous spaces, as noted in \cite{Czech2016}. For homogeneous spaces the Casimir operator of the isometry group is identified with the scalar Laplacian.\footnote{A homogeneous space can be written as the coset space of its isometry group quotiented by the stabilizer subgroup of a point. The Casimir of the isometry group is the scalar Laplacian for the group's Cartan-Killing metric. The same Laplacian is inherited by the coset space when the Casimir acts on functions that are constant on orbits of the stabilizer group. Note that a point in kinematic space is an AdS geodesic, so the stabiliser subgroup of a geodesic in AdS should be used in the quotient.} This was demonstrated for AdS$_3$ in eq. \eqref{eqcasadslaplacian} and for dS$_2\times$dS$_2$ in eq. \eqref{eqopecas}. On the right side of eq. \eqref{eqintertwinedcasimirs} the Casimir acts on a scalar AdS field so the Casimir is in the bulk scalar representation $-\square_{AdS}$. On the left side the Casimir acts on a function of geodesics so it is in the kinematic space representation $-2(\square_{dS}+\bar{\square}_{dS})$. Using the equation of motion for the bulk field and the definition \eqref{eqxray} leads to an equation of motion for the X-ray transform as a scalar field on kinematic space
\begin{equation}\label{eqxrayeom}
2(\square_{dS}+\bar{\square}_{dS})R[\varphi_{AdS}](\Ga)=R[\square_{AdS} \varphi_{AdS}](\Ga)=R[m^2\varphi_{AdS}](\Ga)=m^2R[\varphi_{AdS}](\Ga).
\end{equation}
This shows that free bulk scalars integrated over boundary anchored geodesics are free scalar fields propagating on kinematic space. This is the same equation satisfied by the OPE block \eqref{eqOPEblockeom} of a spin zero quasi-primary of dimension given by $-\De(\De-2)=m^2$. The X-ray transform and OPE block also satisfy the same initial conditions on a Cauchy slice which establishes that they are dual quantities \cite{Czech2016}. 

\paragraph{Conical defect case.} Now let us analyze the conical defect case, again restricting to the quotients AdS$_3/\mathbb{Z}_N$. The fact that conical defects are not homogeneous spaces precludes the possibility of running through the previous argument directly. However, it is possible to use the intertwining relations obtained in the pure case and only then perform the $\mathbb{Z}_N$ quotient with an appropriate prescription for the X-ray transform over conical defect fields.

Consider a massive bulk scalar field $\varphi_{CD}$ on AdS$_3/\mathbb{Z}_N$, and similarly $\varphi_{AdS}$ on pure AdS$_3$, each described by the action 
\begin{equation}
S=-\frac{1}{2}\int d^3 x\sqrt{-g}\left((\pa\varphi)^2+m^2 \varphi^2\right) .
\end{equation}
The Klein-Gordon equation in global coordinates, eq. \eqref{eqgads} for pure AdS, and eq. \eqref{eqCDmetric} for the defect, is
\begin{equation}\label{eqKGcompare}
\square\varphi=-\frac{1}{\cosh^2\rho}\pa^2_t\varphi+\frac{1}{\sinh^2\rho}\pa^2_\phi\varphi +\frac{1}{\cosh\rho\sinh{\rho}}\pa_\rho(\cosh\rho\sinh{\rho}\ \pa_\rho\varphi)=m^2\varphi.
\end{equation}
For $\varphi_{AdS}$ the angular coordinate is $\phi\in(0,2\pi)$, while for $\varphi_{CD}$, $\phi$ should be replaced by $\tilde\phi\in(0,2\pi/N)$. 

In either case the solutions are obtained through separation of variables \cite{Arefeva2016}. For example, in the AdS case solutions are $\varphi_{AdS}(t,\rho,\phi)=e^{i\om t}Y_l(\phi)R(\rho)$, with the circular harmonics
\begin{equation}
Y_{l}( \phi)=e^{i l\phi},\quad  Y_{ l}( \phi+2\pi n)= Y_{  l}( \phi), \quad l,n\in \mathbb{Z}.
\end{equation}
Similarly, $\varphi_{CD}(t,\rho,\tilde\phi)=e^{i\om t}\tilde Y_{m}(\tilde\phi)R(\rho)$. The circular harmonic here is $2\pi/N$ periodic,
\begin{equation}
\tilde Y_m(\tilde\phi)=e^{iNm\tilde\phi},\quad \tilde Y_m\left(\tilde \phi+\frac{2\pi n}{N}\right)=\tilde Y_m(\tilde\phi),\quad \quad m,n\in \mathbb{Z}.
\end{equation}
Therefore the $\varphi_{CD}$ modes are a subset of the $\varphi_{AdS}$ modes with $l=Nm$. They are $\mathbb{Z}_N$ symmetric $\varphi_{AdS}$ modes that are solutions of the conical defect Klein-Gordon equation in each $\mathbb{Z}_N$ wedge of the covering space, reflecting the quotient structure of the defect. Appropriately symmetrized modes of AdS will be denoted $\varphi_{\mathbb{Z}}(\phi)$; any $\varphi_{CD}$ can be obtained by restricting some $\varphi_{\mathbb{Z}}$ to a single $\mathbb{Z}_N$ wedge.

\begin{figure}
    \centering
    \begin{subfigure}[b]{0.3\textwidth}
        \includegraphics[width=\textwidth]{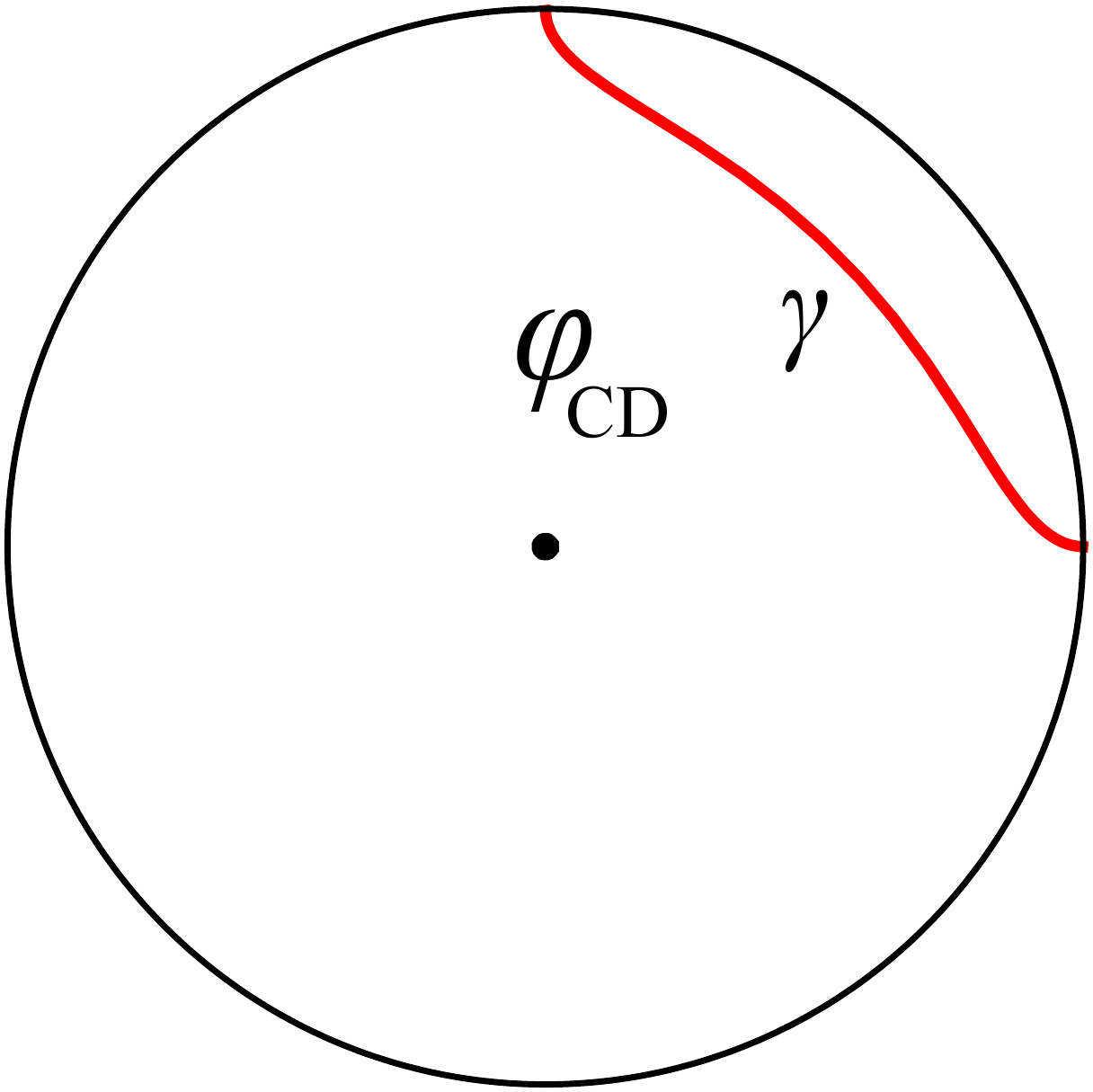}
        \caption{}
        \label{fig:CDGDIntegral}
    \end{subfigure}
     \qquad \qquad 
    \begin{subfigure}[b]{0.3\textwidth}
        \includegraphics[width=\textwidth]{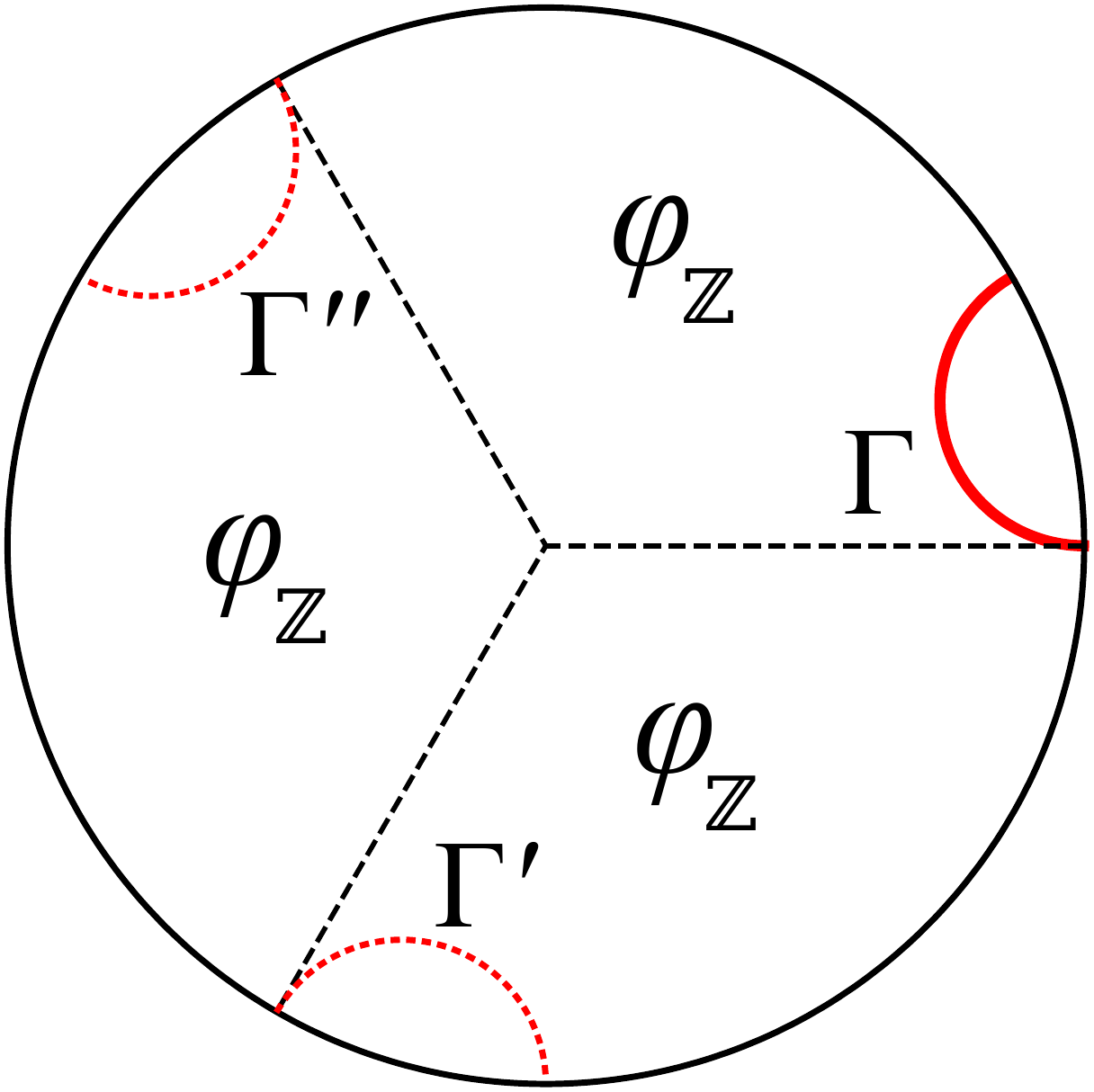}
        \caption{}
        \label{fig:SymGDIntegral}
    \end{subfigure}
    \caption[Conical defect field as symmetrized AdS field]{(a) A conical defect field integrated over a geodesic $\ga$ is the same as (b) a symmetrized AdS$_3$ field integrated over one of the preimages of $\ga$ under the $\mathbb{Z}_N$ quotient. Identifying the edges of any wedge gives the conical defect of (a).}
    \label{fig:6}
\end{figure}

The X-ray transform for the conical defect can then be defined as usual
\begin{equation}
R[\varphi_{CD}](\ga)=\int_\ga ds\ \varphi_{CD}(\tilde x).
\end{equation}
However, this transform acts in a non-homogeneous space and may not share the same invertibility properties as its counterpart eq. \eqref{eqxray}. It is preferable to lift $\varphi_{CD}$ and $\ga$ to the covering space where 
\begin{equation}\label{eqxraycd}
R[\varphi_{CD}](\ga)=\int_\ga ds\ \varphi_{CD}(\tilde x) =\int_\Ga ds \ \varphi_{\mathbb{Z}}( x)=R[\varphi_\mathbb{Z}](\Ga) .
\end{equation}
Instead of integrating $\varphi_{CD}$ over a geodesic $\ga$ in the conical defect spacetime, the corresponding symmetrized AdS field $\varphi_{\mathbb{Z}}$ can be integrated over one of the preimages $\Ga$ of $\ga$ under the $\mathbb{Z}_N$ quotient, see Figure \ref{fig:6}. This prescription works for all boundary anchored $\ga$, minimal or non-minimal, since all conical defect geodesics descend from geodesics $\Ga$ on AdS. 

 Note that in going from $\ga$ to $\Ga$, and $\varphi_{CD}$ to $\varphi_{\mathbb{Z}}$ in eq. \eqref{eqxraycd} there is the freedom to choose one of several identical wedges. The choice of wedge will lead to different coordinate values for $\varphi_{\mathbb{Z}}(\phi)$ and $\Ga(\al,\te)$. This is analogous to the ambiguities encountered throughout this chapter in choosing a fundamental region. For consistency with the previous choice of a vertical strip of kinematic space, see Figure \ref{fig:KSCDVertical}, let $\Ga$ be the preimage of $\ga$ with the smallest centre angle which will always be in the range $\te\in(0,2\pi/N)$. 

By working with the right side of eq. \eqref{eqxraycd}, the properties of homogeneous spaces can be used to find an intertwining relation for the equations of motion. Once again, let $g$ be an infinitesimal isometry of AdS$_3$. The intertwining relation eq. \eqref{eqintertwining} for homogeneous spaces applies as before,
\begin{equation}\label{eqintertwiningcd}
L_{AB}^\Ga R[\varphi_{\mathbb{Z}}](\Ga)=-R[L_{AB}^x \varphi_{\mathbb{Z}}](\Ga),
\end{equation}
and leads to the intertwined Casimirs
\begin{equation}
\mathcal{C}^{\Ga}_2 R[ \varphi_{\mathbb{Z}}](\Ga)=R[\mathcal{C}^{x}_2\varphi_{\mathbb{Z}}(x)](\Ga).
\end{equation}
On the right side the Casimir of AdS isometries becomes the AdS Laplacian which produces the mass eigenvalue. On the left side the Casimir is in the kinematic space representation $-2(\square_{dS}+\bar{\square}_{dS})$ so that
\begin{equation}
2(\square_{dS}+\bar{\square}_{dS}) R[ \varphi_{\mathbb{Z}}](\Ga)=m^2R[\varphi_{\mathbb{Z}}](\Ga).
\end{equation}
On both sides eq. \eqref{eqxraycd} can be used to find the equation of motion for geodesic integrated fields on the conical defect
\begin{equation}\label{eqxraycdeom}
2(\square_{dS/\mathbb{Z}}+\bar{\square}_{dS/\mathbb{Z}}) R[ \varphi_{CD}](\ga)=m^2R[\varphi_{CD}](\ga).
\end{equation} 
The notation $\square_{dS/\mathbb{Z}}$ is to remind that this operator now acts on the subspace of dS$_2$ obtained by restricting to $\te\in(0,2\pi/N)$ with periodic boundary conditions. This is the same behaviour that the symmetrized field exhibits under the quotient, namely $\square_{AdS}\varphi_{AdS}(x)=\square_{CD}\varphi_{CD}(\tilde x)$ within any single wedge.

One might worry that the above argument leading to eq. \eqref{eqintertwiningcd} could break down when $g$ is a boost isometry of AdS under which the conical defect is not invariant. Under the action of a boost, the field $\varphi_{\mathbb{Z}}(g^{-1}\cdot x)$ may no longer be symmetrized around the origin, but the conical defect no longer sits statically at the origin (see \cite{Arefeva2016a} for a relevant discussion). The moving defect is still locally AdS, and can be obtained directly from the covering AdS$_3$ spacetime through an identification along an AdS Killing vector. The identification is no longer a simple angular identification, but shifts time as well as angle. These identifications are given explicitly in \cite{Matschull1999,Balasubramanian1999,Arefeva2015} for example. The moving conical defect solutions can be viewed as global coordinate transformations of the static case, and do not exhibit any different physics compared to stationary ones. On an appropriately boosted timeslice through the moving conical defect spacetime, the transformed field $\varphi_{CD}'(\tilde x)$ can be obtained from $\varphi'_{\mathbb{Z}}( x)$ using the identification that produces the spacetime itself.

The equation of motion for geodesic integrated fields on the conical defect, eq. \eqref{eqxraycdeom}, after taking the equal time limit is the same as the Casimir equation for the base OPE block $\tilde{\mathcal{B}_k}$. The OPE block $\tilde{\mathcal{B}}_k$ represents the contribution to the $\tilde\op_i\tilde\op_j$ OPE  from the conformal family of the quasi-primary $\tilde\op_k$. From the bulk this contribution is obtained by integrating $\varphi$, the dual of $\tilde\op_k$, over \emph{all} geodesics connecting the boundary insertion points of $\tilde\op_i$ and $\tilde\op_j$. This is the well known geodesic approximation which has been used to compute correlation functions \cite{Balasubramanian1999,Arefeva2016a,Goto2017}, and geodesic Witten diagrams \cite{Hijano2016}. Non-minimal geodesics provide a finite number of sub-leading corrections to the minimal geodesic contribution, but can become significant in some regimes.

The connection between bulk and boundary can be made more detailed through the use of kinematic space. Consider the case where $\ga$ is a minimal geodesic. The X-ray transform $R[\varphi_{CD}](\ga_\mathrm{min})$ over a minimal $\ga( \tilde\al,\tilde \te)$, is restricted to $\tilde \al\in(0,\pi/2N)$, $\tilde\te\in(0,2\pi/N)$ with periodicity in the $\tilde \te$ coordinate. The appropriate wave equation \eqref{eqxraycdeom} on the $t=0$ timeslice is
\begin{equation}
4\sin^2 \tilde\al\left(-\frac{\pa^2}{\pa \tilde\al^2}+\frac{\pa^2}{\pa\tilde\te^2}\right)R[ \varphi_{CD}](\ga)=m^2R[\varphi_{CD}](\ga).
\end{equation}
Comparing with eq. \eqref{eqcasequationtildebkn} suggests that the ${\mathcal{B}}_{k,m}$ block with $m=0$ is dual to $R[\varphi_{CD}](\ga_\mathrm{min})$ and represents the contribution to the base OPE from a single class of geodesics, the minimal ones. The duality between ${\mathcal{B}}_{k,0}$ and $R[ \varphi_{CD}](\ga_{\mathrm{min}})$ is established by showing that these quantities satisfy the same initial conditions. The $\tilde\al=0$ Cauchy slice of kinematic space is obtained by taking the coincidence limit of the OPE block, and in the bulk by integrating over a small geodesic that stays near the boundary. These limits are unchanged from the pure AdS case and have been discussed previously \cite{Hijano2016,Czech2016, Boer2016}. In the coincidence limit only the quasi-primary $\op_k$ on the cover, and not its descendants, contributes to the partial OPE block
\begin{equation}
\lim_{\al\to 0}{\mathcal{B}}_{k,0}( \al, \te)=\lim_{\al\to 0} |2 \al|^{\De_k}\op_k(\te),
\end{equation}
while in the conical defect spacetime the behaviour of the dual scalar field near the AdS boundary is given by the extrapolate dictionary
\begin{equation}\label{eqextrapolate}
\lim_{\rho\to \infty}\varphi_{CD}(t=0,\rho,\tilde \phi)=\rho^{-\De_k}\op_k(\tilde \phi),
\end{equation}
so that integrating over a small geodesic localized at $\tilde\phi=\tilde \te$ gives
\begin{equation}
\lim_{\tilde \al\to 0}R[\varphi_{CD}](\ga_{\mathrm{min}}(\tilde\al,\tilde\te))=\lim_{\tilde\al\to 0}\frac{\Ga({\De_k}/{2})^2}{2\Ga(\De_k)}|2\tilde\al|^{\De_k}\op_k(\tilde \te).
\end{equation}
Hence, the initial conditions on kinematic space provide the relative normalization between the dual quantities,

\begin{equation}
R[\varphi_{CD}](\ga_{\mathrm{min}}(\tilde\al,\tilde\te))=\frac{\Ga({\De_k}/{2})^2}{2\Ga(\De_k)}{\mathcal{B}}_{k,0}( \al, \te).
\end{equation}

\begin{figure}
    \centering
    \begin{subfigure}[b]{0.3\textwidth}
        \includegraphics[width=\textwidth]{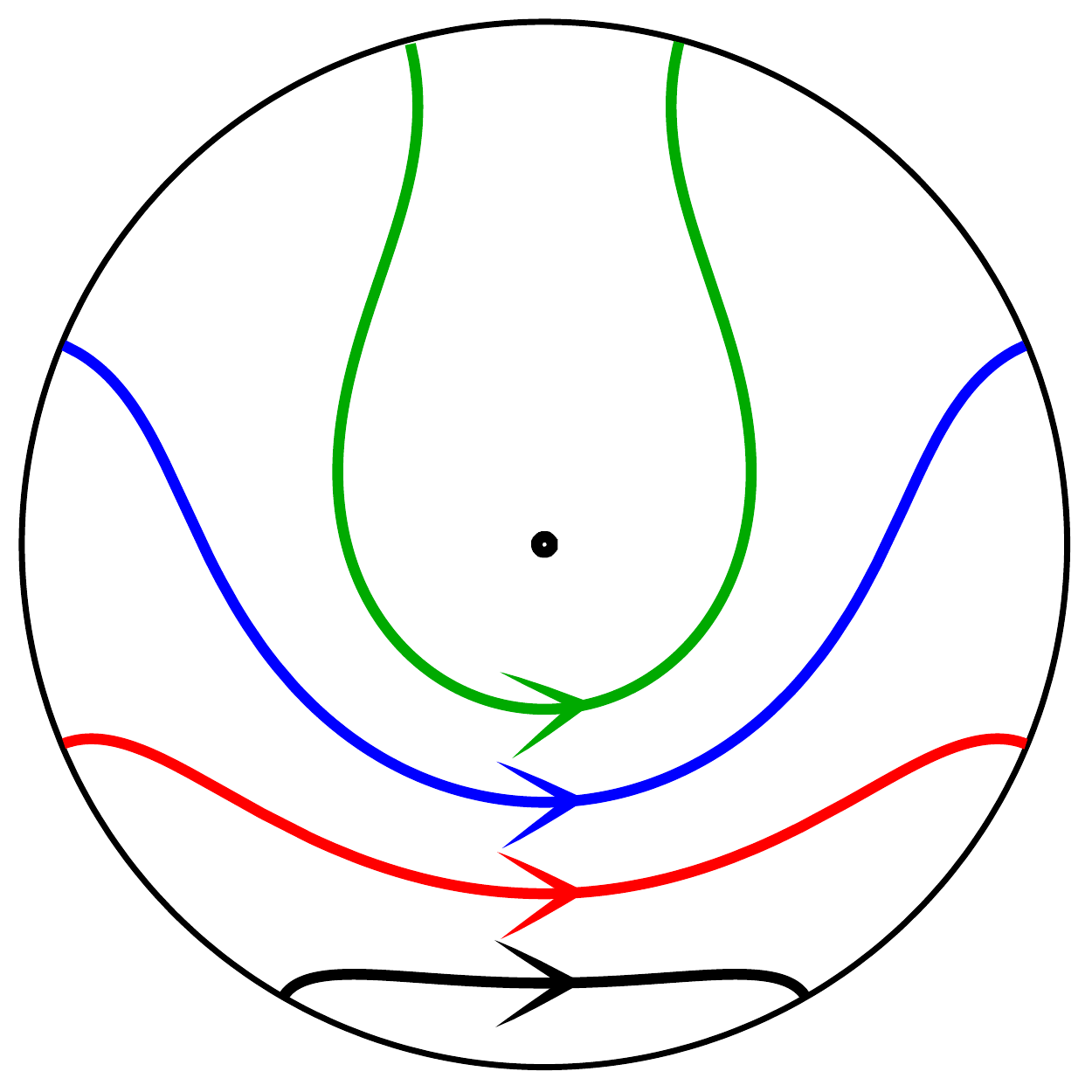}
        \caption{}
        \label{fig:GDContinuity1}
    \end{subfigure}
     \quad 
    \begin{subfigure}[b]{0.3\textwidth}
        \includegraphics[width=\textwidth]{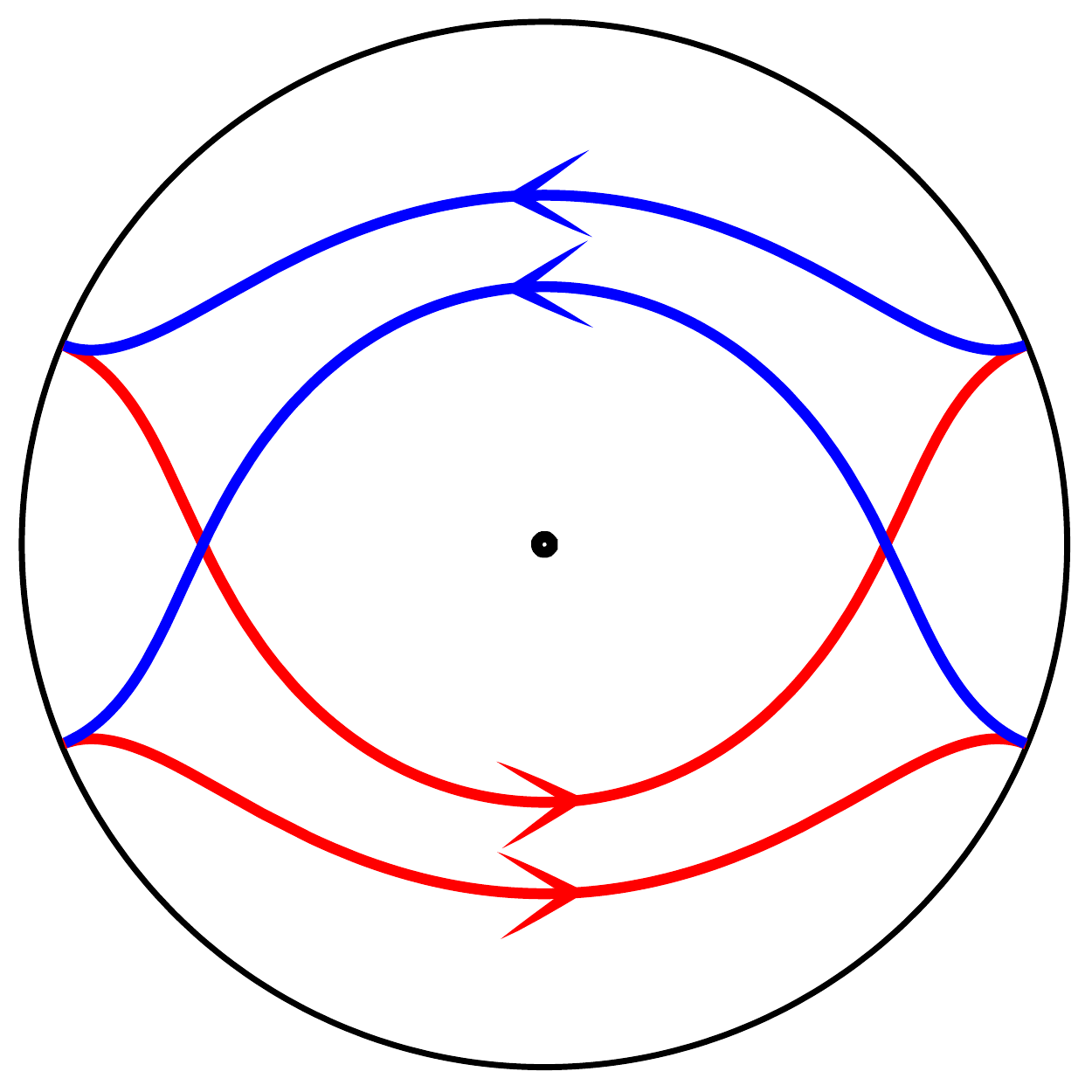}
        \caption{}
        \label{fig:GDContinuity2}
    \end{subfigure}
     \quad 
    \begin{subfigure}[b]{0.3\textwidth}
        \includegraphics[width=\textwidth]{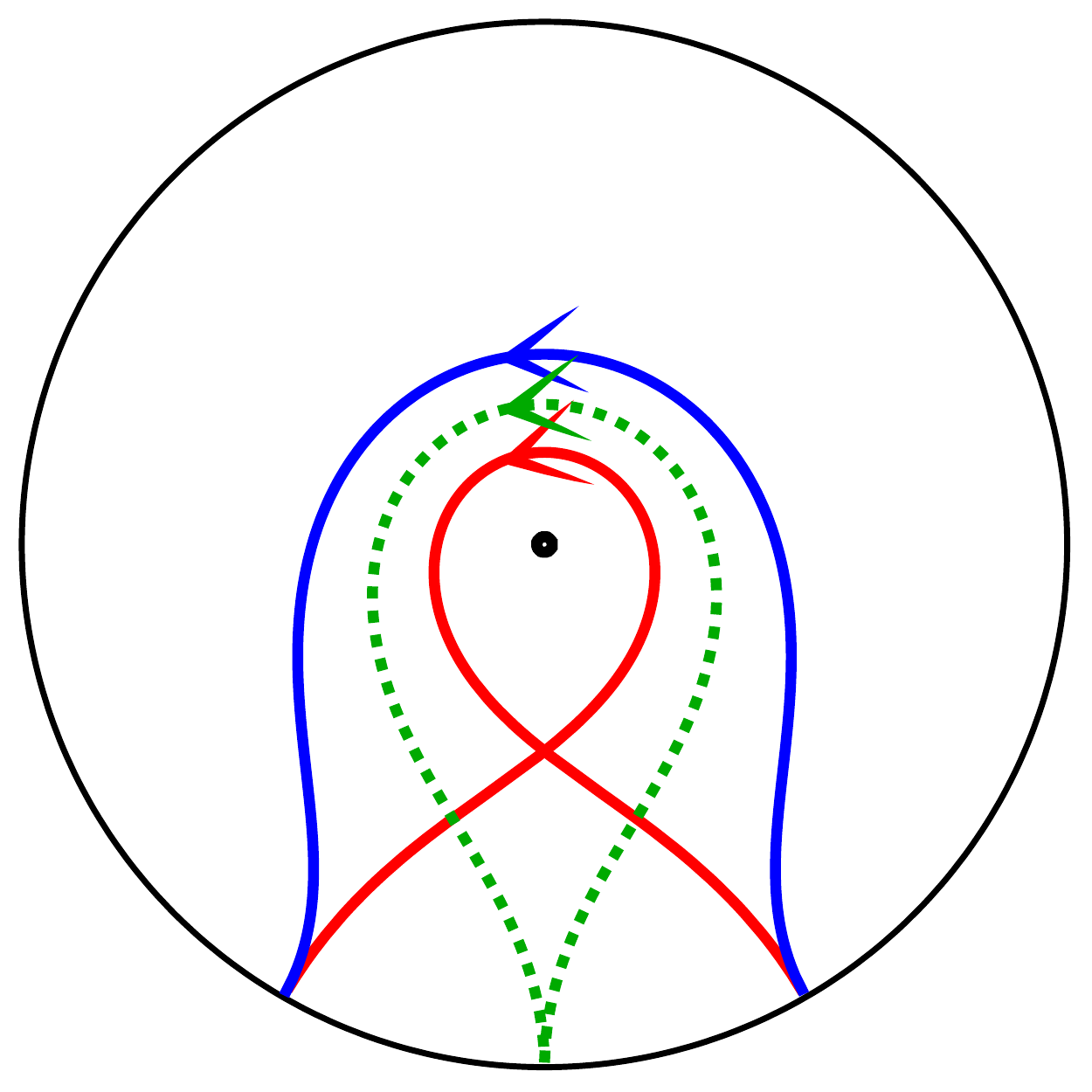}
        \caption{}
        \label{fig:GDContinuity3}
    \end{subfigure}
    \caption[Evolution of oriented geodesics around conical defect]{(a) Oriented geodesics away from the defect are continuous in length and shape as their opening angle is increased. (b) Geodesics with the same endpoints but different orientation cannot be smoothly transformed into one another across the defect. (c) As the opening angle of the blue geodesic increases it reaches the dashed geodesic. The red geodesic also reaches the dashed geodesic as its opening angle decreases, showing continuous behaviour even as the winding number jumps.}
    \label{fig:7}
\end{figure}

In general, the ${\mathcal{B}}_{k,m}$ block represents the contribution to the base OPE from the dual field $\varphi_{CD}$ integrated over geodesics with winding number $n$, where $m$ and $n$ are related by table \eqref{phi1n-odd} or \eqref{phi1n-even}. For the non-minimal cases with $n\geq1$, the geodesics do not stay near the boundary, preventing the use of eq. \eqref{eqextrapolate}. However, the transition between winding numbers is smooth. Away from the defect it is clear that there is no discontinuity in the length or shape of \emph{oriented} geodesics as $\tilde \al$ is increased, even as the winding number jumps, see Figure \ref{fig:7}. This means the X-ray transform $R[\varphi_{CD}](\ga)$ is a continuous and smooth function of the bulk geodesics on the $\tilde \al<\pi/2$ region of kinematic space. Similarly, the partial OPE blocks ${\mathcal{B}}_{k,m}$ blocks defined in eq. \eqref{eqbkn} are continuous in $\tilde \al$ across transitions in the winding number. This is simply because the OPE behaves smoothly as the operator insertions are moved, and it remains convergent for all separations \cite{Pappadopulo2012}.

\begin{figure}
    \centering
    \begin{subfigure}[b]{0.3\textwidth}
        \includegraphics[width=\textwidth]{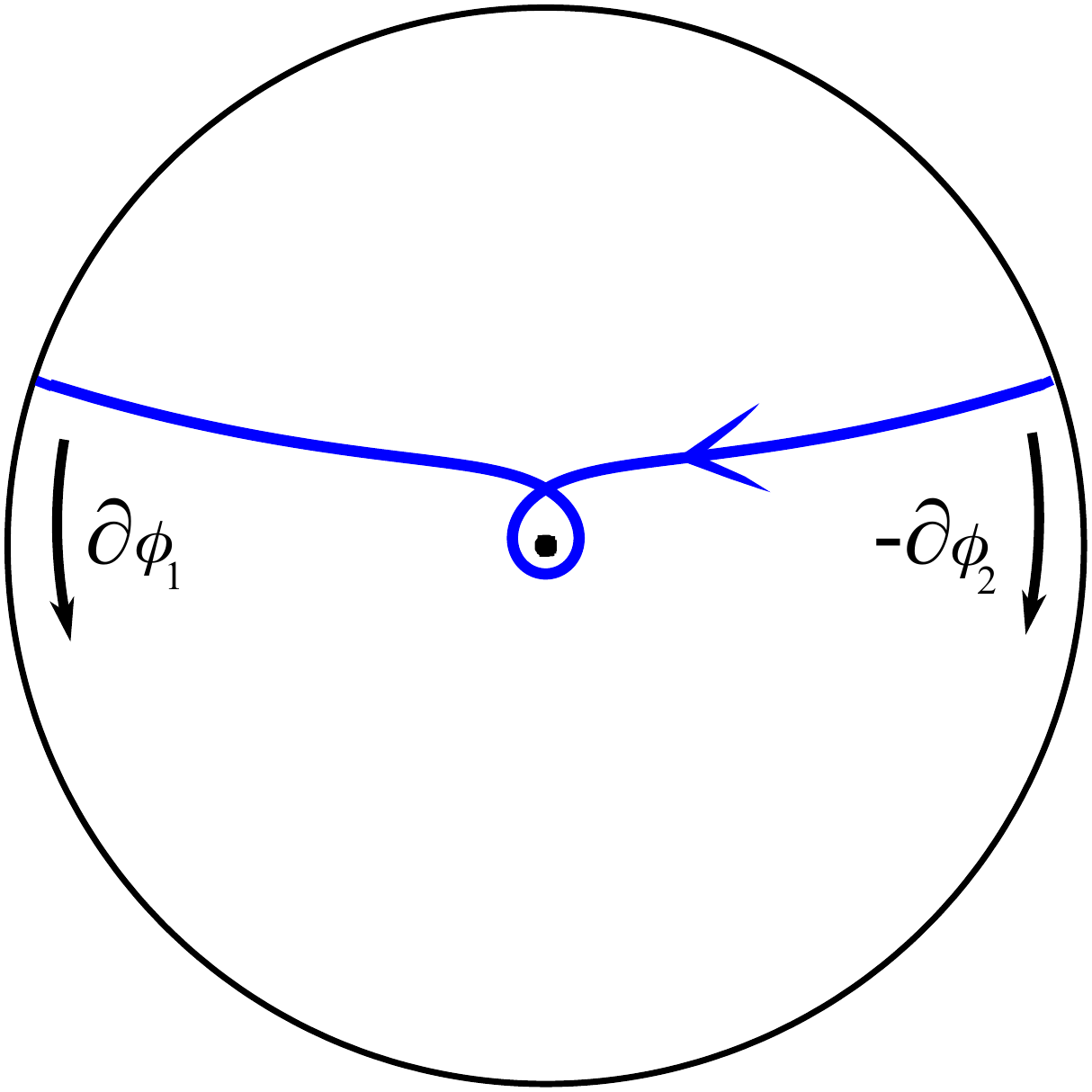}
        \caption{}
        \label{fig:Pinch1}
    \end{subfigure}
     \qquad   \qquad  \qquad
    \begin{subfigure}[b]{0.3\textwidth}
        \includegraphics[width=\textwidth]{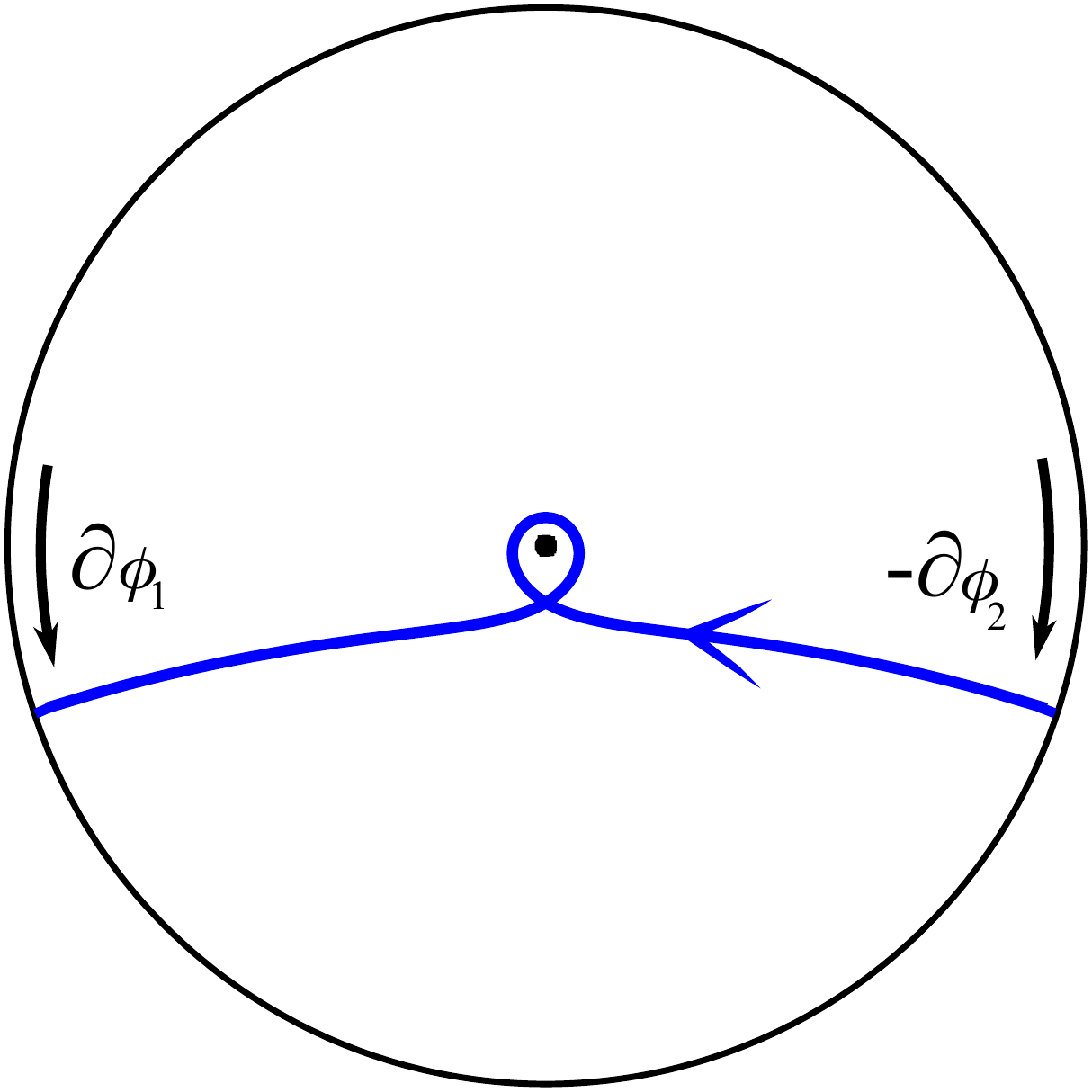}
        \caption{}
        \label{fig:Pinch2}
    \end{subfigure}
    \caption[Evolution of oriented geodesics through conical defect]{Geodesics in AdS$_3$ with $\al=\pi/2$ pass through the origin, and descend to geodesics which touch the conical defect. On AdS$_3$, the behaviour of such geodesics is completely smooth as $\al$ is varied, but on the defect spacetime the endpoints may jump as (a) $\al=\pi/2-\ep$ increases to (b) $\al=\pi/2+\ep$ with $\tilde \te$ held constant. This will be the case when $N$ is not an odd integer. Despite this, the length and shape of such geodesics varies smoothly.}
    \label{fig:pinch}
\end{figure}

There is a potential obstacle to the continuity of $R[\varphi_{CD}](\ga)$ at $\tilde \al=\pi/2$ where geodesics touch the defect. Geodesics in AdS$_3$ with $\tilde\al=\pi/2$ pass through the origin and behave smoothly as $\tilde\al$ is varied, but the corresponding geodesics on the defect spacetime must jump as they pinch in on the defect. As depicted in Figure \ref{fig:pinch}, geodesics with constant center angle $\tilde\te$ jump as $\tilde\al$ is increased past $\pi/2$ and are not homologous across the jump. Despite this, the length and shape of such geodesics varies smoothly which suggests the X-ray transform of $\varphi_{CD}$ will be smooth as well. That this must be the case is easiest to see by using the lifted X-ray transform \eqref{eqxraycd}. There is no discontinuity whatsoever in the transform of lifted geodesics as $\al$ is increased past $\pi/2$.

One can avoid this obstacle entirely by considering an alternative Cauchy slice on the upper half of kinematic space, namely $\tilde\al=\pi$, which also corresponds to near-boundary geodesics and the coincidence limit for the OPE. By the same argument made for $\tilde\al=0$, the X-ray transform $R[\varphi_{CD}](\ga)$ over a minimal geodesic with $\tilde \al\in((1-1/2N)\pi,\pi)$ obeys the same initial conditions as the corresponding partial OPE block, and both are continuous functions on the $\tilde\al>\pi/2$ half of kinematic space. 

Since the initial conditions in the $\tilde\al\to 0,\pi$ limits match between ${\mathcal{B}}_{k,m}$ and $R[\varphi_{CD}](\ga)$, the equations of motion \eqref{eqcasequationtildebkn} and \eqref{eqxraycdeom} along with continuity in $\tilde\al$ establish the duality between OPE blocks and geodesic operators for static conical defects. The base OPE receives contributions from each bulk geodesic, minimal and non-minimal, connecting the boundary insertion points. Each partial OPE block encapsulates the contribution to the base OPE from the dual bulk field integrated over a single geodesic of fixed winding number.

\subsection{Future directions}

The various approaches to kinematic space used in this chapter were adapted to constant time slices of the bulk geometry, equivalently the equal-time limit of the OPE. In each case it was seen that the kinematic space for a quotient spacetime was a quotient of the pure AdS$_3$ kinematic space. The full four dimensional geometry of kinematic space describing the time dependent bulk \cite{Czech2016} should also be obtainable using this quotient. There will be a new ambiguity, in addition to the choice of fundamental regions discussed in this chapter, from the possibility of rotating in time the faces of AdS$_3$ which are identified, see for example Figure 1 from \cite{Arefeva2015}. On neighbouring constant time slices of the AdS$_3$ geometry the wedge representing the conical defect spacetime can have a relative shift in its angular coordinate. The kinematic spaces for subsequent time slices would be vertical strips of dS$_2$ with different ranges of centre angle $\te$. Since the twisted and untwisted identifications of AdS$_3$ produce physically identical conical defect spacetimes, this extra ambiguity can be resolved by making a canonical prescription for an appropriate fundamental region of kinematic space. A complete description of this ambiguity is left for future work.

The conical defect CFT results in this chapter were derived in the special case dual to AdS$_3/\mathbb{Z}_N$, since there is a particularly simple description of this system in terms of a covering CFT in its vacuum state. It is not surprising that the CFT descriptions of the integer and non-integer cases are significantly different when the holographic consequences are kept in mind. The integer defect spacetimes in the bulk have a mild orbifold singularity that does not obstruct the construction of a consistent string theory on this background \cite{Giusto2013,Balasubramanian2003,Son2001}.\footnote{We thank Oleg Lunin for comments on this point.}

Furthermore, this chapter was mainly concerned with static conical defects. These are part of a more general class of moving defects which are produced either by boosting the static solution, or by taking a quotient of AdS$_3$ along a Killing vector with a timelike component \cite{Matschull1999,Balasubramanian1999,Arefeva2015,Arefeva2016a}. It would be interesting to perform this quotient on the AdS$_3$ kinematic space to obtain the kinematic space of a moving defect. Then, using the relation between OPE blocks and geodesic bulk fields it may be possible to use the method of images to relate back to results on the geodesic approximation for correlation functions in those spacetimes.

The partial OPE blocks discussed in this chapter reorganize the operator contributions to the base OPE as compared to the traditional OPE blocks. While a clean CFT interpretation of the operator grouping in terms of conformal families is not obvious from this perspective, we gain a bulk interpretation in terms of the contributions of geodesics with different winding numbers. It would be enlightening to understand better the CFT operator contributions that are represented by partial OPE blocks, and we return to this question in the following chapter. One potential avenue to explore is the superficially similar construction used in \cite{Maloney2017}. Our partial OPE blocks were constructed by first un-gauging a discrete symmetry in going to the covering space description. Gauge invariance is restored by considering symmetrized sums of cover operators under the action of the $\mathbb{Z}_N$ symmetry. The authors of \cite{Maloney2017} studied conformal blocks which give the contribution of a conformal family to a four-point function.  The blocks were approximated by considering only the contribution from light descendants at the cost of modular invariance for the four-point function. Modular invariance was restored by summing over images of the approximate block under the action of modular generators. It may be that these two constructions are related on a deeper level.
The sum over descendants composing an OPE block evinces that they are non-local operators in the CFT.  As such, OPE blocks ${\mathcal{B}_k}(x_1,x_2)$ have a smeared representation where the quasi-primary $\op_k$ they are built from is integrated over a causal diamond defined by the insertion points $x_1, x_2$ \cite{Czech2016,Boer2016}. It was suggested in \cite{Czech2016} that for conical defects the OPE blocks corresponding to winding geodesics should have a smeared representation over diamonds which wrap all the way around the CFT cylinder (See Figure 20 of \cite{Czech2016}). Indeed, our cover OPE blocks have a smeared representation over causal diamonds on the covering CFT cylinder, and so partial OPE blocks can be viewed as symmetrized sums over smeared operators on the cover \eqref{eqbkn}. For the block ${\mathcal{B}}_{k,0}$ representing minimal geodesics, the causal diamonds on the cover are each contained within one $\mathbb{Z}_N$ portion of the cylinder and do not overlap. For blocks representing winding geodesics, the causal diamonds extend over multiple $\mathbb{Z}_N$ portions and can overlap with each other (cf. Figure \ref{fig:3RGreenGDs}). Imposing the $\mathbb{Z}_N$ angular identification on any one of these large causal diamonds produces a diamond which wraps around the cylinder of the base CFT and can overlap on itself. It would be interesting to know if the CFT avatar of entwinement \cite{Balasubramanian2015, Balasubramanian2016} can be cast in terms of partial OPE blocks and wrapping diamonds, and how the bulk can be probed in a more fine-grained fashion using these objects.

Other locally AdS$_3$ geometries and their kinematic spaces have been studied from the bulk and using the differential entropy definition \cite{Czech2015b,Czech2015c,Asplund2016,Zhang2017}, but differences in definitions for kinematic space have led to inconsistent results. For instance, the geometries for the kinematic space of the BTZ black holes described in \cite{Zhang2017} include geodesics of both orientations, while \cite{Asplund2016} and \cite{Czech2015b} do not. Furthermore, the authors of \cite{Asplund2016} chose to include only minimal geodesics in their definition of kinematic space, in contrast to the choice we have made here. In this chapter we have advocated for defining kinematic space from the CFT in terms of OPE blocks, rather than from pairs of points, and have isolated the important contributions of non-minimal geodesics. In the next chapter we will study the newly defined partial OPE blocks in several of the geometries mentioned here.

\chapter{Holographic relations for OPE blocks in excited states}\label{ch:OPE}

\emph{This chapter is based on the paper \cite{Cresswell2018} published in JHEP.}

\section{Introduction}

Since its initial formulation, the AdS/CFT correspondence has opened up many new avenues for studying gravity \cite{Maldacena1999}. It provides a dictionary that can translate unfamiliar gravitational physics into familiar field theory, and vice versa. One of its most powerful aspects is the ability to encode the spatial organization of the bulk as a relationship between the degrees of freedom in the CFT. A particularly useful way of analyzing the geometry of spacetime is through examining the structure of geodesics and extremal surfaces. This has a long history 
in the AdS/CFT context, and an important new theme was begun with the work of \cite{Ryu2006}. Their results in AdS${}_3$ showed that the entanglement entropy of a CFT$_2$ interval is dual to the length of a bulk geodesic anchored at the interval's endpoints. 

The connection between entanglement and geometry \cite{VanRaamsdonk2010} has become of fundamental interest, and has been expanded to many other aspects of quantum information. These include the emergence of gravitational equations of motion from CFT entanglement entropies \cite{Faulkner2017}, bulk gauge freedom interpreted as boundary quantum error correcting codes \cite{Almheiri2015,Mintun2015,Pastawski2015}, the volume of Einstein-Rosen bridges as complexity \cite{Susskind2014}, and the entanglement wedge cross section as CFT entanglement of purification \cite{Umemoto2018}. 

A useful auxiliary space termed kinematic space has been introduced describing the structure of geodesics while also geometrizing entanglement entropy \cite{Czech2015,Czech2016,Boer2016}. Each boundary anchored geodesic, or equivalently each pair of boundary points, is viewed as a single point in kinematic space. One of the major developments discovered through this construction was the holographic dual of a bulk field integrated over a boundary anchored geodesic, namely the OPE block of the corresponding dual operator in the CFT. This is closely related to the duality between conformal blocks in the CFT and geodesic Witten diagrams in the bulk \cite{Hijano2016,Hijano2015}. The properties of OPE blocks themselves have been studied further for defect CFTs \cite{Fukuda2018, Kobayashi2018} and using modular flow \cite{Sarosi2018}.

While these works on kinematic space were thorough, they mainly focused on pure AdS. Followup papers \cite{Czech2016a,Asplund2016,Zhang2017,Karch2017,Cresswell2017,Abt2017,Abt2018} have worked towards extending kinematic space and the OPE block duality to more general AdS spacetimes. We will continue this line of inquiry for AdS${}_3$, where all vacuum solutions to the Einstein equation with negative cosmological constant are locally AdS${}_3$ and can be obtained as quotients. The immediate challenge is that there is no longer a unique geodesic through the bulk between any pair of boundary endpoints. A natural question is to ask how the CFT dual of a geodesic integrated bulk field changes. We will argue that in states dual to quotient geometries, OPE blocks decompose into contributions which are invariant under the quotient action. Each contribution is dual to a bulk field integrated over a single geodesic which may wind around the quotient's fixed points. 

Our arguments are based on the monodromy of maps between pure AdS$_3$ and the quotient geometries. In the bulk the monodromy is responsible for the appearance of non-minimal geodesics, and on the boundary it induces non-analyticities in the OPE blocks. We resolve the latter issue by constructing quotient invariant OPE blocks, and interpret them in terms of winding geodesics. We often utilize the group manifold description of AdS$_3$ and its quotients, in which the structure of geodesics is made clear, and their lengths are easily computable. Throughout, we work with the Euclidean and Lorentzian versions of the construction in parallel to emphasize their differences.

In Section \ref{sec:2} we review the duality between OPE blocks and geodesic integrated bulk fields. Then we introduce the quotient spacetimes of interest and find explicit maps between them and pure AdS$_3$. In Section \ref{sec:3} we use these maps to study the structure of geodesics in the quotient geometries and determine their lengths. In Section \ref{sec:4} we construct quotient invariant OPE blocks, highlighting their relationship to winding geodesics. In Section \ref{sec:5} we conclude with a summary and discussion of remaining open questions.
 
\section{Preliminaries}\label{sec:2}
\subsection{OPE blocks and kinematic space}

In a 2d CFT, the OPE allows us to expand the product of two quasiprimary operators in terms of a basis of local operators at a single location. The OPE can be organized by the contributions from conformal families in the theory, each consisting of a quasiprimary $\mathcal{O}_k$ and its descendants. Considering two scalar operators with the same conformal weight $\Delta$, conformal symmetry dictates that
\begin{equation}\
\mathcal{O}_{i}(x)\mathcal{O}_{j}(0) =  \sum_{k}C_{ijk}\left|x\right|^{\Delta_{k}-2\Delta}\big(1+b_{1}\,x^{\mu}\partial_{\mu}+b_{2}\,x^{\mu}x^{\nu}\partial_{\mu}\partial_{\nu}+\ldots\big)\mathcal{O}_{k}(0)\,,
\end{equation}
with some theory dependent constants $C_{ijk}$, and theory independent constants $b_i$. Since much of this structure is fixed by symmetry, it is convenient to define an OPE block ${\mathcal{B}}^{ij}_k(x_i,x_j)$ associated to each quasiprimary $\mathcal{O}_k$ that repackages the contribution of a conformal family,
\begin{equation}
\mathcal{O}_i(x_i)\mathcal{O}_j(x_j)=x^{-2\Delta}_{12}\sum\limits_kC_{ijk}{\mathcal{B}}^{ij}_k(x_i,x_j)\,.
\end{equation}

Kinematic space has been defined as the space of pairs of CFT points, or equivalently as the space of boundary anchored geodesics in pure AdS \cite{Czech2016,Boer2016}. Since OPE blocks are functions of two boundary points they are fields on kinematic space, and this suggests that they are related to the geodesics of the bulk dual. Indeed, it was shown that for pure AdS the dual of a scalar OPE block is a bulk field integrated over a boundary anchored geodesic,
\begin{equation}\label{Duality}
{\mathcal{B}}^{ij}_k(x_i,x_j)\sim\int_{\gamma_{ij}}ds\ \phi_k(x)\,,
\end{equation}
where $\gamma_{ij}$ is the geodesic with endpoints $(x_i,x_j)$ and $\phi_k$ is the scalar field dual to  $\mathcal{O}_k$. 

The duality between the OPE blocks and geodesic integrated fields was established by showing that both objects behave as fields on kinematic space with the same equation of motion, and the same boundary conditions. Each OPE block built from a scalar quasiprimary $\mathcal{O}_k$ is in an irreducible representation of the conformal group and satisfies an eigenvalue equation under the action of a quadratic conformal Casimir $L^2$, with the eigenvalue induced from $\mathcal{O}_k$,
\begin{equation}
[ L^2,{\mathcal{B}}^{ij}_k(x_i,x_j)]=-\Delta_k(\Delta_k-2){\mathcal{B}}^{ij}_k(x_i,x_j)\,.
\end{equation}
 By expressing the Casimir operator in the differential representation appropriate for ${\mathcal{B}}^{ij}_k$, this becomes a Laplacian on the dS$_2\ \times$ dS$_2$ kinematic space, 
 \begin{equation}\label{OPEblockEoM}
2[\Box_{dS_2}+\bar{\Box}_{dS_2}]{\mathcal{B}}^{ij}_k(x_i,x_j)=-\Delta_k(\Delta_k-2){\mathcal{B}}^{ij}_k(x_i,x_j)\,.
\end{equation}

On the other hand, the bulk scalar field $\phi_k(x)$ dual to $\mathcal{O}_k$ satisfies a wave equation on AdS$_3$, with its mass related to $\Delta_k$ by the holographic dictionary,
\begin{equation}
\Box_{AdS_3} \phi_k(x)=m^2\phi_k(x)=\Delta_k(\Delta_k-2)\phi_k(x)\,.
\end{equation}
Then, the remarkable intertwining property of isometry generators determines the equation of motion for the geodesic integrated field \cite{Czech2016}
\begin{equation}
\int_{\gamma_{ij}}ds\ \Box_{AdS_3} \phi_k(x)=-2[\Box_{dS_2}+\bar{\Box}_{dS_2}]\int_{\gamma_{ij}}ds \ \phi_k(x)\,.
\end{equation}
The conclusion is that the geodesic integrated field obeys the same kinematic space wave equation \eqref{OPEblockEoM} as the OPE block,
\begin{equation}
2[\Box_{dS_2}+\bar{\Box}_{dS_2}]\int_{\gamma_{ij}}ds\ \phi_k(x)=-\Delta_k(\Delta_k-2)\int_{\gamma_{ij}}ds\ \phi_k(x)\,.
\end{equation}
Rounding out the proof requires showing both quantities satisfy the same constraints and the same boundary conditions, which determine the relative normalization omitted in \eqref{Duality}.

For pure AdS there is a one-to-one correspondence between pairs of spacelike separated boundary points and geodesics in the bulk. This makes it simple to identify both the space of pairs of boundary points, and the space of bulk geodesics as the same kinematic space. But for spacetimes that are locally AdS${}_3$, the existence of non-minimal geodesics in the bulk obfuscates this prescription. In such cases it is not \emph{a priori} clear in what sense the duality \eqref{Duality} holds.

This question was addressed for the case of conical defect spacetimes in Chapter \ref{ch:KSCD} (see \cite{Cresswell2017}). Static conical defects are locally AdS$_3$ geometries obtained from AdS$_3$ by a $\mathbb{Z}_{N}$ quotient in the angular direction, leaving a $2\pi/N$ periodic $\tilde \phi$ coordinate. This coordinate parametrizes the one dimensional boundary of a timeslice on which the OPE can be studied. The exact CFT states dual to the conical defect geometries will depend on the system under scrutiny, but in general they can be viewed as the CFT vacuum excited by a heavy operator that sources the defect in the bulk \cite{Balasubramanian1999,Balasubramanian2001,Lunin2002}. In the presence of other operators the OPE does not have an infinite radius of convergence, and it becomes more difficult to study the properties of the OPE blocks directly. Instead, in Chapter \ref{ch:KSCD} the excited CFT states were lifted to vacuum states of a covering space CFT on an $N$-times longer circle parametrized by $\phi$ \cite{Balasubramanian2015}. This process can be seen as removing the discrete $\mathbb{Z}_N$ symmetry of the base CFT states; only appropriately symmetrized quantities on the cover descend to observables on the base \cite{Balasubramanian2016}. 

With this construction, the OPE blocks in the base and cover CFTs can be related. Individual OPE blocks on the cover ${\mathcal{B}}_k(\phi_1, \phi_2)$ are not $\mathbb{Z}_N$ symmetric, but can be combined into gauge invariant observables dubbed partial OPE blocks,
\begin{equation}
{\mathcal{B}}_{k,m}(\alpha_m,\theta)=\frac{1}{N}|2-2\cos(2\alpha_m)|^{-\Delta_k}\sum\limits^{N-1}_{b=0}\exp\left(i\frac{2\pi b}{N}\frac{\partial}{\partial\theta}\right){\mathcal{B}}_k(\alpha_m,\theta)\,.
\end{equation}
 Here, the cover OPE blocks are written in terms of the half opening angle $\alpha=({\phi}_1-{\phi}_2)/2$ and centre angle $\theta=({\phi}_1+{\phi}_2)/2$. The angular distance $\alpha$ between operators is taken to be fixed at $\alpha_m$ while the rotations generated by $\partial/\partial \theta$ implement the symmetrization. The full OPE blocks in the base theory ${\mathcal{B}}'_k$ receive contributions from partial OPE blocks at all allowed angular separations $\alpha_m$ on the cover
\begin{equation}\label{OPEblocksCD-JAW}
{\mathcal{B}}'_k(\alpha,\theta)=\frac{1}{N}\sum\limits_{m=0}^{N-1}\exp\left(i\frac{2\pi m}{N}\frac{\partial}{\partial\phi_1}\right){\mathcal{B}}_{k,m}(\alpha_m,\theta)\,,
\end{equation}
where $\partial/\partial\phi_1$ generates changes in separation. 

Finally, it was shown that the partial OPE blocks individually satisfy duality relations like \eqref{Duality} as fields integrated over minimal or non-minimal geodesics in the conical defect spacetime. The angular separation $\alpha_m$ of the block ${\mathcal{B}}_{k,m}$ is related to the winding number of the geodesic in $\int_{\gamma_m} ds\ \phi_k$. Hence, the new observables ${\mathcal{B}}_{k,m}$ allow us to obtain more fine-grained information about the bulk spacetime that reaches beyond the entanglement shadow limiting minimal geodesics and Ryu-Takayanagi entanglement entropy.

Our approach in this chapter will be similar, but can more readily be applied to the broad class of AdS$_3$ quotient geometries. We will argue that the base OPE blocks for states dual to these geometries can be obtained through the coordinate maps we develop as a sum over partial OPE blocks. The partial blocks are constructed to be invariant under the quotient action. We propose that a partial block is dual to a bulk field integrated over an individual geodesic, which can be minimal or not, as specified by the monodromy under the map. To avoid branch cuts in the full OPE blocks, we identify them as a sum over partial OPE blocks. 

While the bulk interpretation of the partial blocks is clear, they give the contribution to the OPE from individual geodesics or saddlepoints of the path length action \cite{Balasubramanian1999}, our new method also affords a better understanding of the CFT interpretation. Each partial block gives a contribution to the OPE as distinguished by the monodromy around the excited state's heavy operator insertion. To reach these results, we must first develop exact mappings between AdS$_3$ and the quotient geometries that can be used to transform the OPE blocks. We proceed with the Euclidean and Lorentzian cases in turn.

\subsection{AdS${}_3$ quotients}\label{orbifoldsec}
\subsubsection{Euclidean AdS}

One construction of AdS$_3$ is through the $\mathbb{R}^{3,1}$ embedding space. We start with the metric $ds^2=dX_0^2+dX_1^2+dX_2^2-dX_3^2$, with AdS$_3$ defined as the surface $X^2=X_0^2+X^2_1+X_2^2-X_3^2=-\ell^2$. There are a number of different parametrizations of this hyperboloid which give different patches of AdS. 
We focus on the Poincar\'e patch, which only covers part of the hyperboloid. To get the Poincar\'e metric, we implement the coordinates
\begin{gather}\label{EuclideanPoincareEmbeddingCoords}
      \begin{aligned}
         X_0&=\frac{1}{2u}\big(u^2-\ell^2+x^2+t^2\big)\\
	X_1&=\ell \,\frac{x}{u}\\
	X_2&=\ell \,\frac{t}{u}\\
	X_3&=\frac{1}{2u}\big(u^2+\ell^2+x^2+t^2\big) \,,
      \end{aligned}
\end{gather}
which leads to 
\begin{equation}
ds^2=\frac{\ell^2}{u^2}\big(dt^2+dx^2+du^2\big) \,.
\end{equation}
Here, $\ell$ is the AdS radius. We can do a further coordinate transformation by setting $w=x+it$, $\bar w=x-it$, which gives us the metric
\begin{equation}\label{HartmanMetric}
ds^2=\frac{\ell^2}{u^2}\big(dw\,d\bar{w}+du^2\big)\,.
\end{equation}

Boundary anchored geodesics, and especially their lengths, will be very important for understanding the OPE block duality. In Poincar\'e coordinates, the geodesic distance $d$ along the embedding surface between two points $P_1$ and $P_2$ obeys
\begin{gather}
\begin{aligned}\label{generalgdlength}
\cosh \frac{d}{\ell}=-\frac{P_1\cdot P_2}{\ell^2}&=\frac{1}{2u_1 u_2}\left((t_1-t_2)^2+(x_1-x_2)^2+u_1^2+u_2^2\right)\\
&=\frac{1}{2u_1u_2}\left((w_1-w_2)(\bar{w}_1-\bar{w}_2)+u_1^2+u_2^2\right)\,.
\end{aligned}
\end{gather}
 In the limit where both points approach the boundary, such that $u_1,u_2\to0$ with their ratio held fixed, $u_1/u_2\rightarrow 1$, this becomes
\begin{align}
\cosh \frac{d}{\ell}&=1+\frac{1}{2u_1u_2}(w_1-w_2)(\bar w_1-\bar w_2) \,.
\end{align}
The length of a boundary anchored geodesic can then be approximated by
\begin{equation}\label{gdlength}
d\approx\ell\,\log\left(\frac{(w_1-w_2)(\bar w_1-\bar w_2)}{u_1u_2}\right)\,.
\end{equation}

We can also construct Poincar\'e AdS${}_3$ as a group manifold \cite{Banados1993}. This is done by considering each point $g$ in Euclidean AdS${}_3$ as an element of $SL(2,\mathbb{C})/SU(2)$ where, in the embedding coordinates,
\begin{equation}\label{EuclideanEmbeddingGroupElement}
g=\left(\begin{array}{cc}
X_3+X_0 & X_1+iX_2\\
X_1-iX_2 & X_3-X_0
\end{array}\right)\,.
\end{equation} 
For the Euclidean Poincar\'e embedding we have,
\begin{equation}
g=\left(\begin{array}{cc}
u+w\bar{w}/u & \ell w/u\\
\ell\bar{w}/u & \ell^2/u
\end{array}\right)\,.
\end{equation} 
The metric on AdS$_3$ (\ref{HartmanMetric}) is then given by the Cartan-Killing metric $ds^2=\frac{1}{2}\text{Tr}(g^{-1}dgg^{-1}dg)$ which has the correct isometry group for Poincar\'e AdS$_3$, $SL(2,\mathbb{C})/\mathbb{Z}_2$ \cite{Carlip1995}. Other locally AdS${}_3$ solutions are constructed as quotients by a subgroup of the isometry group. The subgroups we study in this chapter are conjugacy classes generated by the elliptic, parabolic, and hyperbolic elements of the form
\begin{equation}\label{conjugacyclasses}
h_{ell}=\left(\begin{array}{cc}
e^{-i\pi\gamma} & 0\\
0 & e^{i\pi\gamma}
\end{array}\right),\qquad
h_{para}=\left(\begin{array}{cc}
1 & \alpha\\
0 & 1
\end{array}\right),\qquad
h_{hyper}=\left(\begin{array}{cc}
e^{\beta/2} & 0\\
0 & e^{-\beta/2}
\end{array}\right)\,,
\end{equation}
where $0<\gamma<1$, $\alpha\in\mathbb{C}$, and $\beta\in\mathbb{R}$. In each case, elements related by conjugation, $g\sim h g h^\dag$, are identified to obtain the quotient manifold. 

Each type of element produces a different locally AdS$_3$ solution. Identification using the elliptic element will give the conical defect, abbreviated `CD', with deficit angle $2\pi(1-\gamma)$. Accounting for the $\mathbb{Z}_2$ quotient of the isometry group, the subgroup generated by an elliptic element is the cyclic group $\mathbb{Z}_N$, where we take $N=1/\gamma\in \mathbb{N}$. The other two elements lead to infinite discrete groups. A quotient using the parabolic element with $\alpha=2\pi $ yields the massless BTZ black hole, which we abbreviate as `0M'. The hyperbolic element with $\beta=2\pi\sqrt{M}$ gives the static BTZ black hole with mass $M$, which we abbreviate as `BTZ'. In summary, the three types of quotient lead to identifications on the Poincar\'e patch as follows,
\begin{align}\label{CDident}
\text{CD:}\quad  & (w,u)\sim(e^{-2\pi i/N}w,u) \,,\\
\label{M=0ident}
\text{0M:}\quad & (w,u)\sim(w+2\pi \ell,u) \,,\\
\label{BTZident}
\text{BTZ:}\quad & (w,u)\sim(e^{2\pi\sqrt{M}}w,e^{2\pi\sqrt{M}}u) \,.
\end{align}
The $N \to\infty$ limit of the CD metric and the $M\to0$ limit of the BTZ metric both produce the 0M metric, but the respective conjugacy classes \eqref{conjugacyclasses} by which elements are identified are not related in this way. Some differences between these limits have been noted in \cite{Chen2018}. For these reasons we treat the 0M solution as a distinct case throughout.

Other locally AdS$_3$ solutions can be obtained using quotients by more complicated subgroups, such as a rotating BTZ black hole using a combination of elliptic and hyperbolic identifications, but we focus on the three archetypal examples above.

Finally, one may wonder if we can consider conical defects where $N>1$ but not an integer. Considering the rational case of $\gamma=m/n$, we find that the subgroup generated by this is $\mathbb{Z}_n$, which is not distinguishable from the integer case. For non-rational $\gamma$ things are worse, as the subgroup generated is no longer finite and the identification one gets is ambiguous. In addition, the validity of non-integer conical defects is suspect in string theory \cite{Lunin2002,Alday2006}, so we will not consider them further.

\subsubsection{Lorentzian AdS}

The Lorentzian case presents a challenge in our approach because the boundary cannot be described by a single complex coordinate. Still, one direct way of approaching Lorentzian AdS using our knowledge of the Euclidean case is to compare them on a timeslice. The $t=0$ slice in embedding coordinates is
\begin{gather}\label{LorentzianPoincareEmbeddingCoords}
      \begin{aligned}
         X_0&=\frac{1}{2u}\big(u^2-\ell^2+x^2\big)\\
	X_1&=\ell \,\frac{x}{u}\\
	X_2&=0\\
	X_3&=\frac{1}{2u}\big(u^2+\ell^2+x^2\big) \,.
      \end{aligned}
\end{gather}
This now satisfies the Lorentzian constraint equation $X^2=X_0^2+X^2_1-X_2^2-X_3^2=-\ell^2$ as well as the Euclidean one, allowing for direct comparison between signatures. On the timeslice the metric is
\begin{equation}
ds^2=\frac{\ell^2(dx^2+du^2)}{u^2}\,,
\end{equation}
which transforms to the upper half plane (UHP) using $s=x+iu$, $\bar{s}=x-iu$
\begin{equation}
ds^2=\frac{-4\ell^2dsd\bar{s}}{(s-\bar{s})^2}\,.
\end{equation}
The upper half plane inherits a $PSL(2,\mathbb{R})$ isometry group from the full $SL(2,\mathbb{R})\times SL(2,\mathbb{R})/\mathbb{Z}_2$ of Lorentzian AdS$_3$ when restricted to the timeslice. Again, we can describe a point $g$ in the timeslice using 
\begin{equation}\label{LorentzianEmbeddingGroupElement}
g=\left(\begin{array}{cc}
X_3+X_0 & X_1-X_2\\
X_1+X_2 & X_3-X_0
\end{array}\right)\,.
\end{equation} 
Then in the group manifold description a point on the UHP is
\begin{equation}
g=\frac{2 i}{s-\bar{s}}\left(\begin{array}{cc}
|s|^2 & \frac{s+\bar{s}}{2}\\
\frac{s+\bar{s}}{2} & 1
\end{array}\right)\,.
\end{equation}
The action of a $PSL(2,\mathbb{R})$ isometry group element
\begin{equation}
\left(\begin{array}{cc}
a & b\\
c & d
\end{array}\right)\,,\quad ad-bc=1\,,
\end{equation}
will transform the UHP coordinate as
\begin{equation}
s\rightarrow\frac{as+b}{cs+d}\,.
\end{equation}

The $PSL(2,\mathbb{R})$ isometry group also has three different types of elements that define conjugacy classes. The elliptic, parabolic, and hyperbolic elements are now given by
\begin{equation}
h_{ell}=\left(\begin{array}{cc}
\cos\theta & -\sin\theta\\
\sin\theta & \cos\theta
\end{array}\right),\qquad
h_{para}=\left(\begin{array}{cc}
1 & \alpha\\
0 & 1
\end{array}\right),\qquad
h_{hyper}=\left(\begin{array}{cc}
e^{\beta/2} & 0\\
0 & e^{-\beta/2}
\end{array}\right)\,,
\end{equation}
where $0<\theta<2\pi$, $\alpha\in\mathbb{R}$, and $\beta\in\mathbb{R}$. Note that there are differences from the $SL(2,\mathbb{C})/\mathbb{Z}_2$ cases we had previously. In particular, the parabolic element involves a real value and the structure of the elliptic element is different. Once again, locally AdS$_3$ spacetimes are obtained as a quotient of $PSL(2,\mathbb{R})$ by subgroups. For the two BTZ cases, the identifications are exactly the same as before with $\alpha=2\pi$ and $\beta=2\pi \sqrt{M}$,
\begin{align}
&\text{0M: }(x,u)\sim(x+2\pi \ell,u),\label{Lor0MIdentification}\\
&\text{BTZ: }(x,u)\sim(e^{2\pi\sqrt{M}}x,e^{2\pi\sqrt{M}}u)\,.\label{LorBTZIdentification}
\end{align}
However, the identification in the elliptic case is significantly more complicated. We take $\theta=\pi/N$ with $N\in\mathbb{N}$ to reproduce the conical defect geometry, and find the identifications
 \begin{align}\label{LorCDIdentification}
\text{CD: }&x\sim\frac{ \ell^2 x \cos(2\pi/N)+\tfrac{\ell}{2}(u^2+x^2-\ell^2)\sin(2\pi/N)}{\ell^2\cos^2(\pi/N)+\ell x\sin(2\pi/N)+(u^2+x^2)\sin^2(\pi/N)}\\
&u\sim\frac{\ell^2 u}{\ell^2\cos^2(\pi/N)+\ell x\sin(2\pi/N)+(u^2+x^2)\sin^2(\pi/N)}\,.
\end{align}
It is simpler in this case to use the complex $s$ coordinate, $s=x+iu$, which is identified as 
\begin{equation}\label{LorCDIdentificationS}
\text{CD: } s\sim\frac{\ell \cos(\pi/N)s-\ell^2 \sin(\pi/N)}{\sin(\pi/N)s+\ell \cos(\pi/N)}\,.
\end{equation}

\subsection{AdS${}_3$ maps and metrics}\label{mapsec}
\subsubsection{Euclidean AdS}

We will be making use of powerful maps that relate pure AdS${}_3$ to other locally AdS${}_3$ geometries \cite{Banados1999a,Roberts2012}. We begin by considering a general AdS${}_3$ solution, written as
\begin{equation}\label{RobertsMetric}
ds^2=\ell^2 \left(-\frac{L}{2}dz^2-\frac{\bar{L}}{2} d\bar{z}^2+\left(\frac{1}{y^2}+\frac{y^2}{16}L\bar{L}\right) dz d\bar{z}+ \frac{dy^2}{y^2}\right)\,.
\end{equation}
We can see that for $L=\bar{L}=0$ this is the usual Poincar\'e metric of pure AdS${}_3$. More generally, we have the relationship 
\begin{equation}
T(z)=\frac{c}{12} L(z)\,,
\end{equation}
where $T(z)$ is the holomorphic stress tensor and $c=3\ell/2G$ is the usual central charge given by the Brown-Henneaux formula \cite{Brown1986}. The analogous relation holds for the anti-holomorphic stress tensor. For what follows, we will set $\ell=1$. 

The transformation of the stress tensor can be exploited to find maps between AdS$_3$ and the quotients. We consider starting with the usual Poincar\'e metric \eqref{HartmanMetric} and implementing the asymptotic relationship $w=f(z)$. The stress tensor transforms as
\begin{equation}
T(z)=\left(\frac{df}{dz}\right)^2T(w)+\frac{c}{12}\{f(z),z\}\,,
\end{equation}
where $\{f(z),z\}$ is the Schwarzian derivative. Since $T(w)=0$ for pure AdS, in the general spacetime \eqref{RobertsMetric} we have
\begin{equation}\label{L=Schw}
L(z)=\{f(z),z\} \,.
\end{equation}
From the CFT point of view, this allows us to get to any background we wish by identifying $f(z)$. Suppose we have the state $|\psi\rangle$ which is excited by an operator with weight $h_{\psi}$. Since
\begin{equation}
\langle\psi|T(z)|\psi\rangle = \frac{h_{\psi}}{z^2}\,,
\end{equation}
we can find the asymptotic map $f(z)$ relating this background to the flat background by solving the differential equation
\begin{equation}\label{hpsi}
\frac{h_{\psi}}{z^2}=\frac{c}{12}\{f(z),z\}\,.
\end{equation}

In turn, the asymptotic map $f(z)$ can be extended into the bulk using \cite{Roberts2012}
\begin{align}
\begin{aligned}\label{fullmaps}
w&=f(z)-\frac{2y^2f'(z)^2\bar{f}''(\bar{z})}{4f'(z)\bar{f}'(\bar{z})+y^2f''(z)\bar{f}''(\bar{z})}\,,\\
\bar{w}&=\bar{f}(\bar{z})-\frac{2y^2\bar{f}'(\bar{z})^2f''(z)}{4f'(z)\bar{f}'(\bar{z})+y^2f''(z)\bar{f}''(\bar{z})}\,,\\
u&=y\,\frac{4(f'(z)\bar{f}'(\bar{z}))^{3/2}}{4f'(z)\bar{f}'(\bar{z})+y^2f''(z)\bar{f}''(\bar{z})}\,,
\end{aligned}
\end{align}
which gives the full map between \eqref{HartmanMetric} and \eqref{RobertsMetric}. In addition, if there is a map $w=f(z)$ that asymptotically implements the transformation, then for any constants $a_1,a_2,a_3$, a more general solution to $\eqref{hpsi}$ is
\begin{equation}
\frac{a_1 f(z)}{1+a_2 f(z)}+a_3\,,
\end{equation}
which comes from $SL(2,\mathbb{C})$ invariance. These maps will give the same metric regardless of the $a_i$ parameters but the corresponding coordinate transformations will differ. For simplicity we take $a_1=1$, $a_2=a_3=0$. 

With this in place, we would like to work out the maps \eqref{fullmaps} for our AdS${}_3$ quotients. The three cases we study correspond in the CFT to states excited by operators with weights 
\begin{gather}
\begin{aligned}\label{3weights}
h_{\text{CD}}&=\frac{c}{24}\left(1-\frac{1}{N^2}\right)\,,\\
h_{\text{0M}}&=\frac{c}{24}\,,\\
h_{\text{BTZ}}&=\frac{c}{24}\big(1+M\big)\,.
\end{aligned}
\end{gather}
In the case of the conical defect, we can see the weight is that of the twist operator and these maps have been looked at before in other contexts \cite{Asplund2015,Anand2018}. The 0M case is the $N\rightarrow\infty$ or $M\rightarrow 0$ limit of the other two. Furthermore, these weights are all non-negative for $N\geq 1$ and $M\geq 0$, as they should be in a unitary CFT. 

These three cases lead to three differential equations \eqref{hpsi}. One can try to solve them using normal methods, or alternatively, one can surmise the form of $f(z)$ from invariance under the identifications \eqref{CDident},\ \eqref{M=0ident},\ \eqref{BTZident} found from the group manifold approach. These identifications suggest the asymptotic maps 
\begin{align}\label{CDasympMap}
f_{\text{CD}}(z) &= z^{-1/N}\,,\\\label{M=0asympMap}
f_{\text{0M}}(z)&=-i\log(z)\,,\\\label{BTZasympMap}
f_{\text{BTZ}}(z)&=\exp{\left(-i\sqrt{M}\log z\right)}\,,
\end{align}
which reproduce the expected weights. As can be seen from the form of the conjugacy classes \eqref{conjugacyclasses}, the $N\to\infty$ and $M\to 0$ limits produce the identity map, rather than the appropriate 0M map, further emphasizing its distinct character. 

Each asymptotic map can be extended into the bulk using \eqref{fullmaps}, which for the conical defect yields the full coordinate transformations
\begin{gather}
\begin{aligned}\label{CDFullTransformation}
w_{\text{CD}}&=\frac{z^{-1/N}((N^2-1)y^2+4N^2z\bar{z})}{((N+1)^2y^2+4N^2z\bar{z})}\,,\\
\bar{w}_{\text{CD}}&=\frac{\bar{z}^{-1/N}((N^2-1)y^2+4N^2z\bar{z})}{((N+1)^2y^2+4N^2z\bar{z})}\,,\\
u_{\text{CD}}&=\frac{4Ny(z\bar{z})^{(N-1)/2N}}{((N+1)^2y^2+4N^2z\bar{z})} \,.
\end{aligned}
\end{gather}
Similarly, for massless BTZ we have the full coordinate transformations
\begin{gather}
\begin{aligned}\label{M=0FullTransformation}
w_{\text{0M}}&=-i\frac{2y^2+(y^2+4z\bar{z})\log z}{y^2+4z\bar{z}}\,,\\
\bar{w}_{\text{0M}}&=i\frac{2y^2+(y^2+4z\bar{z})\log\bar{z}}{y^2+4z\bar{z}}\,,\\
u_{\text{0M}}&=\frac{4y\sqrt{z\bar{z}}}{y^2+4z\bar{z}} \,.
\end{aligned}
\end{gather}
Finally, for massive BTZ the full coordinate transformations are
\begin{gather}
\begin{aligned}\label{BTZFullTransformation}
w_{\text{BTZ}}&=\frac{\big((1-i\sqrt{M})^2y^2+4z\bar{z}\big) \exp{\big(-i\sqrt{M}\log z\big)}}{(1+M)y^2+4z\bar{z}}\,,\\
\bar{w}_{\text{BTZ}}&=\frac{\big((1+i\sqrt{M})^2y^2+4z\bar{z}\big) \exp{\big(i\sqrt{M}\log \bar{z}\big)}}{(1+M)y^2+4z\bar{z}}\,,\\
u_{\text{BTZ}}&=\frac{4y\sqrt{Mz\bar{z}}\,\exp\left(-\frac{i\sqrt{M}}{2}\log (\frac{z}{\bar{z}})\right) }{(1+M)y^2+4z\bar{z}}\,.
\end{aligned}
\end{gather}
Applying these transformations to pure AdS$_3$ yields metrics of the form \eqref{RobertsMetric}, with $L$ and $\bar L$ determined by \eqref{3weights} through \eqref{L=Schw} and \eqref{hpsi},
\begin{gather}
\begin{aligned}\label{3metrics}
ds^2_{\text{CD}}&=\frac{dzd\bar{z}+dy^2}{y^2}-\frac{1}{4}\big(1{-}\frac{1}{N^2}\big)\frac{dz^2}{z^2}-\frac{1}{4}\big(1{-}\frac{1}{N^2}\big)\frac{d\bar{z}^2}{\bar{z}^2}+\frac{1}{16}\big(1{-}\frac{1}{N^2}\big)^2\frac{y^2}{(z\bar{z})^2}dzd\bar{z}\,,\\
ds^2_{\text{0M}}&=\frac{dzd\bar{z}+dy^2}{y^2}-\frac{1}{4}\frac{dz^2}{z^2}-\frac{1}{4}\frac{d\bar{z}^2}{\bar{z}^2}+\frac{1}{16}\frac{y^2}{(z\bar{z})^2}dzd\bar{z}\,,\\
ds^2_{\text{BTZ}}&=\frac{dzd\bar{z}+dy^2}{y^2}-\frac{(1{+}M)}{4}\frac{dz^2}{z^2}-\frac{(1{+}M)}{4}\frac{d\bar{z}^2}{\bar{z}^2}+\frac{(1{+}M)^2}{16}\frac{y^2}{(z\bar{z})^2}dzd\bar{z}\,,
\end{aligned}
\end{gather}
which confirms that the asymptotic maps in \eqref{CDasympMap}-\eqref{BTZasympMap} produce the expected metrics when extended into the bulk. We finish by noting that although the massless BTZ metric can be obtained as a simple limit $N\rightarrow\infty$ or $M\rightarrow 0$ of the conical defect or BTZ metrics respectively, the coordinate transformations are not related in this way.
\subsubsection{Lorentzian AdS}\label{LorentzianMapSec}
The above maps do not generalize straightforwardly to the timeslice. However, we can again use the knowledge that the maps should respect the identifications \eqref{Lor0MIdentification}, \eqref{LorBTZIdentification}, and \eqref{LorCDIdentificationS} to determine 
\begin{align}
s_{\text{CD}}&= i\frac{1+z^{-1/N}}{1-z^{-1/N}}\,,\label{LorSTransformationCD}\\
s_{\text{0M}}&=-i\log(z)\,,\label{LorSTransformation0M}\\
s_{\text{BTZ}}&=\exp{\left(-i\sqrt{M}\log z\right)}\,\label{LorSTransformationBTZ}.
\end{align}
We note that these are full maps on the UHP, not asymptotic ones. The latter two are similar to the asymptotic maps we had before, as the identification on the timeslice is unaffected. The map for the conical defect has a similar piece, but needs to be changed to reflect the change in the elliptic element. In the following, it will be easiest to write the single complex coordinate $z$, which we will call the quotient coordinate for all three cases, as $z=re^{i\theta}$.

In the original $x,u$ coordinates, the map for the conical defect looks like
\begin{gather}\label{LorentzianCDtransform}
	\begin{aligned}
	x_{\text{CD}}&= \frac{2r^{-1/N}\sin(\theta/N)}{1+r^{-2/N}-2r^{-1/N}\cos(\theta/N)}\,,\\
	u_{\text{CD}}&=\frac{1-r^{-2/N}}{1+r^{-2/N}-2r^{-1/N}\cos(\theta/N)}\,.
	\end{aligned}
\end{gather}
For massless BTZ it takes the form
\begin{gather}\label{Lorentzian0Mtransform}
	\begin{aligned}
	x_{\text{0M}}&= \theta\,,\\
	u_{\text{0M}}&=-\log r\,.
	\end{aligned}
\end{gather}
Finally for massive BTZ it looks like
\begin{gather}\label{LorentzianBTZtransform}
	\begin{aligned}
	x_{\text{BTZ}}&=e^{\sqrt{M}\theta}\cos(\sqrt{M}\log r)\,,\\
	u_{\text{BTZ}}&=-e^{\sqrt{M}\theta}\sin(\sqrt{M}\log r)\,.
	\end{aligned}
\end{gather}
In the first two cases the boundary $u=0$ is when $r=1$ in the new coordinates, but for massive BTZ  we have two boundaries, $r=1$ and $r=\exp\left(-\frac{\pi}{\sqrt{M}}\right)$. The identification also produces a horizon at $x=0$ in the Poincar\'e coordinates which interpolates between the boundaries \cite{Fuente2013}. Furthermore, to have $u\geq0$, we need $r>1$ for CD, $r\leq 1$ for 0M, and $\exp\left(-\frac{\pi}{\sqrt{M}}\right)\leq r\leq 1$ for BTZ. Transforming the metric with these maps produces
\begin{align}
ds^2_{\text{CD}}&=\frac{4r^{2/N}}{N^2r^2(r^{2/N}-1)^2}(dr^2+r^2d\theta^2)\,,\\
ds^2_{\text{0M}}&=\frac{1}{r^2\log(r)^2}(dr^2+r^2d\theta^2)\,,\\
ds^2_{\text{BTZ}}&=\frac{M}{r^2\sin^2(\sqrt{M}\log r)}(dr^2+r^2d\theta^2)\,.
\end{align}
We see that the limits $N\to\infty$ and $M\to 0$ reproduce the 0M metric, while taking $N\to 1$ or inserting $M= -1$ gives back pure AdS$_3$. 

Finally, for the CFT analysis, we are interested in the asymptotic maps which are now easily obtained from the full ones
\begin{align}
x_{\text{CD}}&= \cot\left(\frac{\theta}{2N}\right)\,,\label{LorAsyMapCD}\\
x_{\text{0M}}&=\theta,\label{LorAsyMap0M}\\
x_{\text{BTZ}}&=\pm e^{\sqrt{M}\theta}\,.\label{LorAsyMapBTZ}
\end{align}
Note that the sign in the BTZ case will depend on which boundary one considers. We can interpolate between the two boundaries by analytic continuation, $\theta\rightarrow\theta+i\frac{\pi}{\sqrt{M}}$ \cite{Goto2017}. Further, if we interpret $\theta$ to be the complex angle of $z=re^{i\theta}$, the monodromy $z=ze^{2\pi i}$ will implement the identifications \eqref{LorCDIdentificationS}, \eqref{Lor0MIdentification}, and \eqref{LorBTZIdentification}, similarly to the Euclidean case.

\section{Bulk analysis of geodesic structure}\label{sec:3}
\subsection{Euclidean analysis}\label{EuclideanGeodesicSec}

In this section we use the maps between Poincar\'e AdS$_3$ and the quotient geometries to study the resulting structure of geodesics via the group manifold approach. The non-analyticities in the maps allow us to distinguish geodesics with different winding numbers.

Since the geometries \eqref{3metrics} are all locally AdS$_3$, the properties of their geodesics are closely related to those of pure AdS$_3$. More concretely, the lengths of quotient geodesics are given by lengths of AdS$_3$ geodesics whose endpoints are related by the quotient action. We calculate them using the method outlined in \cite{Maxfield2015}. We consider points $p,q$ in the group manifold of AdS$_3$ as in equation (\ref{EuclideanEmbeddingGroupElement}). The length of the geodesic between these points found in \eqref{generalgdlength} is then rewritten as
\begin{equation}
d(p,q)=\cosh^{-1}\left(\frac{\text{Tr}(p^{-1}q)}{2}\right)\,.
\end{equation}
The boundary is represented by singular matrices $p,q$, up to a divergent factor, and the geodesic distance between them diverges. We regulate by considering curves $p(\rho)$, $q(\rho)$ which approach the boundary as $\rho\to\infty$, and which have the property that $\lim_{\rho\to \infty}p(\rho)/\rho=p_{\partial}$, and similarly $q_\partial$, are finite and non-zero. Then in the boundary limit the geodesic length goes to
\begin{equation}\label{RegulatedGeoLength}
d(p_\partial,q_\partial)=\log\rho^2+\log(\text{Tr}(R_{\perp}p^T_{\partial}R_\perp^Tq_{\partial}))+O(1)\,,
\end{equation}
where $R_\perp=(\begin{smallmatrix}0&-1\\1&0\end{smallmatrix})$. The correction term indicates that any rescaling of $\rho$ can give a different finite contribution. In our quotient coordinates, we choose $\rho=1/\epsilon$ where the boundary is cut off at $y=\epsilon$. The radial coordinate is different for each of the different quotient geometries, so the different regulators are labelled.

This approach affords a very clear understanding of non-minimal geodesic lengths. We quotient the AdS$_3$ group manifold by the discrete group generated by one element from \eqref{conjugacyclasses}. The length of the geodesic connecting the boundary points $p_\partial$ and $h q_\partial h^\dag$ is still given by \eqref{RegulatedGeoLength}, 
\begin{equation}\label{NonMinGeoLength}
d(p_\partial,h q_\partial h^\dag)=\log\rho^2+\log(\text{Tr}(R_{\perp}p^T_{\partial}R_\perp^T h q_{\partial}h^\dag))+O(1)\,,
\end{equation}
but in the quotient spacetime $q_\partial$ and $ h q_\partial h^\dag$ are identified. Typically $d(p_\partial,q_\partial)\neq d(p_\partial,h q_\partial h^\dag)$. We now show that non-minimal geodesics can also be identified from monodromies in the asymptotic maps.

We now parametrize the points in the quotient manifold by mapping the embedding coordinates for Poincar\'e, equation (\ref{EuclideanPoincareEmbeddingCoords}), to our quotient coordinates $(z,\bar{z},y)$. Using (\ref{EuclideanEmbeddingGroupElement}) to find the group elements yields
\begin{align}
\text{CD: }&\frac{(z\bar{z})^{\frac{1-N}{2N}}}{4Ny}\left(\begin{array}{cc}
(z\bar{z})^{-\frac{1}{N}}((N-1)^2y^2+4N^2z\bar{z}) & z^{-\frac{1}{N}}((N^2-1)y^2+4N^2z\bar{z})\\
\bar{z}^{-\frac{1}{N}}((N^2-1)y^2+4N^2z\bar{z})& (N+1)^2y^2+4N^2z\bar{z}
\end{array}\right),\\
\text{0M: }&\frac{1}{4y\sqrt{z\bar{z}}}\left(\begin{array}{cc}
2y^2(2+\log(z\bar{z}))+(y^2+4z\bar{z})\log z\log\bar{z}& -i(2y^2+(y^2+4z\bar{z})\log z)\\
i(2y^2+(y^2+4z\bar{z})\log\bar{z})& y^2+4z\bar{z}
\end{array}\right),\\
\text{BTZ: }&\frac{z^{-(1-i\sqrt{M})/2}\bar{z}^{-(1+i\sqrt{M})/2}}{4\sqrt{M}y}\\\nonumber
&\times\left(\begin{array}{cc}
z^{-i\sqrt{M}}\bar{z}^{i\sqrt{M}}((M+1)y^2+4z\bar{z}) & z^{-i\sqrt{M}}((1-i\sqrt{M})^2y^2+4z\bar{z})\\
\bar{z}^{i\sqrt{M}}((1+i\sqrt{M})^2y^2+4z\bar{z})& (M+1)y^2+4z\bar{z}
\end{array}\right)\,.
\end{align}
One can check that conjugation by the elliptic, parabolic, or hyperbolic generators corresponds to taking $z\rightarrow ze^{2\pi i}$ for the respective points. To consider boundary points we take the limit described above resulting in
\begin{align}
\text{CD: }\qquad&\frac{N(z\bar{z})^{\frac{1+N}{2N}}}{\epsilon_{\text{CD}}}\left(\begin{array}{cc}
(z\bar{z})^{-\frac{1}{N}} & z^{-\frac{1}{N}}\\
\bar{z}^{-\frac{1}{N}}& 1
\end{array}\right),\\
\text{0M: }\qquad&\frac{\sqrt{z\bar{z}}}{\epsilon_{\text{0M}}}\left(\begin{array}{cc}
\log z\log\bar{z}& -i\log z\\
i\log\bar{z}& 1
\end{array}\right),\\
\text{BTZ: }\qquad&\frac{z^{(1+i\sqrt{M})/2}\bar{z}^{(1-i\sqrt{M})/2}}{\sqrt{M}\epsilon_{\text{BTZ}}}\left(\begin{array}{cc}
z^{-i\sqrt{M}}\bar{z}^{i\sqrt{M}} & z^{-i\sqrt{M}}\\
\bar{z}^{i\sqrt{M}}& 1
\end{array}\right)\,.
\end{align}
Now we can pick two points, say $z_1$ and $z_2$, and compute the geodesic length using equation (\ref{RegulatedGeoLength}),
\begin{align}\label{EucCDlength}
d_{\text{CD}}&=\log\left[N^2(z_1^{\frac{1}{N}}-z_2^{\frac{1}{N}})(\bar{z}_1^{\frac{1}{N}}-\bar{z}_2^{\frac{1}{N}})\right]+\frac{N-1}{2N}\log z_1\bar{z}_1z_2\bar{z}_2 -2\log \epsilon_{\text{CD}},\\\label{Euc0Mlength}
d_{\text{0M}}&=\log\left[(\log z_1-\log z_2)(\log\bar{z}_1-\log\bar{z}_2)\right]+\frac{1}{2}\log z_1\bar{z}_1z_2\bar{z}_2-2\log{\epsilon_{\text{0M}}},\\\label{EucBTZlength}
d_{\text{BTZ}}&=\log\left[M^{-1}(z_1^{i\sqrt{M}}-z_2^{i\sqrt{M}})(\bar{z}_2^{i\sqrt{M}}-\bar{z}_1^{i\sqrt{M}})\right]+\frac{1-i\sqrt{M}}{2}\log z_1\bar{z}_1z_2\bar{z}_2 -2\log \epsilon_{\text{BTZ}}\,.
\end{align}
Like for the metrics, but unlike for the transformations, the 0M geodesic distance is correctly obtained by taking either $N\to\infty$ or $M\to 0$. Since conjugation by a quotient generator takes $z\to z e^{2\pi i}$, and with reference to \eqref{NonMinGeoLength}, we also obtain winding geodesic lengths from these formulae. This demonstrates how non-analyticities in the asymptotic maps give rise to winding geodesics in the defect geometries.

\subsection{Lorentzian analysis}

We can proceed similarly using our maps (\ref{LorentzianCDtransform}), (\ref{Lorentzian0Mtransform}), and (\ref{LorentzianBTZtransform}) on the embedding coordinates (\ref{LorentzianPoincareEmbeddingCoords}) to find the matrix representations of points in the various quotients as
\begin{align}
\text{CD: }\qquad&\frac{1}{1-r^{-2/N}}\left(\begin{array}{cc}
1+r^{-2/N}+2r^{-1/N}\cos\left(\frac{\theta}{N}\right) & 2r^{-1/N}\sin\left(\frac{\theta}{N}\right)\\
2r^{-1/N}\sin\left(\frac{\theta}{N}\right)& 1+r^{-2/N}-2r^{-1/N}\cos\left(\frac{\theta}{N}\right) 
\end{array}\right),\\
\text{0M: }\qquad&-\frac{1}{\log r}\left(\begin{array}{cc}
\theta^2+\log r ^2 & \theta\\
\theta& 1
\end{array}\right),\\
\text{BTZ: }\qquad&-\frac{1}{\sin(\sqrt{M}\log r)}\left(\begin{array}{cc}
e^{\sqrt{M}\theta} & \cos(\sqrt{M}\log r)\\
\cos(\sqrt{M}\log r)& e^{-\sqrt{M}\theta}
\end{array}\right).
\end{align}
Again, conjugation by the appropriate quotient generator takes $\theta\rightarrow \theta+ 2\pi$. For boundary points we take the limit $r-1=\epsilon\rightarrow0$ in the conical defect case, and $1-r=\epsilon\rightarrow0$ in the massless and massive BTZ cases. This is due to the difference in domains of $r$, as described in Sec. \ref{LorentzianMapSec}. Taking these limits gives the points
\begin{align}
\text{CD: }\qquad&\frac{2N\sin^2\left(\frac{\theta}{2N}\right) }{\epsilon_{\text{CD}}}\left(\begin{array}{cc}
\cot^2\left(\frac{\theta}{2 N}\right) & \cot\left(\frac{\theta}{2N}\right)\\
\cot\left(\frac{\theta}{2N}\right)& 1
\end{array}\right),\\
\text{0M: }\qquad&\frac{1}{\epsilon_{\text{0M}}}\left(\begin{array}{cc}
\theta^2& \theta\\
\theta& 1
\end{array}\right),\\
\text{BTZ: }\qquad&\frac{e^{-\sqrt{M}\theta}}{\sqrt{M}\epsilon_{\text{BTZ}}}\left(\begin{array}{cc}
e^{2\sqrt{M}\theta} & e^{\sqrt{M}\theta}\\
e^{\sqrt{M}\theta}& 1
\end{array}\right).
\end{align}
We can pick two points on the boundary circle, $\theta_1$ and $\theta_2$, to find the geodesic lengths from \eqref{RegulatedGeoLength},
\begin{align}
d_{\text{CD}}&=\log\left[4N^2\sin^2\left(\frac{\theta_1-\theta_2}{2N}\right)\right]-2\log\epsilon_{\text{CD}}\,,\label{LorCDlength}\\
d_{\text{0M}}&=\log[(\theta_1-\theta_2)^2]-2\log\epsilon_{\text{0M}}\,,\label{Lor0Mlength}\\
d_{\text{BTZ}}&=\log\left[\frac{4}{M}\sinh^2\left(\sqrt{M}\frac{\theta_1-\theta_2}{2}\right)\right]-2\log\epsilon_{\text{BTZ}}\,.\label{LorBTZlength}
\end{align}
Again, we see a nice smooth limit between the $N\rightarrow\infty$ and $M\rightarrow0$ limits for the massless BTZ geodesic lengths even though their maps and their embedding coordinates do not have a smooth limit.

In the BTZ expression above we took both points to be on the same boundary $r=1$. Points on the $r=\exp[-\frac{\pi}{\sqrt{M}}]$ boundary are parametrized as
\begin{equation}
\text{BTZ: }\qquad\frac{e^{-\sqrt{M}\theta}}{\sqrt{M}\tilde{\epsilon}_{\text{BTZ}}}\left(\begin{array}{cc}
e^{2\sqrt{M}\theta} & -e^{\sqrt{M}\theta}\\
-e^{\sqrt{M}\theta}& 1
\end{array}\right)\,,
\end{equation}
where we have a different regulator, $\exp[\frac{\pi}{\sqrt{M}}]r-1=\tilde{\epsilon}_{\text{BTZ}}\rightarrow0$. For two points on the $r=\exp[-\frac{\pi}{\sqrt{M}}]$ boundary the distance formula is unchanged, but for horizon crossing geodesics between the two boundaries the lengths are
\begin{equation}
d_{\text{BTZ, crossing}}=\log\left[\frac{4}{M}\cosh^2\left(\sqrt{M}\frac{\theta_1-\theta_2}{2}\right)\right]-\log\epsilon_{\text{BTZ}}\tilde{\epsilon}_{\text{BTZ}}\,.\label{gdlengthsBTZCross}
\end{equation}
Note that this is related to the single sided geodesic length with $\theta\rightarrow\theta+\frac{i\pi}{\sqrt{M}}$.

Once again, in view of \eqref{NonMinGeoLength} and the fact that quotient generators take $\theta\to \theta+2\pi$ we find that non-analyticities in the maps between pure AdS$_3$ and the quotient geometries distinguish boundary anchored geodesics of different windings.

\section{CFT analysis of OPE blocks}\label{sec:4}
\subsection{Euclidean analysis}\label{sec:4.1}

In this section we argue that the non-analyticities in the asymptotic maps between pure AdS$_3$ and the quotient geometries which distinguish winding geodesics also distinguish quotient invariant contributions to OPE blocks. The terms in the OPE block decomposition are in correspondence with the winding geodesics, which suggests a dual relationship. 

We start by mapping vacuum OPE blocks to a non-trivial background using the asymptotic maps from our bulk analysis. Consider a transformation $x\rightarrow x'$ where
\begin{equation}
\Omega(x')=\det\left(\frac{\partial x'^{\mu}}{\partial x^{\nu}}\right)\,.
\end{equation}
An OPE block $B$ of scalar operators will in general transform as \cite{Czech2016}
\begin{equation}\label{OPEBlocktransform}
{{B}}^{ij}_k(x_i,x_j)=\left(\frac{\Omega(x'_i) }{\Omega(x'_j)}\right)^{\Delta_{ij}/2}{{B}}^{ij}_k(x'_i,x'_j)\,,
\end{equation}
where $\Delta_{ij}\equiv \Delta_i -\Delta_j$\,. For simplicity, we will set $\Delta_{ij}=0$. Now we apply \cref{CDasympMap,M=0asympMap,BTZasympMap} for the CD, 0M, and BTZ cases respectively which naively gives the transformation
\begin{equation}
{{B}}^{ij}_k(z_i,\bar{z}_i,z_j,\bar{z}_j) ={{B}}^{ij}_k(w_i,\bar{w}_i,w_j,\bar{w}_j)\,.
\end{equation}
However, we immediately see a problem. All of these maps have a branch cut as we take $z\rightarrow ze^{2\pi i}$, whereas the OPE block should be a single-valued observable. If we wish to remove branch cuts from the OPE block, we should instead consider 
\begin{align}
{\text{CD:}}\quad {{\mathcal{B}}}^{ij}_k(z_i,\bar{z}_i,z_j,\bar{z}_j)
&=\sum_{p_i,p_j}{\mathcal{B}}^{ij}_k(w_ie^{-\frac{2\pi i p_i}{N}},\bar{w}_ie^{\frac{2\pi i p_i}{N}},w_je^{-\frac{2\pi i p_j}{N}},\bar{w}_je^{\frac{2\pi i p_j}{N}})\,,\\
{\text{0M:}}\quad {{\mathcal{B}}}^{ij}_k(z_i,\bar{z}_i,z_j,\bar{z}_j)
&=\sum_{p_i,p_j}{\mathcal{B}}^{ij}_k(w_i+2\pi p_i,\bar{w}_i+2\pi p_i,w_j+2\pi p_j,\bar{w}_j+2\pi p_j)\,,\\
{\text{BTZ:}}\quad {{\mathcal{B}}}^{ij}_k(z_i,\bar{z}_i,z_j,\bar{z}_j) &=\sum_{p_i,p_j} {\mathcal{B}}^{ij}_k(w_ie^{2\pi p_i\sqrt{M}},\bar{w}_ie^{2\pi p_i\sqrt{M}},w_je^{2\pi p_j\sqrt{M}},\bar{w}_je^{2\pi p_j\sqrt{M}}) \,.
\end{align}
These are sums over pre-images of points identified under the maps. Alternatively, these sums can be argued for from the quotient identifications on pure AdS$_3$ in equations (\ref{CDident}), (\ref{M=0ident}), and (\ref{BTZident}) respectively as they are invariant under the boundary action of the quotient. This method of images has been used frequently for describing quotient invariant observables \cite{Balasubramanian1999,Balasubramanian2003,Arefeva2016}.

We now relate these images to geodesics. Fixing one of the points in the vacuum OPE block and taking images of the other point defines a sequence of different geodesics in the pure AdS$_3$ bulk. Under the quotient these all map to geodesics with the same endpoints, but differing by their winding. For conical defects we found in Chapter \ref{ch:KSCD} that fields integrated on each of these winding geodesics have a dual description, the partial OPE block, summarized in equation (\ref{OPEblocksCD-JAW}). Similarly, we can reorganize the sums above, decomposing the full OPE blocks into distinct contributions labelled by $m$,
\begin{align}\label{OPEblocksCD}
{\text{CD}}:\quad {\mathcal{B}}^{ij}_k(z_i,\bar{z}_i,z_j,\bar{z}_j)&=\sum_{m}{\mathcal{B}}^{ij}_{k,m}(w_ie^{-\frac{2\pi i m}{N}},\bar{w}_ie^{\frac{2\pi i m}{N}},w_je^{\frac{-2\pi i m}{N}},\bar{w}_je^{\frac{2\pi im}{N}})\,, \\\label{OPEblocksM=0}
{\text{0M}}:\quad {\mathcal{B}}^{ij}_k(z_i,\bar{z}_i,z_j,\bar{z}_j)&=\sum_{m}{\mathcal{B}}^{ij}_{k,m}(w_i+2\pi m,\bar{w}_i+2\pi m,w_j+2\pi m,\bar{w}_j+2\pi m)\,, \\\label{OPEblocksBTZ}
{\text{BTZ}}:\quad {\mathcal{B}}^{ij}_k(z_i,\bar{z}_i,z_j,\bar{z}_j)&=\sum_{m}{\mathcal{B}}^{ij}_{k,m}(w_ie^{2\pi m\sqrt{M}},\bar{w}_ie^{2\pi m\sqrt{M}},w_je^{2\pi m\sqrt{M}},\bar{w}_je^{2\pi m\sqrt{M}}) \,,
\end{align}
where
\begin{align}\label{PartialOPEblocksCD}
{\text{CD}}:\quad {\mathcal{B}}^{ij}_{k,m}(w_i,\bar{w}_i,w_j,\bar{w}_j)&=\sum_{b}{\mathcal{B}}^{ij}_{k}(w_ie^{\frac{2\pi i (m-b)}{N}},\bar{w}_ie^{\frac{-2\pi i (m-b)}{N}},w_je^{-\frac{2\pi i b}{N}},\bar{w}_je^{\frac{2\pi ib}{N}})\,, \\\label{PartialOPEblocksM=0}
{\text{0M}}:\quad {\mathcal{B}}^{ij}_{k,m}(w_i,\bar{w}_i,w_j,\bar{w}_j)&=\sum_{b}{\mathcal{B}}^{ij}_{k}(w_i+2\pi(b-m),\bar{w}_i+2\pi(b-m),w_j+2\pi b,\bar{w}_j+2\pi b)\,, \\
{\text{BTZ}}:\quad {\mathcal{B}}^{ij}_{k,m}(w_i,\bar{w}_i,w_j,\bar{w}_j)&=\sum_{b}{\mathcal{B}}^{ij}_{k}(w_ie^{2\pi \sqrt{M}(b-m)},\bar{w}_ie^{2\pi \sqrt{M}(b-m)},w_je^{2\pi \sqrt{M}b},\bar{w}_je^{2\pi \sqrt{M}b})\label{PartialOPEblocksBTZ} \,.
\end{align}
Each of the new quantities ${\mathcal{B}}^{ij}_{k,m}$ is invariant under the appropriate quotient action on both coordinates $z_{i,j}$ sending $z\to z e^{2\pi i}$, meaning they are valid observables in the  quotient coordinates. This has been expressed before in terms of invariance under the CFTs discrete gauge symmetry that is induced by the quotient \cite{Balasubramanian2016, Cresswell2017}. 

Our suggestion is that each partial OPE block ${\mathcal{B}}^{ij}_{k,m}(w_i,\bar{w}_i,w_j,\bar{w}_j)$ is dual to the bulk field integrated over a geodesic with winding related to the label $m$. By construction, each partial OPE block depends on pairs of boundary points at a fixed separation determined by $m$. This can be seen from the geodesic distance formulae, \cref{EucCDlength,Euc0Mlength,EucBTZlength}, by acting with the quotient generator $b$ times on point $z_1$, and $b+m$ times on point $z_2$, as dictated by \cref{PartialOPEblocksCD,PartialOPEblocksM=0,PartialOPEblocksBTZ} and \cref{OPEblocksCD,OPEblocksM=0,OPEblocksBTZ}:
\begin{align}
\label{CDgdlength}
d_{CD}(m,b)=&\log\left[N^2(z_1^{\frac{1}{N}}-z_2^{\frac{1}{N}}e^{2\pi m i/N})(\bar{z}_1^{\frac{1}{N}}-\bar{z}_2^{\frac{1}{N}}e^{-2\pi m i/N})\right]\\
\nonumber&+\frac{N-1}{2N}\log z_1\bar{z}_1z_2\bar{z}_2 -2\log \epsilon_{\text{CD}}
,\\
\label{0Mgdlength}
d_{0M}(m,b)=&\log\left[(\log z_1-\log z_2-2\pi m i)(\log\bar{z}_1-\log\bar{z}_2+2 \pi m i)\right]\\
\nonumber&+\frac{1}{2}\log z_1\bar{z}_1z_2\bar{z}_2-2\log{\epsilon_{\text{0M}}},\\
\label{BTZgdlength}
d_{BTZ}(m,b)=&\log\left[M^{-1}(z_1^{i\sqrt{M}}-z_2^{i\sqrt{M}}e^{2\pi \sqrt{M}m})(\bar{z}_2^{i\sqrt{M}}-\bar{z}_1^{i\sqrt{M}}e^{-2\pi \sqrt{M}m})\right]\\
\nonumber&+\frac{1-i\sqrt{M}}{2}\log z_1\bar{z}_1z_2\bar{z}_2 -2\log \epsilon_{\text{BTZ}}.
\end{align}
In each case we find that all dependence on the $b$-sum index drops out. This means that each vacuum OPE block entering ${\mathcal{B}}^{ij}_{k,m}(w_i,\bar{w}_i,w_j,\bar{w}_j)$ defines an AdS$_3$ geodesic, all of which have the same length and become identified under the quotient. Hence, each ${\mathcal{B}}^{ij}_{k,m}(w_i,\bar{w}_i,w_j,\bar{w}_j)$ picks out a unique geodesic in the dual quotient geometry, with winding specified by $m$. Blocks with different $m$ are related by repeated action of the quotient generator on only one of the boundary points in the same way that geodesics with different windings are related, as seen in \eqref{NonMinGeoLength} and the results of Section \ref{sec:3}.

The new quantities ${\mathcal{B}}^{ij}_{k,m}$ are each defined as a sum over vacuum OPE blocks which are known to be convergent inside correlation functions \cite{Pappadopulo2012,Rychkov2016}, but any required normalization has been neglected above. For the conical defect \eqref{PartialOPEblocksCD}, the sum is finite and can be normalized as
\begin{equation}\label{normalizedCD}
{\text{CD}}:\quad {\mathcal{B}}^{ij}_{k,m}(w_i,\bar{w}_i,w_j,\bar{w}_j)=\frac{1}{N}\sum_{b=0}^{N-1}{\mathcal{B}}^{ij}_{k}(w_ie^{\frac{2\pi i (m-b)}{N}},\bar{w}_ie^{\frac{-2\pi i (m-b)}{N}},w_je^{-\frac{2\pi i b}{N}},\bar{w}_je^{\frac{2\pi ib}{N}}).
\end{equation}
 The $b$-sum ensures that ${\mathcal{B}}^{ij}_{k,m}$ is quotient invariant, but does not alter the overall contribution to the OPE. This follows since the $N$ terms in the sum each give equivalent contributions due to conformal symmetry, or from bulk considerations due to the equality of geodesic distances discussed in the previous paragraph.

 For the massless and massive BTZ cases, the $b$-sums are infinite making the normalization appear ambiguous and bringing the convergence of the sum into question. However, we know that the OPE itself is convergent in CFTs, and our ${\mathcal{B}}^{ij}_{k,m}$ represents only a partial contribution to the full OPE. Again, although an infinite number of images are included to ensure invariance under the quotient, each image represents an equivalent contribution by symmetry. We can normalize the operators using a formal limit
\begin{align}
{\text{0M}}:\quad {\mathcal{B}}^{ij}_{k,m}=\lim_{N\to\infty} \frac{1}{2N+1}\sum_{b=-N}^{N}{\mathcal{B}}^{ij}_{k}(w_i+2\pi(b-m),\bar{w}_i+2\pi(b-m),w_j+2\pi b,\bar{w}_j+2\pi b),\label{normalized0M} \\
{\text{BTZ}}:\quad {\mathcal{B}}^{ij}_{k,m}=\lim_{N\to\infty} \frac{1}{2N+1}\sum_{b=-N}^{N}{\mathcal{B}}^{ij}_{k}(w_ie^{2\pi \sqrt{M}(b-m)},\bar{w}_ie^{2\pi \sqrt{M}(b-m)},w_je^{2\pi \sqrt{M}b},\bar{w}_je^{2\pi \sqrt{M}b}).\label{normalizedBTZ}
\end{align}

In contrast, the full OPE blocks in \cref{OPEblocksCD,OPEblocksM=0,OPEblocksBTZ} are not sums over equivalent contributions. By convention we can arrange for the $m=0$ block to correspond to the minimal operator separation, and hence the minimal bulk geodesic. All other $m\neq0$ blocks are subleading since they represent operators at greater separation in the vacuum where there are no complications from the presence of other operators. The fall off with distance can be seen explicitly in the smeared representation for vacuum OPE blocks \cite{Czech2016}. The conical defect sum is finite and can be normalized as in \eqref{normalizedCD}, whereas for the BTZ cases, we see from \eqref{0Mgdlength} and \eqref{BTZgdlength} that the operators become infinitely separated for large $|m|$, and their contribution becomes negligible. This is the mechanism by which similar applications of the method of images for conical defects and BTZ spacetimes produce finite correlators from infinite sums \cite{Keski-Vakkuri1998,Balasubramanian1999,Balasubramanian2013}.


\subsection{Lorentzian analysis}

The Lorentzian case is slightly different because the boundary is not parametrized by a complex coordinate. Still, we can rely on invariance under the quotient action to guide us. OPE blocks in the quotient coordinate $\theta$ transform to vacuum OPE blocks using eq. \eqref{OPEBlocktransform} with the asymptotic maps \eqref{LorAsyMapCD}-\eqref{LorAsyMapBTZ}. For simplicity, we will specialize to $\Delta_i=\Delta_j$. Once again, these maps are not invariant under $\theta\to \theta+2\pi$ meaning there is an ambiguity in the transformation of the naive defect OPE blocks. To define single-valued OPE blocks we sum over images, ensuring consistency with the $u\to0$ boundary limits of \eqref{Lor0MIdentification}-\eqref{LorCDIdentification}.  We then have the following transformations for OPE blocks
\begin{align}
{\text{CD:}}\quad {{\mathcal{B}}}^{ij}_k(\theta_i,\theta_j)
&=\sum_{p_i,p_j}{\mathcal{B}}^{ij}_k\left(\frac{\cos(p_i\pi/N)x_i-\sin(p_i\pi/N)}{\sin(p_i\pi/N)x_i+\cos(p_i\pi/N)},\frac{\cos(p_j\pi/N)x_j-\sin(p_j\pi/N)}{\sin(p_j\pi/N)x_j+\cos(p_j\pi/N)}\right)\,,\\
{\text{0M:}}\quad {{\mathcal{B}}}^{ij}_k(\theta_i,\theta_j)
&=\sum_{p_i,p_j}{\mathcal{B}}^{ij}_k(x_i+2\pi p_i,x_j+2\pi p_j)\,,\\
{\text{BTZ:}}\quad {{\mathcal{B}}}^{ij}_k(\theta_i,\theta_j)&=\sum_{p_i,p_j} {\mathcal{B}}^{ij}_k(x_ie^{2\pi p_i\sqrt{M}},x_je^{2\pi p_j\sqrt{M}}) \,.
\end{align}
For the BTZ case we have written the single sided OPE block above. The OPE block relating operators on different boundaries is related by the analytic continuation of one of the $\theta$ coordinates,
\begin{align}
{\text{BTZ, crossing:}}\quad {{\mathcal{B}}}^{ij}_k(\theta_i+i\pi/\sqrt{M},\theta_j)&=\sum_{p_i,p_j} {\mathcal{B}}^{ij}_k(-x_ie^{2\pi p_i\sqrt{M}},x_je^{2\pi p_j\sqrt{M}}) \,.
\end{align}
This matches nicely with the analytic continuation found both in the coordinate transformations \eqref{LorAsyMapBTZ} and in the geodesic lengths \eqref{gdlengthsBTZCross}. 

As before we can reorganize the sums, writing them as a decomposition into quotient invariant partial OPE blocks 
\begin{align}
&{\text{CD:}}\quad {{\mathcal{B}}}^{ij}_k(\theta_i,\theta_j)
=\sum_{m}{\mathcal{B}}^{ij}_{k,m}\left(\frac{\cos(m\pi/N)x_i-\sin(m\pi/N)}{\sin(m\pi/N)x_i+\cos(m\pi/N)},\frac{\cos(m\pi/N)x_j-\sin(m\pi/N)}{\sin(m\pi/N)x_j+\cos(m\pi/N)}\right)\,,\\
&{\text{0M:}}\quad {{\mathcal{B}}}^{ij}_k(\theta_i,\theta_j)
=\sum_{m}{\mathcal{B}}^{ij}_{k,m}(x_i+2\pi m,x_j+2\pi m)\,,\\
&{\text{BTZ:}}\quad {{\mathcal{B}}}^{ij}_k(\theta_i,\theta_j)=\sum_{m} {\mathcal{B}}^{ij}_{k,m}(x_ie^{2\pi m\sqrt{M}},x_je^{2\pi m\sqrt{M}}) \,,\\
&{\text{BTZ, crossing:}}\quad {{\mathcal{B}}}^{ij}_k(\theta_i+i\pi/\sqrt{M},\theta_j)=\sum_{m} {\mathcal{B}}^{ij}_{k,m}(-x_ie^{2\pi m\sqrt{M}},x_je^{2\pi m\sqrt{M}}) \,.
\end{align}
where 
\begin{align}
&{\text{CD:}}\quad {\mathcal{B}}^{ij}_{k,m}(x_i,x_j)\\\nonumber
&\quad\quad=\sum_{b}{\mathcal{B}}^{ij}_{k}\left(\frac{\cos((b-m)\pi/N)x_i-\sin((b-m)\pi/N)}{\sin((b-m)\pi/N)x_i+\cos((b-m)\pi/N)},\frac{\cos(b\pi/N)x_j-\sin(b\pi/N)}{\sin(b\pi/N)x_j+\cos(b\pi/N)}\right)\,,\\
&{\text{0M:}}\quad {\mathcal{B}}^{ij}_{k,m}(x_i,x_j)
=\sum_{b}{\mathcal{B}}^{ij}_{k}(x_i+2\pi (b-m),x_j+2\pi b)\,,\\
&{\text{BTZ:}}\quad {\mathcal{B}}^{ij}_{k,m}(x_i,x_j)
=\sum_{b} {\mathcal{B}}^{ij}_{k}(x_ie^{2\pi (b-m)\sqrt{M}},x_je^{2\pi b\sqrt{M}}) \,.
\end{align}
For the BTZ partial OPE blocks, the above equations encompass both signs of the $x$ coordinates allowed in  \eqref{LorAsyMapBTZ}. 

The partial OPE blocks $ {\mathcal{B}}^{ij}_{k,m}(x_i,x_j)$ give the contribution to the full OPE block from image operators at a fixed separation in $x$, indicated by the label $m$. Each vacuum OPE block included in the sum gives an identical contribution, as is apparent by the conformal symmetry of the vacuum state, but the sum is necessary for manifest invariance under the quotient. This can be compared with the geodesic distance formulae, \cref{LorCDlength,Lor0Mlength,LorBTZlength} and \eqref{gdlengthsBTZCross}. Acting with the quotient generator $b$ times on point $\theta_1$, and $b+m$ times on point $\theta_2$ gives

 \begin{align}
\label{LorCDlengthbm}
d_{\text{CD}}(b,m)&=\log\left[4N^2\sin^2\left(\frac{\theta_1-\theta_2-2\pi m}{2N}\right)\right]-2\log\epsilon_{\text{CD}}
,\\
\label{Lor0Mlengthbm}
d_{0M}(b,m)&=\log\left[(\theta_1-\theta_2+2\pi m)^2\right]-2\log \epsilon_{0M},\\
\label{LorBTZlengthbm}
d_{\text{BTZ}}(b,m)&=\log\left[\frac{4}{M}\sinh^2\left(\sqrt{M}\frac{\theta_1-\theta_2-2\pi m}{2}\right)\right]-2\log\epsilon_{\text{BTZ}},\\
\label{LorBTZcrossinglengthbm}
d_{\text{BTZ,crossing}}(b,m)&=\log\left[\frac{4}{M}\cosh^2\left(\sqrt{M}\frac{\theta_1-\theta_2-2\pi m}{2}\right)\right]-\log\epsilon_{\text{BTZ}}\tilde\epsilon_{\text{BTZ}}.
\end{align}
In every case the dependence on $b$ drops out, showing a precise matching between the behaviour of geodesics and the structure of ${\mathcal{B}}^{ij}_{k,m}(x_i,x_j)$. Since each term gives an equivalent contribution, the partial OPE blocks can be normalized in the same way as described in Section \ref{sec:4.1}.

Each ${\mathcal{B}}^{ij}_{k,m}(x_i,x_j)$ block is invariant when the quotient acts on both $x_{i,j}$, while blocks with different $m$ are related by repeated action on only one of $x_{i,j}$. Winding or crossing geodesics of different lengths are related by the repeated quotient action on one endpoint, and each is invariant under the action on both endpoints. Hence, we also interpret the $ {\mathcal{B}}^{ij}_{k,m}(x_i,x_j)$ as giving the contribution to the full OPE block from the dual bulk field integrated over a single geodesic, which may be minimal, winding, or horizon crossing as appropriate.

\section{Discussion}\label{sec:5}

In this chapter we have explored generalizing the holographic duality between OPE blocks and geodesic integrated fields to non-trivial locally AdS${}_3$ spacetimes, both in the Euclidean case and for the Poincar\'e disk of Lorentzian AdS$_3$. Such spacetimes can be described as quotients of AdS$_3$ by discrete subgroups of the isometry group. We found that the transformations between AdS$_3$ and its quotients involve non-analyticities which lead to branch cuts in OPE blocks for the dual excited CFT states. We proposed that the branch cuts should be removed by summing over image points of the quotient action, while also noting a natural decomposition of the OPE blocks into quotient invariant contributions. These contributions, partial OPE blocks, are observables in and of themselves, carrying more fine-grained information than the full OPE block. We explained how this decomposition arises from the coordinate transformations, and offered a dual interpretation of the partial OPE blocks as bulk fields integrated over individual winding or crossing geodesics.

On the bulk side we presented coordinate transformations between pure AdS$_3$ and the conical defect, the massless BTZ black hole, and massive BTZ geometries. These maps incorporate the corresponding quotient identifications, which are expressed as a monodromy of the complex coordinate describing the defect spacetime. The identifications map sets of boundary anchored geodesics between distinct pairs of points in pure AdS$_3$ to geodesics with identical endpoints in the new spacetime, differentiated by their winding around the defect. We showed how the lengths of these geodesics transform emphasizing the relation to monodromy. 

In the CFT we showed that branch cuts appear in OPE blocks after the transformation from pure AdS$_3$ to the quotient spacetime. Removing these branch cuts by summing over images led to a new quotient invariant quantity, the partial OPE block. This process can also be seen as requiring the OPE blocks to be invariant under a discrete gauge symmetry induced by the quotient. The various partial OPE blocks are related by applying the quotient generator to one of the insertion points. The same action distinguishes geodesics with different winding. In view of the duality known for pure AdS$_3$, we conjecture that partial OPE blocks are dual to fields integrated over the individual geodesics in the bulk which can be minimal, non-minimal, or even horizon crossing. 

In the case of the conical defect, the discrete quotient group is finite and therefore isomorphic to $\mathbb{Z}_N$. However, for both BTZ cases, the group is infinite and the interpretation of how the orbifold CFT is properly defined is less clear. The idea of orbifolding by these infinite discrete groups is not new \cite{Horowitz1998}, but our interpretation of how these discrete gauge symmetries affect the OPE blocks and their dual is. We have not proven explicitly that the partial OPE blocks are dual to fields integrated over the minimal or non-minimal geodesics, as this would require a greater understanding of the intertwining relation for the Radon transform in non-pure AdS$_3$ \cite{Czech2016}. 

Differences arise between the Euclidean and Lorenztian descriptions for the obvious reason: the monodromy of the $z$ coordinate only exists if $z$ is complex. In Euclidean signature the boundary is naturally described by a complex coordinate and the monodromy affecting OPE blocks is easily understood. In Lorentzian signature we restricted our considerations to the upper half plane description of the Poincar\'e disk to accord with this. In the full Lorentzian case, it is difficult to see how we could reduce the action of the quotient into the monodromy of a complex coordinate as it is unclear what the correct combination of coordinates would be. In addition, for Lorentzian AdS$_3$ there are no geodesics between timelike separated boundary points, whereas OPE blocks for timelike separated insertions remain well-defined. It would be interesting to understand the duality in these cases, and also to find maps analogous to those displayed here for coordinate systems other than Poincar\'e, in both the Euclidean and Lorentzian cases.

There is a superficial similarity of our discussions about the monodromy of OPE blocks with other works that have considered monodromies. Some papers, such as \cite{Banerjee2016,Anous2016,Anous2017}, focus on correlators with large numbers of light operators in the background of two heavy insertions. Monodromy is used to relate the possible OPE channels of the overall correlator. Other papers, such as \cite{Cunha2016,Maloney2017}, use monodromy as a way to pick out different channels of four point functions by switching heavy OPE exchanges with lighter ones. There are two main differences in what we have discussed. First, we are considering a single OPE block, not the full OPE, so the exchanged operators are fixed. All the works mentioned above involve multiple operators, which can fuse in different channels. In contrast the OPE block is a single operator; there is no notion of different fusion channels. Second, we implement sums to conform to the discrete gauge symmetry that is present on the base but not on the cover, which differs from the above works.

\chapter{Interlude}\label{ch:interlude}

In the remainder of this thesis we will shift our focus towards quantum information theory in general, not specifically with a focus on its applications to quantum gravity. We will study the dynamical properties of several measures of quantum information, all of which have been applied in holography but have been more widely used in nearly every branch of quantum physics. We seek to provide analysis for aspects of dynamics in as much generality as possible, so as to leave applications to QFT or holography available, but without discounting the more traditional quantum mechanical uses. Because of the shift in motivation, in this chapter we briefly recount the primary objective of quantum information theory, the quantification of entanglement. Many in-depth reviews of this topic are available elsewhere \cite{Plenio2005,Nielsen2000}.

\section{Entanglement in quantum information theory}

Entanglement is not an observable in quantum theory, but simply a property of a state together with a partition of the theory's Hilbert space. For our purposes it will be sufficient to consider bipartite entanglement, where the Hilbert space $\mathcal{H}$ of the system under consideration is divided into two parts, $\mathcal{H}=\mathcal{H}_A\ot\mathcal{H}_B$. A pure state of the theory $\left | \psi \right\rangle$ is separable if it can be written as a product of states on the subsystems, $\left | \psi \right\rangle = \left | \psi_A \right\rangle\ot \left | \psi_B \right\rangle$. This type of state exhibits no correlations, quantum or classical, between measurements performed on the two subsystems, and hence has no entanglement. If a state cannot be written as a product in any basis, then it is entangled. This definition of entanglement is not practical however, since for any given separable state it may be inconvenient to find a basis that expresses the product nature. Instead we can introduce a measure of entanglement such as the entanglement entropy
\begin{equation}
S(\rho_A)=-\tr_A( \rho_A\log\rho_A), \quad \rho_A=\tr_B(\left | \psi \right\rangle\!\left\langle\psi\right |),
\end{equation}
which is zero for all pure separable states, and positive definite for all entangled states. Instead of searching for a satisfactory basis, one can compute $S(\rho_A)$ in any basis to determine if a state is entangled or not.

For large enough systems, it can still be challenging to compute $\log\rho_A$, since this typically requires a diagonalization of $\rho_A$. It can be easier to work in a preferred basis which diagonalizes both subsystems, usually known as the Schmidt basis,
\begin{equation}\label{eq:Schmidt1}
\left | \psi \right\rangle=\sum_i \sqrt{\la_i} \left | i_A \right\rangle\ot\left | i_B \right\rangle.
\end{equation}
Such a decomposition always exists for pure states, with orthonormal subsystem bases $ \left | i_A \right\rangle$ and $\left | i_B \right\rangle$, and positive real Schmidt coefficients $\la_i$ such that $\sum_i \la_i=1$ for normalized states. One benefit is that the entanglement entropy can be computed simply as $S(\rho_A)=-\sum_i \la_i\log\la_i=S(\rho_B)$. Clearly, separable states have only one non-zero Schmidt coefficient. The number of non-zero coefficients is called the Schmidt rank, and is itself a rudimentary test for entanglement. We also note that $S(\rho_A)$ is maximized when all $\la_i$ are equal, meaning that the reduced density matrix $\rho_A$ is proportional to the identity.

Quantum correlations can only be established in a bipartite system by interacting the two subsystems together. In particular, local unitary operations which act only on one subsystem cannot change the entanglement entropy; from \eqref{eq:Schmidt1} an operation like $U_A\ot\id_B$ with $U_A$ unitary only rotates the $ \left | i_A \right\rangle$ basis, but does not affect the $\la_i$ or $S(\rho_A)$. More generally, entanglement entropy cannot be increased by any local quantum operation $E_A\ot\id_B$ (LO) or by classical communication (CC) between the parties controlling subsystems $A$ and $B$. This singles out LOCC as a useful set of protocol for ordering states according to their amount of entanglement; a state $\rho$ is more entangled than $\si$ if $\rho\to\si$ can be achieved with LOCC, but $\si\to\rho$ cannot. The partial ordering helps to define entanglement as a resource for performing operational tasks, such as quantum teleportation, quantum cryptography, or quantum computation, since states with greater entanglement allow these tasks to be performed better or to a greater extent. These observations will allow us to define what is meant by an entanglement measure, but first there is one additional confounding factor to consider.

When the exact state of the system is unknown, there being some probabilities $p_i$ for the system to be in a number of different states $\left | \psi_i \right\rangle$, it is best described by a density matrix $\rho=\sum_i p_i \left | \psi_i \right\rangle\!\left\langle\psi_i\right |$. Such a state is no longer pure, but mixed, and will exhibit classical correlations in addition to any quantum correlations. Quantities like the entanglement entropy are sensitive to these correlations as well. A state like $\rho=\sum_i p_i\left | \psi_i \right\rangle\!\left\langle\psi_i\right |_A\ot\left | \phi_i \right\rangle\!\left\langle\phi_i\right |_B$ is a mixture of product states, each with no entanglement, yet the overall state has an entropy from the classical distribution, $S=-\sum_i p_i\log p_i$. This already suggests that entanglement entropy has undesirable traits for a mixed state entanglement measure. Even more generally, we could have a mixture of products of mixed states
\begin{equation}\label{eq:separablestate}
\rho=\sum_i p_i \rho^{i}_A\ot\rho^{i}_B,
\end{equation}
 which still has no entanglement, but even larger subsystem entropies. Such states are the most general class of separable states. States which cannot be written in this way are entangled.

With these complications in mind, we can define a number of properties which an entanglement measure $E(\rho)$ should obey. First, $E(\rho)$ should be a positive function on the space of density matrices, and achieve its lower bound of zero if $\rho$ is separable. In addition, $E(\rho)$ should be non-increasing under LOCC\footnote{Some authors use the term \emph{entanglement monotone} for $E(\rho)$ with these properties, referring to monotonicity under LOCC, and insist that entanglement measures for mixed states must reduce to the entanglement entropy for pure states \cite{Plenio2005}.}. Other desirable properties include (sub)additivity, which requires that $E(\rho)$ for a composite system is (less than or) equal to the sum of the entanglement in the subsystems, and convexity which means that $E(\rho)$ is a convex function of the density operator. Some classic examples of entanglement measures include the distillable entanglement \cite{Bennett1996}, and entanglement of formation \cite{Wootters1998}. Notably, entanglement entropy fails to satisfy $(S\circ\tr_B)(\rho)=0$ for mixed separable states, but due to its simplicity for pure states it is still commonly used. 

Two measures in particular will be the focus of our next chapters. First is actually a family of measures introduced briefly before in \eqref{eq:Renyidefinition}, the quantum R\'enyi entropies \cite{Renyi1961},
\begin{equation}\label{eq:Renyidefinition1}
S_\al(\rho_A)=\frac{1}{1-\al}\log \tr_A\rho_A^\al,
\end{equation}
defined for integer values $\al>1$. These constitute an extension of the entanglement entropy, which is obtained as the limit $\lim_{\al\to1} S_\al(\rho_A)$. In general, $S_\al$ is decreasing in $\al$ \cite{Headrick2010}. Individually, they are each positive when acting on density matrices, zero if and only if $\rho$ is overall pure and separable, invariant under local unitaries, non-increasing under LOCC, and for pure states are symmetric $S_\al(\rho_A)=S_\al(\rho_B)$. 

The primary reason to prefer R\'enyi entropies over entanglement entropy alone is that information about the entanglement properties of a state cannot be completely characterized by a single number $S(\rho_A)$. The different measures we have already mentioned are complementary, each having a distinct operational meaning. For instance, two states can have the same entanglement entropy, but if one has less than maximal Schmidt rank, while the other has full Schmidt rank, then only for the latter will the entanglement allow us to express an operation on one subsystem as a related operation on the other subsystem as in \eqref{eq:statecompare} \cite{Cresswell2019a}. This statement can be viewed as a quantum mechanical version of the cyclicity property discussed in the context of the Reeh-Schlieder theorem \cite{Witten2018}. For the R\'enyi entropies, having information about each $S_\al(\rho_A)$ is equivalent to knowing the full eigenvalue distribution of $\rho_A$, the $\la_i$ in the Schmidt decomposition \eqref{eq:Schmidt1} for pure states\footnote{If the dimensionality $d_A$ of $\mathcal{H}_A$ is finite, then only $d_A-1$ R\'enyi entropies are sufficient to determine the spectrum.}. This constitutes much more information than $S(\rho_A)$ alone provides. For pure states, the Schmidt coefficients are the local-unitarily invariant quantities of the state that completely describe its entanglement properties, and the R\'enyi entropies determine them.

The second measure that we will consider is the entanglement negativity, a measure whose origins can roughly be traced
	to 1996, with Peres' Positive Partial Transpose (PPT) condition \cite{Peres1996}: if a
	bipartite state is separable, the transpose taken with respect to
	either subsystem is positive. The PPT condition, stronger and more efficient
	than entropic criteria based on the R\'enyi entropies \cite{Vollbrecht2002},
	was shown by the Horodeckis to be sufficient for the separability
	of $2\otimes2$ and $2\otimes3$ systems \cite{Horodecki1996}. We
	summarize this as follows:
	\begin{theorem*}[Peres-Horodecki Criterion]
		\emph{Let $T_{B}\left(\rho\right)\coloneqq\rho^{T_{B}}\coloneqq \left(\id\otimes T\right)\rho$
			be the }partial transposition map \emph{with respect to system} $B$.
		\emph{Then $\rho\text{ is separable}\implies\rho^{T_{B}}\geq 0.$ Furthermore,
			if $(d_{A},d_{B})\in\{(2,2),(2,3)\}$,
			then $\rho^{T_{B}}\geq0\implies\rho\text{ is separable}$.
		}
		\label{thm:PHC}
	\end{theorem*}
	It is easy to demonstrate the first implication. For a separable state \eqref{eq:separablestate}, the partial transpose acts as
	\begin{equation}
\rho_{\text{sep}}^{T_B}=\sum_i p_i \rho^{i}_A\ot(\rho^{i}_B)^T.
\end{equation}
Now, the transpose map is positive, and in fact acting on each reduced density matrix $\rho_B^i$ preserves its eigenvalues, meaning each $(\rho^{i}_B)^T$ is still positive semidefinite, and hence $\rho^{T_B}$ is as well. In general, transposition is a positive map, but not completely positive, which means the composition we have called partial transposition $T_B=\id\ot T$ is not always positive. For a general state $\rho$, if $\rho^{T_B}$ is found to have negative eigenvalues then by the above argument $\rho$ cannot be separable. However, for systems larger than $2\ot 3$, there exist states which are entangled, yet have $\rho^{T_B}\geq 0$ so that the partial transposition test is inconclusive. States of this type are said to be bound entangled.

	From the Peres-Horodecki Criterion emerges the negativity, $\mathcal{N}$,
	of $\rho$\emph{ }\cite{Zyczkowski1998,Vidal2001}, which encodes
	the degree to which the partial transpose of $\rho$ is negative. Negativity is defined as
	\begin{equation}
	\mathcal{N}\left(\rho\right)\coloneqq\sum_{\lambda<0}\left|\lambda\left(\rho^{T_{B}}\right)\right|,
	\label{eq:negativity-eigvals}
	\end{equation}
	where the $\lambda$'s are the eigenvalues of $\rho^{T_{B}}$. It is worth noting that $\rho^{T_B}=(\rho^{T_A})^T$, so that the choice of transposing $B$ rather than $A$ is not significant. Negativity is a positive function on density matrices, zero for all separable states, monotonic under LOCC, and convex but it is not additive \cite{Plenio2005a}. Although there exist entangled states for which $\mathcal{N}$ vanishes (it is not a faithful measure), negativity has an important advantage over other measures in that it is easily computable, even for mixed states \cite{Huang2014}. For some
	purposes it is useful to define the \emph{logarithmic negativity},
	$E_{\mathcal{N}}\left(\rho\right)\coloneqq\log_{2}\left[2\mathcal{N}\left(\rho\right)+1\right]$, which is monotonic and additive but not convex  \cite{Plenio2005a}. Relationships between the logarithmic negativity, distillable entanglement, entanglement cost, and teleportation capacity have been demonstrated in the literature \cite{Vidal2001, Audenaert2003}.

In the following chapters we will study the evolution of R\'enyi entropies and negativity perturbatively to shed light on how entanglement can be generated in dynamical systems. Our goal is to keep the analysis as broadly applicable as possible. For the R\'enyi entropies, we refrain from specifying any particular Hamitonian to generate the dynamics, instead looking at a general class of initial states to extract universal behaviour. For negativity, no restriction on the type of dynamics is required. Our main outcomes are mathematical tools for handling derivatives with respect to constrained matrices, in particular the Hermitian density matrices, which allows us to write analytical expressions for the time derivatives of negativity.

\chapter{Universal timescale for R\'enyi entropies}\label{ch:Renyi}

\emph{This chapter is based on the paper \cite{Cresswell2017a} published in Phys. Rev. A.}
\section{Introduction}

Composite quantum systems exhibit correlations among subsystems which 
cannot be explained in terms of classical probabilities. For pure states, these 
quantum correlations are known as entanglement. In this chapter, we study how 
entanglement is generated by the mutual interactions among 
subsystems as the overall state evolves in time. 

The time evolution of entanglement has become a focus in a variety of 
research fields. Its early study in quantum optical systems 
\cite{Phoenix1988,Gea-Banacloche1990} has bloomed into a major area of 
research in many-body and condensed-matter systems 
\cite{Calabrese2004,Calabrese2009,Casini2009,Amico2008,
Laflorencie2015}, 
and conformal field theories dual to theories of quantum gravity 
\cite{Maldacena1999,Ryu2006,Hubeny2007}. For some classes of systems, 
general features have been found, including scaling laws 
\cite{Eisert2006,Bravyi2006} and generic linear growth 
\cite{Calabrese2005,Liu2014a,Hartman2013,Bianchi2018}.

The growth of entanglement is especially important in experimental systems 
where entanglement between the system and its environment leads to 
decoherence \cite{Zurek2003}. A complete understanding of the evolution of 
entanglement requires solving the dynamics of the overall state. This is often 
not feasible, including for decoherence where the Hamiltonian describing 
interactions with the environment is not known explicitly.

It is therefore interesting to ask what aspects of entanglement growth, if any, 
are shared by all quantum systems. Broad statements can be made 
in this direction with minimal assumptions about system 
dynamics by relying on special initial conditions instead.

To begin, bipartite entanglement between subsystems must be defined 
with respect to a partition of the system's degrees of 
freedom, represented as a fixed factorization of the Hilbert 
space $\mathcal{H}=\mathcal{H}_A\ot \mathcal{H}_B$. The 
Hamiltonian for the full system can be expressed as
\begin{equation}\label{eqHam}
H=\sum_n A_n\ot B_n,
\end{equation}
where each $A_n$ is an operator acting on subsystem $\mathcal{H}_A$, and each 
$B_n$ acts on $\mathcal{H}_B$. Any number of terms may be 
included as long as $H$ is Hermitian. Since the algebra of 
operators acting on $\mathcal{H}$ is isomorphic to the tensor product of 
subsystem algebras, any Hamiltonian can be 
represented this way \cite{Zanardi2001}.

Recently it was shown by Yang \cite{Yang2017a} (see also the earlier work \cite{Kim1996}) that starting from a 
pure, unentangled state
\begin{equation}\label{eqSt}
\ket{\Psi(0)}=\ket{\psi(0)}_A\ot\ket{\psi(0)}_B,
\end{equation}
 the growth of entanglement under the unitary evolution generated by 
 \eqref{eqHam} is characterized by a universal timescale,
 \small
\begin{equation}\label{eqTS}
T_{\mathrm{ent}}=
\left[\sum_{n,m}\left(\langle A_nA_m \rangle{-}\langle A_n \rangle\langle A_m 
\rangle\right)\left(\langle B_nB_m \rangle{-}\langle B_n \rangle\langle B_m 
\rangle\right)\right]^{-\tfrac{1}{2}}.
\end{equation}
\normalsize
Here the expectation values are taken in the initial state. The timescale is 
universal in the sense that it takes this form for any quantum system that 
satisfies the requirements \eqref{eqHam} and \eqref{eqSt}. The 
entanglement timescale was derived by studying one particular measure of 
the entanglement between subsystems $A$ and $B$, namely, the purity 
$P(\rho_A)=\tr_A \rho_A^2$ of the reduced density matrix $\rho_A=\tr_B 
\rho$. By the assumption \eqref{eqSt}, the purity is initially maximal so that its 
dynamics are governed at lowest order in $t$ by $d^2P/dt^2$. The second 
derivative is proportional to $T_{\mathrm{ent}}^{-2}$ which is entirely 
determined by the expectation values of the interaction Hamiltonian operators 
in the initial state. 

In this chapter, we show that the same entanglement timescale \eqref{eqTS} 
governs the growth of entanglement as measured by the entire family of 
quantum R\'enyi entropies \eqref{eq:Renyidefinition1}. As a family, the R\'enyi entropies 
provide complete information about the eigenvalue distribution of the reduced 
density matrix $\rho_A$, and hence completely characterize the 
entanglement in an overall pure, bipartite state \cite{Li2008,Headrick2010}. 
Therefore, the entanglement timescale \eqref{eqTS} is a universal feature of 
bipartite entanglement. 

The most common measure of entanglement, the entanglement entropy 
{$S(\rho_A)\ {=}-\tr_A( \rho_A\ln \rho_A)$}, corresponds to the $\al\to 1$ limit of 
\eqref{eq:Renyidefinition1}. Its second time derivative can be obtained by an analytic 
continuation in $\al$ from our general results for $\al\geq2$ after which 
\eqref{eqTS} appears with a logarithmically divergent prefactor, reflecting the 
sensitivity of $S(\rho_A)$ to small eigenvalues of the density matrix. We 
provide an example of these results by working with the Jaynes-Cummings 
model \cite{Shore1993}.
 
 In Sec. \ref{sec:3rdOrder} we extend the leading order analysis to the next-to-leading order. At this order, several additional terms appear in the derivatives of the R\'enyi entropies with $\al\geq 3$. One might expect at first glance that the $\al=2$ case should behave differently from the rest of the family. However, we find surprising cancellations between all of the additional terms which leaves the evolution identical for all R\'enyi entropies at this order as well. The next-to-leading order behaviour is characterized by a second universal timescale for all $\al$. This pattern cannot continue to all orders in perturbation theory because the R\'enyi entropies are truly distinct functions. It is likely to break at fourth order around $t=0$, since there are additional classes of new terms which arise for $\al=3$ and $\al\geq 4$, and it is difficult to imagine perfect cancellations between them all. We remark on some of the properties of the third order timescale, and how it relates to the leading order timescale.

Notably, the entanglement timescale can be computed without the need to 
solve for the dynamics of the system. For a given experimental preparation of 
an unentangled state, our results provide an easily calculable estimate of 
when entanglement will become significant. Advances in the optical control of 
atoms have led to the first direct measurement of a R\'enyi entropy in a 
many-body system, and subsequently to measurements of its growth 
\cite{Daley2012,Schachenmayer2013,Islam2015a,Kaufman2016,Elben2018}. 
We return to these measurements for comparison to the entanglement 
timescale in Sec. \ref{dis}.

\section{The entanglement timescale for R\'enyi entropies}\label{sec2}

In this section, we derive an entanglement timescale for the 
R\'enyi entropies of a pure bipartite state \eqref{eqSt} evolving under a 
general Hamiltonian \eqref{eqHam}. Initially the subsystems are pure, $
\rho_A\ {=}\ \rho_A^2$, because \eqref{eqSt} is separable, and therefore 
$S_\al(\rho_A)|_{t=0}=\frac{1}{1-\al} \ln \tr_A \rho_A^\al |_{t=0} =0$. 
As the state evolves, the interactions between subsystems will generate 
entanglement. Starting at a minimum of $S_\al$, the first time derivative is 
initially zero. We will calculate the second derivative to obtain a Taylor 
expansion around $t=0$ of the form 
\begin{equation}
S_\al(\rho_A) =C_\al\frac{t^2}{T_{\mathrm{ent}}^2}+O(t^3).
\end{equation}
We will find that the entanglement timescale 
$T_{\mathrm{ent}}$ takes the same form for all R\'enyi entropies, with 
$C_\al$ a constant.

Since the R\'enyi entropies are initially minimal, their first derivatives must 
vanish. We find 

\begin{equation}
\frac{d}{dt}S_\al(\rho_A)=\frac{\al}{1-\al}
\left(\tr_A\rho_A^\al\right)^{-1}\tr_A\left[(\tr_B\rho)^{\al-1}{\tr_B}{\left(\parder{\rho
}{t}\right)}\right].
\end{equation}
Note that in general, $\left[{\tr_B}{\left(\pa\rho/\pa t\right)},
\tr_B\rho\right]\ {\neq}\ 0$. However, inside the $A$ trace, we can cyclically 
permute each term produced by the derivative into a common ordering as 
shown. Using the von Neumann equation $\pa\rho/\pa t =-i [H,\rho]$ with $
\hbar=1$ and using \eqref{eqHam} and \eqref{eqSt} in the $t=0$ limit, we find 

\begin{equation}
\frac{d}{dt}S_\al(\rho_A)|_{t=0}=\frac{i\al}
{\al-1}{\left(\tr_A\rho_A^\al\right)^{-1}}{\sum_n}\tr_B(\rho_B B_n)
\tr_A\big(\rho_A^{\al-1}A_n\rho_A-\rho_A^\al A_n\big)=0.
\end{equation}

The leading order of the time evolution comes from the second derivative,
\begin{align}\label{eq2nd}
\begin{aligned}
\frac{d^2}{dt^2}S_\al(\rho_A)\ =\frac{1}{1-\al}
\bigg(\left(\tr_A\rho_A^\al\right)^{-1}\tr_A\bigg\{\frac{d^2}{dt^2}
[\tr_B\rho(t)]^\al\bigg\}-\left(\tr_A\rho_A^\al\right)^{-2}\bigg\{\tr_A\frac{d}{dt}
[\tr_B\rho(t)]^\al\bigg\}^2\bigg).
\end{aligned}
\end{align}
 The second term vanishes when the $t\to0$ limit is taken; this was the result 
of the first derivative calculation. We are left with the first term of 
\eqref{eq2nd} for which we find
\begin{align}\label{eq2}
\begin{aligned}
\tr_A\bigg\{\frac{d^2}{dt^2}
[\tr_B\rho(t)]^\al\bigg\}=\al\tr_A\left[(\tr_B\rho)^{\al-1}\tr_B\parder{^2\rho}{t^2}+
\sum_{\be=0}^{\al-2}(\tr_B\rho)^{\be}\tr_B\parder{\rho}{t}(\tr_B\rho)^{\al-2-\be}
\tr_B\parder{\rho}{t}\right].
\end{aligned}
\end{align}
 The $\be$ sum keeps track of the non-commuting factors which cannot be 
permuted into a common ordering. Applying the von Neumann equation 
leads to 
\begin{align}\label{eq3}
\frac{d^2}{dt^2}&S_\al(\rho_A)\big|_{t=0}\, =\frac{\al}
{\al-1}\left(\tr_A\rho_A^\al\right)^{-1}\sum_{n,m}\bigg[\tr_B(B_nB_m\rho_B)
\tr_A\left(2A_nA_m\rho_A^\al-2A_m\rho_AA_n\rho_A^{\al-1}\right)\\
&+\tr_B(B_n\rho_B)\tr_B(B_m\rho_B){\sum_{\be=0}
^{\al-2}}{\tr_A}{\left(2\rho_A^{\be+1}A_n\rho_A^{\al-\be-1}A_m-\rho_A^\be 
A_n\rho_A^{\al-\be}A_m-\rho_A^{\be+2}A_n\rho_A^{\al-2-\be}A_m\right)}
\bigg].\nonumber
\end{align}

Before simplifying \eqref{eq3} for general $\al$, it is useful to look at the 
unique case of $\al=2$ which corresponds to the purity studied in 
\cite{Yang2017a}. In this case, the $\be$ sum contains only a single term. 
Using the assumption of purity at $t=0$ allows us to write
\begin{align}\label{eq5}
\begin{aligned}
\frac{d^2}{dt^2}S_2(\rho_A)\big|_{t=0} =4\sum_{n,m}
\left[\tr_B(B_nB_m\rho_B)-\tr_B(B_n\rho_B)\tr_B(B_m\rho_B)\right]
\left[\tr_A(A_nA_m\rho_A)-\tr_A(A_m\rho_AA_n\rho_A)\right].
\end{aligned}
\end{align}
Note that we have \emph{not} assumed that $[A_n,A_m]=0$. Instead, we 
have used the symmetry of\\ $\tr_B(B_n\rho_B)\tr_B(B_m\rho_B)$ in the $n$, $m$ indices to exchange $A_n$ and $A_m$. Indeed, \eqref{eq5} exactly 
matches the main result of \cite{Yang2017a} when we account for the 
difference in the definitions of the purity and R\'enyi entropy. Defining the $
\al$-purity, $P_\al(\rho_A)=\tr_A\rho_A^\al$, we have under our assumptions $
\frac{d^2}{dt^2}S_\al(\rho_A)|_{t=0}=\frac{1}{1-\al}\frac{d^2}{dt^2}
P_\al(\rho_A)|_{t=0}.$

Returning to the general case, it is possible to greatly simplify \eqref{eq3} by 
using the idempotency of $\rho_A(t=0)$, and $\rho_A^0=\id_A$ where $
\id_A$ is the identity operator for subsystem $A$. The special case of $
\rho_A^0=\id_A$ only occurs in the $\be$ sum when $\be$ takes on its 
extreme values of 0 and $\al-2$. Each other term in the sum vanishes. The 
general result for $\al>2$ is
\begin{align}\label{eq6}
\frac{d^2}{dt^2}S_{\al}(\rho_A)\big|_{t=0} &=\frac{2\al}{\al-1}\sum_{n,m}
\left[\tr_B(B_nB_m\rho_B)-\tr_B(B_n\rho_B)\tr_B(B_m\rho_B)\right]
\left[\tr_A(A_nA_m\rho_A)-\tr_A(A_m\rho_AA_n\rho_A)\right]\nonumber \\
&=\frac{2\al}{\al-1}\sum_{n,m}\left[\langle B_nB_m \rangle-\langle B_n 
\rangle\langle B_m \rangle\right]\left[\langle A_nA_m \rangle-\langle A_n 
\rangle\langle A_m \rangle\right]=\frac{2\al}{\al-1}T_{\mathrm{ent}}^{-2},
\end{align}
 where we have used the simplification $
\tr_A(A_m\rho_AA_n\rho_A)=\tr_A(A_m\rho_A)\tr_A(A_n\rho_A)$ for pure $
\rho_A$ as shown in \cite{Yang2017a}.

Equation \eqref{eq6} is our main result and shows that the second derivative 
of every R\'enyi entropy for $\al>2$ is of the same universal form as the $
\al=2$ case studied previously. In fact, the coefficient incorporates the $\al=2$ 
case in Eq. \eqref{eq5} as well. The only remaining case is $\al=1$, which we 
turn to now.

The entanglement entropy $S(\rho_A)=-\tr_A(\rho_A\ln\rho_A)$ is the most 
widely used entanglement measure in the literature. It corresponds to the $
\al\to1^+$ limit of $S_\al(\rho_A)$ after an analytic continuation in $\al$ 
\cite{Calabrese2004,Calabrese2009}. Inserting $\al=1$ at intermediate steps 
in the derivation leading to \eqref{eq6} produces ill-defined quantities since 
the density matrix $\rho_A(t=0)$ is pure, and therefore singular. 
Nevertheless, we emphasize that inverse powers of $\rho_A$ do not appear 
in the final result \eqref{eq6}. The prefactor $2\al/(\al-1)$ can be analytically 
continued in $\al$ and is analytic away from the simple pole at $\al=1$. 
Taking the limit of $2\al/(\al-1)$ as $\al\to1^+$ along the real axis shows that 
$d^2S(\rho_A)/dt^2|_{t=0}$ is proportional to the entanglement timescale 
with a divergent prefactor. This reflects the entanglement entropy's sensitivity 
to small eigenvalues of $\rho_A$ via the logarithm.

 To make this point more clear, let $p_i(t)$ be the eigenvalues of $\rho_A$ 
such that $p_1(0)=1$ and $p_j(0)=0$ ($j\neq 1$). Then the second derivative 
of the entanglement entropy, $S(\rho_A)=-\sum (p_i \ln p_i)$, in the $t\to 0$ 
limit is
\begin{equation}\label{eqEET}
\frac{d^2 S}{dt^2}=-\frac{d^2p_1}{dt^2}-\sum_{j\neq1}\left[( \ln 
p_j+3)\frac{d^2p_j}{dt^2}\right].
\end{equation}
 Generically, $\lim_{t\to0}(d^2p_j/dt^2) \ln p_j$ is divergent since $d^2p_j/
dt^2$ is not required to be zero initially. Still, the divergence of $d^2 S/dt^2$ 
at $t=0$ does not imply that the entanglement entropy itself diverges; on the 
contrary, $S(\rho_A)$ is strictly bounded above by the dimension of the 
Hilbert space of subsystem $A$. Rather, $d^2 S/dt^2$ appears in the Taylor 
series as the coefficient of $t^2$ which tames the logarithmic divergence. It 
should be noted that higher derivatives also diverge logarithmically at $t=0$, 
but are suppressed by higher powers of $t$.

\section{Example - Jaynes-Cummings model}

Equation \eqref{eqEET} shows that the divergence of $d^2S/dt^2$ at $t=0$ 
for an initially pure product state found in \eqref{eq6} is not an artifact of the 
analytic continuation in $\al$. This is the generic behavior of the 
entanglement entropy for an initially separable state. To explore the physical 
significance of the entanglement timescale, and to check the divergence of 
$d^2S/dt^2|_{t=0}$, we work with the Jaynes-Cummings model (JCM) of a 
two-level atom interacting with a quantized radiation field 
\cite{Jaynes1963,Shore1993}. This system has been extensively studied in 
quantum optics because of its interesting entanglement properties 
\cite{Phoenix1988,Quesada2013} and quantum revivals 
\cite{Eberly1980,Pimenta2016}. In this section, we calculate the entanglement 
timescale for initially separable states, first by finding an analytic solution for 
the R\'enyi entropies at all times, and then by studying the expectation values 
of the interaction terms in the initial state as dictated by \eqref{eq6}. We 
explicitly show that the divergence of $d^2S/dt^2|_{t=0}$ is only logarithmic.

In the rotating-wave approximation, the JCM Hamiltonian is \cite{Shore1993}
\begin{equation}
\frac{H}{\hbar}=\frac{\om_0}{2}\si_z+\om a^\dag a+\la(a^\dag \si_-+a\si_+).
\end{equation}
\n Here, $\om_0$ is the atomic transition frequency, $\om$ is the characteristic 
field frequency, and $\la$ is a coupling constant. For simplicity, we impose the 
resonance condition $\om=\om_0$ and set $\hbar=1$. The Pauli operators 
can be written in terms of the atomic ground state $\ket{g}$ and excited state 
$\ket{e}$ as $\si_z=\ket{e}\bra{e}-\ket{g}\bra{g}$, $\si_-=\ket{g}\bra{e}$, and $
\si_+=\ket{e}\bra{g}$. The field mode has a Fock basis $\ket{n}$ on which the 
creation and annihilation operators $a^\dag$, $a$ act in the usual way. Notice 
that this Hamiltonian is of the assumed product form \eqref{eqHam} and is 
time independent.

Let the overall initial state be the product of an arbitrary atomic state 
$\ket{\psi}_A=C_g\ket{g}+C_e\ket{e}$ and field state $\ket{\psi}_F{=}\sum_{n=0}
^\infty C_n \ket{n}$. Then the overall state at any time is 
\cite{Gea-Banacloche1990}
\small
\begin{align}\begin{aligned}\label{eqjcmstate}
\ket{\Psi(t)}&={\sum_{n=0}^\infty}\{[C_e{C_n} \cos( {\la} \sqrt{n+1} t 
) {-}iC_g C_{n+1}\sin( {\la}\sqrt{n+1} t )] \ket{e}\\
 &\ \ {+}[ -iC_eC_{n-1}\sin( {\la}\sqrt{n}t)+C_g C_n\cos( {\la}\sqrt{n}t 
)]\ket{g}\}\ket{n},
\end{aligned}\raisetag{\baselineskip}\end{align}
\normalsize
 which is entangled for most times. Since the exact solution for the state is 
available, the R\'enyi entropies can be calculated directly for either 
subsystem after a partial trace. When the atom is initially excited $(C_e{=}\ 1,\ 
C_g{=}\ 0)$,
\small
\begin{align}\begin{aligned}\label{eqjcme}
\frac{d^2}{dt^2}S_\al(\rho_A)\bigg|_{t=0}&=\frac{2\al}{\al-1} {\lambda ^2}\left[ \sum_{n=0}^{\infty } (n+1) \left| C_n\right| ^2\right. - \left. {\sum _{n,m=0}^{\infty}} \sqrt{m+1} \sqrt{n+1} C_{n+1}^* C_{n} C_{m+1}C_{m}^*\right].
\end{aligned}\end{align}
\normalsize
 For comparison, if the atom is initially in the ground state, then the result in 
\eqref{eqjcme} changes slightly by the replacement $\left| C_n\right| ^2\to\left| 
C_{n+1}\right| ^2$ in the first sum.

\begin{figure*}
    \centering
    \begin{subfigure}[b]{0.49\textwidth}
        \includegraphics[width=0.8\textwidth]{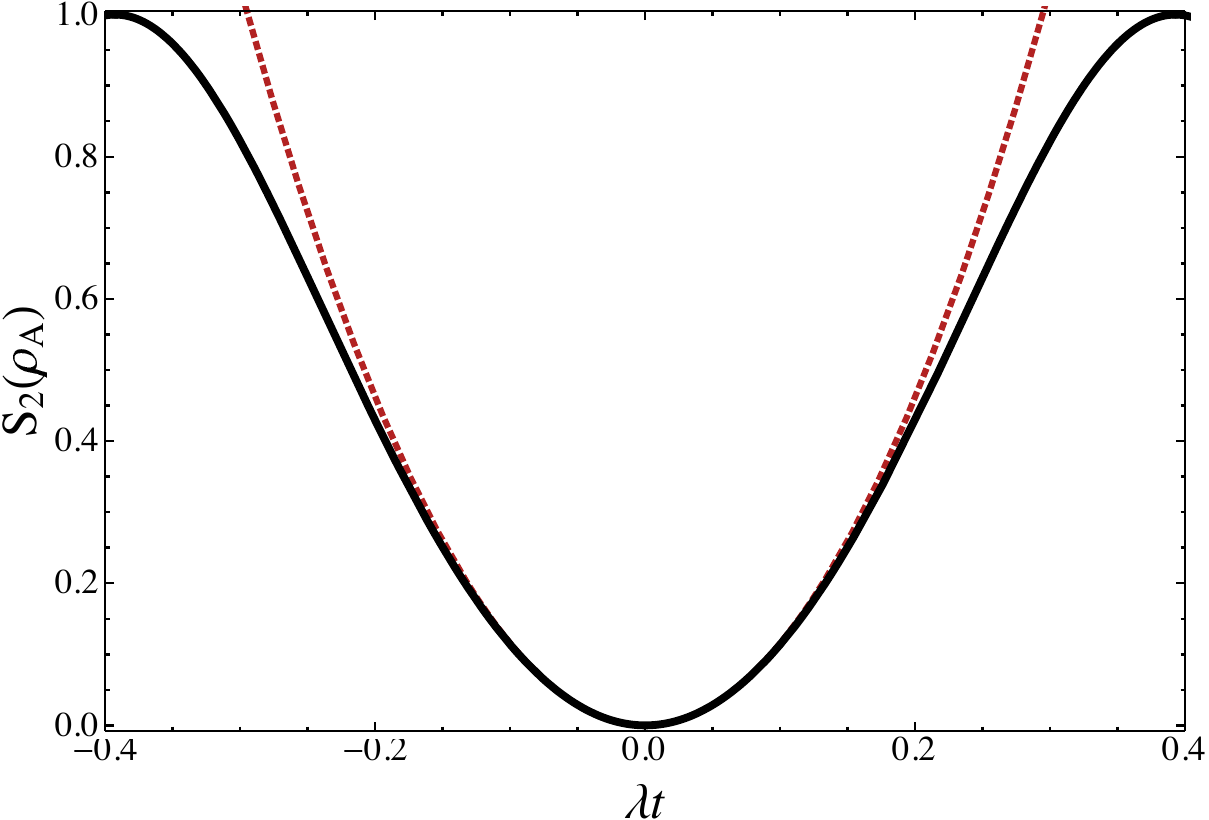}
        \label{fig:focka}
    \end{subfigure}
    \begin{subfigure}[b]{0.49\textwidth}
        \includegraphics[width=0.8\textwidth]{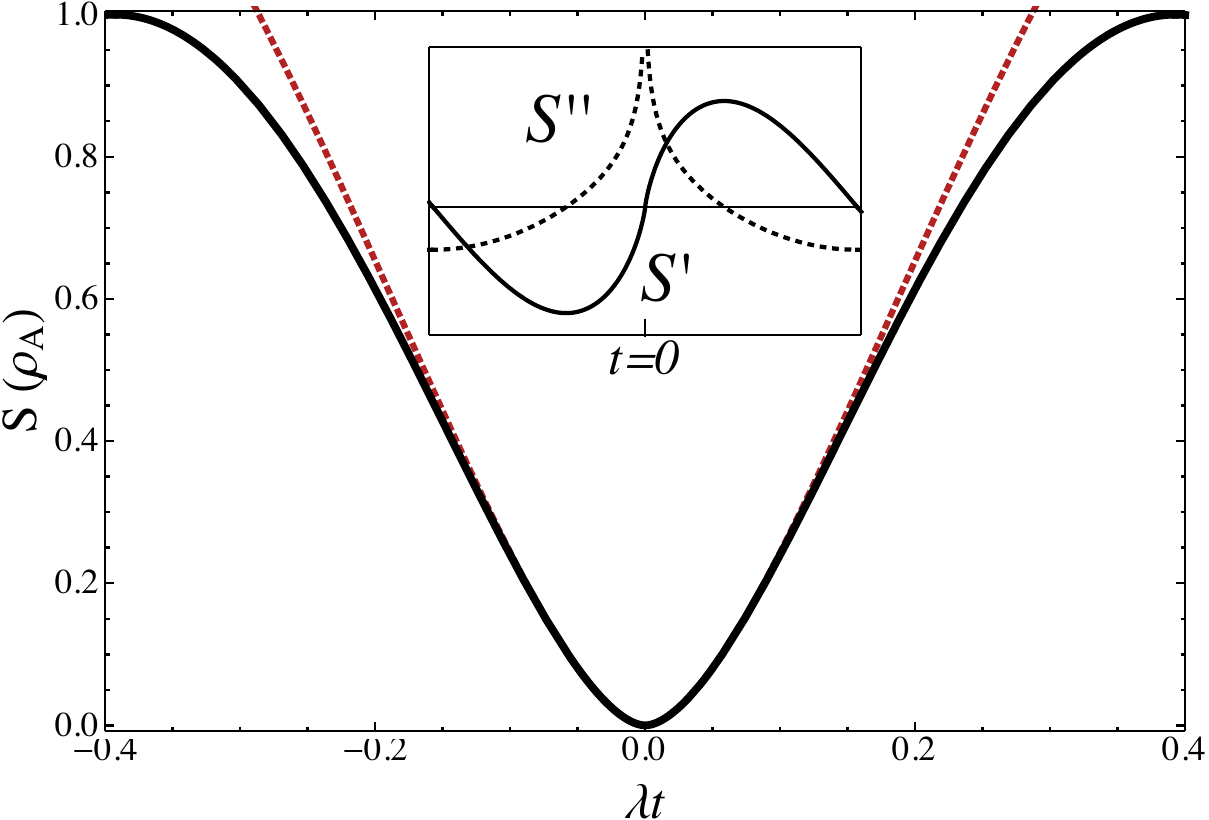}
         \label{fig:fockb}
    \end{subfigure}
    \caption[Entropy growth from Fock state in Jaynes-Cummings model]{
  (a) $S_2(\rho_A)$ for the Fock state with $N=3$ and 
        $C_e=1$ is sinusoidal and $C^\infty$ smooth. $S_2$ is compared to 
        the quadratic approximation with 
        timescale $\la T_{\mathrm{ent},e}=1/4$ (dashed red line). (b) $S(\rho_A)$ 
        for the same state is differentiable, but $d^2 S/dt^2$ is discontinuous 
        at $t=0$ (inset, dashed line). Units of $\ln(2)$ are used in all figures. 
         \normalsize	
        }
        \label{fig:fock}
\end{figure*}

The entanglement timescale can alternatively be computed from the 
Hamiltonian and initial state by using the definition in \eqref{eq6}. This is 
much simpler because it does not require solving for the time evolution of the 
system. When the atom is initially excited, the only nonzero term in 
\eqref{eq6} is
\small
\begin{align}\label{eqTe}\begin{aligned}
T_{\mathrm{ent},e}^{-2}&=\la^2(\langle a a^\dag  \rangle-\langle a 
\rangle\langle a^\dag \rangle)(\langle \si_+ \si_- \rangle-\langle \si_+ 
\rangle\langle \si_-\rangle)\\
&=\lambda ^2\left[ \sum _{n=0}^{\infty } (n+1) \left| C_n\right| ^2\right. \left. - \sum 
_{n,m=0}^{\infty } \sqrt{m+1} \sqrt{n+1} C_{n+1}^* C_{n} C_{m+1}C_{m}
^*\right]\geq1.
\end{aligned}\end{align}
\normalsize
 Similarly for the ground-state case, we find a single nonzero term,
\begin{equation}
T_{\mathrm{ent},g}^{-2}=\la^2(\langle a^\dag a\rangle-\langle a^\dag 
\rangle\langle a \rangle)(\langle \si_- \si_+ \rangle-\langle \si_- 
\rangle\langle \si_+\rangle)\geq0,
\end{equation}
which~is~like~{\eqref{eqTe}}~but~with~${| C_n|^2}{\to}{| C_{n{+}1}| 
^2}$~in~the~first~sum.

The growth of entanglement is always controlled by the strength of the 
coupling $\la$ between subsystems. Indeed, it was pointed out in early 
studies of the JCM that $\la^{-1}$ is proportional to the time period over 
which the reduced states remain approximately pure 
\cite{Gea-Banacloche1990}. The positivity of R\'enyi entropies requires that 
$T_{\mathrm{ent}}^{-2}$ is positive. This is ensured by the results of 
\cite{Yang2017a}, but can be seen here as a consequence of the 
Cauchy-Schwarz inequality which implies 
$\langle a^\dag a\rangle\geq\langle a^\dag \rangle\langle a \rangle$, etc.

From these general expressions, we can easily examine the growth of 
entanglement for some common field states. Consider when the field is 
initially in a Fock state, $\ket{\psi}_F=\ket{N}$. For the initially excited state, we 
find $T_{\mathrm{ent},e}=({\la\sqrt{N+1}})^{-1}$ and for the ground state, 
$T_{\mathrm{ent},g}=({\la\sqrt{N}})^{-1}$. Figure \ref{fig:fock} shows 
$S_2(\rho_A)$ and $S(\rho_A)$ for $C_e=1,\ C_g=0$, and $N=3$, 
along with the quadratic timescale approximation. Whereas $S_\al(\rho_A)$ for 
$\al\geq2$ is $C^\infty$ smooth in this example, we see that 
$d^2 S(\rho_A)/dt^2$ diverges at $t=0$ as expected, while $dS(\rho_A)/dt$ 
is continuous at $t=0$.

\begin{figure*}
    \centering
    \begin{subfigure}[b]{0.49\textwidth}
        \includegraphics[width=0.8\textwidth]{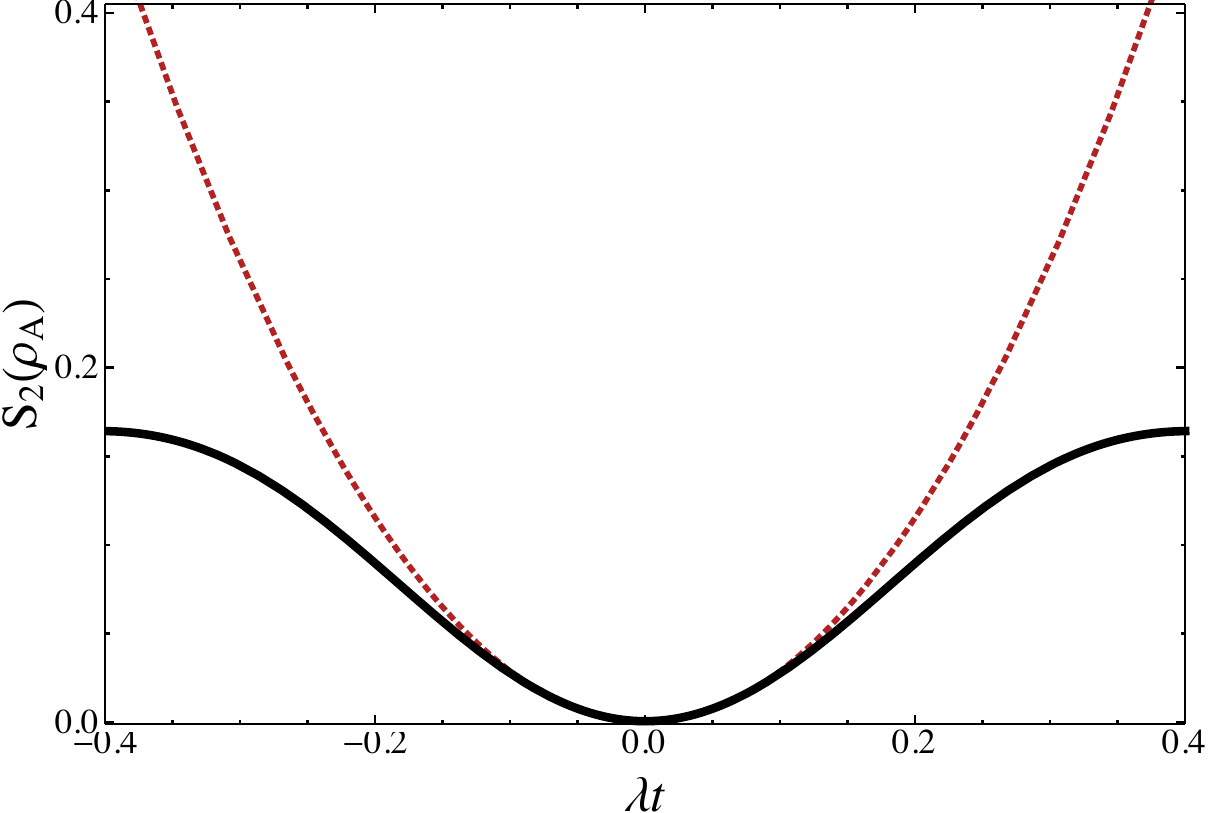}
         \label{fig:cohEa}
    \end{subfigure}
    \begin{subfigure}[b]{0.49\textwidth}
        \includegraphics[width=0.8\textwidth]{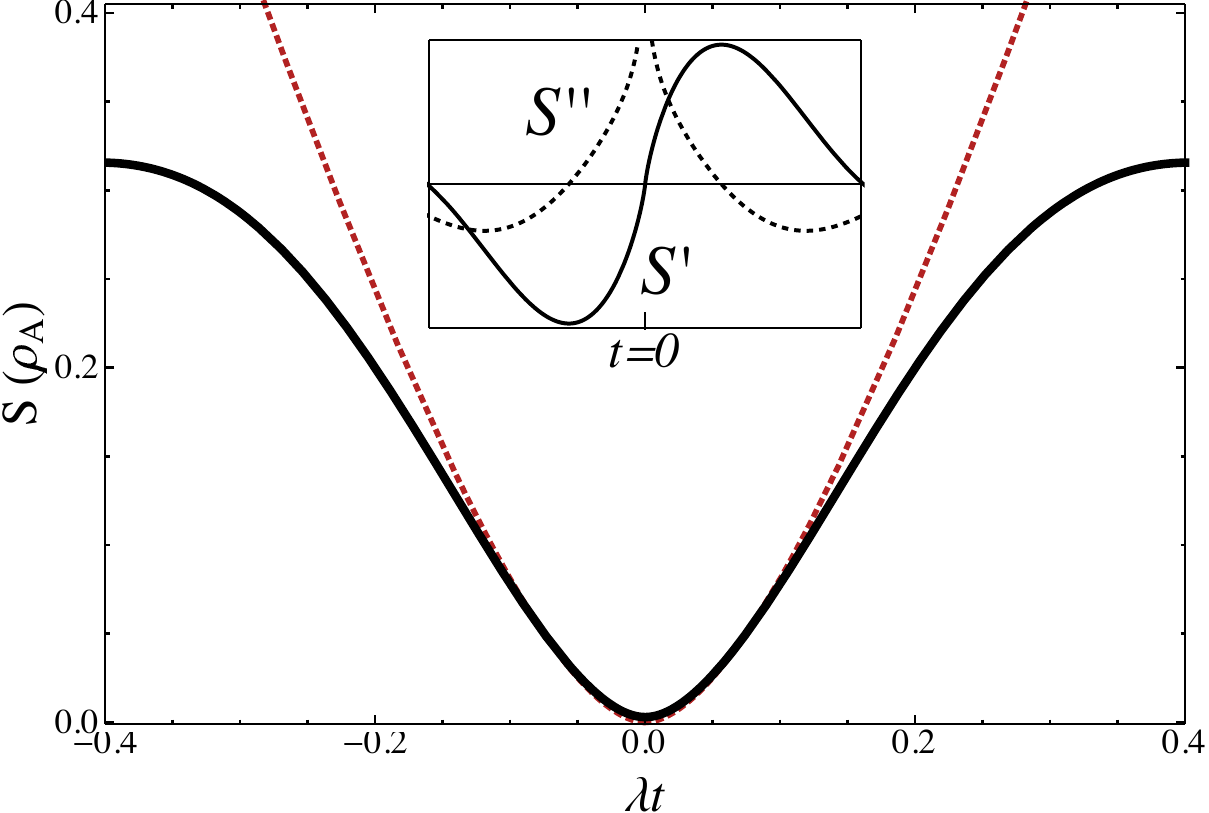}
         \label{fig:cohEb}
    \end{subfigure}
    \caption[Entropy growth from coherent state in Jaynes-Cummings model]{\small (a) $S_2(\rho_A)$ for the coherent state with $\nu=3$ and 
        $C_e=1$. The small-$t$ behavior is independent of $\nu$ and described 
        by the quadratic timescale $\la T_{\mathrm{ent},e}=1$ (dashed red line). 
        (b) $S(\rho_A)$ 
        for the same state is differentiable, but $d^2 S/dt^2$ is discontinuous 
        at $t=0$ (inset dashed line).
    \normalsize}\label{fig:cohE}
\end{figure*}

 Instead, if the field starts in a coherent state,
\begin{equation}
\ket{\psi}_F=e^{-\tfrac{1}{2}|\nu|^2}\sum_{n=0}^\infty \frac{\nu^n}{\sqrt{n!}}
\ket{n},\quad a\ket{\psi}_F=\nu\ket{\psi}_F,
\end{equation}
 then the excited state timescale is $T_{\mathrm{ent},e}{=}\ {1/\la}$, whereas for 
the ground state, $T_{\mathrm{ent},g}^{-1}{=}\ 0$. Notably, these timescales are 
independent of $\nu$. Figure \ref{fig:cohE} shows $S_2(\rho_A)$ and 
$S(\rho_A)$ for the coherent state with $\nu\ {=} \ 3$ and $C_e\ {=}\ 1$,~$C_g\ {=}\ 0$. 
Once again, $d^2 S(\rho_A)/dt^2$ diverges at $t\ {=}\ 0$, while $dS(\rho_A)/dt$ is continuous at $t=0$. 

\begin{figure*}
    \centering
    \begin{subfigure}[b]{0.49\textwidth}
        \includegraphics[width=0.8\textwidth]{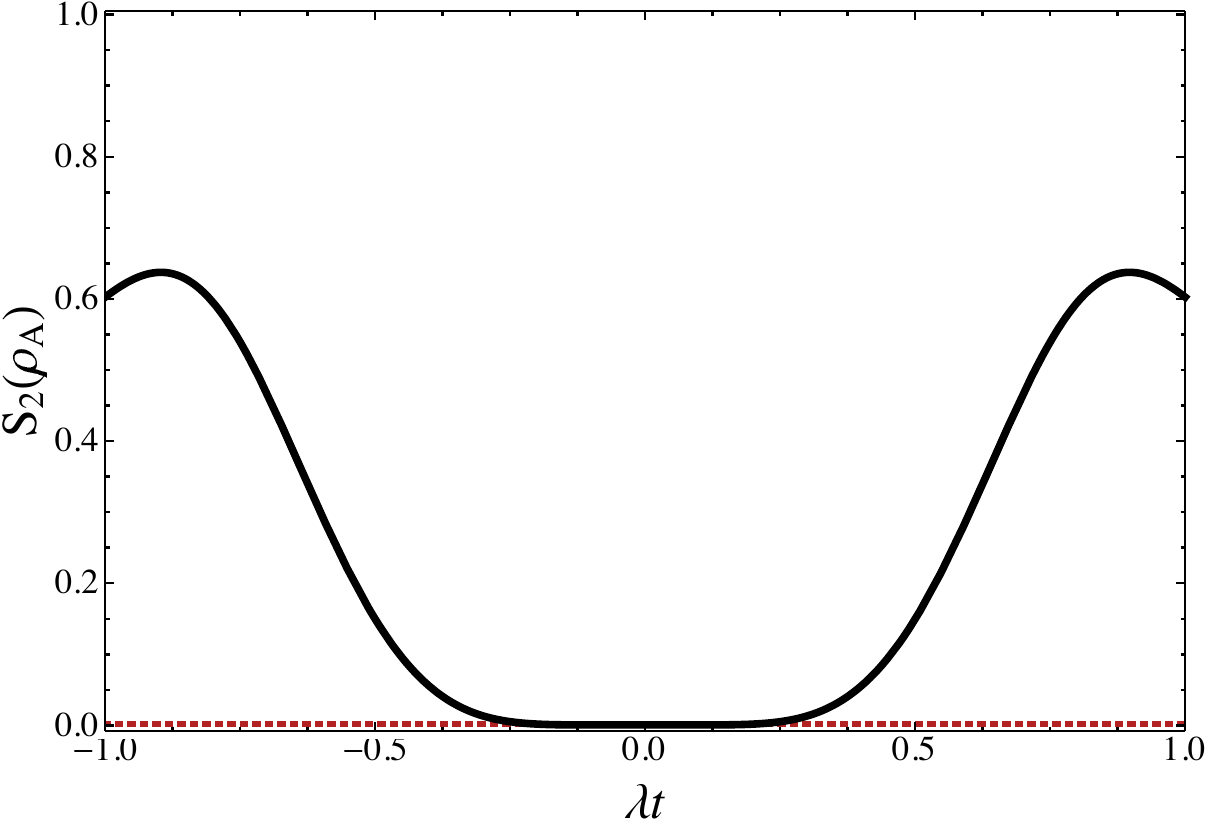}
         \label{fig:cohGa}
    \end{subfigure}
    \begin{subfigure}[b]{0.49\textwidth}
        \includegraphics[width=0.8\textwidth]{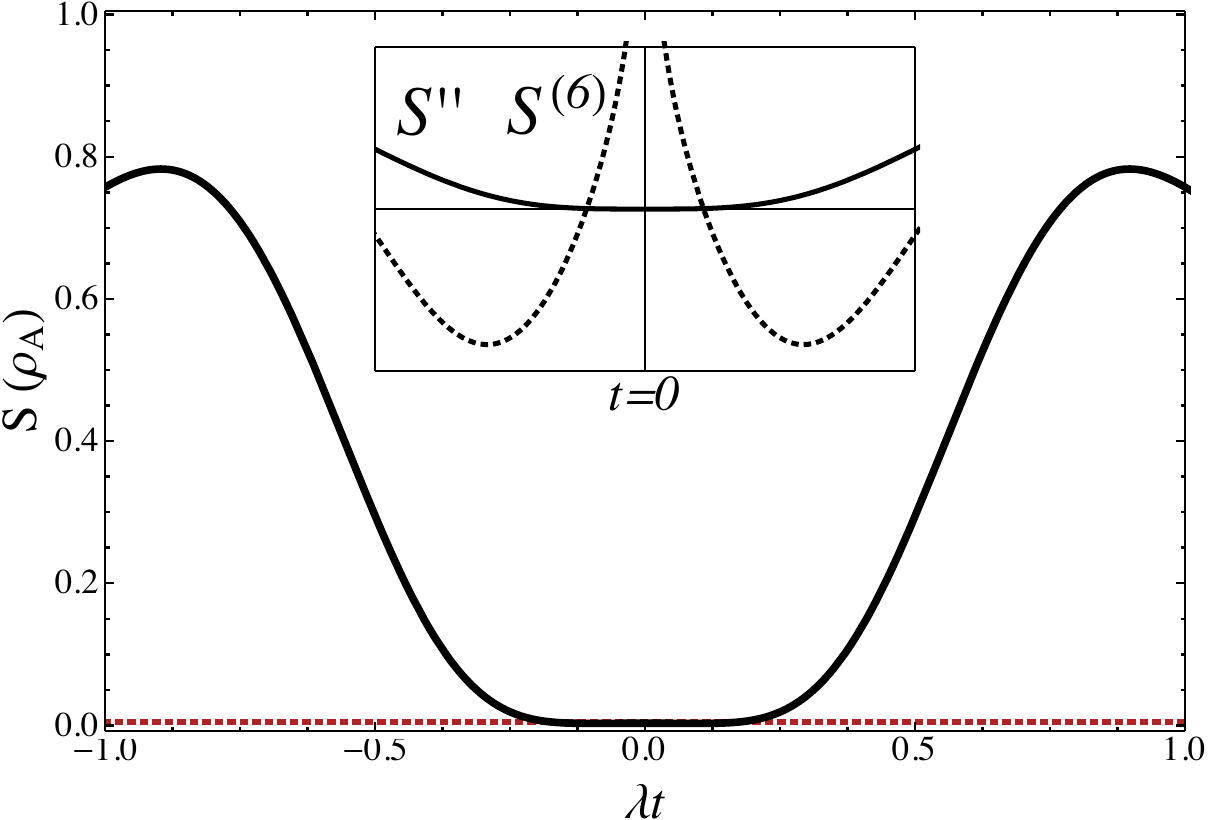}
       \label{fig:cohGb}
    \end{subfigure}
     \caption[Higher order entropy growth from coherent state in Jaynes-Cummings model]{\small (a) $S_2(\rho_A)$ for the coherent state with $\nu=3$ and 
        $C_g=1$, where $T_{\mathrm{ent},g}^{-1}=0$
        indicates that the state remains effectively separable for a significant time. 
        The leading behavior 
        around $t=0$ is sixth order in $t$. (b) $S(\rho_A)$ 
        for the same state is $C^5$ smooth, with $d^2 S/dt^2|_{t=0}=0$ 
        (inset, solid line), and $d^6 S/dt^6$ discontinuous 
        at $t=0$ (inset, dashed line).\normalsize}\label{fig:cohG}
\end{figure*}

For comparison, the coherent state with $\nu=3$ and $C_e=0$, $C_g=1$ 
remains effectively separable for some time, as shown in 
Fig. \ref{fig:cohG}. The divergence of the 
entanglement timescale in this case means one must look to higher orders in 
the Taylor expansion of $S_\al(t)$ to see the growth of entanglement. This is 
one example of an initial state where the correlated quantum uncertainty 
defined in \cite{Yang2017a} vanishes. 

Equation \eqref{eqjcme} shows that the second time derivative of the 
entanglement entropy typically will be divergent in separable states. 
This is not a flaw of  taking the $\al\to1$ limit of the R\'enyi entropy, 
but is the actual behavior of the 
entanglement entropy. From the state \eqref{eqjcmstate}, we can calculate 
the entanglement entropy directly for all times by diagonalizing the reduced 
density matrix of the atom $\rho_A(t)$ and finding its eigenvalues,
$p_{1}(t)=\frac{1}{2}(1+|\vec s(t)|)$, and $p_{2}(t)=\frac{1}{2}(1-|\vec s(t)|)$ in 
terms of the Bloch vector $\vec s(t)$ \cite{Gerry2005}. For instance, starting 
with the atom in its excited state, we find $d^2p_1/dt^2|_{t=0}
=-2T_{\mathrm{ent},e}^{-2}=-d^2p_2/dt^2|_{t=0}.$ Using \eqref{eqEET} leads to the logarithmically divergent result,
\small
\begin{align}\begin{aligned}
 \frac{d^2 S}{dt^2}\bigg|_{t=0} =2\left\{{-2+\ln 2-\lim_{t\to0}{\ln}{\left[1-
{\sum_{n=0}^\infty} |C_{n}|^2\cos^2(\la \sqrt{n+1}  t)\right]}}\right\}T_{\mathrm{ent},e}^{-2}.
\end{aligned}\end{align}
\normalsize
A similar logarithmic divergence occurs for the atom initially in its ground 
state. 

\section{Universal growth at third order}\label{sec:3rdOrder}

A natural extension of the results so far is to consider the growth of R\'enyi entropies at third order around pure, separable states. Although the leading order behaviour is physically identical for all $\al$, we would not expect this to be the case at all orders. The measures in this family are, in general, independent of each other. Surprisingly, a universal timescale also emerges at third order in $t$ around the initial state. We proceed in the same manner as Sec. \ref{sec2}.

Acting with another derivative on \eqref{eq2nd} produces four distinct terms. As a result of the first order calculation, any of these terms proportional to a single derivative of $ [\tr_B\rho(t)]^\al$ will vanish in the $t\to 0$ limit. This leaves only the term with a third order derivative,
\begin{align}\label{eq:3rdDeriv}
\frac{d^3}{dt^3}&S_\al(\rho_A)|_{t=0}=\frac{1}{1-\al}\left[\left(\tr_A(\tr_B\rho(t))^\al\right)^{-1}\tr_A\left(\frac{d^3}{dt^3}(\tr_B\rho(t))^\al\right)\right]_{t=0}\nonumber\\
&=\frac{1}{1-\al}\tr_A\left(\al\frac{d^2}{dt^2}\left[(\tr_B\rho(t))^{\al-1}\tr_B\parder{\rho}{t}\right]\right)_{t=0}\\
&=\frac{\al}{1-\al}\tr_A\left(\frac{d}{dt}\left[\Te(\al-2)\sum_{\be=0}^{\al-2}(\tr_B\rho)^{\be}\tr_B\parder{\rho}{t}(\tr_B\rho)^{\al-2-\be}\tr_B\parder{\rho}{t}+(\tr_B\rho(t))^{\al-1}\tr_B\parder{^2\rho}{t^2}\right]\right)_{t=0}.\nonumber
\end{align}
For small values of $\al$ there are not enough factors of $\rho$ to produce terms with multiple factors of $\pa \rho/\pa t$. In order to treat all values of $\al$ simultaneously we introduce step functions $\Te(x)$ which equal unity if $x\geq0$, and are zero otherwise.

Acting with the third derivative appears to produce six distinct types of terms. However, by relabeling the summation indices and cyclically permuting terms we realize that there are only three truly distinct possibilities,
\begin{align}
\begin{aligned}\label{eq:3terms}
\frac{d^3}{dt^3}S_\al&(\rho_A)|_{t=0}=\frac{\al}{1-\al}\tr_A\bigg[(\tr_B\rho)^{\al-1}\tr_B\frac{\pa^3\rho}{\pa t^3}+3\Te(\al-2)\sum_{\be=0}^{\al-2}(\tr_B\rho)^{\be}\tr_B\parder{\rho}{t}(\tr_B\rho)^{\al-2-\be}\tr_B\frac{\pa^2\rho}{\pa t^2}\\
&+2\Te(\al-3)\sum_{\be=0}^{\al-2}\Te(\be-1)\sum_{\ga=0}^{\be-1}\left( (\tr_B\rho)^\ga\tr_B\parder{\rho}{t}(\tr_B\rho)^{\be-1-\ga} \right)\tr_B\parder{\rho}{t}(\tr_B\rho)^{\al-2-\be}\tr_B\parder{\rho}{t}\bigg]_{t=0}.
\end{aligned}
\end{align}
At this point we can apply the von Neumann equation for each factor $\tfrac{\pa^n \rho}{\pa t^n}$. For instance, the first term involves $\frac{\pa^3\rho}{\pa t^3}=(-i)^3(H^3\rho-3H^2\rho H+3 H \rho H^2-\rho H^3)$ and simplifies to 
\begin{align}
\begin{aligned}
\tr_A\left[(\tr_B\rho)^{\al-1}\tr_B\frac{\pa^3\rho}{\pa t^3}\right]_{t=0}=3i\sum_{n,m,l}\tr_B(B_nB_m B_l\rho_B)\tr_A\left[\rho_A^{\al-1}A_l\rho_AA_n A_m-\rho_A^{\al-1} A_m A_l \rho_AA_n)\right].
\end{aligned}
\end{align}
Upon using the simplification that the state is initially pure, this term reduces to 
\begin{equation}
3i\sum_{n,m,l}\tr_B(B_nB_m B_l\rho_B)\left[\tr_A(\rho_AA_l)\tr_A(\rho_AA_n A_m)-\tr_A( \rho_AA_n)\tr_A(\rho_A A_m A_l)\right].
\end{equation}
We note that already dependence on $\al$ has dropped out. The second type of term in \eqref{eq:3terms} goes similarly, except that the $\al=2$ and $\al>2$ cases must be treated separately when applying the initial conditions. This is one place where we could begin to see differing behaviour for different R\'enyi entropies. However, due to some remarkable cancellations, the result ends up being the same, and no dependence on $\al$ remains:
\begin{align}
\begin{aligned}
&\tr_A\left[3\Te(\al{-}2)\sum_{\be=0}^{\al-2}(\tr_B\rho)^{\be}\tr_B\parder{\rho}{t}(\tr_B\rho)^{\al{-}2{-}\be}\tr_B\frac{\pa^2\rho}{\pa t^2}\right]_{t=0}=3i\Te(\al{-}2)\sum_{n,m,l}\tr_B(B_n\rho_B)\tr_B(B_mB_l\rho_B)\\
&\tr_A\left[  2\tr_A(\rho_A A_m)\tr_A(\rho_AA_nA_l)-2 \tr_A(\rho_AA_l)\tr_A(\rho_A A_mA_n)+ \tr_A(\rho_AA_m A_lA_n) -\tr_A(\rho_AA_nA_mA_l) \right].
\end{aligned}
\end{align}
Finally, the third type of term in \eqref{eq:3terms} only exists for $\al\geq3$, so we may expect that R\'enyi entropies with $\al\geq3$ will have deviations in their third order growth compared to the $\al=2$ measure due to these contributions. The cases $\al=3$ and $\al>3$ should be treated separately, since in the former the sums collapse to a single term. In both cases, after a sequence of seemingly magical cancellations due to the initial conditions, we find that \emph{this type of term vanishes identically},
\begin{equation}
\tr_A\left[2\Te(\al-3)\sum_{\be=1}^{\al-2}\sum_{\ga=0}^{\be-1}\left( (\tr_B\rho)^\ga\tr_B\parder{\rho}{t}(\tr_B\rho)^{\be-1-\ga} \right)\tr_B\parder{\rho}{t}(\tr_B\rho)^{\al-2-\be}\tr_B\parder{\rho}{t}\right]_{t=0}=0.
\end{equation}

Despite numerous opportunities for new contributions to appear for $\al>2$, every possible difference cancels out perfectly leaving the final result for the third order growth of R\'enyi entropies around pure, separable states
\begin{align}\label{third}
&\frac{d^3}{dt^3}S_\al(\rho_A)|_{t=0}=\frac{6i\al}{1-\al}\bigg\{\\
&\sum_{n,m,l}[\tr_B(B_n B_mB_l\rho_B){-}\tr_B(B_m\rho_B)\tr_B(B_n B_l\rho_B)][\tr_A(A_l\rho_A)\tr_A(A_nA_m\rho_A){-}\tr_A(A_n\rho_A)\tr_A(A_mA_l\rho_A)]{+}\nonumber\\
&\sum_{n,m,l}[\tr_A(A_n A_mA_l\rho_A){-}\tr_A(A_m\rho_A)\tr_A(A_n A_l\rho_A)][\tr_B(B_l\rho_B)\tr_B(B_nB_m\rho_B){-}\tr_B(B_n\rho_B)\tr_B(B_mB_l\rho_B)]\bigg\}.\nonumber
\end{align}
The only dependence on $\al$ is through the normalization factor, but the dynamics is completely universal for all R\'enyi entropies\footnote{For the entanglement entropy we can take the $\al\to1^+$ limit after an analytic continuation and find the same dynamical behaviour, with a divergent prefactor. This exactly mirrors the situation found for the second order derivative discussed in Sec. \ref{sec2}}. We also note that the result is totally symmetric in $A\leftrightarrow B$, and does not receive contributions from self-energy terms in the Hamiltonian, only from non-local interaction terms. Furthermore, since the R\'enyi entropies are bounded below by zero, and the $t^3$ term in the Taylor expansion is odd, this result must vanish if the second order timescale vanishes. Indeed this can be verified to occur by going to the basis where $\rho_A^{ij}=\rho_B^{ij}=\de^{1i}\de^{1j}$ initially \cite{Yang2017a}.

At fourth order the remarkable cancellations leading to universal behaviour will almost certainly not continue. Here, and for any higher order, there will be more than one term in the initial step similar to \eqref{eq2nd} that does not contain a term proportional to a single derivative of $ [\tr_B\rho(t)]^\al$. Furthermore, in the step similar to \eqref{eq:3terms} there will be terms which only appear for $\al\geq 3$ or $\al\geq 4$, etc. Universality at higher orders would require each of these types of terms to vanish identically, which becomes increasingly unlikely to imagine. In any case, the R\'enyi entropies are truly independent functions so the common behaviour must cease at some order. Our result in this section shows that non-universal behaviour does not appear at the lowest possible order.

\section{Discussion}\label{dis}

The main result of \cite{Yang2017a} showed that for any unentangled pure 
bipartite state evolving under an arbitrary Hamiltonian, the growth of 
entanglement is characterized by a timescale which takes the universal form
 \small
\begin{equation}
T_{\mathrm{ent}}{=}
\left[\sum_{n,m}\left(\langle A_nA_m \rangle{-}\langle A_n \rangle\langle A_m 
\rangle\right)\left(\langle B_nB_m \rangle{-}\langle B_n \rangle\langle B_m 
\rangle\right)\right]^{-\tfrac{1}{2}},
\end{equation}
\normalsize
 where entanglement is measured by the purity of subsystems. In this chapter, 
we have shown that the same timescale characterizes the growth of 
entanglement as measured by any R\'enyi entropy. Since the family of R\'enyi 
entropies constitutes a complete determination of the entanglement in a pure 
bipartite system, the entanglement timescale universally describes the initial 
growth of bipartite entanglement. A universal timescale with similar properties
also governs the next-to-leading order growth as a result of several non-trivial
cancellations between terms that appear for $\al\geq 3$ but not $\al=2$. 

It is easy to prove that the entanglement timescale obeys several properties 
expected of the R\'enyi entropy. As shown in \cite{Yang2017a}, 
$T_{\mathrm{ent}}^2$ is a manifestly positive quantity so that the R\'enyi 
entropies initially increase from their minimum value. It is also symmetric 
between the subsystems $A$ and $B$ which reflects the symmetry 
$S_\al(\rho_A)=S_\al(\rho_B)$ for overall pure states. Furthermore, the 
coefficient ${2\al}/(\al-1)$ in \eqref{eq6} is monotonically decreasing in $\al$, 
which is required by the general condition $\pa S_\al/\pa \al\leq 0$.

R\'enyi entropies are widely used theoretically and have recently been 
measured in isolated many-body systems \cite{Islam2015a}, including their 
time dependence after an interaction is turned on \cite{Kaufman2016}. The 
first such measurement was performed on a Bose-Einstein condensate 
trapped in an optical lattice and evolving under the Bose-Hubbard 
Hamiltonian in one dimension,
\begin{equation}\label{eqBH}
H=-J\sum_{\langle i,j \rangle} a^\dag_i a_j +\frac{U}{2}\sum_{i} a^\dag_i 
a_i(a^\dag_i a_i-1).
\end{equation}
 The first sum is over nearest-neighbor pairs and represents tunneling 
between neighboring sites at a rate $J$. The second sum over each lattice 
site represents the attractive energy among bosons sharing a site. In the 
experiment \cite{Kaufman2016}, a product of one-particle Fock states was 
prepared on six adjacent lattice sites with a barrier on each end. After a 
quench in which the interaction in \eqref{eqBH} was turned on, the second 
R\'enyi $S_2(\rho_A)$ was measured in time for all unique partitions of the 
six sites.

The only interaction term in \eqref{eqBH} that couples $A$ to $B$ is $-
J(a^\dag_i a_{i+1}+a_i a_{i+1}^\dag)$, where sites $i$ and $i {+} 1$ are 
neighbors across the partition. Thus, for any nontrivial partitioning, the 
entanglement timescale is the same, $T_{\mathrm{ent},BH}^{-2}
={J^2}{\langle1|a_i^\dag a_i|1\rangle}{\langle{1|a_{i{+}1} a_{i{+}1}^\dag|1}\rangle}~{+}~{J^2}{\langle{1|a_i a_i^\dag|1}\rangle}{\langle{1|a_{i{+}1}^\dag a_{i{+}1}|1}\rangle}=$ $4 J^2$. Using the 
experimental value of $J/2\pi=66$ Hz, we can estimate that the entanglement 
will become significant within a time $T_{\mathrm{ent},BH}=1.2\ \mathrm{ms}
$, which agrees with the experimental result displayed in Fig. 3 of Ref. 
\cite{Kaufman2016}. This comparison is only approximate since the actual 
initial states prepared in the experiment were not free of entanglement.

The original motivation to determine the entanglement timescale was to 
estimate how quickly a generic quantum system will decohere due to 
entanglement with gravitational degrees of freedom 
\cite{Zurek2003,Baker2017,Yang2017}. This question is relevant to the black-hole information problem \cite{Almheiri2013,Harlow2016}, where the 
Hawking quanta escaping from the black-hole horizon region may entangle 
with the geometry itself. To make any concrete statements about 
entanglement with gravitational degrees of freedom, one needs to work with 
quantum field theory or, better yet, quantum gravity. Since our derivation of 
the entanglement timescale assumes that the initial state is pure and 
unentangled, it is difficult to generalize these results to quantum field theory, 
where typical states are highly entangled on all scales 
\cite{Bombelli1986,Srednicki1993,Holzhey1994}. UV divergent 
entanglements can be avoided by considering the entanglement difference 
between states, for example with the relative entropy, which lends hope for 
our analysis of $d^2S_\al/dt^2$ \cite{Lashkari2014,Ruggiero2017}. One can 
otherwise avoid divergences by considering causally separated subregions, 
but this comes at the cost of losing purity for the combined system 
\cite{Hollands2017}. Moreover, for gauge field theories, or in the algebraic approach to QFT, the Hilbert space 
does not factorize across spatial boundaries, invalidating our assumptions 
\cite{Casini2014,Donnelly2014}. Still, the growth of entanglement in quantum field theory 
states is a major area of research in many-body, condensed-matter, and high-energy physics \cite{Avery2014,Liu2014,Belin2013,Cresswell2015,Caputa2014}, and it 
would be interesting to develop an entanglement timescale in these regimes.

\chapter{Perturbative expansion of negativity using patterned matrix calculus}\label{ch:neg}

\emph{This chapter is based on the paper \cite{Cresswell2019} published in Phys. Rev. A.}

	\section{Introduction}\label{sec:intro}

	In 1935, Einstein, Podolsky, and Rosen imagined a composite quantum
	state that did not admit a complete local description. 
	In this kind of state, outcomes of measurements performed on the subsystems were perfectly anti-correlated regardless of the chosen measurement basis. Schr\"{o}dinger
	shortly gave this remarkable feature of the quantum formalism the name
	\emph{entanglement}, and the notion continues to be the subject of
	extensive theoretical study and experiment \cite{Horodecki2009}, with applications like quantum teleportation \cite{Bennett1993,Bouwmeester1997,Barrett2004,Riebe2004}, quantum-enhanced metrology \cite{Mitchell2004, Pezze2009,giovannetti2011advances,gross2012nonlinear}, and quantum cryptography \cite{Ekert1991,Gisin2002,Curty2004,Tang2014}.
	We continue this line of study in this chapter which will provide insight
	into how entanglement evolves as the state of a bipartite system is varied.

	The dynamics of entanglement, studied initially in quantum-optical systems, has become an active area of research
	in many-body systems \cite{Audenaertetal2002,Anders2008,EislerZimboras2014}, condensed-matter physics \cite{HelmesWessel2014,EislerZimboras2015,Shermanetal2016,Shimetal2018}, and quantum field
	theories \cite{Calabreseetal2012,Coser2014,Chaturvedietal2018}. Perturbative approaches to entanglement dynamics have revealed a universal timescale characterizing the growth of entanglement
	in initially pure, separable states under unitary evolution as measured by the purity of the reduced density matrix \cite{Kim1996, Yang2017a}, and by the R\'enyi entropies \cite{Cresswell2017a}.
	
	In this paper we focus on
	the \emph{negativity} \eqref{eq:negativity-eigvals}. Although negativity has been studied for two decades, there is, to our knowledge, no general perturbative expansion available. This omission is likely owed to three nuances in the differentiation of negativity: the matrix representation of the partial transposition map, issues of non-differentiability of the trace norm, and the computation of the trace norm's derivative, which requires a careful consideration of the calculus of complex matrices with patterns. 
	
	To expand on these difficulties, we recast negativity in terms of the \emph{trace norm}\footnote{The trace norm, also known as the \textit{nuclear norm}, is one of the \textit{Schatten norms} and one of the \textit{Ky Fan norms}.} $\left\Vert X\right\Vert _{1}\coloneqq\text{Tr}\sqrt{X^{\dagger}X}$ of a matrix, $X$, so that
	\begin{equation}\label{eq:negativitydef}
	\mathcal{N}\left(\rho\right)\coloneqq\frac{1}{2}\left(\left\Vert \rho^{T_{B}}\right\Vert _{1}-1\right).
	\end{equation}
	A perturbative expansion of $\mathcal{N}\left(\rho\right)$ as $\rho(\mu)$ varies with respect to some parameter $\mu$ requires a derivative that we schematically write as 
	\begin{equation}\label{eq:SchematicDerivative}
	\frac{d\mathcal{N}}{d \mu}=\frac{1}{2}\frac{\partial\left\Vert \rho^{T_{B}}\right\Vert _{1}}{\partial\rho^{T_{B}}}\frac{\partial\rho^{T_{B}}}{\partial\rho}\frac{\partial\rho}{\partial \mu}.
	\end{equation}
	In the following sections we will be more precise about how these derivatives are defined and multiplied. The last factor, ${\partial\rho}/{\partial \mu}$, will depend on the application at hand, but for concreteness we will mainly consider time evolution, with ${\partial\rho}/{\partial t}$ governed by the von Neumann dynamics of a closed system or the Lindblad dynamics of an open system. It should be understood that more general variations can be treated in the same way. The middle factor requires an explicit expression for the action of the partial transposition map on the density matrix. Some forms are available in the literature, but in what follows we work in the vectorized representation where $\rho$ and $\rho^{T_B}$ are vectors in $\mathbb{C}^{(d_Ad_B)^2}$. The linear map $T_{B}=\id\otimes T$ has a simple representation in this vector space that is easy to implement numerically, while also providing a form for ${\partial\rho^{T_{B}}}/{\partial\rho}$. 
	
	The more challenging factor to understand is the derivative of the trace norm with respect to its argument. One complication is that the trace norm is only differentiable if its argument is invertible \cite{Watson1992}. At singular arguments one has the notion of a \emph{subdifferential}, to which we return in the discussion. Otherwise, we assume invertibility of $\rho^{T_B}$ throughout. More importantly, the derivative ${\partial\left\Vert \rho^{T_{B}}\right\Vert _{1}}/{\partial\rho^{T_{B}}}$ is taken with respect to a Hermitian matrix whose elements are not independent variables. This is perilous; it is necessary to represent $\rho^{T_B}$ in terms of a set of independent variables before differentiating, which is the main notion behind patterned matrix calculus \cite{Hjorungnes2011}. Generally, we refer to any matrix whose elements are not independent variables as a \emph{patterned} matrix, some other examples of which are symmetric, unitary, or diagonal matrices. Patterned matrix calculus is an underexplored branch of mathematics that we find to be crucial in the perturbative analysis of negativity.

	In this chapter, we address the nuances just mentioned to further our understanding of entanglement
	dynamics by way of negativity.
	In Section \ref{sec:Dynamics},
	we provide a means by which to compute the perturbative expansion of negativity.
	To this end, we offer new matrix representations of the partial transposition
	map, along with explicit computations of the first and second derivatives
	of the trace norm with respect to complex, patterned arguments. The techniques we develop can be straightforwardly carried out to any order in the expansion for negativity.
	In Section \ref{sec:systems}, we apply our results to several physical
	systems with illustrative differences and compare them with the behaviour of other entanglement
	measures. Section \ref{sec:QSL} explores how patterned derivatives can be used more broadly in quantum information theory with an application to quantum speed limits and bounds on entanglement dynamics. 
	Then, in Section \ref{sec:formalcalculus} we use a more formal version of patterned matrix calculus to explain why it is necessary when studying some quantities, like negativity, but not for others, like the R\'enyi entropies that were the focus of the previous chapter. Our result is a theorem describing a class of matrix functions for which patterned derivatives are equal to their unpatterned counterparts, and hence these subtleties can be safely ignored.
Finally, in Section \ref{sec:Discussion}, we discuss in greater
	detail the challenges of our approach, the validity of our assumptions, and the benefits and
	limitations of our results.

	\section{Perturbative expansion of negativity \label{sec:Dynamics}}
	Let us suppose that the density matrix, $\rho(t)$, of a quantum state and some number of its derivatives are known at a given time, $t_0$. To understand how the entanglement between two subsystems changes near $t_0$ due to the evolution of $\rho(t)$ we can expand the negativity as
	\begin{equation}\label{eq:expansion}
	\mathcal{N}(t)=\mathcal{N}(t_0)+\left.\frac{d \mathcal{N}}{d t}\right|_{t_0} (t-t_0)+\frac{1}{2}\left.\frac{d^2 \mathcal{N}}{dt^2}\right|_{t_0} (t-t_0)^2+\cdots .
	\end{equation}
	Our goal is to provide general expressions for the derivatives of negativity in this expansion.\footnote{The logarithmic negativity, $E_{\mathcal{N}}\left(\rho\right)\coloneqq\log_{2}\Vert\rho^{T_B}\Vert_{1}$, can be treated analogously.} In this section we offer explicit expressions for the first and second derivatives using a method that can be carried out systematically to any desired order of precision.
	
	\subsection{Vectorization formalism and the partial transposition map}
	\label{subsec:ptranspose}
	
	Expressions like \eqref{eq:SchematicDerivative} are cumbersome to work with since $\partial \rho^{T_B}/\partial \rho$ represents a four-dimensional array that must be contracted against two matrices to produce a scalar. To avoid such complications we prefer to work in the vectorization formalism, where we represent the state $\rho$ as a column vector, $\ve\rho$. In general, the vectorization operation, $\text{vec}$, stacks the columns of an $m\times n$ matrix into an $mn\times 1$ vector. It admits two useful identities that we rely on, namely
	\begin{equation}
	\ve\left(ABC\right) =\left(C^{T}\otimes A\right)\ve B\label{eq:vecABC},
	\end{equation}
	where $A,B,C$ are any three compatible matrices, and
	\begin{equation}
	\tr\left(A^T B\right) =(\ve ^T A) \ve B,\label{eq:TrAB}
	\end{equation}
	where $A,B$ are the same size and $\ve ^T A \coloneqq (\ve A)^T$ \cite{Magnus1985, Magnus2007}. In this formalism we can rewrite the first derivative of negativity as
	\begin{equation}\label{eq:FirstDerivative}
	\frac{d\mathcal{N}}{d t}=\frac{1}{2}\frac{\partial\left\Vert \rho^{T_{B}}\right\Vert _{1}}{\partial\,\veT\rho^{T_{B}}}\frac{\partial\,\ve\rho^{T_{B}}}{\partial\,\veT\rho}\frac{\partial\,\ve\rho}{\partial t}.
	\end{equation}
	With this simple change in notation the middle factor is an ordinary matrix multiplied by two vectors. To demonstrate the convenience of this formalism we first consider how the standard matrix transposition map can be represented, before applying it to the partial transposition and determining $\partial\,\ve\rho^{T_{B}}/{\partial\,\veT\rho}$.

	Let $\phi$ be a superoperator, i.e., a linear map from the space of $m\times n$ matrices to the space of $q\times r$ matrices. Due to linearity, $\phi$ can always be written as
	\begin{equation}\label{eq:superopmap}
	\phi(X)=\sum_i A_i X B_i,
	\end{equation}
	where $X$, the $A_i$ and the $B_i$ are $m \times n$, $q\times m$, and $n\times r$ matrices, respectively. If $X$ is instead vectorized as $\ve X$, then there is a related operator, $M_\phi$, acting on the space of $mn\times 1$ vectors, and represented by the $qr\times mn$ matrix
	\begin{equation}\label{eq153}
	M_\phi=\sum_i  B_i^T\otimes A_i.
	\end{equation}
	There is no restriction on the number of terms in each sum, so these representations will not be unique. The $M_\phi$ operator can be viewed as implementing the linear map $\phi$ on the vectorized space of matrices,
	\begin{equation}\label{eq:vectorizedsupop}
	\ve \phi(X)= M_\phi \ve X,
	\end{equation}
	which follows from identity \eqref{eq:vecABC}. This demonstrates the existence of an isomorphism\footnote{This is distinct from the Choi-Jamiolkowski isomorphism.} between the space of superoperators acting on $\mathbb{C}^{m,n}$ and those acting on $\mathbb{C}^{mn}$ \cite{Havel2002, Zyczkowski2004}.
	
	Matrix transposition is a linear operation, meaning we can define a \emph{commutation matrix}, $K_{mn}$, such that
	\begin{equation}
	\ve X^{T}=K_{mn}\ve X,\label{eq:kmat}
	\end{equation}
	where $K_{mn}$ has dimensions $mn \times mn$. $K_{mn}$ is a symmetric permutation matrix satisfying the useful identities
	\begin{align} 
	(K_{mn})^{2} &= \id,\\
	\label{Kcommid}
	K_{mq}\left(X\otimes Y\right) &= \left(Y\otimes X\right)K_{nr},
	\end{align}
	where $Y$ is $q\times r$. In the following we only deal with square matrices and hence use $K \coloneqq K_{nn}$ for ease of notation. The commutation matrix has a simple representation of the form \eqref{eq153} in the standard basis of matrices $\left\{ J^{ij}_{n}\right\}$, where $J^{ij}_n$ is the $n\times n$ \emph{single-entry matrix} defined through
	$\left(J^{ij}_n\right)_{kl}=\delta_{ik}\delta_{jl}$. In terms of these elements
	\begin{equation}\label{eq:commutation}
	K=\sum_{i,j=1}^n J^{ji}_n\otimes J^{ij}_n.
	\end{equation}

	In the same vein as \eqref{eq:kmat}, we can represent the partial transposition map as a linear superoperator acting on the space of $\left(d_{A}d_{B}\right)^{2}\times 1$ column vectors,
	\begin{equation}\label{eq:vecrhoTb}
	\ve \rho^{T_{B}}=K_B\ve \rho ,
	\end{equation}
	where the subscript indicates the subsystem to be transposed. We call $K_{B}$ a \emph{partial commutation matrix}. It is $\left(d_{A}d_{B}\right)^{2}\times \left(d_{A}d_{B}\right)^{2}$ and is self-inverse. The advantage of this formalism is that we can immediately identify the middle factor in \eqref{eq:FirstDerivative} as
	\begin{equation}
	\frac{\partial\,\ve \rho^{T_{B}}}{\partial\,\veT \rho}=K_B,\label{eq:drhoTBdrho}
	\end{equation}
	which follows when we observe that the partial commutation matrix is constant. To establish this we can investigate the form of $K_B$, again working with standard basis elements. We find
	\begin{equation}
	K_B=\sum_{i,j=1}^{d_{B}}\left(\id_{d_{A}}\otimes J_{d_{B}}^{ji}\right)\otimes\left(\id_{d_{A}}\otimes J_{d_{B}}^{ij}\right).\label{eq:supop}
	\end{equation}
	We note that $K_B$ is a constant matrix that depends only on the dimensions $d_A$, and $d_B$. Furthermore, when $d_A=1$, then $K_B=K$, as expected. In Sec. \ref{sec:superoperator} we show how \eqref{eq:supop} can be obtained through the action of the partial transposition map on the standard basis, and we also present a convenient form for $K_B$ by identifying its eigenvectors.  For other representations of the transposition and partial transposition maps, see, for example, \cite{Jaroslaw2011}.
	
	With this brief introduction to aspects of algebra on vectorized matrices, we now turn to calculus in order to identify ${\partial\left\Vert \rho^{T_{B}}\right\Vert _{1}}/{\partial\,\veT\rho^{T_{B}}}$, the remaining factor in \eqref{eq:FirstDerivative}.

	\subsection{First derivative of the trace norm \label{subsec:trnormder}}
	
	In this section we use $X$ to represent an unpatterned $n\times n$ matrix of complex variables. The unpatterned derivatives of a scalar function $g(X,X^*)$,
	\begin{equation}
	D_X g\coloneqq \frac{\partial g}{\partial\,\veT X},\quad \text{and} \quad D_{X^*} g\coloneqq \frac{\partial g}{\partial\,\veT X^*},
	\end{equation}
	are found by expressing the differential of $g$ in the form
	\begin{equation}\label{eq:differentialTrNorm}
	d g =(D_X g) d\ve X+(D_{X^*} g) d\ve X^*,
	\end{equation}
	and reading off the prefactors of $d\ve X$ and $ d\ve X^*$. The differentials $d\ve X$ and $d\ve X^*$ are taken to be independent, and when $g$ is differentiable the unpatterned derivatives are unique \cite{Magnus1985,Hjorungnes2007}.

	As discussed in Sec.\ref{sec:intro}, $\rho^{T_B}$ is Hermitian, meaning its matrix elements are interdependent and special care must be taken in defining a derivative with respect to it. With some effort, the derivative of a scalar function with respect to a Hermitian argument $A$ can be found in terms of the unpatterned derivatives as in \cite{Hjorungnes2011}:
	\begin{equation}\label{eq:DTrNorm}
	D_{A} g\coloneqq\frac{\partial g}{\partial\,\veT A}=(D_X g)|_{X=A}+ (D_{X^*} g)|_{X=A} K,
	\end{equation}
	where $(D_X g)|_{X=A}$ means the Hermitian pattern is applied after the unpatterned derivative has been computed, and $K$ is the commutation matrix from \eqref{eq:commutation}. Sec. \ref{sec:Acomp_pat} contains an abstract summary of our general approach to patterned derivatives {which can be used to establish \eqref{eq:DTrNorm} rigourously. For a briefer derivation, we note that the differentials of $A$ are not independent, since $d\ve A^{*}=d\ve A^{T}=Kd\ve A$. Hence, when we apply the Hermitian pattern, \eqref{eq:differentialTrNorm} becomes
		\begin{equation}
		dg=
		\left[D_{X}g+\left(D_{X^{*}}g\right)K\right]_{X=A}d\ve A,
		\end{equation}
		and we identify the patterned derivative in \eqref{eq:DTrNorm}. We note that $D_{A^*}g$ can be found similarly, but is not independent from $D_{A}g$ since $D_{A^*}g=(D_{A}g) K$.}

	The trace norm $\Vert \rho^{T_B}\Vert_{1}$ is a scalar function of a Hermitian matrix, so in order to find its derivative we start by computing the differential of $\Vert X\Vert_{1}$ as in \eqref{eq:differentialTrNorm}. We take $X$ to be invertible so that $\Vert X\Vert_{1}$ is differentiable, and define $\left|X\right|\coloneqq\sqrt{X^{\dagger}X}$, so that
	$\left\Vert X\right\Vert _{1}=\text{Tr}\left|X\right|$. Then, by definition,
	$\left|X\right|\left|X\right|=X^{\dagger}X$, and we can take the
	differential of both sides to obtain
	\begin{equation}
	\left(d\left|X\right|\right)\left|X\right|+\left|X\right|d\left|X\right|=X^{\dagger}dX+\left(dX^{\dagger}\right)X\label{eq:firstdif}.
	\end{equation}
	Multiplying by $\left|X\right|^{-1}$ on the left and taking the trace allows us to isolate $\tr \left(d\left\vert X\right\vert\right)$
	\begin{equation}
	2\tr\left(d\left|X\right|\right)=\tr\left(\left\vert X\right\vert^{-1} X^\dag dX \right)+\tr\left(X \left\vert X\right\vert^{-1} dX^\dag\right).
	\end{equation}
	The differential operator $d$ commutes with both the trace and $\text{vec}$ operations, so this equation actually gives the differential of the trace norm, $\tr\left(d\left\vert X\right\vert\right)=d \Vert X\Vert_{1}$. Identity \eqref{eq:TrAB} allows us to express $d\Vert X\Vert_{1}$ in the required form \eqref{eq:differentialTrNorm},
	\begin{equation}
	d \Vert X\Vert_{1}=\frac{1}{2}\veT\left[X^{*}\left(\left|X\right|^{-1}\right)^{T}\right]d\ve X+\frac{1}{2}\veT\left[\left(\left|X\right|^{-1}\right)^{T}X^{T}\right]Kd\ve X^{*}\label{eq:firstdifferential},
	\end{equation}
	where we have also used \eqref{eq:kmat} for $d\ve X^\dag$. Now we can identify the derivatives with respect to $X$ and $X^*$ as
	\begin{align}
	    \begin{aligned}
	    D_X \Vert X\Vert_{1}&=\frac{1}{2}\veT\left[X^{*}\left(\left|X\right|^{-1}\right)^{T}\right], \\ D_{X^*} \Vert X\Vert_{1}&=\frac{1}{2}\veT\left[\left(\left|X\right|^{-1}\right)^{T}X^{T}\right]K.\label{eq:unpatder}
	    \end{aligned}
	    \end{align}
	The derivative with respect to a Hermitian argument follows by substituting \eqref{eq:unpatder} in \eqref{eq:DTrNorm}, and recalling $K^2=\id$:
	\begin{align}
	\begin{aligned}
	D_A \Vert A\Vert_{1}&=\frac{1}{2}\veT\left[X^{*}\left(\left|X\right|^{-1}\right)^{T}+\left(\left|X\right|^{-1}\right)^{T}X^{T}\right]_{X=A}\\
	&=\veT\left[A^T\left(\left|A\right|^{-1}\right)^T\right]=\veT\left[A\left\vert A\right\vert^{-1}\right]K\label{eq:drhopatterned},
	\end{aligned}
	\end{align}
	where we note that $A$ commutes with $\left\vert A\right\vert^{-1}$. The matrix $A\left\vert A\right\vert^{-1}$ appearing in the derivative is the matrix extension of the sign function, defined such that $A=\text{sign}(A)|A|$ \cite{Higham2008}.

	The derivative simplifies even further when we use the eigendecomposition for $\rho^{T_B}=U\Lambda U^\dag$, where $\Lambda$ contains the eigenvalues of $\rho^{T_B}$ in decreasing order, 
	\begin{equation}
	\frac{\partial\left\Vert \rho^{T_{B}}\right\Vert _{1}}{\partial\,\veT\rho^{T_{B}}}=D_{\rho^{T_B}}\left\Vert \rho^{T_{B}}\right\Vert _{1}=\veT\left( U^{*}\text{sign}\left(\Lambda\right)U^{T}\right).
	\label{eq:patder}
	\end{equation}
	
	We would like to emphasize that the derivative taken with respect to a Hermitian argument \eqref{eq:patder} is twice the unpatterned derivative \eqref{eq:unpatder}, which shows the indispensability of patterned matrix calculus for understanding derivatives of negativity. Our results \eqref{eq:supop} and \eqref{eq:patder} can be combined to give the first derivative of negativity \eqref{eq:FirstDerivative}. Next, we compute the second derivative explicitly, and then proceed to show how the perturbative expansion can be carried out to any order.

	\subsection{Second derivative of the trace norm \label{subsec:trnormhessian}}
	
	Taking another derivative of \eqref{eq:FirstDerivative} and noting that $K_B$ is constant gives
	\begin{align}\begin{aligned}\label{eq:secondDerivative}
	\frac{d^2 \mathcal{N}}{dt^2}=&\, \frac{1}{2}\left(\frac{\partial\,\ve \rho^{T_B}}{\partial\,\veT \rho}\frac{\partial\,\ve\rho}{\partial t}\right)^T\frac{\partial}{\partial\,\veT \rho^{T_B}}\left(\frac{\partial\left\Vert\rho^{T_B}\right\Vert_{1}}{\partial\,\veT \rho^{T_B}}\right)^T
	\left(\frac{\partial\,\ve \rho^{T_B}}{\partial\,\veT \rho}\frac{\partial\,\ve\rho}{\partial t}\right)\\
	&+\frac{1}{2}\frac{\partial\left\Vert\rho^{T_B}\right\Vert_{1}}{\partial\,\veT \rho^{T_B}}\frac{\partial\,\ve \rho^{T_B}}{\partial\,\veT \rho}\frac{\partial^2\,\ve\rho}{\partial t^2}.
	\end{aligned}\end{align}
	The second line can be computed using the above results if $\partial^2 \rho/\partial t^2$ is known, but the first line involves the Hessian of $\Vert\rho^{T_B}\Vert_{1}$ with respect to a Hermitian argument,
	
	\begin{equation}\label{eq:HermHessian}
	\hess_{\rho^{T_B},\rho^{T_B}}\left(\Vert\rho^{T_B}\Vert_{1}\right)\coloneqq \frac{\partial}{\partial\,\veT \rho^{T_B}}\left(\frac{\partial\left\Vert\rho^{T_B}\right\Vert_{1}}{\partial\,\veT \rho^{T_B}}\right)^T.
	\end{equation}
	To our knowledge, Hessians with respect to patterned matrices have not been discussed in the literature. Therefore, in the following we present a detailed discussion of such Hessians in general before applying our new methods to the trace norm to compute \eqref{eq:HermHessian}. 
	
	As with Jacobians \eqref{eq:differentialTrNorm}, unpatterned Hessians are defined through the differential,\footnote{We follow the conventions in \cite{Hjorungnes2011}.}
	\eq{\label{eq:seconddifferentialTrNorm}
	d^2 g = & \, d\veT X \left[ \hess_{X,X}(g) \right] d\ve X+d\veT X^*
	\left[ \hess_{X,X^*}(g) \right] d\ve X\\
	&+d\veT X \left[ \hess_{X^*,X}(g) \right] d\ve X^*
	+d\veT X^*  \left[ \hess_{X^*,X^*}(g) \right] d\ve X^*,}
	where we take 
	
	\begin{equation}
	\hess_{X^*,X}(g)\coloneqq D_{X^*}(D_X g)^T\coloneqq\frac{\partial}{\partial\,\veT X^*}\left[\frac{\partial g}{\partial\,\veT X}\right]^T,
	\end{equation}
		 etc., and note that second differentials of the variables $X$ and $X^{*}$ are zero by definition. Since the four Hessians here are not independent, there is some freedom to exchange terms in the form \eqref{eq:seconddifferentialTrNorm}. To compute the Hessians, one must write the second differential of $g(X,X^*)$ as
	\begin{align}\begin{aligned}\label{eq:Aexpansion}
	d^2 g = & \, d\veT X (B_{10})\ d\ve X+d\veT X^* (B_{00})\ d\ve X\\
	&+d\veT X (B_{11})\ d\ve X^*+d\veT X^* (B_{01})\ d\ve X^*.
	\end{aligned}\end{align} 
	Since partial derivatives commute, the $\hess_{X,X}$ and $\hess_{X^*,X^*}$ Hessians should be symmetric, while $\hess_{X,X^*}$ and $\hess_{X^*,X}$ should be transposes of one another. One can conclude that
	\begin{align}
	\hess_{X,X}(g)&=\frac{1}{2}\left(B_{10}+B_{10}^T\right),\label{eq:unpatHess1}\\
	\hess_{X^*,X^*}(g)&=\frac{1}{2}\left(B_{01}+B_{01}^T\right),\label{eq:unpatHess2}\\
	\hess_{X,X^*}(g)&=\frac{1}{2}\left(B_{00}+B_{11}^T\right)=[\hess_{X^*,X}(g)]^T.\label{eq:unpatHess3}
	\end{align}

	When the argument of $g(A,A^*)$ is a Hermitian matrix, this procedure must be modified to account for the interdependence of $A$ and $A^\dag$. We can define a Hessian with respect to a Hermitian argument by noting
	that the patterned Jacobian for scalar $g$, Eq. \eqref{eq:DTrNorm}, also applies to a vector function $\boldsymbol{g}$,
	since it applies elementwise:
	\begin{equation}
	D_{A}\boldsymbol{g}=\left[D_{X}\boldsymbol{g}+\left(D_{X^{*}}\boldsymbol{g}\right)K\right]_{X=A}.\label{eq:hermvecder}
	\end{equation}
	Then, the Hessian of $g\left(A,A^{*}\right)$ with respect to Hermitian $A$ can be found by applying \eqref {eq:hermvecder} to the Jacobian $\boldsymbol{g}=\left(D_{A}g\right)^{T}$, resulting in
	\eq{\label{eq:hessscal}
	\hess_{A,A}(g)  :=D_{A}\boldsymbol{g} =&\left[D_{X}\left(D_{X}g\right)^{T}+K D_{X}\left(D_{X^{*}}g\right)^{T}
	+D_{X^{*}}\left(D_{X}g\right)^{T}K+KD_{X^{*}}\left(D_{X^{*}}g\right)^{T}K\right]_{X=A} 
	\\
	 =&\, \hess_{X,X}(g)\big\vert_{X=A}+K\hess_{X,X^{*}}(g)\big\vert_{X=A}
	+\hess_{X^{*}X}(g)\big\vert_{X=A}K+K\hess_{X^{*}X^{*}}(g)\big\vert_{X=A}K.
	}
	
	One way to gauge the correctness of this result is to use expression \eqref{eq:seconddifferentialTrNorm}
	for the second differential of a scalar function $g(X,X^*)$. Then, letting $X\to A$ and recalling $d\ve A^{*}=Kd\ve A$ gives
	\begin{align}
	\begin{aligned}
	d^{2}g  =d\veT A\left[\hess_{X,X}(g)\vert_{X=A}+K\hess_{X,X^{*}}(g)\vert_{X=A}+\hess_{X^{*}X}(g)\vert_{X=A}K+K\hess_{X^{*}X^{*}}(g)\vert_{X=A}K\right]d\ve A. \label{eq:hessalt}
	\end{aligned}
	\end{align}
	We can see that the part of expression \eqref{eq:hessalt} in square brackets \textendash{}
	defined to be $\hess_{A,A}(g)$ \textendash{} matches Eq. \eqref{eq:hessscal}. This result can also be confirmed using the formal calculus described in Sec. \ref{sec:Acomp_pat}. Whereas there are three independent Hessians of $g$ with respect to combinations of $\{X,X^*\}$, there is only one independent Hessian in the Hermitian case. The Hessians are related by
	\begin{equation}
	\hess_{A,A^*}(g)=K\hess_{A,A}(g),\quad \hess_{A^*,A}(g)=\hess_{A,A}(g)K,\quad \hess_{A^*,A^*}(g)=K\hess_{A,A}(g)K.
	\end{equation}

	We now use our result \eqref{eq:hessscal} to compute \eqref{eq:HermHessian} from the unpatterned Hessians of the trace norm. Taking the differential of both sides of Eq. \eqref{eq:firstdif} and noting $d^{2}X=0=d^{2}X^{\dagger}$ gives us
	\begin{equation}
	\left(d^{2}\left|X\right|\right)\left|X\right|+\left|X\right|d^{2}\left|X\right|=2dX^{\dagger}dX-2\left(d\left|X\right|\right)^2.
	\label{eq:doublediff}
	\end{equation}
	Once again we left-multiply by $|X|^{-1}$ and take the trace,
	\begin{equation}
	{\tr}( d^{2}\left|X\right|)=\tr[(dX^{\dagger}){\left(dX\right)}\left\vert X\right\vert^{-1}]-\tr[\left(d\left\vert X\right\vert\right)\left|X\right|^{-1}d\left\vert X\right\vert],
	\end{equation}
	which becomes
	\eq{d^{2}\left\Vert X\right\Vert_{1}=d\veT\left(X^{*}\right)\left[\left(\left\vert X\right\vert^{-1}\right)^{T}\otimes\id\right]d\ve X
	-d\veT\left(\left|X\right|^T\right)\left[\id\otimes\left|X\right|^{-1}\right]d\ve \left|X\right|.\label{eq:d2tr}}
	To find $d\ve\left|X\right|$, we may vectorize both sides of Eq. \eqref{eq:firstdif} and use identity \eqref{eq:vecABC} which results in
	\begin{align}
	 \begin{aligned}
	   (\left|X\right|^{T}\otimes\id+\id\otimes\left|X\right|) d\ve \left|X\right|
	=\left(\id\otimes X^{\dagger}\right)d\ve X+\left(X^{T}\otimes\id\right)d\ve X^{\dagger}.
	  \end{aligned}
	\end{align}
	For compact notation, let us introduce the \emph{Kronecker sum} $A\oplus B \coloneqq A\otimes\id+\id\otimes B$ and define \eq{X_{\oplus}\coloneqq X^{T}\otimes\id+\id\otimes X=X^{T}\oplus X.} Then, since $|X|_{\oplus}$ is invertible,\footnote{Notice
		that $
		\text{det}|X|_{\oplus}\geq\text{det}\left(\left|X\right|^{T}\otimes\id\right)+\text{det}\left(\id\otimes\left|X\right|\right)=2\left(\text{det}\left|X\right|\right)^{n}>0$,
		where we have used the fact that $\left|X\right|^{T}\otimes\id$ and
		$\id\otimes\left|X\right|$ are positive semidefinite and $\left|X\right|$
		is nonsingular. Hence $|X|_{\oplus}$ is invertible. Eq. \eqref{eq:firstdif} is a \emph{Sylvester equation} for which solvability conditions are known and met in our case. 
	}
	\eq{d\ve \left|X\right|=\left(|X|_{\oplus}\right)^{-1}\left(\id\otimes X^{\dagger}\right)d\ve X
	+\left(|X|_{\oplus}\right)^{-1}\left(X^{T}\otimes\id\right)Kd\ve X^{*}.\label{eq:dvec|X|}}
	Inserting this in Eq. \eqref{eq:d2tr} brings us to the desired form \eqref{eq:Aexpansion} from which we can read off the $B$ matrices, and combine them to form the Hessians in Eqs. \eqref{eq:unpatHess1} to \eqref{eq:unpatHess3}. 
	
	All that remains is to merge the unpatterned Hessians as in \eqref{eq:hessscal} to obtain the Hessian with respect to a Hermitian variable. 
	We show these computations in more detail in Sec. \ref{sec:simphess}. The result is
	\eq{\label{eq:patternedHessian}
	{\hess_{\rho^{T_{B}},\rho^{T_{B}}}}{\left(\left\Vert \rho^{T_{B}}\right\Vert _{1}\right)}=\frac{1}{2}K\left[{\left(\left|\rho^{T_{B}}\right|^{-1}\right)_{\oplus}}-\rho_{\oplus}^{T_{B}}\left(\left\vert\rho^{T_{B}}\right\vert_{\oplus}\right)^{-1}
	\left(\left\vert\rho^{T_{B}}\right\vert^{-1}\right)_{\oplus}\left(\left\vert\rho^{T_{B}}\right\vert_{\oplus}\right)^{-1}\rho_{\oplus}^{T_{B}}\right].
	}
	For computational efficiency we can simplify this expression in terms of the  eigendecomposition of $\rho^{T_{B}}$, as we did for the first derivative. We find
	\eq{\hess_{\rho^{T_{B}},\rho^{T_{B}}}\left(\left\Vert \rho^{T_{B}}\right\Vert _{1}\right) 
	=K(U^*\otimes U)\left[
	\id-\text{sign}\Lambda\right.\left.
	\otimes\ \text{sign}\Lambda\right]\left(\left|\Lambda\right|_{\oplus}\right)^{-1}(U^T\otimes U^\dag).\label{eq:spectHessian2}}
	This form provides additional insight into the behaviour of negativity since the Hessian vanishes when the eigenvalues of $\rho^{T_B}$ are all positive. This Hessian, along with the results of Secs. \ref{subsec:ptranspose} and \ref{subsec:trnormder}, allows the second derivative of negativity \eqref{eq:secondDerivative} to be written in terms of the density matrix and derived quantities.

	\subsection{Summary of method \label{subsec:summary}}
	
	In this section we summarize our results in an algorithm for computing the perturbative expansion for negativity \eqref{eq:expansion} to second order:
	
	\renewcommand{\theenumi}{\Roman{enumi}}

	(1) Determine the derivatives of the density matrix at $t=t_0$, e.g. from an equation of motion.
		
		(2) Construct the commutation matrix $K$ from \eqref{eq:commutation} and partial commutation matrix $K_B$ from \eqref{eq:supop} appropriate for subsystem dimensions $d_A,d_B$ from the basis of single-entry matrices $J_n^{ij}$.
		
		(3) Compute the eigendecomposition for the initial state $\rho^{T_B}(t_0)=U \Lambda U^\dag$.
		
		(4) Use the above in the first derivative of negativity, found from \eqref{eq:FirstDerivative} using \eqref{eq:drhoTBdrho} and \eqref{eq:patder}:
		\begin{align}
		\left.\frac{d\mathcal{N}}{d t}\right|_{t=t_0}&=\frac{1}{2}\veT\left[U^{*}\text{sign}\left(\Lambda\right)U^{T}\right]K_B\ve\dot{\rho}\left(t_0\right).
		\label{eq:first derivative}
		\end{align}
		
		(5) Use the above in the second derivative of negativity, found from \eqref{eq:secondDerivative} using \eqref{eq:drhoTBdrho}, \eqref{eq:patder}, and additionally \eqref{eq:spectHessian2} for the patterned Hessian of the trace norm $\hess_{\rho^{T_{B}},\rho^{T_{B}}}\left(\left\Vert \rho^{T_{B}}\right\Vert _{1}\right)$. We summarize it as
		\begin{align}
		    \begin{aligned}
		    \label{eq:second derivative}
		{\left.\frac{d^{2}\mathcal{N}}{d t^{2}}\right|_{t=t_0}}{=} & \, \frac{1}{2}{\left[K_B\ve\dot{\rho}\left(t_0\right)\right]^{T}}{\hess_{\rho^{T_{B}},\rho^{T_{B}}}}{\left(\left\Vert \rho^{T_{B}}\right\Vert _{1}\right)}{K_B\ve\dot{\rho}\left(t_0\right)}\\
		& +\frac{1}{2}\veT\left[U^{*}\text{sign}\left(\Lambda\right)U^{T}\right]K_B\ve\ddot{\rho}\left(t_0\right).
		    \end{aligned}
		\end{align}

	In light of Sections \ref{subsec:trnormder} and \ref{subsec:trnormhessian}, it is clear that higher differentials $d^{n} \Vert X\Vert_{1}$, with $n\geq 3$, can be computed iteratively by solving the equation $ d^n(|X| |X|)=d^n(X^\dag X)=0$ for the differentials $d^n\ve |X|$. Each such equation takes the form
	\begin{equation}
	(d^n\left\vert X\right\vert) \left\vert X\right\vert+\left\vert X\right\vert d^n\left\vert X\right\vert=C_n,\end{equation}
	where $C_n$ only contains differentials of order less than $n$, and each equation can be solved as in \eqref{eq:dvec|X|},
	\begin{equation}
	d^n\ve \left\vert X\right\vert=\left(\left\vert X\right\vert_{\oplus}\right)^{-1}\ve C_n.
	\end{equation}
	In terms of lower-order differentials of $\left\vert X\right\vert$, then, 
	\begin{equation}\label{eq:ndifferential}
	d^n\left\Vert X\right\Vert_{1}=\frac{1}{2} \tr (\left\vert X\right\vert^{-1} C_n).
	\end{equation}
	Finally, the form of higher-order derivatives of $\left\Vert \rho^{T_B}\right\Vert_{1}$ with respect to its Hermitian argument can be generalized from the methods we will present in Sec. \ref{sec:Acomp_pat} and read off from \eqref{eq:ndifferential}.
	This extends the steps in Section \ref{subsec:trnormhessian} for a perturbative expansion of negativity to any order. 
	
	We conclude this section by noting that, for certain classes of systems,  all terms involving higher derivatives of the trace norm vanish. In these cases we have
	\begin{equation}
	\frac{d^{n}\mathcal{N}}{d t^{n}}=\frac{1}{2}\veT\left[U^{*}\text{sign}\left(\Lambda\right)U^{T}\right]K_B\ve\left.\frac{\partial^{n}\rho}{\partial t^{n}}\right\vert_{t=t_0},
	\end{equation}
	and the expansion \eqref{eq:expansion} resums to
	\eq{\label{eq:specialexpansion}
	\mathcal{N}\left(t\right)=\mathcal{N}\left(t_{0}\right)+\frac{1}{2}\left.\left(\text{vec}^{T}\left[U^{*}\text{sign}\left(\Lambda\right)U^{T}\right]K_B\right)\right\vert_{t=t_{0}}\ve\left[\rho\left(t\right)-\rho\left(t_{0}\right)\right].}
	The simplified Eq. \eqref{eq:specialexpansion} holds for a number of systems, but we have yet to find an \textit{a priori} condition that guarantees its validity. Necessary and sufficient conditions for the vanishing of terms containing higher derivatives of the trace norm merit further study.
	
	\section{Negativity growth in various systems \label{sec:systems}}
	The derivatives of negativity can now be calculated by knowledge of the density matrix and its derivatives at a specific instant in time $t_0$. This is often much simpler than computing $\rho^{T_B}\left(t\right)$ for all times, finding all of the eigenvalues, and then differentiating the sum in Eq. \eqref{eq:negativity-eigvals}. Moreover, the latter method can be difficult to implement numerically, as it relies on derivatives of absolute value functions, which can lead to spurious results if not treated carefully. Here we introduce some physical systems to exemplify the usefulness and robustness of our method.
	
	\subsection{Jaynes-Cummings model}
	A commonly used model in quantum optics is the Jaynes-Cummings model (JCM), which characterizes a two-level atom interacting with a single quantized mode of a bosonic field. The JCM has been the subject of much theoretical and experimental work \cite{Shore1993,Finketal2008}, including recent theoretical studies of its entanglement properties \cite{Quesada2013,Cresswell2017a}. The JCM Hamiltonian is given in units of $\hbar=1$ by
	\begin{equation}
		{H_\text{JCM}}=\omega\id\otimes{\had\ha} +\left({\omega}-{\Delta}\right)\hat{\sigma}^\dagger\hat{\sigma}\otimes\id-\iu g\left(\hat{\sigma}\otimes\had-\hat{\sigma}^\dagger\otimes\ha\right),
	\end{equation}
	where $\ha$ is the bosonic annihilation operator for the field, $\hat{\sigma}=\ket{g}\bra{e}$ lowers the atom from the excited state $\ket{e}$ to the ground state $\ket{g}$, $\omega$ is the frequency of the bosonic mode, $\Delta$ is the detuning between the mode and the atomic transition frequency, and $g$ is a coupling constant
	\cite{Quesada2013}. The Hamiltonian conserves total excitation number $\id\otimes\had\ha+\hat{\sigma}^\dagger\hat{\sigma}\otimes\id$, restricting the dynamics to systems of size $2\times N$, where $N$ is the number of Fock states of the bosonic mode that are coupled to by the initial conditions. This subsumes systems for which the PPT criterion is sufficient ($N=2,\,3$), as well as systems that can have vanishing negativity yet remain entangled ($N>3$). We use this model to explore both types of systems, as delineated by the \hyperref[thm:PHC]{Peres-Horodecki Criterion}.
	
\begin{figure}
	\centering\includegraphics[trim={0cm 1.3cm 0cm 3.8cm
		},clip,width=0.7\columnwidth]{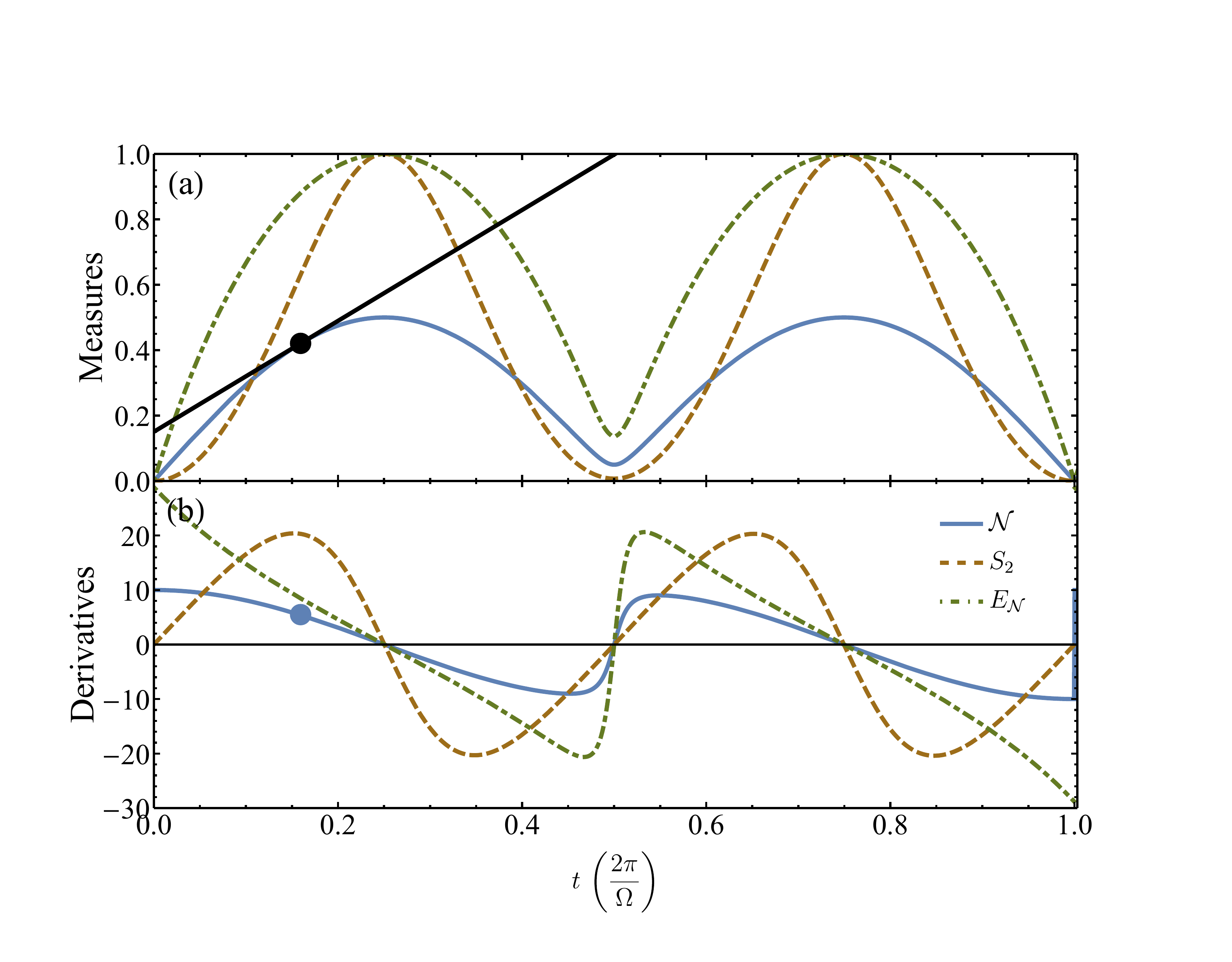}
		\caption[Comparison of entanglement measures in the Jaynes-Cummings model]{Evolution of the entanglement measures negativity $\mathcal{N}$ (blue solid lines), second-order R\'enyi entropy $S_2$ (brown dashed lines), and logarithmic negativity $E_{\mathcal{N}}$ (green dot-dashed lines) in the Jaynes-Cummings model for the initial state $\ket{e}\otimes\ket{3}$ and parameters $\left(\omega,\Delta,g\right)=\left(10,1,5\right)$. \textbf{(a)} Entanglement measures and \textbf{(b)} time-derivatives of entanglement measures versus time, with time in units of $2\pi/\Omega$. The time derivatives involving negativity are calculated using Eq. \eqref{eq:first derivative}. The measures agree in regions where entanglement is increasing, decreasing, maximal, minimal, and absent. Notably, negativity is {initially} more sensitive than R\'enyi entropies to growth in entanglement from the initial, separable state; the R\'enyi entropy does not grow linearly in time around $t=0$. We exemplify the success of  Eq. \eqref{eq:first derivative} by plotting a tangent to the negativity curve in (a) with slope found from the derivative curve in (b).\normalsize}
		\label{fig:JCM-2by2-measures}
	\end{figure}

	As a first example we choose the initial conditions ${\rho}(t_0)=\ket{\psi_0}\bra{\psi_0},\,\ket{\psi_0}=\ket{e}\otimes\ket{3}$; the atom is in its excited state and the field has three excitations. The state of the system for all time is given by $\rho(t)=\ket{\psi\left(t\right)}\bra{\psi\left(t\right)}$, where 
	\begin{align}
	    \begin{aligned}
	    {\ket{\psi\left(t\right)}}=\eu^{-\iu H_\text{JCM}t}\ket{\psi_0}
		=\frac{\eu^{\iu t\left(\Delta-8\omega\right)/2}}{\Omega}\left[4g\sin\left(\frac{\Omega t}{2}\right)\ket{g}
		\otimes\ket{4}
		+\left(\Omega\cos\frac{\Omega t}{2}-\iu \Delta\sin\frac{\Omega t}{2}\right)\ket{e}\otimes\ket{3}\right]
		,
	    \end{aligned}
	\end{align}
	and we have defined the Rabi frequency through $\Omega^2\coloneqq\Delta^2+\left(4g\right)^2$. The negativity can be calculated analytically for this system, which has an effective dimension of $2\times2$ at all times; the atom Hilbert space is spanned by $\ket{g}$ and $\ket{e}$ while the field Hilbert space is spanned by $\ket{3}$ and $\ket{4}$, with excitations trading between the two subsystems.
	We find, using \eqref{eq:first derivative}, that \eq{\frac{d\mathcal{N}}{d t}
		&=\frac{2g\sin\left(\Omega t\right)\left[\Delta^2+\left(4g\right)^2\cos\Omega t\right]}{\Omega\sqrt{\Delta^2\left(1-\cos\Omega t\right)^2+\Omega^2\sin^2\Omega t}},} which agrees with the result obtained by differentiating \eqref{eq:negativity-eigvals} with respect to time.
	We plot the negativity, second-order R\'enyi entropy, and logarithmic negativity for this system versus time for some fiducial parameters in Fig. \ref{fig:JCM-2by2-measures}. All of these quantities act as entanglement measures for the $2\times2$ system, as guaranteed by the PPT criterion. Of note, the measures involving negativity are initially more sensitive than the R\'enyi entropy, as the former grow linearly with time from separable states while the latter grows only quadratically \cite{Cresswell2017a}.
	
\begin{figure}\centering\includegraphics[width=0.7\columnwidth]{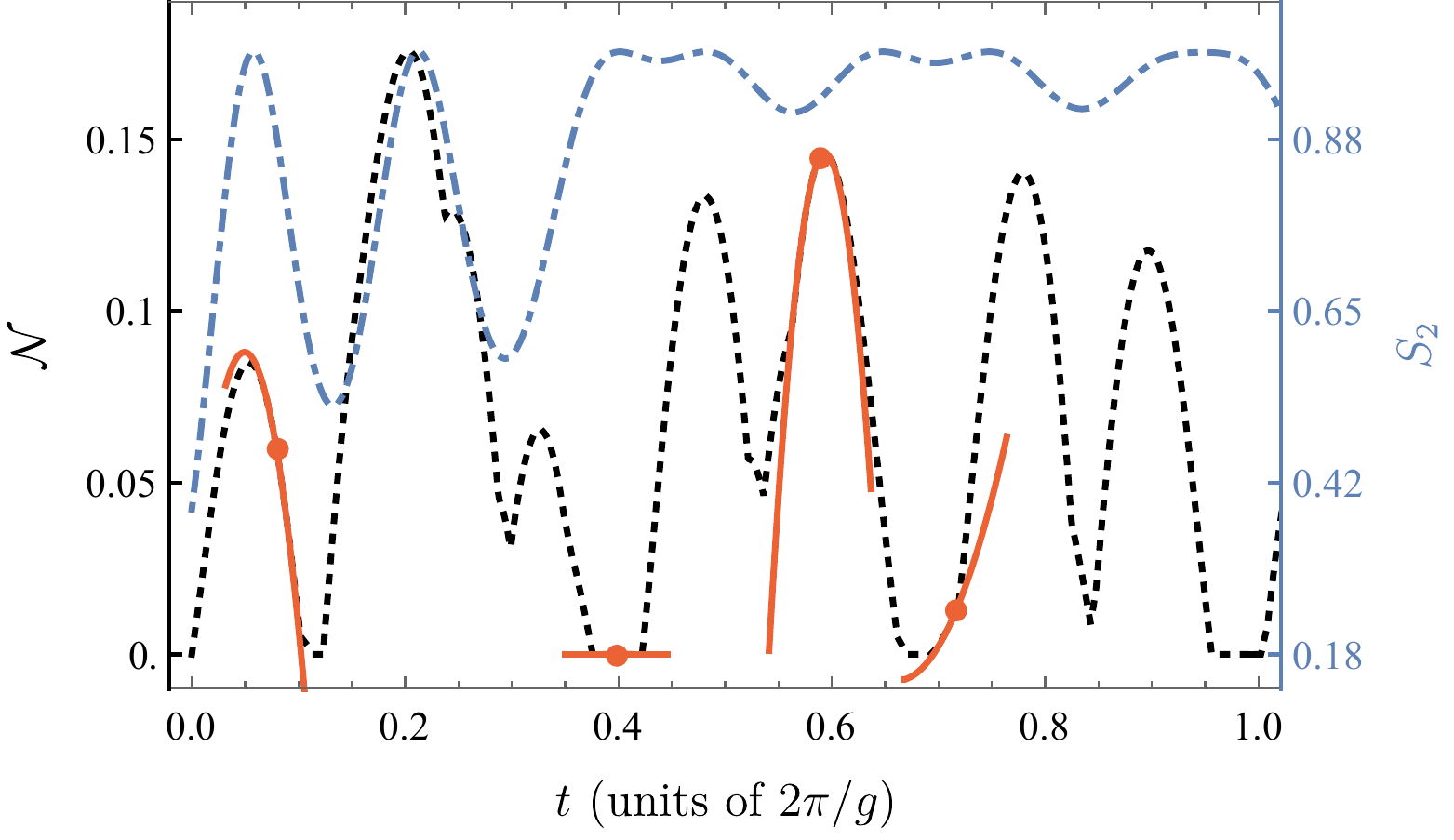}
		\caption[Quadratic approximations of negativity in Jaynes-Cummings model]{ Evolution of entanglement monotones in the Jaynes-Cummings model for the initially bound-entangled state given by Eq. \eqref{eq:JCM-bound-initial} and parameters $\left(\omega,\Delta,g\right)=\left(10,1,5\right)$. The negativity (dotted black curve) and R\'enyi entropy (dot-dashed blue curve) no longer oscillate with a single frequency, so we measure time in units of $2\pi/g$. We plot the second-order expansion from Eqs. \eqref{eq:first derivative} and \eqref{eq:second derivative} about various time points (solid orange parabolas); these agree with numerically-calculated derivatives of Eq. \eqref{eq:negativity-eigvals} in regions where the graph of negativity is both concave and convex as a function of time. Moreover, at times when the R\'enyi entropy is changing yet negativity is constant, our method successfully captures the dynamics of negativity.
			\normalsize}	
		\label{fig:negativity-JCM-bound-renyi}
	\end{figure}

	To investigate a system for which the PPT criterion is not sufficient, let us choose the initial state
	\eq{
		\label{eq:JCM-bound-initial}
		\nonumber  \rho\propto &
		\ket{g}\bra{g}\otimes\left[4\left(\ket{0}\bra{0}+\ket{1}\bra{1}\right)+9\left(\ket{2}\bra{2}+\ket{3}\bra{3}\right)\right]\\
		&+\ket{e}\bra{e}\otimes\left(\ket{0}\bra{0}+\ket{1}\bra{1}+\ket{2}\bra{2}+\ket{3}\bra{3}\right)
		\\
		&+\left\lbrace\ket{g}\bra{e}\otimes\left[2\left(\ket{1}\bra{0}+\ket{2}\bra{1}\right)+3\ket{3}\bra{2}\right]+\text{H.c.}
		\right\rbrace,
	}
	which was shown in Ref. \cite{Quesada2013} to have zero negativity while remaining entangled (these states are `bound' entangled \cite{Horodeckietal1998Bound,Lavoieetal2010}).
	The negativity and second-order R\'enyi entropy are plotted in Fig. \ref{fig:negativity-JCM-bound-renyi} for the same fiducial parameters as in Fig. \ref{fig:JCM-2by2-measures}. There are distinct regions in which the negativity fails to witness entanglement, i.e., in which negativity is zero and R\'enyi entropy is nonzero (such as $t=0$). The first and second derivatives, given by Eqs. \eqref{eq:first derivative} and Eq. \eqref{eq:second derivative}, agree to machine precision with the results obtained by differentiating \eqref{eq:negativity-eigvals}. We also plot in Fig. \ref{fig:negativity-JCM-bound-renyi} the second-order expansion found using Eqs. \eqref{eq:first derivative} and \eqref{eq:second derivative} about an assortment of time points to show that our equations capture the negativity dynamics even in regions where negativity is constant,  in intervals when $\rho^{T_B}$ is positive semi-definite, and in the presence of bound entanglement. One may also use our method to analyze how negativity changes with respect to the system parameters $\Delta$ and $g$ in order to explore how entanglement in the JCM is sensitive to the entire parameter landscape.
	
	\subsection{Open system dynamics: entangled cavity photons}
	\label{sec:open system}
	
	The perturbation theory developed above admits variations of negativity  with respect to any parameter $\mu$, given the derivatives $\partial^n\rho/\partial\mu^n$. In the JCM examples, we used time as the perturbation parameter, with the unitary evolution equation $\partial\rho/\partial t=-\iu\left[H_\text{JCM},\rho\right]$. A natural extension of our method is to parametrize non-unitary evolution; we can ask how negativity changes with time in systems whose dynamics are coupled to other, external systems. Sometimes the external systems themselves are responsible for the entanglement generated with time \cite{BenattiFloreaniniPiani2003}. We can also ask how negativity changes with respect to other parameters, including dynamical parameters and initial conditions. In this section we exhibit the versatility of our method in another quantum-optical context.
	
\begin{figure}\centering\includegraphics[trim={0cm 2.0cm 0cm 3.4cm
		},clip,width=0.7\columnwidth]{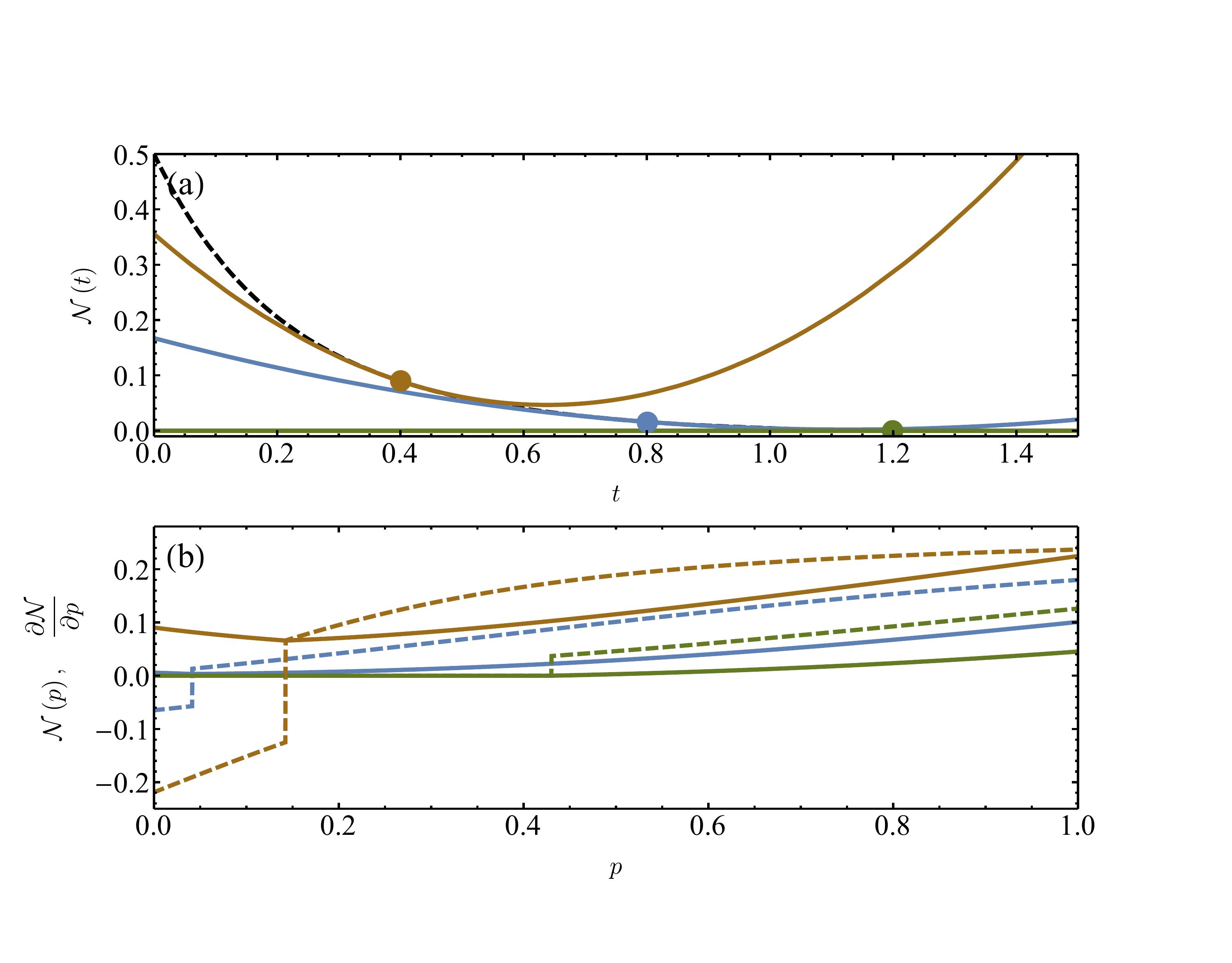}
		\caption[Dependence of negativity on system parameters in an open quantum system]{
			Dependence of negativity on system parameters in an open quantum system. \textbf{(a)} Evolution of negativity with respect to time (black, dashed curve), and perturbative expansions to second order using Eqs. \eqref{eq:first derivative} and \eqref{eq:second derivative} at time points $t=0.4$ (brown), $t=0.8$ (blue), and $t= 1.2$ (green). The initial mixing parameter is set to $p=0.35$. Negativity decays to 0 in finite time, beyond which all of the derivatives vanish, as successfully captured by our perturbation theory. Time is measured in units of the decay constant defined in Ref. \cite{Wangetal2016}. \textbf{(b)}
			Negativity (solid lines) and its derivatives with respect to the initial mixing parameter $p$ (dashed lines), for the same three time points as in (a) ($t=0.4$ is kinked at $p\approx 0.14$, $t=0.8$ at $p\approx 0.04$, and $t=1.2$ at $p\approx 0.43$). Regardless of time, negativity reaches a minimum for a particular value of $p$. When $t$ is small, negativity reaches a minimum for some $p=p_0$ between $0$ and $1$; $p_0$ is not monotonic with $t$. When $t$ is sufficiently large, negativity vanishes for all $p$ below a critical value, as seen in the $t=1.2$ curves. The derivatives are again calculated using Eq. \eqref{eq:first derivative}, and agree numerically with those found by differentiating $\mathcal{N}\left(p\right)$.\normalsize
		}
		\label{fig:arxiv}
	\end{figure}

	Negativity has recently been studied in the open system of a pair of cavities coupled to a pair of reservoirs with a flat spectrum {\cite{Lopezetal2008,Wangetal2016}}. The authors of Ref. \cite{Wangetal2016} showed that an initial mixture of maximally-entangled pairs of cavities, with states given by 
	\eq{\label{eq:cavity state}\rho_\text{cav}(t=0;&\,p)=p\ket{\psi}\bra{\psi}+\left(1-p\right)\ket{\phi}\bra{\phi},\quad 0\leq p\leq 1\\ &\ket{\psi}\propto\ket{0}\otimes\ket{0}+\ket{1}\otimes\ket{1},\\ &\ket{\phi}\propto\ket{0}\otimes\ket{2}+\ket{1}\otimes\ket{3},}
	can exhibit \emph{entanglement sudden death}; viz., negativity can decay to zero in finite time \cite{Almeidaetal2007,AlQasimiJames2008}. Furthermore, the cavity states coupled to by the dynamics are of dimension $2\times4$, so the PPT criterion does not hold in this system. The authors supply an analytic expression for $\rho_\text{cav}\left(t;\,p\right)$ (see Sec. \ref{app:open system dynamics} below), which can be compared to our perturbation theory method (Fig. \ref{fig:arxiv}). 
	
	In Fig. \ref{fig:arxiv}(a) we see a perfect agreement between our perturbation theory and the evolution of negativity with respect to time in this open system for a particular value of $p$. The dynamics are fully captured, including the time beyond which negativity decays to 0 and remains unchanged.
	
	Fig. \ref{fig:arxiv}(b) shows negativity and its derivatives from Eq. \eqref{eq:first derivative} with respect to the initial mixing parameter $p$ at various time points. The derivatives again match those found by differentiating Eq. \eqref{eq:negativity-eigvals} to machine precision. They give insight into the entanglement sudden death phenomenon, showing its dependence on initial conditions, as studied in depth in Ref. \cite{Wangetal2016}. Depending on the amount of initial mixing between the two entangled states $\ket{\psi_1}$ and $\ket{\psi_2}$, negativity decays at different rates with respect to time. For time evolution, negativity eventually reaches zero and remains there. With respect to $p$, negativity exhibits another sudden death feature: it decays to zero with shrinking $p$ at sufficiently long times. However, at shorter times, negativity reaches a minimum at intermediate values $p=p_0$, then grows again for increasing $\left|p-p_0\right|$, where the $p_0$ values are are highly sensitive to the time at which they are being evaluated. Our perturbation theory is an excellent tool for probing these complex phenomena or the dependence of negativity on any parameter $\mu$ in all systems for which $\partial\rho/\partial\mu$ is known.

	\section{Quantum speed limits and bounds on negativity growth}\label{sec:QSL}
	
	The techniques we have developed can also be readily adapted to other functions of quantum states and observables. For example, our perturbation theory can immediately be applied to any dynamics involving the trace norm. This includes the trace distance $\tfrac{1}{2}\left\Vert \rho-\sigma\right\Vert_{1}$ between two states, which has been used, for instance, to investigate non-Markovian systems \cite{Breueretal2009,Rivasetal2010,Aaronsonetal2013,Deffner2013}. It has been shown that non-Markovianity holds when the trace distance between two states undergoing the same dynamics increases over time \cite{Breueretal2009}, a condition that can now be investigated using our matrix calculus techniques. The fidelity between quantum states $F(\rho,\sigma)=\left\Vert \sqrt{\rho}\sqrt{\sigma}\right\Vert_1$, the Hilbert-Schmidt distance $D_{HS}(\rho,\si)=\Vert \rho-\si\Vert ^2_{HS}=\tr[(\rho-\si)^\dag(\rho-\si)]$, and other norm-based functions of the density matrix are all matrix functions for which the argument's patterns must be considered, as we will prove in the following section. In this section we give an example of how our techniques can be applied more broadly in quantum physics by studying quantum speed limits, which usually involve bounds on $\tfrac{d}{dt}\left\Vert \rho(t)-\rho(0)\right\Vert$ for some norm (see \cite{Deffner2017} for a recent review).
	
	The original quantum speed limit, the Mandelstam-Tamm bound, applied a generalized version of the Heisenberg uncertainty principle to find a lower bound on the time $\tau$ for any pure state of a quantum system $| \psi(0)\rangle$ to evolve into an orthogonal state $\langle \psi(\tau)|\psi(0)\rangle=0$:
	\begin{equation}\label{eq:MTbound}
\tau\geq \tau_{\text{QSL}}=\frac{\pi}{2}\frac{\hbar}{\De H}.
\end{equation}
As the Mandelstam-Tamm bound was an attempt at formalizing the energy-time uncertainty relation $\De t\De E\gtrsim \hbar$, which is not the consequence of any canonical commutation relation as opposed to $\De x\De p\gtrsim \hbar$, it naturally is phrased in terms of the uncertainty in the Hamiltonian, $\De H$.  While this bound always holds, it is not always tight. A second approach to the derivation of the minimal time for orthogonal evolution used the time-dependent Schr\"odinger equation and resulted in the slightly different Margolus-Levitin bound,
	\begin{equation}\label{eq:MLbound}
\tau\geq \tau_{\text{QSL}}=\frac{\pi}{2}\frac{\hbar}{\langle H\rangle}.
\end{equation}
In fact both bounds hold true, and it can be show that the unified bound is tight,
\begin{equation}\label{eq:unifiedbound}
 \tau_{\text{QSL}}=\max\left(\frac{\pi}{2}\frac{\hbar}{\De H},\frac{\pi}{2}\frac{\hbar}{\langle H\rangle}\right).
\end{equation}

When extending quantum speed limits to the case of mixed states one can no longer simply consider orthogonality as the obvious endpoint of evolution. In same spirit, though, we can instead consider measures of distinguishability, and determine the time required for a state to evolve to a certain distinctness. Orthogonal pure states are perfectly distinguishable, but we cannot expect perfect distinguishability for generic mixed state evolutions. Even so, there are many valid measures of distinguishability, such as the trace distance, fidelity, or Hilbert-Schmidt distance mentioned above. For our purposes, we notice that all of these measures can be reduced to matrix norms in the representative form $\Vert\rho(t)-\rho(0)\Vert$. The rate of change of this type of quantity is then telling us about the speed of quantum evolution through the space of density matrices, according to a distance measure determined by the choice of norm \cite{Deffner2017a}. An upper bound on this evolution speed, for any appropriate norm, would represent a fundamental limitation on the speed at which a quantum system can evolve, 
	\begin{equation}\label{eq:vQSL}
v_{\text{QSL}}\geq v = \frac{d}{dt}\left\Vert \rho(t)-\rho(0)\right\Vert,
\end{equation} and, if desired, a speed limit $v_{\text{QSL}}$ could be turned into a speed limit time by averaging, 
	\begin{equation}
\tau_{\text{QSL}}=\frac{\tau}{\int_0^\tau dt\ v_{\text{QSL}}}.
\end{equation}
As we have pointed out, explicit computation of derivatives of this type must be carried out with careful attention to the Hermitian pattern of $\rho$. 
	
	The patterned derivatives for the trace norm from \eqref{eq:drhopatterned} already leads to an interesting bound of the form \eqref{eq:vQSL} on the evolution of quantum systems, but in the following we will consider the much more general set of Schatten $p$-norms
	\begin{equation}
\Vert X\Vert _p=[\tr(|X|^p)]^{1/p}.
\end{equation}
By computing the patterned derivatives of this norm, we will be able to bound the evolution speed $\tfrac{d}{dt} \left\Vert \rho(t)-\rho(0)\right\Vert_p$ in terms of $\left\Vert \dot \rho\right\Vert_p$ which is directly connected to the equations of motion of the system. 

Following the same procedure as before we start with the unpatterned derivatives. The differential is
\begin{equation}
d\Vert X\Vert _p=(\Vert X\Vert _p)^{1-p}\tr(|X|^{p-1}d|X|),
\end{equation}
and from $d(|X| |X|)=d(X^\dag X)$ this can be reexpressed as
\begin{equation}
d\Vert X\Vert _p=\frac{1}{2}(\Vert X\Vert _p)^{1-p}\tr(|X|^{p-1}(X^\dag dX+dX^\dag X)).
\end{equation}
At this point it is convenient to vectorize,
\begin{equation}
d\Vert X\Vert _p=\frac{1}{2}(\Vert X\Vert _p)^{1-p}[\veT(|X|^{p-2} X^\dag)Kd\ve X+\veT(X|X|^{p-2}) d\ve X^*],
\end{equation}
which allows us to read off the derivatives
\begin{equation}
D_X\Vert X\Vert _p=\frac{1}{2}(\Vert X\Vert _p)^{1-p}\veT(|X|^{p-2} X^\dag)K,\quad D_{X^*}\Vert X\Vert _p=\frac{1}{2}(\Vert X\Vert _p)^{1-p}\veT(X|X|^{p-2}).
\end{equation}

For the application to quantum systems, we require the Hermitian derivative for $A^\dag=A$ as in \eqref{eq:DTrNorm} which simplifies to
\begin{equation}
D_A\Vert A\Vert _p=(\Vert A\Vert _p)^{1-p}\veT(A|A|^{p-2})K.
\end{equation}
As a check, putting $p=1$ reproduces the trace norm results \eqref{eq:unpatder} and \eqref{eq:drhopatterned}:
\begin{equation}
D_X\Vert X\Vert _1=\frac{1}{2}\veT(|X|^{-1} X^\dag)K,\quad D_{X^*}\Vert X\Vert _1=\frac{1}{2}\veT(X|X|^{-1}),
\end{equation}
\begin{equation}
D_A\Vert A\Vert _1=\veT(A|A|^{-1})K.
\end{equation}
Another interesting intermediate result comes from setting $p=2$, which gives the derivatives of the Hilbert-Schmidt norm
\begin{equation}
D_X\Vert X\Vert _2=\frac{1}{2\Vert X\Vert _2}\veT( X^*),\quad D_{X^*}\Vert X\Vert _2=\frac{1}{2\Vert X\Vert _2}\veT(X),
\end{equation}
\begin{equation}
D_A\Vert A\Vert _2=(\Vert A\Vert _2)^{-1}\veT(A)K.
\end{equation}

The desired quantity $\tfrac{d}{dt}\left\Vert \rho(t)-\rho(0)\right\Vert_p$ is now found through a simple application of identity \eqref{eq:TrAB}, 
\begin{align}\begin{aligned}
\frac{d\left\Vert \rho(t)-\rho(0)\right\Vert_p}{dt}&=D_\rho\Vert\rho(t)-\rho(0)\Vert_p\frac{\pa \ve\rho(t)}{\pa t}\\
&=(\Vert\rho(t)-\rho(0)\Vert_p)^{1-p}\tr\{[\rho(t)-\rho(0)] |\rho(t)-\rho(0)|^{p-2}\dot \rho\}.
\end{aligned}\end{align}
We can bound the evolution speed in terms of $\left\Vert \dot \rho\right\Vert_p$ by first using the operator inequality $|\tr(A)|\leq \tr(|A|)$, leading to
 \begin{equation}
\frac{d\left\Vert \rho(t)-\rho(0)\right\Vert_p}{dt}\leq\left\vert \frac{d\left\Vert \rho(t)-\rho(0)\right\Vert_p}{dt}\right\vert \leq (\Vert\rho(t)-\rho(0)\Vert_p)^{1-p}\tr \left\vert[\rho(t)-\rho(0)]|\rho(t)-\rho(0)|^{p-2}|\dot \rho\right\vert.
\end{equation}
Next we can apply H\"older's inequality for the Schatten norms
\begin{equation}\label{eq:Holder}
    \Vert YZ\Vert _1=\tr|YZ|\leq\Vert Y\Vert _{q}\Vert Z\Vert _{q^*},
\end{equation}
where $q,q^*$ are chosen such that the norms are dual, $1/q+1/q^*=1$. With the judicious choices $Y=[\rho(t)-\rho(0)]|\rho(t)-\rho(0)|^{p-2}$, $Z=\dot\rho$, $q=\tfrac{p}{p-1}$, and $q^*=p$, H\"older's inequality precisely dictates that
 \begin{equation}\label{eq:vQSLbound}
\frac{d\left\Vert \rho(t)-\rho(0)\right\Vert_p}{dt} \leq \Vert\dot\rho\Vert_p=v_{\text{QSL}}.
\end{equation}
This relation was derived without reference to the particular dynamics of the system $\dot\rho$, so it will be valid for open or closed systems and for von Neumann, Lindbladian, or even non-Markovian dynamics. Simply put, $\Vert \dot\rho\Vert_p$ is a fundamental quantum speed limit for the evolution velocity through state space measured with respect to the Schatten $p$-norm distance.

We now turn to another bound on quantum dynamics that can be proven as an extension of our results on negativity in this chapter. This time the focus will be on the rate of entanglement generation rather than the state evolution itself. In particular, we will bound the rate of change of negativity in terms of $\Vert\dot\rho\Vert_1$, similar to the quantum speed limit above.

Since negativity is, broadly, just the trace norm of the partial transpose, we will focus on the representation presented in Sec. \ref{subsec:ptranspose}. With the partial transpose map \eqref{eq:supop} acting as in \eqref{eq:vecrhoTb}, we find from \eqref{eq153} the necessary matrix superoperator \eqref{eq:superopmap}
\begin{equation}
\rho^{T_B}=\sum_{ij}(\id_A\ot J^{ij}_{d_B})\rho(\id_A\ot J^{ij}_{d_B}).
\end{equation}
The trace norm of this expression can be bounded by applying the triangle inequality for each term in the sums,
\begin{equation}
\left\Vert\rho^{T_B}\right\Vert_1\leq\sum_{ij}\left\Vert(\id_A\ot J^{ij}_{d_B})\rho(\id_A\ot J^{ij}_{d_B})\right\Vert_1,
\end{equation}
and then with two applications of H\"older's inequality \eqref{eq:Holder} with $q=1$ and $q^*=\infty$,
\begin{equation}
\left\Vert\rho^{T_B}\right\Vert_1\leq \sum_{ij} \left\Vert  \left(\id_{d_{A}}\otimes J_{d_{B}}^{ij}\right) \right\Vert_{\infty} \left\Vert \rho \right\Vert_1 \left\Vert  \left(\id_{d_{A}}\otimes J_{d_{B}}^{ij}\right)\right\Vert_{\infty}.
\end{equation}
The operator norm $\Vert \cdot\Vert_\infty$ appearing here gives the largest singular value of its argument, as opposed to the trace norm which gives the sum of all singular values. This is a useful characterization since singular values combine simply under the Kronecker product. If $A$ has non-zero singular values $\si^A_i$ for $i=1\dots r_A$, and $B$ has $\si^B_j$ for $j=1\dots r_B$, then $A\ot B$ will have rank $r_A r_B$, and its singular values are all the possible combinations $\si^A_i \si^B_j$. Given that singular values are non-negative and listed in decreasing order, the operator norm of the Kronecker product is $\Vert A\ot B\Vert_\infty=\si^A_1 \si^B_1=\Vert A\Vert_\infty \Vert B\Vert_\infty$. The previous line then simplifies greatly to 
\begin{equation}
\left\Vert\rho^{T_B}\right\Vert_1\leq \left\Vert \id_{d_{A}}\right\Vert_{\infty}^2\left\Vert \rho \right\Vert_1\sum_{ij} \left\Vert J_{d_{B}}^{ij}\right\Vert_{\infty}^2.
\end{equation}
These factors are quite manageable since $\left\Vert \id_{d_{A}}\right\Vert_{\infty}^2 = 1$, while $\sum_{ij} \big\Vert J_{d_{B}}^{ij}\big\Vert_{\infty}^2 = \sum_i \left\Vert J_{d_{B}}^{ii}\right\Vert_{\infty}^2=d_B$, since the eigenvalues of $J^{ij}$ are 0 for $i\neq j$ and 1 for $i=j$. Our explicit representation for the partial transpose has allowed us to determine \begin{equation}
\left\Vert \rho^{T_{B}} \right\Vert_1 \leq d_B \left\Vert \rho \right\Vert_1.
\end{equation}

Before introducing dynamics, we can quickly improve this bound by noting that $\rho^{T_{A}}=\left(\rho^{T_{B}}\right)^T$, which means $\left\Vert \rho^{T_{A}} \right\Vert_1 =\left\Vert \rho^{T_{B}} \right\Vert_1 $. Hence, the bound should be symmetric in $d_A$ and $d_B$. Indeed, repeating the above computation for the partial transpose with respect to the other subsystem
\begin{equation}
\rho^{T_{A}} = \sum_{ij}^{d_A} \left(J_{d_{A}}^{ij}\otimes \id_{d_B} \right) \rho \left(J_{d_{A}}^{ij}\otimes \id_{d_B} \right),
\end{equation}
gives 
\begin{equation}
\left\Vert \rho^{T_{B}} \right\Vert_1 \leq d_A \left\Vert \rho \right\Vert_1.
\end{equation}
This proves the stricter bound
\begin{equation}\label{eq:ptransrhobound}
\left\Vert \rho^{T_{B}} \right\Vert_1 \leq \min(d_A,d_B) \left\Vert \rho \right\Vert_1.
\end{equation}
In fact this bound is tight, since maximally entangled states saturate it. Take the generalized Bell state between two $d$-dimensional qudits,
\begin{equation}
\left | \Psi \right\rangle = \frac{1}{\sqrt{d}}\sum_{j=0}^d \left | j, j \right \rangle.
\end{equation}
Its density matrix $\rho=\left| \Psi\right \rangle \left\langle \Psi \right |$ is normalized to unity in the trace norm, whereas the partial transpose $\rho^{T_B}=\tfrac{1}{d}\sum_{ij=0}^d\left| i,j\right \rangle \left\langle j,i \right |$ has $d^2$ singular values equal to $\tfrac{1}{d}$. Hence $\Vert \rho^{T_B}\Vert_1=d \Vert \rho \Vert_1$ for this class of states.

We can now make use of our first main result of this section, \eqref{eq:vQSLbound}, to find a bound on the rate of change of our proxy for negativity, $\Vert \rho^{T_B}\Vert_1$, from \eqref{eq:ptransrhobound}:
\begin{equation}
    \left\vert\frac{\partial\left\Vert\rho^{T_B}\right\Vert_1}{\partial t}\right\vert\leq \min(d_A,d_B)\left |\frac{\partial\left\Vert\rho\right\Vert_1}{\partial t}\right | \leq \min(d_A,d_B) \left\Vert\dot{\rho}\right\Vert_1.
\end{equation}
The quantum speed limit \eqref{eq:vQSLbound} has allowed us to bound the growth of negativity directly in terms of the system's dynamics. As an example, the basic case of unitary evolution under Hamiltonian $H$ could be treated by first rewriting the dynamical term as
\begin{equation}\left\Vert\dot{\rho}\right\Vert_1 = \left\Vert H \rho-\rho H\right\Vert_1 \leq \left\Vert H \rho\right\Vert_1+\left\Vert\rho H\right\Vert_1 \leq 2\left\Vert H \right\Vert_1.
\end{equation}
These steps use the triangle inequality, sub-multiplicitivity which all Schatten $p$-norms obey, and the normalization of $\rho$. If we also reinstate the negativity according to its definition \eqref{eq:negativitydef} we have a new bound on  entanglement dynamics
\begin{equation}\label{eq:neggrowthbound}
    \left\vert\dot{\mathcal{N}}(t)\right\vert \leq \min(d_A,d_B) \left\Vert H \right\Vert_1.
\end{equation}
We note that self-energy terms in the Hamiltonian will not produce entanglement, so one would expect that a bound on entanglement dynamics should involve only the interaction terms of $H$. To resolve this issue we note that constant terms in $H$ do not contribute to entanglement either, but do still affect its trace norm. Therefore this upper bound \eqref{eq:neggrowthbound} can be lowered by adding a constant multiple of the identity to $H$ such that $\Vert H+\lambda\id\Vert_1$ is minimized. Doing so can be thought of as cancelling out the influence of any self-energy terms on the $\left\Vert H \right\Vert_1$ side of the bound, while not affecting $\left\vert\dot{\mathcal{N}}(t)\right\vert$.

\section{When is patterned matrix calculus required?}\label{sec:formalcalculus}

It may seem inconsistent that this chapter has emphasized the importance of taking into account the Hermiticity of the density matrix for derivatives of negativity, while in Chapter \ref{ch:Renyi} this issue was ignored completely. Indeed, the subtleties of patterned matrix calculus have been ignored in many calculations in the literature, yet are unavoidable when studying negativity as done here. In this section we explain the major difference between negativity and, for example, R\'enyi entropies, which entails that patterns can be ignored for the latter, but not the former when taking derivatives. Our result is a theorem stating that for complex analytic matrix functions, that is, for functions $G(X)$ that do not depend explicitly on the complex conjugate $X^*$, there is no functional difference between patterned and unpatterned derivatives. First, we recount the calculus of complex, patterned matrices developed in \cite{ Hjorungnes2007,Tracy1988,Hjorungnes2008,Hjorungnes2008a} and summarized in \cite{Hjorungnes2011}. {This calculus was used in \cite{Hjorungnes2011} to rigorously compute the derivative of a scalar function with respect to a Hermitian argument \eqref{eq:DTrNorm}, and also provides a rigourous derivation of our result on patterned Hessians \eqref{eq:hessscal}. After this introduction to the formal calculus, we leverage it to prove a theorem on the types of functions for which patterned and unpatterned derivatives are not equal.}
	
	\subsection{Formal calculus of complex patterned matrices \label{sec:Acomp_pat}}
	
	Consider $F\left(P,W,W^{*}\right)$,
	a differentiable, complex matrix-valued function of a real matrix variable $P$,
	a complex matrix variable \textbf{$W$} and its complex conjugate
	$W^{*}$. The differential of such a function is given by
	\begin{align}\label{eq:differential}
	\begin{aligned}
	d\ve F=&\left(D_{P}F\right)d\ve P+\left(D_{W}F\right)d\ve W+\left(D_{W^{*}}F\right)d\ve W^{*},
	\end{aligned}
	\end{align}
	where the differentials of $P$, $W$ and $W^{*}$ are independent,
	and the Jacobian $D_{P}F$, for example, is
	\begin{align}
	D_{P}F & \coloneqq\frac{\partial\,\ve F}{\partial\,\veT P}.
	\end{align}
	Fortuitously, the differential commutes with vectorization, tracing,
	transposition, and conjugation:
	\begin{align}
	\begin{aligned}
	d\left(\ve X\right)&=\ve\left(dX\right),\quad d\left(\tr X\right)=\tr dX,\\
	d\left(X^{T}\right)&=\left(dX\right)^{T},\quad d\left(X^{*}\right)=\left(dX\right)^{*}.
	\end{aligned}
	\end{align}
	Also, derivatives in this formalism satisfy a chain rule; for a composite function
	\begin{equation}\label{eq:chainrule}
	H\left(P,W,W^{*}\right)=G\left[F\left(P,W,W^{*}\right),F^{*}\left(P,W,W^{*}\right)\right],
	\end{equation}
	we have
	\begin{align}
	D_{P}H&=\left(D_{F}G\right)\left(D_{P}F\right)+\left(D_{F^{*}}G\right)\left(D_{P}F^{*}\right),\\
	D_{W}H&=\left(D_{F}G\right)\left(D_{W}F\right)+\left(D_{F^{*}}G\right)\left(D_{W}F^{*}\right),\\
	D_{W^*}H&=\left(D_{F}G\right)\left(D_{W^*}F\right)+\left(D_{F^{*}}G\right)\left(D_{W^*}F^{*}\right).
	\end{align}

	We must employ a careful strategy for taking derivatives with respect to a matrix if there are any elements in that matrix which are (possibly constant) functions of the other elements. Such an approach
	was developed in \cite{Tracy1988, Hjorungnes2008, Hjorungnes2008a} and we summarize it here for a differentiable function $G(A,A^*)$ of a complex patterned matrix $A$:
	
		(1) Let $F$ be a function that acts on a set of unpatterned matrices $[P,W,W^{*}]$
		to make a patterned matrix $A=F\left(P,W,W^{*}\right)$. This function
		must be differentiable with respect to $P,$ $W$, and $W^{*}$,
		and a diffeomorphism between the sets of patterned and unpatterned
		matrices. That is, $F$ must be a smooth, bijective function whose inverse is also
		smooth. The number of independent parameters contained in $P,W,W^{*}$ that fully parametrize
		the set of patterned matrices should be minimal.
	
		(2) Let $X$ be an unpatterned matrix with the same size as $A$. Extend $G$ to act on unpatterned matrices and find its derivatives $D_{X}G\left(X,X^{*}\right)$ and $D_{X^*}G\left(X,X^{*}\right)$. Use the chain rule for $G(A,A^*)=H\left(P,W,W^{*}\right)$ as in \eqref{eq:chainrule} to find
		\begin{equation}
		{D_{P}H} ={D_{X}G}{\left({X}{,}{X^{*}}\right)}{\vert_{{X}={A}}}D_{P}F+D_{X^{*}}G{\left({X}{,}{X^{*}}\right)}{\vert_{{X}={A}}}D_{P}F^{*}, \label{eq:patapp}
		\end{equation}
		etc., where patterns are applied after differentiation.

	(3) The derivative of $G(A,A^*)$ with respect to the patterned matrix $A$ is given by
		\begin{equation}
		D_{A}G=\left[D_{P}H,D_{W}H,D_{W^{*}}H\right]D_{A}F^{-1}.\label{eq:finalpatapp}
		\end{equation}

	In other words, one should find a minimal basis to represent the set of patterned matrices $A$, compute derivatives in this basis, then transform back to the standard basis. The diffeomorphism $F(P,W,W^*)$ represents the transformation from the minimal basis to the standard basis, while its inverse $F^{-1}(A)$ produces a vector of matrices $[P,W,W^{*}]^T$.

	\subsection{Analytic functions of matrices} \label{sec:conjindcase}
	In this section we discuss a sufficient condition for when consideration of matrix patterns is unnecessary in taking derivatives of functions with respect to those matrices. Let $G(X,X^{*})$ be a matrix-valued function that is differentiable in matrices $X$. Suppose that $G(X,X^{*})$ is analytic in the sense that it is independent of $X^{*}$, i.e., $D_{X^{*}}G=0$. Let $A$ be a patterned matrix with the same size as $X$, and $F$ an appropriate diffeomorphism acting on the minimal set of matrix parameters $\left[P,W,W^{*}\right]$ such that $A=F(P,W,W^{*})$. Define $H$ so that $H(P,W,W^{*})=G(A,A^{*})$ as in the beginning of Sec. \ref{sec:Acomp_pat}. We have, from Eq. \eqref{eq:patapp}, that
	\begin{equation}
	D_{P}H  =D_{X}G\left(X,X^{*}\right)\vert_{X=A}D_{P}F,
	\end{equation}
	and similarly for $D_{W}H$ and $D_{W^{*}}H$. This means that Eq. \eqref{eq:finalpatapp} now reads
	\begin{equation}
	D_{A}G = D_{X}G \vert _{X=A} [D_{P}F,D_{W}F,D_{W^{*}}F]D_{A}F^{-1}.
	\end{equation}
	But, by construction, the diffeomorphism $F$ satisfies,
	\begin{equation}
	[D_{P}F,D_{W}F,D_{W^{*}}F]D_{A}F^{-1}=\id,
	\end{equation}
	since this amounts to changing from the standard basis to the minimal basis and back \cite{Hjorungnes2011}.
	Thus we can see that
	\begin{equation}
	D_{A}G = D_{X}G \vert _{X=A}.
	\end{equation}
	
	The conclusion we draw is as follows: for functions independent of the complex conjugate of their argument, taking the patterned derivative is equivalent to differentiating with respect to the unpatterned argument and evaluating it at the patterned matrix. The Schatten $p$-norms, including the trace norm, do not obey this condition, and we found their patterned derivatives to have a different form compared to the unpatterned counterparts.

\section{Discussion \label{sec:Discussion}}
	
In the preceding sections, we have provided a means of computing the
perturbative expansion of the entanglement negativity. The complete
expressions require knowledge of the partial transpose $\rho^{T_{B}}$
at the expansion point; of the partial commutation matrix $K_B$, which we have presented explicitly in various convenient forms; and of the dynamics of $\rho$.
	
Because the trace norm is not differentiable at points where $\rho^{T_{B}}$ is singular, we assumed that $\rho^{T_{B}}$ was invertible in our discussion. It would be interesting to know the conditions for $\rho^{T_B}$ to be invertible based on properties of $\rho$. There is, to our knowledge, no straightforward relationship between the rank of a general density matrix and the rank of its partial transpose. On the other hand, we can make some conclusions by considering pure states. Let $\sigma=\ket{\psi}\bra{\psi}$ be a pure state in a bipartite Hilbert space with dimensions $d_{A}\times d_{B}$ and let $r$ be the \emph{Schmidt rank} of $\ket{\psi}$, the number of non-zero coefficients in its Schmidt decomposition. From \cite{Johnston2018}, we know that the matrix rank of $\sigma^{T_{B}}$ is $r^{2}$. Since $r$ is bounded from above by the minimum of $\left\{ d_{A},d_{B}\right\}$, $\sigma^{T_{B}}$ having maximal rank implies that $r=d_{A}=d_{B}$. For pure states, then, $\sigma^{T_{B}} = \left(\ket{\psi}\bra{\psi}\right)^{T_{B}} $ is invertible if and only if the Schmidt rank of $\ket{\psi}$ is maximal and the dimensions of the subsystems are equal.
	
For singular $\rho^{T_{B}}$  it might be possible to apply our analysis to the evolution of $\rho$ in an $r^{2}$- dimensional subspace where $\rho^{T_{B}}$ is supported. Otherwise, the trace norm still has a well defined subdifferential because it is convex \cite{Watson1992}. It may be possible to optimize over the set of subgradients given extra input, such as the global bound $\mathcal{N}\geq 0$, to determine the evolution of $\mathcal{N}(t)$ around singular points and find a one-sided derivative. We leave these ideas for future exploration.

Many calculations have been presented in the literature involving functions of density matrices that seemingly did not require patterned matrix calculus, including those of the previous chapter. We addressed this intriguing discrepancy by showing that negativity and R\'enyi entropies belong to very different classes of functions from the point of view of patterned matrix calculus. We proved that analytic matrix functions, those that do not explicitly depend on the complex conjugate of the argument, will have equivalent patterned and unpatterned derivatives. It is only for non-analytic functions that one needs to recruit these more subtle techniques. This coincidence allows one to gloss over the patterns of the density matrix when studying common functions like R\'enyi entropies $S_\alpha(\rho)=\tfrac{1}{1-\alpha}\log \tr \left(\rho^\alpha\right)$ \cite{Kim1996, Cresswell2017a}. In contrast, the trace norm $\left\Vert X\right\Vert_{1}=\tr \sqrt{X^\dag X}$ explicitly depends on the complex conjugate $X^*$, so that the negativity is not analytic in this sense. The most common examples of non-analytic matrix functions are matrix norms, like the family of Schatten $p$-norms studied in Sec. \ref{sec:QSL}.
		
The primary challenge in our perturbative expansion was in the correct application of patterned matrix calculus to the problem. The salient pattern was the Hermiticity of the density operator, which implies Hermiticity of its partial transpose. We were able to extend the approach taken in \cite{Hjorungnes2011} to compute the first and second derivatives of the trace norm with respect to a Hermitian argument. However, Hermiticity may not be the only pattern at play. Density matrices are also normalized to have unit trace, and evolution may conspire to endow additional structure to the partial transpose. 
	
As we have discussed, patterned derivatives can be found by first computing unpatterned derivatives, and subsequently imposing patterns on the result. Hermiticity is a strong condition which, as we showed in Sec. \ref{subsec:trnormder}, destroys the independence of the complex differentials $d\ve \rho^{T_B}$ and $d\ve \left(\rho^{T_B}\right)^*$. Hence, imposing Hermiticity greatly alters the functional form of the derivatives, resulting in \eqref{eq:DTrNorm}.

By contrast, the unit trace condition for $\rho^{T_{B}}$ introduces some dependencies among the diagonal elements, but this structure does not affect the patterned derivatives. The unit trace condition is a numerical constraint that does not change the functional form of the derivatives, and can simply be applied to the unpatterned derivative. For this reason we have not endeavoured to treat it with the same rigour as Hermiticity.
	
Additionally, some readers may also have been perturbed to notice that no consideration was given to the patterns of $\rho$ when computing $\partial\,\ve \rho^{T_{B}}/\partial\,\veT \rho$ in \eqref{eq:drhoTBdrho}. One way to explain this is that the function that maps $\ve \rho$ to its partial transpose $\ve \rho^{T_{B}}=K_B\ve \rho$ is constant, depending only on the dimensions of $\rho$, and hence is functionally independent of $\rho^{*}$. It is an analytic matrix function, and Sec. \ref{sec:conjindcase} shows that its patterned and unpatterned derivatives must be equivalent.

Since we allow for general dynamics of $\rho(t)$, in theory its evolution might induce patterns on the partial transpose beyond Hermiticity. For example, one can conceive of a Hamiltonian that keeps $\rho^{T_{B}}$ positive semidefinite for some time interval, indicating a protracted separability or bound entanglement. In such a scenario the patterned derivatives may be functionally different from the Jacobian \eqref{eq:DTrNorm} or Hessian \eqref{eq:hessscal}, and would need to be treated on a case-by-case basis. However, we have seen in Section \ref{sec:open system} that our expansion correctly predicts zero evolution of the negativity when $\rho^{T_B}$ is positive semidefinite for an example system.
	
Our analysis can be applied to probe changes in negativity in a broad assortment of physical systems, and the techniques we employ can be readily adapted to other functions of quantum states and observables. Studies of phenomena as disparate as phase transitions \cite{Lu2018}, quantum quenches \cite{Coser2014,EislerZimboras2014}, and beam propagation \cite{Fedorovetal2018} can harness our methods in their investigations of negativity. In Sec. \ref{sec:QSL} we demonstrated two distinct uses of our techniques to bound the rate of evolution of general quantum systems with a quantum speed limit, as well as providing a bound on entanglement dynamics through negativity. The first main result, eq. \eqref{eq:vQSLbound}, limits the rate at which any quantum state can evolve in state space, according to the distance measure induced by the Schatten $p$-norm. This hearkens back to the original quantum speed limits for a pure state to evolve to an orthogonal state. Our quantum speed limit also allowed us to also provide a bound on the growth of negativity \eqref{eq:neggrowthbound} in terms of the Hamiltonian for closed systems.

One especially interesting application of our calculus is to the linear, or nearly linear, growth of entanglement observed in a large class of many-body systems using entanglement entropy \cite{Calabrese2005,Bianchi2018} and, more recently, negativity \cite{Calabreseetal2012,Coser2014}. In critical systems, quasi-particles produced by a quench spread at a uniform velocity, leading to an emergent lightcone-like behaviour and exactly linear growth of logarithmic negativity. For more general systems, quasi-particles can propagate at varying speeds, leading to an approximate linear growth that has been studied numerically \cite{Coser2014,EislerZimboras2014}. Our approach to the derivatives of negativity provides a new avenue to analytically explore the conditions under which second and higher derivatives of the negativity will vanish. 
	
Our techniques can even be employed for classical applications of complex patterned matrices, such as analyzing the condition number for Mueller matrices \cite{Layden2012}, whose patterns are discussed in \cite{Simon2010}. Understanding the evolution of entanglement and other functions of complex patterned matrices will have ramifications for an expansive range of fields in the near future.

The remainder of this Chapter contains technical details.

\section{Vectorized representation of the partial transposition map\label{sec:superoperator}}

Here we derive Eq. \eqref{eq:supop}, where the partial transposition map $T_B=\id\otimes T$ is recast to act on vectorized $d_{A}d_{B}\times d_{A}d_{B}$ matrices as in \eqref{eq:vectorizedsupop}, and takes the form of \eqref{eq153}, namely $K_B=\sum_{i}B^T_{i}\otimes A_{i}$. This is accomplished by finding the action of $T_B$ on each element of the standard basis of matrices, and then vectorizing. 

The standard basis consists of single-entry matrices $(J^{ij})_{kl}=\delta_{ik}\delta_{jl}$ with the following ordering (we reserve $J^{ij}$ with no subscript for the $d_{A}d_{B}\times d_{A}d_{B}$ case):
\begin{equation}
\left\{J^{1,1} ,J^{2,1},\cdots,J^{d_{A}d_{B}-1,d_{A}d_{B}} ,J^{d_{A}d_{B},d_{A}d_{B}}\right\}.
\end{equation}
If we parametrize $i$ and $j$ by
\begin{equation}
i=(a_{i}-1)d_{B}+b_{i},\quad j=(a_{j}-1)d_{B}+b_{j},
\end{equation}
with $1\leq a_{i},a_{j}\leq d_{A}$ and $1\leq b_{i},b_{j}\leq d_{B}$, then we can decompose $J^{ij}$ as
\begin{equation}
J^{ij}=J^{(a_{i}-1)d_{B}+b_{i},(a_{j}-1)d_{B}+b_{j}}=J_{d_{A}}^{a_{i},a_{j}}\otimes J_{d_{B}}^{b_{i},b_{j}},
\end{equation}
using $m\times m$ single-entry matrices $J_{m}^{ij}$. From this we can read off the action of the partial transposition transformation on the basis elements
\begin{equation}\label{eq:TBonMatrices}
T_{B}\left(J^{ij}\right)=J_{d_{A}}^{a_{i},a_{j}}\otimes J_{d_{B}}^{b_{j},b_{i}}=J^{(a_{i}-1)d_{B}+b_{j},(a_{j}-1)d_{B}+b_{i}}.
\end{equation}
In the vectorized representation we use the basis $\{\ve J^{1,1} ,\ve J^{2,1},{\cdots},\ve J^{d_{A}d_{B}-1,d_{A}d_{B}} ,\ve J^{d_{A}d_{B},d_{A}d_{B}}\}$, and the action of $T_B$ on basis elements is determined by vectorizing both sides of \eqref{eq:TBonMatrices} such that $ \ve T_{B}\left(J^{ij}\right) = K_B \ve J^{ij}$. $K_B$ is a $\left(d_Ad_B\right)^2\times \left(d_Ad_B\right)^2$ permutation matrix whose elements can be expressed in terms of the single-entry matrices $J^{qr}_{\left(d_Ad_B\right)^2}$, with $1\leq q,r\leq \left(d_Ad_B\right)^2$. Note that these matrices are larger than the matrices $J^{ij}$ with no subscripts. The $\ve J^{ij}$ basis element has its non-zero entry in position $r=(j-1)d_{A}d_{B}+i$. Hence, the $r^{\text{th}}$ column of $K_B$ is equal to $\ve T_B\left(J^{ij}\right)$. From \eqref{eq:TBonMatrices} we see that partial transposition takes $i\to i^\prime=\left(a_i-1\right)d_B+b_j$ and $j\to j^\prime=\left(a_j-1\right)d_B+b_i$, so the $r^{\text{th}}$ column only has a non-zero entry in the $q^\text{th}$ row, where $q=\left(j^\prime-1\right)d_{A}d_{B}+i^\prime$. This non-zero element of $K_B$ can be expressed as
\begin{align}
\begin{aligned}
I(i,j)=&\, J_{\left(d_{A}d_{B}\right)^{2}}^{(j^\prime-1)d_{A}d_{B}+i^\prime,(j-1)d_{A}d_{B}+i}=J^{j^\prime,j}\otimes J^{i^\prime,i}=J^{(a_{j}-1)d_{B}+b_{i},(a_{j}-1)d_{B}+b_{j}}\otimes J^{(a_{i}-1)d_{B}+b_{j},(a_{i}-1)d_{B}+b_{i}}.
\end{aligned}
\end{align}
Every vectorized basis element $\ve J^{ij}$ matches with an element $I\left(i,j\right)$, so that $K_B$ is the sum of all such elements:
\begin{align}
\begin{aligned}
K_B=& \sum_{a_{i},a_{j}=1}^{d_{A}}\sum_{b_{i},b_{j}=1}^{d_{B}}J^{(a_{j}-1)d_{B}+b_{i},(a_{j}-1)d_{B}+b_{j}}\otimes J^{(a_{i}-1)d_{B}+b_{j},(a_{i}-1)d_{B}+b_{i}}\\
=& \sum_{b_{i},b_{j}=1}^{d_{B}}\left(\sum_{a=1}^{d_{A}}J^{(a-1)d_{B}+b_{i},(a-1)d_{B}+b_{j}}\right)\otimes\left(\sum_{a=1}^{d_{A}}J^{(a-1)d_{B}+b_{j},(a-1)d_{B}+b_{i}}\right)\\
=& \sum_{b_{i},b_{j}=1}^{d_{B}}\left(\sum_{a=1}^{d_{A}}J_{d_{A}}^{a,a}\otimes J_{d_{B}}^{b_{i},b_{j}}\right)\otimes\left(\sum_{a=1}^{d_{A}}J_{d_{A}}^{a,a}\otimes J_{d_{B}}^{b_{j},b_{i}}\right)\\
=& \sum_{b_{i},b_{j}=1}^{d_{B}}\left(\id_{d_{A}}\otimes J_{d_{B}}^{b_{i},b_{j}}\right)\otimes\left(\id_{d_{A}}\otimes J_{d_{B}}^{b_{j},b_{i}}\right).
\end{aligned}
\end{align}
This is the form we presented in \eqref{eq:supop}.

Another representation of $K_B$ involves a more optimal basis choice. Consider a symmetric matrix, $E$, an antisymmetric matrix, $O$, and an arbitrary matrix, $X$. Notice that
\begin{align}
T_{B}\left(X\otimes E\right) & =X\otimes E^{T}=X\otimes E,\\
T_{B}\left(X\otimes O\right) & =X\otimes O^{T}=-X\otimes O.
\end{align}
Therefore the partial transposition map has as its eigenoperators matrices of the form $X\otimes E$ (eigenvalue 1) and $X\otimes O$ (eigenvalue -1). With this in mind, we can define bases $\mathbb{E}_{\text{S}}$ and
$\mathbb{E}_{\text{A}}$ of the symmetric and antisymmetric matrices, respectively:
\begin{align}
\mathbb{E}_{\text{S}} & =\left\{ J^{ij}_{d_B}+J^{ji}_{d_B}-J^{ij}_{d_B}J^{ij}_{d_B}\right\}, \\
\mathbb{E}_{\text{AS}} & =\left\{ J^{ij}_{d_B}-J^{ji}_{d_B}\vert i \neq j \right\},
\end{align}
which gives us a basis for the combined system
\begin{equation}
\mathbb{E}=\left\{ J^{ij}_{d_A}\right\} \otimes\left(\mathbb{E}_{\text{S}}\cup\mathbb{E}_{\text{AS}}\right).
\end{equation}
$K_B$ is diagonal in the (vectorized) basis $\mathbb{E}$; if we assume the basis matrices are all normalized by their Frobenius norm and order the basis so that the symmetric matrices come first, it has the form
\begin{equation}
K_B^{\mathbb{E}}=\begin{bmatrix}\id_{k} & 0\\
0 & -\id_{l}
\end{bmatrix},
\end{equation}
where $\id_{k}$ is the $k \times k$ identity matrix with $k = d^{2}_{A}\left|\mathbb{E}_{\text{S}}\right|=\frac{1}{2}d^{2}_{A}d_{B}\left(d_{B}+1\right)$, and $\id_{l}$ is the $l \times l$ identity matrix with $l = d^{2}_{A}\left|\mathbb{E}_{\text{AS}}\right|=\frac{1}{2}d^{2}_{A}d_{B}\left(d_{B}-1\right)$.
We can thus write
\begin{equation}
K_B=V^{\mathbb{E}}K_B^{\mathbb{E}}\left(V^{\mathbb{E}}\right)^{T},
\end{equation}
where $V^{\mathbb{E}}$ has as its column vectors the vectorized matrices from $\mathcal{\mathbb{E}}$.

\section{Simplifying the Hessian of the trace norm}\label{sec:simphess}
To obtain the simplified expressions for the trace norm Hessian presented in Eqs. \eqref{eq:patternedHessian} and \eqref{eq:spectHessian2}, we use identity \eqref{Kcommid} as well as a commutation rule for the matrix $K$ and the inverse of a Kronecker sum. Supposing $X \oplus Y$ is invertible, and remembering that $K$ is self-inverse, we have
\begin{equation}
K(X \oplus Y)^{-1} = [(X \oplus Y)K]^{-1} =[K(Y \oplus X)]^{-1} = (Y \oplus X)^{-1}K.
\end{equation}
We can now write down the $B$ matrices introduced in Eq. \eqref{eq:Aexpansion} using \eqref{eq:dvec|X|} in \eqref{eq:d2tr},
\begin{align}
\begin{aligned}
B_{00}=&\left(\left|X\right|^{-1}\right)^{T}\otimes\id-\left(\id\otimes X\right)\left(\left|X\right|^{-1}\right)_{\oplus}\left(\id\otimes\left|X\right|^{-1}\right)\left(\left|X\right|^{-1}\right)_{\oplus}\left(\id\otimes X^{\dagger}\right),\\
B_{01}=&-K\left(X\otimes\id\right)\left(\left|X\right|^{-1}\right)_{\oplus}^{T}\left(\left|X\right|^{-1}\otimes\id\right)\left(\left|X\right|^{-1}\right)_{\oplus}^{T}\left(\id\otimes X^{T}\right),\\
B_{10}=&-K\left(X^{*}\otimes\id\right)\left(\left|X\right|^{-1}\right)_{\oplus}\left(\id\otimes\left|X\right|^{-1}\right)\left(\left|X\right|^{-1}\right)_{\oplus}\left(\id\otimes X^{\dagger}\right),\\
B_{11}=&-\left(\id\otimes X^{*}\right)\left(\left|X\right|^{-1}\right)_{\oplus}^{T}\left(\left|X\right|^{-1}\otimes\id\right)\left(\left|X\right|^{-1}\right)_{\oplus}^{T}\left(\id\otimes X^{T}\right).
\end{aligned}
\end{align}
The unpatterned Hessians are then
\begin{align}
&\hess_{X,X}\left(\left\Vert X\right\Vert _{1}\right)=-\frac{1}{2}K\left(X^{*}\otimes\id\right)\left(|X|_{\oplus}\right)^{-1}\left(\left|X\right|^{-1}\right)_{\oplus}\left(|X|_{\oplus}\right)^{-1}\left(\id\otimes X^{\dagger}\right),\nonumber\\
&\hess_{X^{*},X^{*}}\left(\left\Vert X\right\Vert _{1}\right)=-\frac{1}{2}K\left(X\otimes\id\right)\left(|X|_{\oplus}^{T}\right)^{-1}\left(\left|X\right|^{-1}\right)_{\oplus}^{T}\left(|X|_{\oplus}^{T}\right)^{-1}\left(\id\otimes X^{T}\right),\\
\hess_{X,X^{*}}\left(\left\Vert X\right\Vert _{1}\right)=&\frac{1}{2}\left(\left|X\right|^{-1}\right)^{T}\otimes\id-\frac{1}{2}\left(\id\otimes X\right)\left(|X|_{\oplus}\right)^{-1}\left(\left|X\right|^{-1}\right)_{\oplus}\left(|X|_{\oplus}\right)^{-1}\left(\id\otimes X^{\dagger}\right)=\left(\hess_{X^{*},X}\left\Vert X\right\Vert _{1}\right)^{T}.\nonumber
\end{align}
These are combined to form the patterned Hessian with respect to a Hermitian matrix $A$ \eqref{eq:hessscal},
\begin{equation}
\hess_{A,A}\left(\left\Vert A\right\Vert _{1}\right) = \frac{1}{2}K\left[\left(\left|X\right|^{-1}\right)_{\oplus}-X_{\oplus}\left(\left|X\right|_{\oplus}\right)^{-1}\left(\left|X\right|^{-1}\right)_{\oplus}\left(\left|X\right|_{\oplus}\right)^{-1}X_{\oplus}^{\dagger}\right]_{X=A},
\end{equation}
which was presented in \eqref{eq:patternedHessian}. We can simplify this equation with the eigendecomposition $A=U\Lambda U^{\dagger}$ by noting
\begin{align}
\begin{aligned}
A_\oplus&=A^{T}\oplus A=\left(U^{*}\otimes U\right)\left(\Lambda\oplus\Lambda\right)\left(U^{T}\otimes U^{\dagger}\right),\\
\left\vert A\right\vert_\oplus&=\left|A\right|^{T}\oplus\left|A\right|=\left(U^{*}\otimes U\right)\left(\left|\Lambda\right|\oplus\left|\Lambda\right|\right)\left(U^{T}\otimes U^{\dagger}\right),
\end{aligned}
\end{align}
and so on. Continuing in this manner, the patterned Hessian can be written solely in terms of the eigendecomposition as
\begin{align}\label{eq:altPatternedHessian}
\begin{aligned}
\hess_{A,A}\left(\left\Vert A\right\Vert _{1}\right)=&\frac{1}{2}K(U^*\otimes U)\bigg[(\vert \Lambda\vert^{-1}\oplus \vert \Lambda\vert^{-1})\\
&-(\Lambda\oplus \Lambda)\left(\vert \Lambda\vert\oplus\vert \Lambda\vert\right)^{-1}(\vert \Lambda\vert^{-1}\oplus\vert \Lambda\vert^{-1})\left(\vert \Lambda\vert\oplus\vert \Lambda\vert\right)^{-1}(\Lambda\oplus \Lambda)\bigg](U^T\otimes U^\dag)\\
=&\frac{1}{2}K(U^*\otimes U)\left[\left(\vert \Lambda\vert\oplus\vert \Lambda\vert\right)^{2}-(\Lambda\oplus \Lambda)^2\right]\left(\vert \Lambda\vert\oplus\vert \Lambda\vert\right)^{-2}(\vert \Lambda\vert^{-1}\oplus \vert \Lambda\vert^{-1})(U^T\otimes U^\dag)\\
=&K(U^*\otimes U)\left[\left|\Lambda\right|\otimes\left|\Lambda\right|-\Lambda\otimes\Lambda\right]\left(\vert \Lambda\vert\oplus\vert \Lambda\vert\right)^{-2}(\vert \Lambda\vert^{-1}\oplus \vert \Lambda\vert^{-1})(U^T\otimes U^\dag)\\
=&K(U^*\otimes U)\left[\id-\text{sign}\Lambda\otimes\text{sign}\Lambda\right]\left(\left|\Lambda\right|\otimes\left|\Lambda\right|\right)\left(\vert \Lambda\vert\oplus\vert \Lambda\vert\right)^{-2}(\vert \Lambda\vert^{-1}\oplus \vert \Lambda\vert^{-1})(U^T\otimes U^\dag)\\
=&K(U^*\otimes U)\left[\id-\text{sign}\Lambda\otimes\text{sign}\Lambda\right]\left(\vert \Lambda\vert\oplus\vert \Lambda\vert\right)^{-1}(U^T\otimes U^\dag),
\end{aligned}
\end{align} 
Here, we use that all the matrices involving $\Lambda$ are diagonal and commute to simplify the Hessian. The final line was presented in \eqref{eq:spectHessian2}.
	
\section{Details of open system dynamics}\label{app:open system dynamics}
As per Ref. \cite{Wangetal2016}, the initial state of the pair of cavities is given by Eq. \eqref{eq:cavity state}, and each cavity is coupled to a reservoir with $N\to\infty$ modes. Defining two amplitudes $\xi\left(t\right)=\eu^{-t/2}$ and $\chi\left(t\right)=\sqrt{1-\eu^{-t}}$, where $t$ is measured in units of some dissipative constant, the state evolves to
	\eq{\rho\left(t;\,p\right)=\left(
		\begin{array}{cccccccc}
			a_{11} & 0 & 0 & 0 & 0 & a_{16} & 0 & 0 \\
			0 & a_{22} & 0 & 0 & 0 & 0 & a_{27} & 0 \\
			0 & 0 & a_{33} & 0 & 0 & 0 & 0 & a_{38} \\
			0 & 0 & 0 & a_{44} & 0 & 0 & 0 & 0 \\
			0 & 0 & 0 & 0 & a_{55} & 0 & 0 & 0 \\
			a_{16} & 0 & 0 & 0 & 0 & a_{66} & 0 & 0 \\
			0 & a_{27} & 0 & 0 & 0 & 0 & a_{77} & 0 \\
			0 & 0 & a_{38} & 0 & 0 & 0 & 0 & a_{88} \\
		\end{array}
		\right),
	}
	in the $\left\{\ket{0}\otimes\ket{0},\ket{0}\otimes\ket{1},\ket{0}\otimes\ket{2},\ket{0}\otimes\ket{3},\ket{1}\otimes\ket{0},\ket{1}\otimes\ket{1},\ket{1}\otimes\ket{2},\ket{1}\otimes\ket{3}\right\}$ basis, where the matrix elements are given by
	\begin{align}
	\begin{aligned}
		a_{11}=&\left(p+\chi^4+\chi^8-\chi^8\right)/2, \
		a_{22}=\xi^2\chi^2\left[2-p+3\left(1-p\right)\chi^4\right]/2, \
		a_{33}=\left(1-p\right)\xi^4\left(1+3\chi^4\right)/2, \\
		a_{44}=&\left(1-p\right)\xi^6\chi^2 / 2, \
		a_{55}=\xi^2\chi^2\left(p+\chi^4-p\chi^4\right)/2, \
		a_{66}=\xi^4\left[p+3\left(1-p\right)\chi^4\right]/2, \
		a_{77}=3\left(1-p\right)\xi^6\chi^2 / 2, \\
		a_{88}=&\left(1-p\right)\xi^8/2, \
		a_{16}=\xi^2\left[p+\sqrt{3}\left(1-p\right)\chi^4\right]/2, \
		a_{27}=\sqrt{3/2}\left(1-p\right)\xi^4\chi^2, \
		a_{38}=\left(1-p\right)\xi^6/2.
	\end{aligned}
	\end{align}
This can be used to calculate $\partial\rho/\partial t$ and $\partial\rho/\partial p$ in Eqs. \eqref{eq:first derivative} and \eqref{eq:second derivative}, and can also be used to explicitly calculate the eigenvalues of $\rho$ for use in Eq. \eqref{eq:negativity-eigvals}.

\chapter{Conclusion}\label{ch:conc}

In this thesis we have focused on applications of quantum information theory to quantum gravity through the AdS/CFT correspondence, with additional work on the dynamics of entanglement in general quantum systems. 

The main results of Chapters \ref{ch:KSCD} and \ref{ch:OPE} describe the refinement of a holographic duality that was originally established for the maximally symmetric case of pure AdS, dual to the conformally invariant vacuum state of a CFT \cite{Czech2016}. The duality allows OPE blocks, contributions to the OPE from a conformal primary and its descendants, to be expressed as an integral of a dual bulk field over a geodesic. The OPE block is a bilocal operator, depending on the two boundary points where the OPE is applied, and this feature can be traced as the origin of the geodesic integrated operator's diffeomorphism invariance. This was a significant advance in our understanding of how the CFT can encode the diffeomorphism invariance of a gravitational theory, as most constructions of bulk fields from boundary data focus on local fields \cite{Hamilton2006,Kabat2017} which are not diffeomorphism invariant observables \cite{Donnelly2015,Donnelly2016}. The major shortcoming of this duality was its heavy reliance on the symmetry of the vacuum state to determine the metric on kinematic space and the equations of motion for OPE blocks. In the bulk, the same symmetry ensures that there is a unique geodesic connecting the two spacelike separated boundary insertion points. This leaves the duality in an uncertain state for any less symmetric setting, as the kinematic space will not be entirely determined by symmetry, while in the bulk there will often be several geodesics connecting pairs of boundary points. 

We have argued that the duality continues to hold with some modifications in quotients of AdS$_3$, dual to CFT$_2$ states excited by the insertion of heavy primary operators. We chose to work in quotient spacetimes as they have a rich spectrum of non-minimal geodesics which can wind around singularities, or cross through black hole horizons, while the spacetimes still retain some symmetries of AdS. In the CFT we explained how OPE blocks decompose into more fine-grained, quotient-invariant observables which we termed partial OPE blocks. These new observables compute the contribution to the OPE from a conformal family but each differ from the others by the monodromy of the OPE around the heavy primary operator insertion in the state. We presented two different arguments that each partial OPE block is dual to a geodesic integrated bulk field, where the geodesic can be minimal, non-minimal, or even horizon crossing. In Section \ref{secdis}, utilizing the residual symmetries of quotient spacetimes (the conical defects in particular) we were able to show that geodesic integrated fields obey a wave equation on a region of kinematic space, in a one-to-one correspondence with the partial OPE blocks. More generally, in Chapter \ref{ch:OPE} we showed that the coordinate maps taking us from pure AdS$_3$ to various quotient spacetimes have non-analyticities which induce non-minimal geodesics; acting with these transformations asymptotically on OPE blocks and requiring observables to be single-valued induces the decomposition into partial OPE blocks. Having the same origin, we argued that these quantities will in general be dual in AdS$_3$/CFT$_2$. 

The significance of the partial OPE block / non-minimal geodesic integrated field duality is in its ability to access information deep in the bulk of AdS. As mentioned in the Introduction, one major goal of the AdS/CFT program is to reconstruct the gravitational theory entirely in terms of CFT data without assuming anything about the bulk \emph{ab initio}. Entanglement entropy data has proven to be a useful starting point as the metric of the bulk can be extracted using  the Ryu-Takayangi relation \eqref{eq:RyuTakayanagi} wherever boundary anchored minimal surfaces reach. We noted, however, that typical CFT states will have a dual geometry with entanglement shadows, regions where minimal surfaces do not reach. The CFT$_2$ observable called entwinement \cite{Balasubramanian2015,Balasubramanian2016,Balasubramanian2018} has been proposed to be dual to the length of non-minimal boundary anchored geodesics present in non-pure AdS$_3$ spacetimes. As we have seen in several classes of locally AdS$_3$ spacetimes, non-minimal geodesics can reach all parts of the spacetime. Hence, entwinement is a potential path to reconstructing the entire bulk.

 Unlike the entanglement entropy of a boundary subregion, which is a measure of correlations among spatially organized degrees of freedom, entwinement measures correlations among internal, discretely gauged degrees of freedom. This is very similar to the way we constructed partial OPE blocks in Ch. \ref{ch:KSCD}, first removing a discrete gauge symmetry by lifting operators to the covering CFT, considering the OPE in the cover, and finally projecting down to gauge invariant observables.

In a similar manner to how entwinement allows reconstruction of the metric deep in the bulk, our proposal could allow reconstruction of bulk fields beyond the entanglement wedge. In the case of pure AdS$_3$ the reconstruction of fields from OPE blocks has already been established \cite{Czech2016}. The OPE blocks are first related to geodesic integrated fields, after which the geodesic integral can be inverted to leave the bare AdS field smeared over boundary operators, the same representation as established by other methods \cite{Hamilton2006}. The geodesic integral in this case is a Radon transform on the hyperbolic disk, and explicit inversion formulae are known for this highly symmetric scenario \cite{Rubin2002,Lin2015}. When the bulk geometry is not pure AdS, it may be possible to achieve bulk reconstruction within the entanglement shadow by starting with partial OPE blocks. This would require inverting geodesic integrals in more complicated, locally AdS backgrounds, a mathematical problem not yet solved in general. Another issue encountered for general spacetimes in three dimensions is that some regions never contain geodesic turning points, a situation analogous to entanglement shadows called entanglement shade \cite{Espindola2018}. This potentially blocks reconstruction of these regions from entanglement or entwinement information in the CFT, but has been resolved by using entanglement of purification \eqref{eq:EoP} and its dual, the area of entanglement wedge cross sections \eqref{eq:EWCS}. It is not yet known if entanglement of purification or its dual has a useful manifestation in kinematic space.

Other considerations of non-extremal bulk surfaces in higher dimensional theories have been adressed in \cite{Balasubramanian2018a}, where differential entropy \eqref{eqCrofton} was extended by including shape derivative information. It would be interesting to extend the higher dimensional kinematic space program to less symmetric spacetimes using these ideas. 

To date, kinematic space has mainly been used to examine general features of AdS/CFT that can be constrained by conformal symmetry alone. It would be interesting to apply the kinematic space proposal to a particular realization of AdS/CFT. For instance, one could study a chiral primary state in the D1-D5 CFT in the limit where it is dual to a conical defect in the bulk \cite{Lunin2003}. This theory could be used to study the behaviour of holographic complexity in black hole models through kinematic space \cite{Abt2017,Abt2018}.

In Chapters \ref{ch:Renyi} and \ref{ch:neg} we studied the dynamics of entanglement through the R\'enyi entropies and negativity. For R\'enyi entropies we considered arbitrary von Neumann type dynamics starting from initially pure, unentangled states, and showed that each R\'enyi entropy exhibits the same leading order dynamics characterized by a timescale \eqref{eq6}. Since the family of R\'enyi entropies is sufficient to completely characterize the entanglement of pure states, this timescale is a universal feature of bipartite entanglement. We also showed similar results for next-to-leading order dynamics that only arise due to surprising cancellations of terms that appear for a subset of the family of measures. 

For pure states, negativity can be considered as the R\'enyi entropy for the parameter $\al=\tfrac{1}{2}$ \footnote{Technically logarithmic negativity has this property, not negativity.} \cite{Vidal2001,Rangamani2014}. One may then wonder if the results of Ch. \ref{ch:Renyi} also apply to negativity, but a review of the assumptions made in our derivations shows that $\al=\tfrac{1}{2}$ is explicitly excluded. Indeed, negativity has drastically different behaviour around pure separable states compared to the standard R\'enyi measures. For one thing, negativity grows at first order in $t$ whereas R\'enyi growth started at second order, which could be interpreted as negativity being a more sensitive measure of entanglement around separable states. In order to compare these measures, we sought a perturbative expansion of negativity, but failed to find analytic expressions for its derivatives presented in the literature.

 In Ch. \ref{ch:neg} we explained why this may have been overlooked previously, and why the calculus of patterned matrices was necessary to solve the problem\footnote{We find it interesting that the mathematical theory of patterned matrix calculus that we extended and applied to the problem  \cite{Hjorungnes2008,Hjorungnes2008a} was developed several years after negativity was introduced \cite{Vidal2001}.}. We identified that an expansion of negativity would involve derivatives of a matrix norm with respect to a Hermitian argument which does not constitute a set of independent variables as usually required in matrix calculus. After developing mathematical tools for computing higher derivatives of matrix functions with respect to Hermitian arguments, we applied them to the negativity and presented analytic expressions for its derivatives. We also proved a condition for when the subtleties of patterned matrix calculus can be ignored, as had been done in much of the quantum information literature; matrix functions which do not explicitly depend on the complex conjugate of their argument (analytic matrix functions) will have patterned derivatives that equal their unpatterned counterparts. R\'enyi entropies are examples of analytic matrix functions, whereas negativity is not analytic due to the matrix norm used in its definition. 
 
 Finally, we applied patterned matrix calculus techniques to the larger class of Schatten $p$-norms in the context of quantum speed limits. $p$-norms can be used as distance measures on Hilbert space to quantify how quickly evolving states $\rho(t)$ become distinguishable from the initial state $\rho(0)$. We produced a bound on the rate of change of distinguishability in terms of the $p$-norm of $\dot \rho(t)$ \eqref{eq:vQSLbound}. We then extended this result to a bound on entanglement generation by considering the $p=1$ case relevant to negativity. Using our explicit expressions for the partial transpose operation we bounded the rate of change of negativity in terms of the Hamiltonian's norm \eqref{eq:neggrowthbound}.

To end this thesis we will present a question to be answered in the future. A different avenue to explore which is concerned with the entanglement structure of holographic theories involves the inequalities satisfied by Ryu-Takayanagi entropies. For ordinary quantum systems entanglement entropies satisfy a number of inequalities with important physical meanings. These include subadditivity \eqref{eq:subadditivity1} and, related by purification symmetry, the Araki-Lieb inequality $S(\rho_{AB})\geq |S(\rho_A)-S(\rho_{B})|$ for two subregions, as well as strong subadditivity \eqref{eq:SSA1} and its equivalents for three subregions. To see the structure behind these inequalities, it is helpful to introduce an abstract entropy space, where each direction corresponds to the entropy value of states for a certain subsystem \cite{Pippenger2003}. Then, the entropy inequalities restrict what vectors in this space can be realized by the entropies of a quantum state. 

To make this picture concrete, consider the entropy space for two subregions $A$ and $B$. This space is $\mathbb{R}^3$ with axes corresponding to $S(\rho_A)$, $S(\rho_B)$, and $S(\rho_{AB})$. Subadditivity and the Araki-Lieb inequality imply that there are three hyperplanes constraining attainable entropy vectors, explicitly
\begin{equation} 
S(\rho_A)+S(\rho_B)-S(\rho_{AB})\geq0,\quad S(\rho_B)+S(\rho_{AB})-S(\rho_A)\geq0,\quad S(\rho_{AB})+S(\rho_A)-S(\rho_B)\geq0.
\end{equation}
Allowed entropy vectors lie within the convex cone delimited by these hypersurfaces in entropy space. For three subregions $A$, $B$, and $C$ the entropy space is $\mathbb{R}^7$, and valid entropy vectors lie within a cone defined by upliftings of the two-party inequalities, as well as instances of strong subadditivity. Generally, for $N$ regions, entropy space is $\mathbb{R}^{2^N-1}$, but for $N\geq 4$ \emph{the inequalities defining the entropy cone are not completely known}. This is a major shortcoming of our current understanding of entanglement in general quantum systems.

Slightly more can be said for holographic entanglement entropies which also satisfy subadditivity, strong subadditivity, and inequalities related by symmetry \cite{Headrick2007}. In addition, they satisfy the three-party inequality called monogamy of mutual information (MMI) \eqref{eq:MMI1} \cite{Hayden2013}. While the two-party holographic entropy cone is identical to its general counterpart, MMI further constrains the three party holographic entropy cone \cite{Bao2015}. Going to four parties, the known inequalities are sufficient in holography, whereas for five parties there are five additional types of inequality satisfied by the Ryu-Takayanagi entropy which are not always obeyed in general quantum systems \cite{Bao2015}. Very recently a complete description of the holographic entropy cone for five regions was presented \cite{Hernandez2019}. For even larger numbers of subsystems some new inequalities for holographic entropy are known, but it is undetermined if there are additional unknown inequalities or not. 

Holographic entropy inequalities directly put constraints on the geometries which can emerge through entanglement in AdS/CFT. We can learn about the limitations of emergent spacetimes by studying the extreme edges of allowed entropy space. Spacetimes with Ryu-Takayanagi entropies that saturate the entropy inequalities are on the verge of failing to be holographic, so if we are able to construct such geometries we may be able to extract insights about quantum gravity.

For subadditivity it is extremely simple to classify all the states which saturate the inequality: only product states of the form $\rho_{AB}=\rho_{A}\ot\rho_{B}$ have $I(A:B)=S(\rho_A)+S(\rho_B)-S(\rho_{AB})=0$, since these are the only states with no correlations between $A$ and $B$. This pattern of entropies appears for subsystems of holographic states on one side of entanglement phase transitions as the minimal area bulk surface jumps between connected and disconnected configurations\footnote{In quantum field theories spatial subregions are never completely uncorrelated, and the vacuum sector of Hilbert space does not factorize, points we belaboured when discussing the Reeh-Schlieder theorem in Ch. \ref{ch:intro}. The Ryu-Takayanagi entropy produces $I(A:B)=0$ for widely separated subregions, and hence implies a factorization structure, only to leading order in the $1/G_N$ expansion, which is the regime we will discuss. Quantum corrections from bulk fields ensure that expectations from the Reeh-Schlieder theorem are not violated \cite{Faulkner2013}. } \cite{Headrick2010,Rangamani2016}. 

It is much more difficult to establish the class of states which saturate strong subadditivity \cite{Ruskai2002,Hayden2004}, but it is not to troublesome to describe them: states $\rho_{ABC}$ saturate the inequality \eqref{eq:SSA1} when the Hilbert space of subsystem $B$ factorizes as $\mathcal{H}_B=\bigoplus_j \mathcal{H}_{b^L_i}\otimes \mathcal{H}_{b^R_j}$, and the state takes the form
\begin{equation}
\rho_{ABC}=\bigoplus_j p_j\rho_{A b^L_j}\otimes \rho_{b^R_j C},
\end{equation}
 where $p_j$ is a probability distribution, $\rho_{A b^L_j}\in\mathcal{H}_A\otimes\mathcal{H}_{b^L_i}$ and $\rho_{b^R_j C}\in\mathcal{H}_{b^R_j}\otimes\mathcal{H}_{C}$. This type of state is known as a quantum Markov chain, since the subsystem $B$ mediates all correlations between $A$ and $C$. When $B$ is traced out, we are left with the uncorrelated product $\rho_{AC}=\rho_A\ot \rho_C$. Non-trivial quantum Markov chains cannot be realized in holographic systems because they are ruled out by the stronger inequality MMI \eqref{eq:MMI1}.

 We are then lead to ask, \emph{what class of states saturates MMI, and what geometrical interpretation do they have in holography}? This is an unsolved question, as pointed out in \cite{Casini2008,Rota2016}, but we can present some expectations for its resolution. It will be convenient to introduce the quantity $I_3(A:B:C)$ called tripartite information which is constrained to be negative by MMI,
 \begin{equation}\label{eq:MMI2}
I_3(A:B:C)= S(\rho_{A})+S(\rho_{B})+S(\rho_{C})-S(\rho_{AB})-S(\rho_{AC})- S(\rho_{BC})+S(\rho_{ABC})\leq 0.
\end{equation}
First of all, every pure three party state $\left| \psi \right\rangle_{ABC}$ has $I_3(A:B:C)=0$, as does the maximally mixed state $\rho_{ABC}=\id_{ABC}$, so we only consider sub-maximally mixed states. Secondly, the question is meaningless for general quantum systems since MMI does not hold for entanglement entropy. There are even theories where every state has $I_3=0$ \cite{Casini2005}. In any case, one can consider two states with $I_3(\rho^1_{ABC})=-I_3(\rho^2_{A'B'C'})$, and form the product $ \rho^1_{ABC}\ot\rho^2_{A'B'C'}$ such that the combined state has $I_3(AA':BB':CC')=0$. Hence, there is no special structure for states with $I_3=0$ in general. Conveniently the MMI inequality rules out this trivializing example, so we can still hope for interesting structure holographically.

To build some intuition, we can consider the analogous question in classical probability theory. Defining $I_3(X:Y:Z)$ in the same manner as \eqref{eq:MMI2}, but with Shannon entropies $H(X)=-\sum_x p(x)\log p(x)$, for a three variable joint probability distribution $p(x,y,z)$ the tripartite information is
\begin{equation}\label{eq:I3classical}
I_3(X:Y:Z)=-\sum_{x,y,z}p(x,y,z)\log\frac{p(x,y,z)p(x)p(y)p(z)}{p(x,y)p(y,z)p(z,x)}.
\end{equation}
Marginal probability distributions are defined as $p(x)=\sum_y p(x,y)$ and all are normalized. Each term in the sum in \eqref{eq:I3classical} can be positive or negative, meaning that in general there is no special structure of distributions with $I_3=0$ for the same reasons as above. Motivated by MMI, we can consider only the subset of distributions for which the quantity $\log\frac{p(x,y,z)p(x)p(y)p(z)}{p(x,y)p(y,z)p(z,x)}$ is positive. It is then not hard to see that within this class, the only distributions with $I_3=0$ are of the form
\begin{equation}\label{eq:classicalstate}
p(x,y,z)=p(x|y)p(y|z)p(z|x)=p(y|x)p(z|y)p(x|z).
\end{equation}
Here we have introduced the conditional probability distributions $p(x|y)=p(x,y)/p(y)$ and noted a symmetry\footnote{This structure expresses that the level of correlation between any two variables is unaffected by the value of the third variable, and has sometimes been called a uniformly associative distribution \cite{Pearl1988,Rodriguez2007}. This class of distributions has appeared in statistics literature, but only rarely as it cannot be expressed as a Bayesian network, nor uniquely as a random Markov field; see p. 87ff of \cite{Koller2009} for a discussion.}.

We conjecture that, to leading order in the $1/G_N$ expansion, subsystem density matrices $\rho_{ABC}$ of holographic states with $I_3(A:B:C)=0$ always have the structure
\begin{equation}\label{eq:conjecturedstate}
\rho_{ABC}=\rho_{A_1 B_1}\ot \rho_{B_2 C_2}\ot \rho_{ C_3A_3},
\end{equation}
 which requires the Hilbert spaces to factorize as $\mathcal{H}_A=\mathcal{H}_{A_1}\ot \mathcal{H}_{A_3}$ etc. The class \eqref{eq:conjecturedstate} shares similarities with \eqref{eq:classicalstate} in its cyclic structure and interpretation.  This class of states has no shared three-party correlations, but has arbitrary two-party correlations among any pair of regions. States of this form were considered for independent reasons in \cite{Hubeny2018}. In that work it was shown that the structure \eqref{eq:conjecturedstate} appears in holographic settings on one side of entanglement phase transitions as the bulk Ryu-Takayanagi surface jumps from disconnected components to a multi-legged connected surface joining $ABC$. 

The form \eqref{eq:conjecturedstate} is implied by another recent conjecture concerning the form of holographic states in general \cite{Cui2018}. Using an alternative approach to holographic entanglement entropy called bit-threads, which are based on maximally packed vector fields rather than minimal surfaces, the authors posited that any pure 3-party holographic state should decompose as 
\begin{equation}
\ket{\psi}_{ABC}=\ket{\psi_1}_{A_1 B_1}\ot \ket{\psi_2}_{A_2 C_2}\ot\ket{\psi_{3}}_{B_3 C_3},
\end{equation}
with only bipartite entanglement. Furthermore, for a pure state on four subregions the conjectured holographic state decomposition is
\begin{equation}\label{eq:headrickconjecture}
\ket{\psi}_{ABCD}=\ket{\psi_1}_{A_1B_1}\ot \ket{\psi_2}_{A_2 C_2}\ot\ket{\psi_{3}}_{A_3D_3}\ot\ket{\psi_{4}}_{B_4 C_4}\ot\ket{\psi_{5}}_{B_5D_5}\ot\ket{\psi_{6}}_{C_6D_6}\ot\ket{PT}_{A_7B_7C_7D_7}.
\end{equation}
Each two-party $\ket{\psi}$ is bipartite entangled, while the state $\ket{PT}$ is a four-party perfect tensor defined such that the reduced state on any two subsystems is maximally mixed. It can be shown that all perfect tensor states have $I_3<0$ between any three parties. Hence, the only way for the conjectured state \eqref{eq:headrickconjecture} to have $I_3=0$ after tracing out one subsystem is if the perfect tensor factor is trivial. Removing the perfect tensor factor and tracing out any of the 4 parties reproduces our conjectured state \eqref{eq:conjecturedstate}. However, the conjectured form \eqref{eq:headrickconjecture} has stronger connotations than \eqref{eq:conjecturedstate} alone, and we note that the implied Hilbert space factorizations like $\h_A=\h_{A_1}\ot\h_{A_2}\ot\h_{A_3}\ot\h_{A_7}$ do not necessarily correspond to spatial factorizations. That is, the conjecture \eqref{eq:headrickconjecture} does not imply that each Hilbert space factor $\h_{A_i}$, for example, corresponds to some disjoint spatial subregion of $A$, but instead refers to some internal organization of degrees of freedom \cite{Cui2018}.

Although \eqref{eq:conjecturedstate} is implied by the conjecture in \cite{Cui2018}, it will likely be much easier to prove \eqref{eq:conjecturedstate} directly than to prove the form \eqref{eq:headrickconjecture}. One promising approach that leverages the properties of holographic theories follows as an extension of work disseminated in \cite{Bao2015}. In that paper Ryu-Takayanagi surface configurations are encoded into a graph model, and entropy inequalities are expressed in terms of cuts of the graph. This model can be used to prove MMI and the other holographic inequalities, but it can also be used to study the conditions for inequalities to be saturated. Work in this direction is in progress.

\addcontentsline{toc}{chapter}{Bibliography}
\bibliographystyle{utphys}
\bibliography{thesis}

\providecommand{\href}[2]{#2}\begingroup\raggedright\begin{thebibliography}{100}

\bibitem{Maldacena1999}
J.~Maldacena, ``The large-{N} limit of superconformal field theories and
  supergravity,'' \href{http://dx.doi.org/10.1023/A:1026654312961}{{\em
  International Journal of Theoretical Physics} {\bfseries 38} no.~4, (Apr,
  1999) 1113--1133},  \href{http://arxiv.org/abs/hep-th/9711200}{{\ttfamily
  hep-th/9711200}}.

\bibitem{Hartnoll2009}
S.~A. Hartnoll, ``Lectures on holographic methods for condensed matter
  physics,'' \href{http://dx.doi.org/10.1088/0264-9381/26/22/224002}{{\em
  Classical and Quantum Gravity} {\bfseries 26} no.~22, (Oct, 2009) 224002},
  \href{http://arxiv.org/abs/0903.3246}{{\ttfamily 0903.3246}}.

\bibitem{McGreevy2009}
J.~McGreevy, ``Holographic duality with a view toward many-body physics,''
  \href{http://dx.doi.org/10.1155/2010/723105}{{\em Advances in High Energy
  Physics} {\bfseries 2010} (2010) },
  \href{http://arxiv.org/abs/0909.0518}{{\ttfamily 0909.0518}}.

\bibitem{Liu2014b}
J.~Casalderrey-Solana, H.~Liu, D.~Mateos, K.~Rajagopal, and U.~A. Wiedemann,
  \href{http://dx.doi.org/10.1017/CBO9781139136747}{{\em Gauge/String Duality,
  Hot QCD and Heavy Ion Collisions}}.
\newblock Cambridge University Press, 2014.
\newblock
\newblock \href{http://arxiv.org/abs/1101.0618}{{\ttfamily 1101.0618}}.

\bibitem{Bhattacharyya2008}
S.~Bhattacharyya, S.~Minwalla, V.~E. Hubeny, and M.~Rangamani, ``Nonlinear
  fluid dynamics from gravity,''
  \href{http://dx.doi.org/10.1088/1126-6708/2008/02/045}{{\em Journal of High
  Energy Physics} {\bfseries 2008} no.~02, (Feb, 2008) 045--045},
  \href{http://arxiv.org/abs/0712.2456}{{\ttfamily 0712.2456}}.

\bibitem{Haehl2016}
F.~M. Haehl, R.~Loganayagam, and M.~Rangamani, ``The fluid manifesto: emergent
  symmetries, hydrodynamics, and black holes,''
  \href{http://dx.doi.org/10.1007/JHEP01(2016)184}{{\em Journal of High Energy
  Physics} {\bfseries 2016} no.~1, (Jan, 2016) 184},
  \href{http://arxiv.org/abs/1510.02494}{{\ttfamily 1510.02494}}.

\bibitem{Francesco1997}
P.~Francesco, P.~Mathieu, and D.~S\'en\'echal, {\em Conformal Field Theory}.
\newblock Springer-Verlag New York,
\newblock 1997.

\bibitem{Belavin1984}
A.~Belavin, A.~Polyakov, and A.~Zamolodchikov, ``Infinite conformal symmetry in
  two-dimensional quantum field theory,''
  \href{http://dx.doi.org/https://doi.org/10.1016/0550-3213(84)90052-X}{{\em
  Nuclear Physics B} {\bfseries 241} no.~2, (1984) 333 -- 380}.

\bibitem{Pappadopulo2012}
D.~Pappadopulo, S.~Rychkov, J.~Espin, and R.~Rattazzi, ``Operator product
  expansion convergence in conformal field theory,''
  \href{http://dx.doi.org/10.1103/PhysRevD.86.105043}{{\em Phys. Rev. D}
  {\bfseries 86} (Nov, 2012) 105043},
  \href{http://arxiv.org/abs/1208.6449}{{\ttfamily 1208.6449}}.

\bibitem{Rychkov2016}
S.~Rychkov and P.~Yvernay, ``{Remarks on the Convergence Properties of the
  Conformal Block Expansion},''
  \href{http://dx.doi.org/10.1016/j.physletb.2016.01.004}{{\em Phys. Lett.}
  {\bfseries B753} (2016) 682--686},
\href{http://arxiv.org/abs/1510.08486}{{\ttfamily arXiv:1510.08486 [hep-th]}}.

\bibitem{VanRaamsdonk2010}
M.~Van~Raamsdonk, ``Building up spacetime with quantum entanglement,''
  \href{http://dx.doi.org/10.1007/s10714-010-1034-0}{{\em General Relativity
  and Gravitation} {\bfseries 42} no.~10, (Oct, 2010) 2323--2329},
  \href{http://arxiv.org/abs/1005.3035}{{\ttfamily 1005.3035}}. [Int. J. of
  Mod. Phys. D19, 2429-2435 (2010)].

\bibitem{Reeh1961}
H.~Reeh and S.~Schlieder, ``{Bemerkungen zur Unita\"ar\"aquivalenz von
  Lorentzinvarienten Feldern},''  {\em Nuovo Cimento} {\bfseries 22} (1961)
  1051.

\bibitem{Witten2018}
E.~Witten, ``{APS Medal for Exceptional Achievement in Research: Invited
  article on entanglement properties of quantum field theory},''
  \href{http://dx.doi.org/10.1103/RevModPhys.90.045003}{{\em Rev. Mod. Phys.}
  {\bfseries 90} (Oct, 2018) 045003},
  \href{http://arxiv.org/abs/1803.04993}{{\ttfamily 1803.04993}}.

\bibitem{Plenio2005}
M.~B. Plenio and S.~Virmani, ``An introduction to entanglement measures,'' {\em
  Quantum Information \& Computation} {\bfseries 7} no.~1, (2007) 1--51,
  \href{http://arxiv.org/abs/quant-ph/0504163v3}{{\ttfamily
  quant-ph/0504163v3}}.

\bibitem{Casini2014}
H.~Casini, M.~Huerta, and J.~A. Rosabal, ``Remarks on entanglement entropy for
  gauge fields,'' \href{http://dx.doi.org/10.1103/PhysRevD.89.085012}{{\em
  Phys. Rev. D} {\bfseries 89} (Apr, 2014) 085012},
  \href{http://arxiv.org/abs/1312.1183}{{\ttfamily 1312.1183}}.

\bibitem{Ghosh2015}
S.~Ghosh, R.~M. Soni, and S.~P. Trivedi, ``On the entanglement entropy for
  gauge theories,'' \href{http://dx.doi.org/10.1007/JHEP09(2015)069}{{\em
  Journal of High Energy Physics} {\bfseries 2015} no.~9, (Sep, 2015) 69},
  \href{http://arxiv.org/abs/1501.02593}{{\ttfamily 1501.02593}}.

\bibitem{Lin2018}
J.~Lin and D.~Radicevic, ``Comments on defining entanglement entropy,''
  \href{http://arxiv.org/abs/1808.05939}{{\ttfamily 1808.05939}}.

\bibitem{Calabrese2004}
P.~Calabrese and J.~Cardy, ``Entanglement entropy and quantum field theory,''
  {\em Journal of Statistical Mechanics: Theory and Experiment} {\bfseries
  2004} no.~06, (2004) P06002,
  \href{http://arxiv.org/abs/hep-th/0405152}{{\ttfamily hep-th/0405152}}.

\bibitem{Calabrese2009}
P.~Calabrese and J.~Cardy, ``Entanglement entropy and conformal field theory,''
  {\em Journal of Physics A: Mathematical and Theoretical} {\bfseries 42}
  no.~50, (2009) 504005,  \href{http://arxiv.org/abs/0905.4013}{{\ttfamily
  0905.4013}}.

\bibitem{Srednicki1993}
M.~Srednicki, ``Entropy and area,''
  \href{http://dx.doi.org/10.1103/PhysRevLett.71.666}{{\em Phys. Rev. Lett.}
  {\bfseries 71} (Aug, 1993) 666--669},
  \href{http://arxiv.org/abs/hep-th/9303048}{{\ttfamily hep-th/9303048}}.

\bibitem{Holzhey1994}
C.~Holzhey, F.~Larsen, and F.~Wilczek, ``{Geometric and renormalized entropy in
  conformal field theory},''
  \href{http://dx.doi.org/10.1016/0550-3213(94)90402-2}{{\em Nucl. Phys.}
  {\bfseries B424} (1994) 443--467},
\href{http://arxiv.org/abs/hep-th/9403108}{{\ttfamily arXiv:hep-th/9403108
  [hep-th]}}.

\bibitem{Alcaraz2011}
F.~C. Alcaraz, M.~I. nez Berganza, and G.~Sierra, ``Entanglement of low-energy
  excitations in conformal field theory,''
  \href{http://dx.doi.org/10.1103/PhysRevLett.106.201601}{{\em Phys. Rev.
  Lett.} {\bfseries 106} (May, 2011) 201601},
  \href{http://arxiv.org/abs/1101.2881}{{\ttfamily 1101.2881}}.

\bibitem{Berganza2012}
M.~I. Berganza, F.~C. Alcaraz, and G.~Sierra, ``Entanglement of excited states
  in critical spin chains,''
  \href{http://dx.doi.org/10.1088/1742-5468/2012/01/p01016}{{\em Journal of
  Statistical Mechanics: Theory and Experiment} {\bfseries 2012} no.~01, (Jan,
  2012) P01016},  \href{http://arxiv.org/abs/1109.5673}{{\ttfamily 1109.5673}}.

\bibitem{Bombelli1986}
L.~Bombelli, R.~K. Koul, J.~Lee, and R.~D. Sorkin, ``Quantum source of entropy
  for black holes,''  \href{http://dx.doi.org/10.1103/PhysRevD.34.373}{{\em
  Phys. Rev. D} {\bfseries 34} (Jul, 1986) 373--383}.

\bibitem{Haag1996}
R.~Haag, {\em Local Quantum Physics}.
\newblock Springer-Verlag Berlin Heidelberg,
\newblock 1996.

\bibitem{Knight1961}
J.~M. Knight, ``Strict localization in quantum field theory,''
  \href{http://dx.doi.org/10.1063/1.1703731}{{\em Journal of Mathematical
  Physics} {\bfseries 2} no.~4, (1961) 459--471},
  \href{http://arxiv.org/abs/https://doi.org/10.1063/1.1703731}{{\ttfamily
  https://doi.org/10.1063/1.1703731}}.

\bibitem{Cresswell2019a}
I.~Tzitrin, A.~Z. Goldberg, and J.~C. Cresswell, ``Operational symmetries of
  entangled states,''  \href{http://arxiv.org/abs/1906.07731}{{\ttfamily
  1906.07731}}.

\bibitem{Araki1976}
H.~Araki, ``{Relative entropy of States of von Neumann Algebras},''  {\em Publ.
  RIMS Kokyuroku} {\bfseries 11} (1976) 809--833.

\bibitem{Ibinson2007}
B.~Ibinson, N.~Linden, and A.~Winter, ``All inequalities for the relative
  entropy,'' \href{http://dx.doi.org/10.1007/s00220-006-0081-6}{{\em
  Communications in Mathematical Physics} {\bfseries 269} no.~1, (Jan, 2007)
  223--238},  \href{http://arxiv.org/abs/quant-ph/0511260}{{\ttfamily
  quant-ph/0511260}}.

\bibitem{Witten1998}
E.~Witten, ``Anti de {Sitter} space and holography,'' {\em
  Adv.Theor.Math.Phys.} {\bfseries 2:} (1998) 253--291,
  \href{http://arxiv.org/abs/hep-th/9802150}{{\ttfamily hep-th/9802150}}.

\bibitem{Gubser1998}
S.~Gubser, I.~Klebanov, and A.~Polyakov, ``Gauge theory correlators from
  non-critical string theory,''
  \href{http://dx.doi.org/http://dx.doi.org/10.1016/S0370-2693(98)00377-3}{{\em
  Physics Letters B} {\bfseries 428} no.~1-2, (May, 1998) 105--114},
  \href{http://arxiv.org/abs/hep-th/9802109}{{\ttfamily hep-th/9802109}}.

\bibitem{Hooft1993}
G.~'t~Hooft, ``Dimensional reduction in quantum gravity,''
  \href{http://arxiv.org/abs/gr-qc/9310026}{{\ttfamily gr-qc/9310026}}.

\bibitem{Susskind1995}
L.~Susskind, ``The world as a hologram,''
  \href{http://dx.doi.org/10.1063/1.531249}{{\em Journal of Mathematical
  Physics} {\bfseries 36} no.~11, (Apr., 1995) 6377--6396},
  \href{http://arxiv.org/abs/hep-th/9409089}{{\ttfamily hep-th/9409089}}.

\bibitem{Heemskerk2009}
I.~Heemskerk, J.~Penedones, J.~Polchinski, and J.~Sully, ``Holography from
  conformal field theory,''
  \href{http://dx.doi.org/10.1088/1126-6708/2009/10/079}{{\em Journal of High
  Energy Physics} {\bfseries 2009} no.~10, (Oct, 2009) 079--079},
  \href{http://arxiv.org/abs/0907.0151}{{\ttfamily 0907.0151}}.

\bibitem{deHaro2001}
S.~de~Haro, K.~Skenderis, and S.~N. Solodukhin, ``{Holographic Reconstruction
  of Spacetime and Renormalization in the AdS/CFT Correspondence},''
  \href{http://dx.doi.org/10.1007/s002200100381}{{\em Communications in
  Mathematical Physics} {\bfseries 217} no.~3, (Mar, 2001) 595--622},
  \href{http://arxiv.org/abs/hep-th/0002230}{{\ttfamily hep-th/0002230}}.

\bibitem{Hamilton2006}
A.~Hamilton, D.~Kabat, G.~Lifschytz, and D.~A. Lowe, ``Holographic
  representation of local bulk operators,''
  \href{http://dx.doi.org/10.1103/PhysRevD.74.066009}{{\em Phys. Rev. D}
  {\bfseries 74} (Sep, 2006) 066009},
  \href{http://arxiv.org/abs/hep-th/0606141}{{\ttfamily hep-th/0606141}}.

\bibitem{Heemskerk2012}
I.~Heemskerk, ``Construction of bulk fields with gauge redundancy,''
  \href{http://dx.doi.org/10.1007/JHEP09(2012)106}{{\em Journal of High Energy
  Physics} {\bfseries 2012} no.~9, (Sep, 2012) 106},
  \href{http://arxiv.org/abs/1201.3666}{{\ttfamily 1201.3666}}.

\bibitem{Almheiri2013}
A.~Almheiri, D.~Marolf, J.~Polchinski, and J.~Sully, ``Black holes:
  complementarity or firewalls?,''
  \href{http://dx.doi.org/10.1007/JHEP02(2013)062}{{\em Journal of High Energy
  Physics} {\bfseries 2013} no.~2, (Feb, 2013) 62},
  \href{http://arxiv.org/abs/1207.3123}{{\ttfamily 1207.3123}}.

\bibitem{Harlow2016}
D.~Harlow, ``Jerusalem lectures on black holes and quantum information,''
  \href{http://dx.doi.org/10.1103/RevModPhys.88.015002}{{\em Rev. Mod. Phys.}
  {\bfseries 88} (Feb, 2016) 015002},
  \href{http://arxiv.org/abs/1409.1231}{{\ttfamily 1409.1231}}.

\bibitem{Banks1998}
T.~Banks, M.~R. Douglas, G.~T. Horowitz, and E.~Martinec, ``{AdS} dynamics from
  conformal field theory,'' {\em NSF-ITP-98-082, EFI-98-30, RU-98-xx} (1998) ,
  \href{http://arxiv.org/abs/hep-th/9808016}{{\ttfamily hep-th/9808016}}.

\bibitem{Balasubramanian1999c}
V.~Balasubramanian and P.~Kraus, ``Spacetime and the holographic
  renormalization group,''
  \href{http://dx.doi.org/10.1103/PhysRevLett.83.3605}{{\em Phys. Rev. Lett.}
  {\bfseries 83} (Nov, 1999) 3605--3608},
  \href{http://arxiv.org/abs/hep-th/9903190}{{\ttfamily hep-th/9903190}}.

\bibitem{Harlow2011}
D.~Harlow and D.~Stanford, ``Operator dictionaries and wave functions in
  {AdS/CFT} and {dS/CFT},''  \href{http://arxiv.org/abs/1104.2621}{{\ttfamily
  1104.2621}}.

\bibitem{DHoker2002}
E.~{D'Hoker} and D.~Z. Freedman, {\em Supersymmetric Gauge Theories and the
  AdS/CFT Correspondence},
  \href{http://dx.doi.org/10.1142/9789812702821_0001}{pp.~3--159}.
\newblock World Scientific, 2002.
\newblock
\newblock \href{http://arxiv.org/abs/hep-th/0201253}{{\ttfamily
  hep-th/0201253}}.

\bibitem{Skenderis2002}
K.~Skenderis, ``Lecture notes on holographic renormalization,''
  \href{http://dx.doi.org/10.1088/0264-9381/19/22/306}{{\em Classical and
  Quantum Gravity} {\bfseries 19} no.~22, (Nov, 2002) 5849--5876},
  \href{http://arxiv.org/abs/hep-th/0209067}{{\ttfamily hep-th/0209067}}.

\bibitem{Ryu2006}
S.~Ryu and T.~Takayanagi, ``Holographic derivation of entanglement entropy from
  the anti-de {S}itter space/conformal field theory correspondence,''
  \href{http://dx.doi.org/10.1103/PhysRevLett.96.181602}{{\em Phys. Rev. Lett.}
  {\bfseries 96} (May, 2006) 181602},
  \href{http://arxiv.org/abs/hep-th/0603001}{{\ttfamily hep-th/0603001}}.

\bibitem{Hubeny2007}
V.~E. Hubeny, M.~Rangamani, and T.~Takayanagi, ``A covariant holographic
  entanglement entropy proposal,'' {\em Journal of High Energy Physics}
  {\bfseries 2007} no.~07, (2007) 062,
  \href{http://arxiv.org/abs/0705.0016}{{\ttfamily 0705.0016}}.

\bibitem{Lewkowycz2013}
A.~Lewkowycz and J.~Maldacena, ``Generalized gravitational entropy,''
  \href{http://dx.doi.org/10.1007/JHEP08(2013)090}{{\em Journal of High Energy
  Physics} {\bfseries 2013} no.~8, (Aug, 2013) 90},
  \href{http://arxiv.org/abs/1304.4926}{{\ttfamily 1304.4926}}.

\bibitem{Dong2016a}
X.~Dong, A.~Lewkowycz, and M.~Rangamani, ``Deriving covariant holographic
  entanglement,'' \href{http://dx.doi.org/10.1007/JHEP11(2016)028}{{\em Journal
  of High Energy Physics} {\bfseries 2016} no.~11, (Nov, 2016) 28},
  \href{http://arxiv.org/abs/1607.07506}{{\ttfamily 1607.07506}}.

\bibitem{Brown1986}
J.~D. Brown and M.~Henneaux, ``Central charges in the canonical realization of
  asymptotic symmetries: An example from three dimensional gravity,''
  \href{http://dx.doi.org/10.1007/BF01211590}{{\em Communications in
  Mathematical Physics} {\bfseries 104} no.~2, (Jun, 1986) 207--226}.

\bibitem{Nishioka2009}
T.~Nishioka, S.~Ryu, and T.~Takayanagi, ``Holographic entanglement entropy: an
  overview,'' \href{http://dx.doi.org/10.1088/1751-8113/42/50/504008}{{\em
  Journal of Physics A: Mathematical and Theoretical} {\bfseries 42} no.~50,
  (Dec, 2009) 504008},  \href{http://arxiv.org/abs/0905.0932}{{\ttfamily
  0905.0932}}.

\bibitem{Headrick2010}
M.~Headrick, ``Entanglement {R}\'enyi entropies in holographic theories,''
  \href{http://dx.doi.org/10.1103/PhysRevD.82.126010}{{\em Phys. Rev. D}
  {\bfseries 82} (Dec, 2010) 126010},
  \href{http://arxiv.org/abs/1006.0047}{{\ttfamily 1006.0047}}.

\bibitem{Headrick2014}
M.~Headrick, ``General properties of holographic entanglement entropy,''
  \href{http://dx.doi.org/10.1007/JHEP03(2014)085}{{\em Journal of High Energy
  Physics} {\bfseries 2014} no.~3, (Mar, 2014) 85},
  \href{http://arxiv.org/abs/1312.6717}{{\ttfamily 1312.6717}}.

\bibitem{Freedman2017}
M.~Freedman and M.~Headrick, ``Bit threads and holographic entanglement,''
  \href{http://dx.doi.org/10.1007/s00220-016-2796-3}{{\em Communications in
  Mathematical Physics} {\bfseries 352} no.~1, (May, 2017) 407--438},
  \href{http://arxiv.org/abs/1604.00354}{{\ttfamily 1604.00354}}.

\bibitem{Headrick2007}
M.~Headrick and T.~Takayanagi, ``Holographic proof of the strong subadditivity
  of entanglement entropy,''
  \href{http://dx.doi.org/10.1103/PhysRevD.76.106013}{{\em Phys. Rev. D}
  {\bfseries 76} (Nov, 2007) 106013},
  \href{http://arxiv.org/abs/0704.3719}{{\ttfamily 0704.3719}}.

\bibitem{Hayden2013}
P.~Hayden, M.~Headrick, and A.~Maloney, ``Holographic mutual information is
  monogamous,'' \href{http://dx.doi.org/10.1103/PhysRevD.87.046003}{{\em Phys.
  Rev. D} {\bfseries 87} (Feb, 2013) 046003},
  \href{http://arxiv.org/abs/1107.2940}{{\ttfamily 1107.2940}}.

\bibitem{Bao2015}
N.~Bao, S.~Nezami, H.~Ooguri, B.~Stoica, J.~Sully, and M.~Walter, ``The
  holographic entropy cone,''
  \href{http://dx.doi.org/10.1007/JHEP09(2015)130}{{\em Journal of High Energy
  Physics} {\bfseries 2015} no.~9, (Sep, 2015) 130},
  \href{http://arxiv.org/abs/1505.07839}{{\ttfamily 1505.07839}}.

\bibitem{Hubeny2018}
V.~E. Hubeny, M.~Rangamani, and M.~Rota, ``Holographic entropy relations,''
  \href{http://dx.doi.org/10.1002/prop.201800067}{{\em Fortschr. Phys.}
  {\bfseries 66} no.~11-12, (Apr., 2019) 1800067},
  \href{http://arxiv.org/abs/1808.07871}{{\ttfamily 1808.07871}}.

\bibitem{Hubeny2018a}
V.~E. Hubeny, M.~Rangamani, and M.~Rota, ``The holographic entropy
  arrangement,''  \href{http://arxiv.org/abs/1812.08133}{{\ttfamily
  1812.08133}}.

\bibitem{Cui2018}
S.~X. Cui, P.~Hayden, T.~He, M.~Headrick, B.~Stoica, and M.~Walter, ``Bit
  threads and holographic monogamy,''
  \href{http://arxiv.org/abs/1808.05234}{{\ttfamily 1808.05234}}.

\bibitem{Rangamani2014}
M.~Rangamani and M.~Rota, ``Comments on entanglement negativity in holographic
  field theories,'' \href{http://dx.doi.org/10.1007/JHEP10(2014)060}{{\em
  Journal of High Energy Physics} {\bfseries 2014} no.~10, (Oct, 2014) 60},
  \href{http://arxiv.org/abs/1406.6989}{{\ttfamily 1406.6989}}.

\bibitem{Chaturvedietal2018}
P.~Chaturvedi, V.~Malvimat, and G.~Sengupta, ``Holographic quantum entanglement
  negativity,'' \href{http://dx.doi.org/10.1007/JHEP05(2018)172}{{\em Journal
  of High Energy Physics} {\bfseries 2018} no.~5, (May, 2018) 172},
  \href{http://arxiv.org/abs/1609.06609}{{\ttfamily 1609.06609}}.

\bibitem{Lashkari2016}
N.~Lashkari and M.~Van~Raamsdonk, ``Canonical energy is quantum fisher
  information,'' \href{http://dx.doi.org/10.1007/JHEP04(2016)153}{{\em Journal
  of High Energy Physics} {\bfseries 2016} no.~4, (Apr, 2016) 153},
  \href{http://arxiv.org/abs/1508.00897}{{\ttfamily 1508.00897}}.

\bibitem{Almheiri2015}
A.~Almheiri, X.~Dong, and D.~Harlow, ``{Bulk Locality and Quantum Error
  Correction in AdS/CFT},''
  \href{http://dx.doi.org/10.1007/JHEP04(2015)163}{{\em JHEP} {\bfseries 04}
  (2015) 163},
\href{http://arxiv.org/abs/1411.7041}{{\ttfamily arXiv:1411.7041 [hep-th]}}.

\bibitem{Pastawski2015}
F.~Pastawski, B.~Yoshida, D.~Harlow, and J.~Preskill, ``{Holographic quantum
  error-correcting codes: Toy models for the bulk/boundary correspondence},''
  \href{http://dx.doi.org/10.1007/JHEP06(2015)149}{{\em JHEP} {\bfseries 06}
  (2015) 149},
\href{http://arxiv.org/abs/1503.06237}{{\ttfamily arXiv:1503.06237 [hep-th]}}.

\bibitem{Mintun2015}
E.~Mintun, J.~Polchinski, and V.~Rosenhaus, ``{Bulk-Boundary Duality, Gauge
  Invariance, and Quantum Error Corrections},''
  \href{http://dx.doi.org/10.1103/PhysRevLett.115.151601}{{\em Phys. Rev.
  Lett.} {\bfseries 115} no.~15, (2015) 151601},
\href{http://arxiv.org/abs/1501.06577}{{\ttfamily arXiv:1501.06577 [hep-th]}}.

\bibitem{Harlow2017}
D.~Harlow, ``The {Ryu-Takayanagi} formula from quantum error correction,''
  \href{http://dx.doi.org/10.1007/s00220-017-2904-z}{{\em Communications in
  Mathematical Physics} {\bfseries 354} no.~3, (Sep, 2017) 865--912},
  \href{http://arxiv.org/abs/1607.03901}{{\ttfamily 1607.03901}}.

\bibitem{Jafferis2016}
D.~L. Jafferis, A.~Lewkowycz, J.~Maldacena, and S.~J. Suh, ``Relative entropy
  equals bulk relative entropy,''
  \href{http://dx.doi.org/10.1007/JHEP06(2016)004}{{\em Journal of High Energy
  Physics} {\bfseries 2016} no.~6, (Jun, 2016) 4},
  \href{http://arxiv.org/abs/1512.06431}{{\ttfamily 1512.06431}}.

\bibitem{Faulkner2016}
T.~Faulkner, R.~G. Leigh, O.~Parrikar, and H.~Wang, ``Modular {H}amiltonians
  for deformed half-spaces and the averaged null energy condition,''
  \href{http://dx.doi.org/10.1007/JHEP09(2016)038}{{\em Journal of High Energy
  Physics} {\bfseries 2016} no.~9, (Sep, 2016) 38},
  \href{http://arxiv.org/abs/1605.08072}{{\ttfamily 1605.08072}}.

\bibitem{Faulkner2017a}
T.~Faulkner and A.~Lewkowycz, ``Bulk locality from modular flow,''
  \href{http://dx.doi.org/10.1007/JHEP07(2017)151}{{\em Journal of High Energy
  Physics} {\bfseries 2017} no.~7, (Jul, 2017) 151},
  \href{http://arxiv.org/abs/1704.05464}{{\ttfamily 1704.05464}}.

\bibitem{Faulkner2018}
T.~Faulkner, M.~Li, and H.~Wang, ``A modular toolkit for bulk reconstruction,''
   \href{http://arxiv.org/abs/1806.10560}{{\ttfamily 1806.10560}}.

\bibitem{Umemoto2018}
K.~Umemoto and T.~Takayanagi, ``Entanglement of purification through
  holographic duality,'' {\em Nature Physics} {\bfseries 14} no.~6, (June,
  2018) 573--577,  \href{http://arxiv.org/abs/1708.09393}{{\ttfamily
  1708.09393}}.

\bibitem{Nguyen2018}
P.~Nguyen, T.~Devakul, M.~G. Halbasch, M.~P. Zaletel, and B.~Swingle,
  ``Entanglement of purification: from spin chains to holography,''
  \href{http://dx.doi.org/10.1007/JHEP01(2018)098}{{\em Journal of High Energy
  Physics} {\bfseries 2018} no.~1, (Jan, 2018) 98},
  \href{http://arxiv.org/abs/1709.07424}{{\ttfamily 1709.07424}}.

\bibitem{Terhal2002}
B.~M. Terhal, M.~Horodecki, D.~W. Leung, and D.~P. DiVincenzo, ``The
  entanglement of purification,''
  \href{http://dx.doi.org/10.1063/1.1498001}{{\em Journal of Mathematical
  Physics} {\bfseries 43} no.~9, (2002) 4286--4298},
  \href{http://arxiv.org/abs/quant-ph/0202044}{{\ttfamily quant-ph/0202044}}.

\bibitem{Bao2018}
N.~Bao and I.~F. Halpern, ``Holographic inequalities and entanglement of
  purification,'' \href{http://dx.doi.org/10.1007/JHEP03(2018)006}{{\em Journal
  of High Energy Physics} {\bfseries 2018} no.~3, (Mar, 2018) 6},
  \href{http://arxiv.org/abs/1710.07643}{{\ttfamily 1710.07643}}.

\bibitem{Bao2019}
N.~Bao, A.~Chatwin-Davies, and G.~N. Remmen, ``Entanglement of purification and
  multiboundary wormhole geometries,''
  \href{http://dx.doi.org/10.1007/JHEP02(2019)110}{{\em Journal of High Energy
  Physics} {\bfseries 2019} no.~2, (Feb, 2019) 110},
  \href{http://arxiv.org/abs/1811.01983}{{\ttfamily 1811.01983}}.

\bibitem{Caputa2019}
P.~Caputa, M.~Miyaji, T.~Takayanagi, and K.~Umemoto, ``Holographic entanglement
  of purification from conformal field theories,''
  \href{http://dx.doi.org/10.1103/PhysRevLett.122.111601}{{\em Phys. Rev.
  Lett.} {\bfseries 122} (Mar, 2019) 111601},
  \href{http://arxiv.org/abs/1812.05268}{{\ttfamily 1812.05268}}.

\bibitem{Dutta2019}
S.~Dutta and T.~Faulkner, ``A canonical purification for the entanglement wedge
  cross-section,''  \href{http://arxiv.org/abs/1905.00577}{{\ttfamily
  1905.00577}}.

\bibitem{Bhattacharyya2019}
A.~Bhattacharyya, A.~Jahn, T.~Takayanagi, and K.~Umemoto, ``Entanglement of
  purification in many body systems and symmetry breaking,'' {\em YITP-19-05,
  IPMU19-0014} ,
  \href{http://arxiv.org/abs/http://arxiv.org/abs/1902.02369v1}{{\ttfamily
  http://arxiv.org/abs/1902.02369v1}}.

\bibitem{Nielsen2006}
M.~A. Nielsen, M.~R. Dowling, M.~Gu, and A.~C. Doherty, ``Quantum computation
  as geometry,'' \href{http://dx.doi.org/10.1126/science.1121541}{{\em Science}
  {\bfseries 311} no.~5764, (2006) 1133--1135},
  \href{http://arxiv.org/abs/quant-ph/0603161}{{\ttfamily quant-ph/0603161}}.

\bibitem{Brown2018}
A.~R. Brown and L.~Susskind, ``Second law of quantum complexity,''
  \href{http://dx.doi.org/10.1103/PhysRevD.97.086015}{{\em Phys. Rev. D}
  {\bfseries 97} (Apr, 2018) 086015},
  \href{http://arxiv.org/abs/1701.01107}{{\ttfamily 1701.01107}}.

\bibitem{Jefferson2017}
R.~A. Jefferson and R.~C. Myers, ``Circuit complexity in quantum field
  theory,'' \href{http://dx.doi.org/10.1007/JHEP10(2017)107}{{\em Journal of
  High Energy Physics} {\bfseries 2017} no.~10, (Oct, 2017) 107},
  \href{http://arxiv.org/abs/1707.08570}{{\ttfamily 1707.08570}}.

\bibitem{Maldacena2001}
J.~Maldacena, ``Eternal black holes in anti-de {S}itter,''
  \href{http://dx.doi.org/10.1088/1126-6708/2003/04/021}{{\em Journal of High
  Energy Physics} {\bfseries 2003} no.~04, (Apr, 2003) 021--021},
  \href{http://arxiv.org/abs/hep-th/0106112}{{\ttfamily hep-th/0106112}}.

\bibitem{Sekino2008}
Y.~Sekino and L.~Susskind, ``Fast scramblers,''
  \href{http://dx.doi.org/10.1088/1126-6708/2008/10/065}{{\em Journal of High
  Energy Physics} {\bfseries 2008} no.~10, (Oct, 2008) 065--065},
  \href{http://arxiv.org/abs/0808.2096}{{\ttfamily 0808.2096}}.

\bibitem{Susskind2014}
L.~Susskind, ``Computational complexity and black hole horizons,''
  \href{http://dx.doi.org/10.1002/prop.201500092}{{\em Fortschritte der Physik}
  {\bfseries 64} no.~1, (2016) 24--43},
  \href{http://arxiv.org/abs/1402.5674}{{\ttfamily 1402.5674}}.

\bibitem{Stanford2014}
D.~Stanford and L.~Susskind, ``Complexity and shock wave geometries,''
  \href{http://dx.doi.org/10.1103/PhysRevD.90.126007}{{\em Phys. Rev. D}
  {\bfseries 90} (Dec, 2014) 126007},
  \href{http://arxiv.org/abs/1406.2678}{{\ttfamily 1406.2678}}.

\bibitem{Alishahiha2015}
M.~Alishahiha, ``Holographic complexity,''
  \href{http://dx.doi.org/10.1103/PhysRevD.92.126009}{{\em Phys. Rev. D}
  {\bfseries 92} (Dec, 2015) 126009},
  \href{http://arxiv.org/abs/1509.06614}{{\ttfamily 1509.06614}}.

\bibitem{Brown2016a}
A.~R. Brown, D.~A. Roberts, L.~Susskind, B.~Swingle, and Y.~Zhao, ``Holographic
  complexity equals bulk action?,''
  \href{http://dx.doi.org/10.1103/PhysRevLett.116.191301}{{\em Phys. Rev.
  Lett.} {\bfseries 116} (May, 2016) 191301},
  \href{http://arxiv.org/abs/1509.07876}{{\ttfamily 1509.07876}}.

\bibitem{Brown2016}
A.~R. Brown, D.~A. Roberts, L.~Susskind, B.~Swingle, and Y.~Zhao, ``Complexity,
  action, and black holes,''
  \href{http://dx.doi.org/10.1103/PhysRevD.93.086006}{{\em Phys. Rev. D}
  {\bfseries 93} (Apr, 2016) 086006},
  \href{http://arxiv.org/abs/1512.04993}{{\ttfamily 1512.04993}}.

\bibitem{Carmi2017}
D.~Carmi, R.~C. Myers, and P.~Rath, ``Comments on holographic complexity,''
  \href{http://dx.doi.org/10.1007/JHEP03(2017)118}{{\em Journal of High Energy
  Physics} {\bfseries 2017} no.~3, (Mar, 2017) 118},
  \href{http://arxiv.org/abs/1612.00433}{{\ttfamily 1612.00433}}.

\bibitem{Susskind2016}
L.~Susskind, ``The typical-state paradox: diagnosing horizons with
  complexity,'' \href{http://dx.doi.org/10.1002/prop.201500091}{{\em Fortschr.
  Phys.} {\bfseries 64} no.~1, (Apr., 2016) 84--91},
  \href{http://arxiv.org/abs/1507.02287}{{\ttfamily 1507.02287}}.

\bibitem{Maldacena2013}
J.~Maldacena and L.~Susskind, ``Cool horizons for entangled black holes,''
  \href{http://dx.doi.org/10.1002/prop.201300020}{{\em Fortschr. Phys.}
  {\bfseries 61} no.~9, (Apr., 2013) 781--811},
  \href{http://arxiv.org/abs/1306.0533}{{\ttfamily 1306.0533}}.

\bibitem{Bianchi2014}
E.~Bianchi and R.~C. Myers, ``On the architecture of spacetime geometry,''
  \href{http://dx.doi.org/10.1088/0264-9381/31/21/214002}{{\em Classical and
  Quantum Gravity} {\bfseries 31} no.~21, (Oct, 2014) 214002},
  \href{http://arxiv.org/abs/1212.5183}{{\ttfamily 1212.5183}}.

\bibitem{Maldacena2017}
J.~Maldacena, D.~Simmons-Duffin, and A.~Zhiboedov, ``Looking for a bulk
  point,'' \href{http://dx.doi.org/10.1007/JHEP01(2017)013}{{\em Journal of
  High Energy Physics} {\bfseries 2017} no.~1, (Jan, 2017) 13},
  \href{http://arxiv.org/abs/1509.03612}{{\ttfamily 1509.03612}}.

\bibitem{Engelhardt2016}
N.~Engelhardt and G.~T. Horowitz, ``Towards a reconstruction of general bulk
  metrics,'' \href{http://dx.doi.org/10.1088/1361-6382/34/1/015004}{{\em
  Classical and Quantum Gravity} {\bfseries 34} no.~1, (Dec, 2016) 015004},
  \href{http://arxiv.org/abs/1605.01070}{{\ttfamily 1605.01070}}.

\bibitem{Kabat2018}
D.~Kabat and G.~Lifschytz, ``Emergence of spacetime from the algebra of total
  modular {H}amiltonians,''  \href{http://arxiv.org/abs/1812.02915}{{\ttfamily
  1812.02915}}.

\bibitem{Balasubramanian2014}
V.~Balasubramanian, B.~D. Chowdhury, B.~Czech, J.~de~Boer, and M.~P. Heller,
  ``Bulk curves from boundary data in holography,''
  \href{http://dx.doi.org/10.1103/PhysRevD.89.086004}{{\em Phys. Rev. D}
  {\bfseries 89} (Apr, 2014) 086004},
  \href{http://arxiv.org/abs/1310.4204}{{\ttfamily 1310.4204}}.

\bibitem{Myers2014}
R.~C. Myers, J.~Rao, and S.~Sugishita, ``Holographic holes in higher
  dimensions,'' \href{http://dx.doi.org/10.1007/JHEP06(2014)044}{{\em JHEP}
  {\bfseries 2014} no.~6, (Jun, 2014) 44},
  \href{http://arxiv.org/abs/1403.3416v2}{{\ttfamily 1403.3416v2}}.

\bibitem{Headrick2014a}
M.~Headrick, R.~C. Myers, and J.~Wien, ``Holographic holes and differential
  entropy,'' \href{http://dx.doi.org/10.1007/JHEP10(2014)149}{{\em Journal of
  High Energy Physics} {\bfseries 2014} no.~10, (Oct, 2014) 149},
  \href{http://arxiv.org/abs/1408.4770}{{\ttfamily 1408.4770}}.

\bibitem{Freivogel2014}
B.~Freivogel, R.~A. Jefferson, L.~Kabir, B.~Mosk, and I.-S. Yang, ``Casting
  shadows on holographic reconstruction,''
  \href{http://dx.doi.org/10.1103/PhysRevD.91.086013}{{\em Phys. Rev. D}
  {\bfseries 91} (Apr, 2015) 086013},
  \href{http://arxiv.org/abs/1412.5175}{{\ttfamily 1412.5175}}.

\bibitem{Faulkner2013}
T.~Faulkner, A.~Lewkowycz, and J.~Maldacena, ``Quantum corrections to
  holographic entanglement entropy,''
  \href{http://dx.doi.org/10.1007/JHEP11(2013)074}{{\em Journal of High Energy
  Physics} {\bfseries 2013} no.~11, (Nov, 2013) 74},
  \href{http://arxiv.org/abs/1307.2892}{{\ttfamily 1307.2892}}.

\bibitem{Dong2014}
X.~Dong, ``Holographic entanglement entropy for general higher derivative
  gravity,'' \href{http://dx.doi.org/10.1007/JHEP01(2014)044}{{\em Journal of
  High Energy Physics} {\bfseries 2014} no.~1, (Jan, 2014) 44},
  \href{http://arxiv.org/abs/1310.5713}{{\ttfamily 1310.5713}}.

\bibitem{Balasubramanian2015}
V.~Balasubramanian, B.~D. Chowdhury, B.~Czech, and J.~de~Boer, ``Entwinement
  and the emergence of spacetime,''
  \href{http://dx.doi.org/10.1007/JHEP01(2015)048}{{\em JHEP} {\bfseries 2015}
  no.~1, (2015) 48},  \href{http://arxiv.org/abs/1406.5859v2}{{\ttfamily
  1406.5859v2}}.

\bibitem{Balasubramanian2016}
V.~Balasubramanian, A.~Bernamonti, B.~Craps, T.~De~Jonckheere, and F.~Galli,
  ``Entwinement in discretely gauged theories,''
  \href{http://dx.doi.org/10.1007/JHEP12(2016)094}{{\em JHEP} {\bfseries 2016}
  no.~12, (Dec, 2016) 94},  \href{http://arxiv.org/abs/1609.03991}{{\ttfamily
  1609.03991}}.

\bibitem{Kabat2017}
D.~Kabat and G.~Lifschytz, ``Local bulk physics from intersecting modular
  {H}amiltonians,'' \href{http://dx.doi.org/10.1007/JHEP06(2017)120}{{\em
  Journal of High Energy Physics} {\bfseries 2017} no.~6, (Jun, 2017) 120},
  \href{http://arxiv.org/abs/1703.06523}{{\ttfamily 1703.06523}}.

\bibitem{Wall2014}
A.~C. Wall, ``Maximin surfaces, and the strong subadditivity of the covariant
  holographic entanglement entropy,''
  \href{http://dx.doi.org/10.1088/0264-9381/31/22/225007}{{\em Classical and
  Quantum Gravity} {\bfseries 31} no.~22, (Nov, 2014) 225007},
  \href{http://arxiv.org/abs/1211.3494}{{\ttfamily 1211.3494}}.

\bibitem{Headrick2014b}
M.~Headrick, V.~E. Hubeny, A.~Lawrence, and M.~Rangamani, ``Causality {\&}
  holographic entanglement entropy,''
  \href{http://dx.doi.org/10.1007/JHEP12(2014)162}{{\em Journal of High Energy
  Physics} {\bfseries 2014} no.~12, (Dec, 2014) 162},
  \href{http://arxiv.org/abs/1408.6300}{{\ttfamily 1408.6300}}.

\bibitem{Dong2016}
X.~Dong, D.~Harlow, and A.~C. Wall, ``Reconstruction of bulk operators within
  the entanglement wedge in gauge-gravity duality,''
  \href{http://dx.doi.org/10.1103/PhysRevLett.117.021601}{{\em Phys. Rev.
  Lett.} {\bfseries 117} (Jul, 2016) 021601},
  \href{http://arxiv.org/abs/1601.05416}{{\ttfamily 1601.05416}}.

\bibitem{Czech2012}
B.~Czech, J.~L. Karczmarek, F.~Nogueira, and M.~V. Raamsdonk, ``The gravity
  dual of a density matrix,''
  \href{http://dx.doi.org/10.1088/0264-9381/29/15/155009}{{\em Classical and
  Quantum Gravity} {\bfseries 29} no.~15, (Jul, 2012) 155009},
  \href{http://arxiv.org/abs/1204.1330}{{\ttfamily 1204.1330}}.

\bibitem{Faulkner2014}
T.~Faulkner, M.~Guica, T.~Hartman, R.~C. Myers, and M.~Van~Raamsdonk,
  ``Gravitation from entanglement in holographic {CFTs},''
  \href{http://dx.doi.org/10.1007/JHEP03(2014)051}{{\em Journal of High Energy
  Physics} {\bfseries 2014} no.~3, (Mar, 2014) 51},
  \href{http://arxiv.org/abs/1312.7856}{{\ttfamily 1312.7856}}.

\bibitem{Lashkari2014a}
N.~Lashkari, M.~B. McDermott, and M.~Van~Raamsdonk, ``Gravitational dynamics
  from entanglement ``thermodynamics'',''
  \href{http://dx.doi.org/10.1007/JHEP04(2014)195}{{\em Journal of High Energy
  Physics} {\bfseries 2014} no.~4, (Apr, 2014) 195},
  \href{http://arxiv.org/abs/1308.3716}{{\ttfamily 1308.3716}}.

\bibitem{Faulkner2017}
T.~Faulkner, F.~M. Haehl, E.~Hijano, O.~Parrikar, C.~Rabideau, and
  M.~Van~Raamsdonk, ``Nonlinear gravity from entanglement in conformal field
  theories,'' \href{http://dx.doi.org/10.1007/JHEP08(2017)057}{{\em Journal of
  High Energy Physics} {\bfseries 2017} no.~8, (Aug, 2017) 57},
  \href{http://arxiv.org/abs/1705.03026}{{\ttfamily 1705.03026}}.

\bibitem{Lashkari2015}
N.~Lashkari, C.~Rabideau, P.~Sabella-Garnier, and M.~Van~Raamsdonk,
  ``Inviolable energy conditions from entanglement inequalities,''
  \href{http://dx.doi.org/10.1007/JHEP06(2015)067}{{\em Journal of High Energy
  Physics} {\bfseries 2015} no.~6, (Jun, 2015) 67},
  \href{http://arxiv.org/abs/1412.3514}{{\ttfamily 1412.3514}}.

\bibitem{Lashkari2016a}
N.~Lashkari, J.~Lin, B.~Stoica, H.~Ooguri, and M.~Van~Raamsdonk,
  ``{Gravitational positive energy theorems from information inequalities},''
  \href{http://dx.doi.org/10.1093/ptep/ptw139}{{\em Progress of Theoretical and
  Experimental Physics} {\bfseries 2016} no.~12, (12, 2016) },
  \href{http://arxiv.org/abs/1605.01075}{{\ttfamily 1605.01075}}.

\bibitem{Bousso2016}
R.~Bousso, Z.~Fisher, S.~Leichenauer, and A.~C. Wall, ``Quantum focusing
  conjecture,'' \href{http://dx.doi.org/10.1103/PhysRevD.93.064044}{{\em Phys.
  Rev. D} {\bfseries 93} (Mar, 2016) 064044},
  \href{http://arxiv.org/abs/1506.02669}{{\ttfamily 1506.02669}}.

\bibitem{Bousso2015}
R.~Bousso, Z.~Fisher, J.~Koeller, S.~Leichenauer, and A.~C. Wall, ``Proof of
  the quantum null energy condition,''
  \href{http://dx.doi.org/10.1103/PhysRevD.93.024017}{{\em Phys. Rev. D}
  {\bfseries 93} (Jan, 2016) 024017},
  \href{http://arxiv.org/abs/1509.02542}{{\ttfamily 1509.02542}}.

\bibitem{Hartman2017}
T.~Hartman, S.~Kundu, and A.~Tajdini, ``Averaged null energy condition from
  causality,'' \href{http://dx.doi.org/10.1007/JHEP07(2017)066}{{\em Journal of
  High Energy Physics} {\bfseries 2017} no.~7, (Jul, 2017) 66},
  \href{http://arxiv.org/abs/1610.05308}{{\ttfamily 1610.05308}}.

\bibitem{Wall2017a}
A.~C. Wall, ``Lower bound on the energy density in classical and quantum field
  theories,'' \href{http://dx.doi.org/10.1103/PhysRevLett.118.151601}{{\em
  Phys. Rev. Lett.} {\bfseries 118} (Apr, 2017) 151601},
  \href{http://arxiv.org/abs/1701.03196}{{\ttfamily 1701.03196}}.

\bibitem{Balakrishnan2017}
S.~Balakrishnan, T.~Faulkner, Z.~U. Khandker, and H.~Wang, ``A general proof of
  the quantum null energy condition,''
  \href{http://arxiv.org/abs/1706.09432}{{\ttfamily 1706.09432}}.

\bibitem{Cresswell2017}
J.~C. Cresswell and A.~W. Peet, ``{Kinematic space for conical defects},''
  \href{http://dx.doi.org/10.1007/JHEP11(2017)155}{{\em JHEP} {\bfseries 11}
  (2017) 155},  \href{http://arxiv.org/abs/1708.09838}{{\ttfamily
  arXiv:1708.09838 [hep-th]}}.

\bibitem{Cresswell2018}
J.~C. Cresswell, I.~T. Jardine, and A.~W. Peet, ``Holographic relations for
  {OPE} blocks in excited states,''
  \href{http://dx.doi.org/10.1007/JHEP03(2019)058}{{\em Journal of High Energy
  Physics} {\bfseries 2019} no.~3, (Mar, 2019) 58},
  \href{http://arxiv.org/abs/1809.09107}{{\ttfamily 1809.09107}}.

\bibitem{Cresswell2017a}
J.~C. Cresswell, ``Universal entanglement timescale for {R}\'{e}nyi
  entropies,'' \href{http://dx.doi.org/10.1103/PhysRevA.97.022317}{{\em Phys.
  Rev. A} {\bfseries 97} (Feb, 2018) 022317},
  \href{http://arxiv.org/abs/1709.10064}{{\ttfamily 1709.10064}}.

\bibitem{Cresswell2019}
J.~C. Cresswell, I.~Tzitrin, and A.~Z. Goldberg, ``Perturbative expansion of
  entanglement negativity using patterned matrix calculus,''
  \href{http://dx.doi.org/10.1103/PhysRevA.99.012322}{{\em Phys. Rev. A}
  {\bfseries 99} (Jan, 2019) 012322},
  \href{http://arxiv.org/abs/1809.07772}{{\ttfamily 1809.07772}}.

\bibitem{Ferrara1971}
S.~Ferrara, A.~F. Grillo, and R.~Gatto, ``Manifestly conformal covariant
  operator-product expansion,''
  \href{http://dx.doi.org/10.1007/BF02770435}{{\em Lettere al Nuovo Cimento
  (1971-1985)} {\bfseries 2} no.~26, (1971) 1363--1369}.

\bibitem{Ferrara1972}
S.~Ferrara, A.~Grillo, G.~Parisi, and R.~Gatto, ``Covariant expansion of the
  conformal four -point function,''
  \href{http://dx.doi.org/http://dx.doi.org/10.1016/0550-3213(72)90587-1}{{\em
  Nuclear Physics B} {\bfseries 49} (1972) 77 -- 98}.

\bibitem{Hijano2015}
E.~Hijano, P.~Kraus, and R.~Snively, ``Worldline approach to semi-classical
  conformal blocks,'' \href{http://dx.doi.org/10.1007/JHEP07(2015)131}{{\em
  JHEP} {\bfseries 2015} no.~7, (2015) 131},
  \href{http://arxiv.org/abs/1501.02260}{{\ttfamily 1501.02260}}.

\bibitem{Hijano2016}
E.~Hijano, P.~Kraus, E.~Perlmutter, and R.~Snively, ``Witten diagrams
  revisited: the {AdS} geometry of conformal blocks,''
  \href{http://dx.doi.org/10.1007/JHEP01(2016)146}{{\em JHEP} {\bfseries 2016}
  no.~1, (2016) 146},  \href{http://arxiv.org/abs/1508.00501}{{\ttfamily
  1508.00501}}.

\bibitem{Fukuda2018}
M.~Fukuda, N.~Kobayashi, and T.~Nishioka, ``Operator product expansion for
  conformal defects,'' \href{http://dx.doi.org/10.1007/JHEP01(2018)013}{{\em
  Journal of High Energy Physics} {\bfseries 2018} no.~1, (Jan, 2018) 13},
  \href{http://arxiv.org/abs/1710.11165}{{\ttfamily 1710.11165}}.

\bibitem{Kobayashi2018}
N.~Kobayashi and T.~Nishioka, ``Spinning conformal defects,'' {\em UT-18-11,
  IPMU18-0084} (2018) ,  \href{http://arxiv.org/abs/1805.05967}{{\ttfamily
  1805.05967}}.

\bibitem{DAS2019397}
S.~Das, ``Comments on spinning {OPE} blocks in {AdS$_3$/CFT$_2$},''
  \href{http://dx.doi.org/https://doi.org/10.1016/j.physletb.2019.03.058}{{\em
  Physics Letters B} {\bfseries 792} (2019) 397 -- 405},
  \href{http://arxiv.org/abs/1811.09375}{{\ttfamily 1811.09375}}.

\bibitem{Prudenziati2019}
A.~Prudenziati, ``A geodesic witten diagram description of holographic
  entanglement entropy and its quantum corrections,''
  \href{http://arxiv.org/abs/1902.10161}{{\ttfamily 1902.10161}}.

\bibitem{Czech2015}
B.~Czech, L.~Lamprou, S.~McCandlish, and J.~Sully, ``Integral geometry and
  holography,'' \href{http://dx.doi.org/10.1007/JHEP10(2015)175}{{\em JHEP}
  {\bfseries 2015} no.~10, (2015) 175},
  \href{http://arxiv.org/abs/1505.05515v1}{{\ttfamily 1505.05515v1}}.

\bibitem{Czech2016}
B.~Czech, L.~Lamprou, S.~McCandlish, B.~Mosk, and J.~Sully, ``A stereoscopic
  look into the bulk,'' \href{http://dx.doi.org/10.1007/JHEP07(2016)129}{{\em
  JHEP} {\bfseries 2016} no.~7, (2016) 129},
  \href{http://arxiv.org/abs/1604.03110v2}{{\ttfamily 1604.03110v2}}.

\bibitem{Boer2015}
J.~de~Boer, M.~P. Heller, R.~C. Myers, and Y.~Neiman, ``Holographic de {S}itter
  geometry from entanglement in conformal field theory,''
  \href{http://dx.doi.org/10.1103/PhysRevLett.116.061602}{{\em Phys. Rev.
  Lett.} {\bfseries 116} (Feb, 2016) 061602},
  \href{http://arxiv.org/abs/1509.00113}{{\ttfamily 1509.00113}}.

\bibitem{Donnelly2015}
W.~Donnelly and S.~B. Giddings, ``Diffeomorphism-invariant observables and
  their nonlocal algebra,''
  \href{http://dx.doi.org/10.1103/PhysRevD.93.024030}{{\em Phys. Rev. D}
  {\bfseries 93} (Jan, 2016) 024030},
  \href{http://arxiv.org/abs/1507.07921}{{\ttfamily 1507.07921}}.

\bibitem{Donnelly2016}
W.~Donnelly and S.~B. Giddings, ``Observables, gravitational dressing, and
  obstructions to locality and subsystems,''
  \href{http://dx.doi.org/10.1103/PhysRevD.94.104038}{{\em Phys. Rev. D}
  {\bfseries 94} (Nov, 2016) 104038},
  \href{http://arxiv.org/abs/1607.01025}{{\ttfamily 1607.01025}}.

\bibitem{Bhowmick2017}
S.~Bhowmick, S.~Das, and B.~Ezhuthachan, ``Entanglement entropy and kinematic
  space in {BCFT},''  \href{http://arxiv.org/abs/1703.01759}{{\ttfamily
  1703.01759}}.

\bibitem{Banados1992}
M.~Ba\~nados, C.~Teitelboim, and J.~Zanelli, ``Black hole in three-dimensional
  spacetime,'' \href{http://dx.doi.org/10.1103/PhysRevLett.69.1849}{{\em Phys.
  Rev. Lett.} {\bfseries 69} (Sep, 1992) 1849--1851},
  \href{http://arxiv.org/abs/hep-th/9204099}{{\ttfamily hep-th/9204099}}.

\bibitem{Banados1993}
M.~Ba\~nados, M.~Henneaux, C.~Teitelboim, and J.~Zanelli, ``Geometry of the 2+1
  black hole,'' \href{http://dx.doi.org/10.1103/PhysRevD.48.1506}{{\em Phys.
  Rev. D} {\bfseries 48} (Aug, 1993) 1506--1525},
  \href{http://arxiv.org/abs/gr-qc/9302012}{{\ttfamily gr-qc/9302012}}.

\bibitem{Deser1984}
S.~Deser, R.~Jackiw, and G.~'t~Hooft, ``Three-dimensional {E}instein gravity:
  Dynamics of flat space,''
  \href{http://dx.doi.org/http://dx.doi.org/10.1016/0003-4916(84)90085-X}{{\em
  Annals of Physics} {\bfseries 152} no.~1, (1984) 220 -- 235}.

\bibitem{Zhang2017}
J.-d. Zhang and B.~Chen, ``Kinematic space and wormholes,''
  \href{http://dx.doi.org/10.1007/JHEP01(2017)092}{{\em JHEP} {\bfseries 2017}
  no.~1, (2017) 92},  \href{http://arxiv.org/abs/1610.07134v2}{{\ttfamily
  1610.07134v2}}.

\bibitem{Espindola2017}
R.~Esp{\'i}ndola, A.~G{\"u}ijosa, A.~Landetta, and J.~F. Pedraza, ``What's the
  point? {Hole-ography in Poincar{\'e} AdS},''
  \href{http://dx.doi.org/10.1140/epjc/s10052-018-5563-0}{{\em The European
  Physical Journal C} {\bfseries 78} no.~1, (Jan, 2018) 75},
  \href{http://arxiv.org/abs/1708.02958}{{\ttfamily 1708.02958}}.

\bibitem{Balasubramanian1999}
V.~Balasubramanian and S.~F. Ross, ``Holographic particle detection,''
  \href{http://dx.doi.org/10.1103/PhysRevD.61.044007}{{\em Phys. Rev. D}
  {\bfseries 61} (Jan, 2000) 044007},
  \href{http://arxiv.org/abs/hep-th/9906226v1}{{\ttfamily hep-th/9906226v1}}.

\bibitem{Asplund2014}
C.~T. Asplund, A.~Bernamonti, F.~Galli, and T.~Hartman, ``Holographic
  entanglement entropy from 2d {CFT}: heavy states and local quenches,''
  \href{http://dx.doi.org/10.1007/JHEP02(2015)171}{{\em JHEP} {\bfseries 2015}
  no.~2, (2015) 171},  \href{http://arxiv.org/abs/1410.1392}{{\ttfamily
  1410.1392}}.

\bibitem{Czech2014}
B.~Czech and L.~Lamprou, ``Holographic definition of points and distances,''
  \href{http://dx.doi.org/10.1103/PhysRevD.90.106005}{{\em Phys. Rev. D}
  {\bfseries 90} (Nov, 2014) 106005},
  \href{http://arxiv.org/abs/1409.4473v1}{{\ttfamily 1409.4473v1}}.

\bibitem{Asplund2016}
C.~T. Asplund, N.~Callebaut, and C.~Zukowski, ``Equivalence of emergent de
  {S}itter spaces from conformal field theory,''
  \href{http://dx.doi.org/10.1007/JHEP09(2016)154}{{\em JHEP} {\bfseries 2016}
  no.~9, (2016) 154},  \href{http://arxiv.org/abs/1604.02687}{{\ttfamily
  1604.02687}}.

\bibitem{Boer2016}
J.~de~Boer, F.~M. Haehl, M.~P. Heller, and R.~C. Myers, ``Entanglement,
  holography and causal diamonds,''
  \href{http://dx.doi.org/10.1007/JHEP08(2016)162}{{\em JHEP} {\bfseries 2016}
  no.~8, (2016) 1--83},  \href{http://arxiv.org/abs/1606.03307}{{\ttfamily
  1606.03307}}.

\bibitem{Karch2017}
A.~Karch, J.~Sully, C.~F. Uhlemann, and D.~G.~E. Walker, ``Boundary kinematic
  space,'' \href{http://dx.doi.org/10.1007/JHEP08(2017)039}{{\em JHEP}
  {\bfseries 2017} no.~8, (Aug, 2017) 39},
  \href{http://arxiv.org/abs/1703.02990v1}{{\ttfamily 1703.02990v1}}.

\bibitem{Dolan2004}
F.~Dolan and H.~Osborn, ``Conformal partial waves and the operator product
  expansion,''
  \href{http://dx.doi.org/http://dx.doi.org/10.1016/j.nuclphysb.2003.11.016}{{\em
  Nuclear Physics B} {\bfseries 678} no.~1, (2004) 491 -- 507},
  \href{http://arxiv.org/abs/hep-th/0309180}{{\ttfamily hep-th/0309180}}.

\bibitem{Maldacena1998}
J.~Maldacena and A.~Strominger, ``{AdS$_3$} black holes and a stringy exclusion
  principle,'' \href{http://dx.doi.org/10.1088/1126-6708/1998/12/005}{{\em
  JHEP} {\bfseries 1998} no.~12, (1998) 005},
  \href{http://arxiv.org/abs/hep-th/9804085}{{\ttfamily hep-th/9804085}}.

\bibitem{Lunin2003}
O.~Lunin, S.~D. Mathur, and A.~Saxena, ``What is the gravity dual of a chiral
  primary?,''
  \href{http://dx.doi.org/https://doi.org/10.1016/S0550-3213(03)00081-6}{{\em
  Nuclear Physics B} {\bfseries 655} no.~1â2, (2003) 185 -- 217},
  \href{http://arxiv.org/abs/hep-th/0211292}{{\ttfamily hep-th/0211292}}.

\bibitem{Boer2011}
J.~de~Boer, M.~M. Sheikh-Jabbari, and J.~Sim\'on, ``Near-horizon limits of
  massless {BTZ} and their {CFT} duals,''
  \href{http://dx.doi.org/10.1088/0264-9381/28/17/175012}{{\em Classical and
  Quantum Gravity} {\bfseries 28} no.~17, (2011) 175012},
  \href{http://arxiv.org/abs/1011.1897v2}{{\ttfamily 1011.1897v2}}.

\bibitem{Balasubramanian2003}
V.~Balasubramanian, A.~Naqvi, and J.~Sim\'on, ``A multi-boundary {AdS} orbifold
  and {DLCQ} holography: a universal holographic description of extremal black
  hole horizons,'' \href{http://dx.doi.org/10.1088/1126-6708/2004/08/023}{{\em
  JHEP} {\bfseries 2004} no.~08, (2004) 023},
  \href{http://arxiv.org/abs/hep-th/0311237}{{\ttfamily hep-th/0311237}}.

\bibitem{Arefeva2016}
I.~Y. Aref'eva and M.~A. Khramtsov, ``{AdS/CFT} prescription for angle-deficit
  space and winding geodesics,''
  \href{http://dx.doi.org/10.1007/JHEP04(2016)121}{{\em JHEP} {\bfseries 2016}
  no.~4, (2016) 121},  \href{http://arxiv.org/abs/1601.02008v2}{{\ttfamily
  1601.02008v2}}.

\bibitem{Arefeva2016a}
I.~Y. Aref'eva, M.~A. Khramtsov, and M.~D. Tikhanovskaya, ``Improved image
  method for a holographic description of conical defects,''
  \href{http://dx.doi.org/10.1134/S0040577916110106}{{\em Theoretical and
  Mathematical Physics} {\bfseries 189} no.~2, (2016) 1660--1672},
  \href{http://arxiv.org/abs/1604.08905}{{\ttfamily 1604.08905}}.

\bibitem{Arefeva2017}
I.~Y. Aref'eva, M.~A. Khramtsov, and M.~D. Tikhanovskaya, ``Thermalization
  after holographic bilocal quench,''
  \href{http://dx.doi.org/10.1007/JHEP09(2017)115}{{\em JHEP} {\bfseries 2017}
  no.~9, (Sep, 2017) 115},  \href{http://arxiv.org/abs/1706.07390}{{\ttfamily
  1706.07390}}.

\bibitem{Czech2015c}
B.~Czech, L.~Lamprou, S.~McCandlish, and J.~Sully, ``Tensor networks from
  kinematic space,'' \href{http://dx.doi.org/10.1007/JHEP07(2016)100}{{\em
  JHEP} {\bfseries 2016} no.~7, (2016) 100},
  \href{http://arxiv.org/abs/1512.01548}{{\ttfamily 1512.01548}}.

\bibitem{Cunha2016}
B.~C. da~Cunha and M.~Guica, ``Exploring the {BTZ} bulk with boundary conformal
  blocks,''  \href{http://arxiv.org/abs/1604.07383}{{\ttfamily 1604.07383}}.

\bibitem{Guica2016}
M.~Guica, ``Bulk fields from the boundary {OPE},'' {\em Nordita-2016-144}
  no.~144, (2016) ,  \href{http://arxiv.org/abs/1610.08952v2}{{\ttfamily
  1610.08952v2}}.

\bibitem{Lin2015}
J.~Lin, M.~Marcolli, H.~Ooguri, and B.~Stoica, ``Locality of gravitational
  systems from entanglement of conformal field theories,''
  \href{http://dx.doi.org/10.1103/PhysRevLett.114.221601}{{\em Phys. Rev.
  Lett.} {\bfseries 114} (Jun, 2015) 221601},
  \href{http://arxiv.org/abs/1412.1879}{{\ttfamily 1412.1879}}.

\bibitem{Matschull1999}
H.-J. Matschull, ``Black hole creation in 2 + 1 dimensions,''
  \href{http://dx.doi.org/10.1088/0264-9381/16/3/032}{{\em Classical and
  Quantum Gravity} {\bfseries 16} no.~3, (1999) 1069},
  \href{http://arxiv.org/abs/gr-qc/9809087}{{\ttfamily gr-qc/9809087}}.

\bibitem{Arefeva2015}
I.~Aref\'eva, A.~Bagrov, P.~S\"aterskog, and K.~Schalm, ``Holographic dual of a
  time machine,'' \href{http://dx.doi.org/10.1103/PhysRevD.94.044059}{{\em
  Phys. Rev. D} {\bfseries 94} (Aug, 2016) 044059},
  \href{http://arxiv.org/abs/1508.04440}{{\ttfamily 1508.04440}}.

\bibitem{Goto2017}
K.~Goto and T.~Takayanagi, ``{CFT} descriptions of bulk local states in the
  {AdS} black holes,'' \href{http://dx.doi.org/10.1007/JHEP10(2017)153}{{\em
  JHEP} {\bfseries 2017} no.~10, (Oct, 2017) 153},
  \href{http://arxiv.org/abs/1704.00053v1}{{\ttfamily 1704.00053v1}}.

\bibitem{Giusto2013}
S.~Giusto, O.~Lunin, S.~D. Mathur, and D.~Turton, ``{D}1-{D}5-{P} microstates
  at the cap,'' \href{http://dx.doi.org/10.1007/JHEP02(2013)050}{{\em JHEP}
  {\bfseries 2013} no.~2, (Feb, 2013) 50},
  \href{http://arxiv.org/abs/1211.0306}{{\ttfamily 1211.0306}}.

\bibitem{Son2001}
J.~Son, ``String theory on {AdS}$_3$/{Z}$_n$,'' {\em HUTP-01/A034} (2001) ,
  \href{http://arxiv.org/abs/hep-th/0107131v1}{{\ttfamily hep-th/0107131v1}}.

\bibitem{Maloney2017}
A.~Maloney, H.~Maxfield, and G.~S. Ng, ``A conformal block farey tail,''
  \href{http://dx.doi.org/10.1007/JHEP06(2017)117}{{\em JHEP} {\bfseries 2017}
  no.~6, (Jun, 2017) 117},  \href{http://arxiv.org/abs/1609.02165}{{\ttfamily
  1609.02165}}.

\bibitem{Czech2015b}
B.~Czech, G.~Evenbly, L.~Lamprou, S.~McCandlish, X.-l. Qi, J.~Sully, and
  G.~Vidal, ``Tensor network quotient takes the vacuum to the thermal state,''
  \href{http://dx.doi.org/10.1103/PhysRevB.94.085101}{{\em Phys. Rev. B}
  {\bfseries 94} (Aug, 2016) 085101},
  \href{http://arxiv.org/abs/1510.07637}{{\ttfamily 1510.07637}}.

\bibitem{Sarosi2018}
G.~S{\'a}rosi and T.~Ugajin, ``Modular hamiltonians of excited states, {OPE}
  blocks and emergent bulk fields,''
  \href{http://dx.doi.org/10.1007/JHEP01(2018)012}{{\em Journal of High Energy
  Physics} {\bfseries 2018} no.~1, (Jan, 2018) 12},
  \href{http://arxiv.org/abs/1705.01486}{{\ttfamily 1705.01486}}.

\bibitem{Czech2016a}
B.~Czech, L.~Lamprou, S.~McCandlish, B.~Mosk, and J.~Sully, ``Equivalent
  equations of motion for gravity and entropy,''
  \href{http://dx.doi.org/10.1007/JHEP02(2017)004}{{\em JHEP} {\bfseries 2017}
  no.~2, (2017) 4},  \href{http://arxiv.org/abs/1608.06282v1}{{\ttfamily
  1608.06282v1}}.

\bibitem{Abt2017}
R.~Abt, J.~Erdmenger, H.~Hinrichsen, C.~M. Melby-Thompson, R.~Meyer, C.~Northe,
  and I.~A. Reyes, ``Topological complexity in {AdS$_3$/CFT$_2$},''
  \href{http://dx.doi.org/10.1002/prop.201800034}{{\em Fortschritte der Physik}
  {\bfseries 66} no.~6, (2018) 1800034},
  \href{http://arxiv.org/abs/1710.01327}{{\ttfamily 1710.01327}}.

\bibitem{Abt2018}
R.~Abt, J.~Erdmenger, M.~Gerbershagen, C.~M. Melby-Thompson, and C.~Northe,
  ``Holographic subregion complexity from kinematic space,''
  \href{http://dx.doi.org/10.1007/JHEP01(2019)012}{{\em Journal of High Energy
  Physics} {\bfseries 2019} no.~1, (Jan, 2019) 12},
  \href{http://arxiv.org/abs/1805.10298}{{\ttfamily 1805.10298}}.

\bibitem{Balasubramanian2001}
V.~Balasubramanian, J.~de~Boer, E.~Keski-Vakkuri, and S.~F. Ross,
  ``Supersymmetric conical defects: Towards a string theoretic description of
  black hole formation,''
  \href{http://dx.doi.org/10.1103/PhysRevD.64.064011}{{\em Phys. Rev. D}
  {\bfseries 64} (Aug, 2001) 064011},
  \href{http://arxiv.org/abs/hep-th/0011217}{{\ttfamily hep-th/0011217}}.

\bibitem{Lunin2002}
O.~Lunin, J.~M. Maldacena, and L.~Maoz, ``{Gravity solutions for the D1-D5
  system with angular momentum},''
\href{http://arxiv.org/abs/hep-th/0212210}{{\ttfamily arXiv:hep-th/0212210
  [hep-th]}}.

\bibitem{Carlip1995}
S.~Carlip and C.~Teitelboim, ``Aspects of black hole quantum mechanics and
  thermodynamics in 2+1 dimensions,''
  \href{http://dx.doi.org/10.1103/PhysRevD.51.622}{{\em Phys. Rev. D}
  {\bfseries 51} (Jan, 1995) 622--631},
  \href{http://arxiv.org/abs/gr-qc/9405070}{{\ttfamily gr-qc/9405070}}.

\bibitem{Chen2018}
C.-B. Chen, W.-C. Gan, F.-W. Shu, and B.~Xiong, ``Quantum information metric of
  conical defect,'' \href{http://dx.doi.org/10.1103/PhysRevD.98.046008}{{\em
  Phys. Rev. D} {\bfseries 98} (Aug, 2018) 046008},
  \href{http://arxiv.org/abs/1804.08358}{{\ttfamily 1804.08358}}.

\bibitem{Alday2006}
L.~F. Alday, J.~de~Boer, and I.~Messamah, ``{The gravitational description of
  coarse grained microstates},''
  \href{http://dx.doi.org/10.1088/1126-6708/2006/12/063}{{\em JHEP} {\bfseries
  12} (2006) 063},
\href{http://arxiv.org/abs/hep-th/0607222}{{\ttfamily arXiv:hep-th/0607222
  [hep-th]}}.

\bibitem{Banados1999a}
M.~Ba\~nados, ``{Three-dimensional quantum geometry and black holes},''
  \href{http://dx.doi.org/10.1063/1.59661}{{\em AIP Conf. Proc.} {\bfseries
  484} no.~1, (1999) 147--169},
\href{http://arxiv.org/abs/hep-th/9901148}{{\ttfamily arXiv:hep-th/9901148
  [hep-th]}}.

\bibitem{Roberts2012}
M.~M. Roberts, ``{Time evolution of entanglement entropy from a pulse},''
  \href{http://dx.doi.org/10.1007/JHEP12(2012)027}{{\em JHEP} {\bfseries 12}
  (2012) 027},
\href{http://arxiv.org/abs/1204.1982}{{\ttfamily arXiv:1204.1982 [hep-th]}}.

\bibitem{Asplund2015}
C.~T. Asplund, A.~Bernamonti, F.~Galli, and T.~Hartman, ``Entanglement
  scrambling in 2d conformal field theory,''
  \href{http://dx.doi.org/10.1007/JHEP09(2015)110}{{\em Journal of High Energy
  Physics} {\bfseries 2015} no.~9, (Sep, 2015) 110},
  \href{http://arxiv.org/abs/1506.03772}{{\ttfamily 1506.03772}}.

\bibitem{Anand2018}
N.~Anand, H.~Chen, A.~L. Fitzpatrick, J.~Kaplan, and D.~Li, ``{An Exact
  Operator That Knows Its Location},''
  \href{http://dx.doi.org/10.1007/JHEP02(2018)012}{{\em JHEP} {\bfseries 02}
  (2018) 012},
\href{http://arxiv.org/abs/1708.04246}{{\ttfamily arXiv:1708.04246 [hep-th]}}.

\bibitem{Fuente2013}
A.~de~la Fuente and R.~Sundrum, ``{Holography of the BTZ Black Hole, Inside and
  Out},'' \href{http://dx.doi.org/10.1007/JHEP09(2014)073}{{\em JHEP}
  {\bfseries 09} (2014) 073},
\href{http://arxiv.org/abs/1307.7738}{{\ttfamily arXiv:1307.7738 [hep-th]}}.

\bibitem{Maxfield2015}
H.~Maxfield, ``{Entanglement entropy in three dimensional gravity},''
  \href{http://dx.doi.org/10.1007/JHEP04(2015)031}{{\em JHEP} {\bfseries 04}
  (2015) 031},
\href{http://arxiv.org/abs/1412.0687}{{\ttfamily arXiv:1412.0687 [hep-th]}}.

\bibitem{Keski-Vakkuri1998}
E.~Keski-Vakkuri, ``Bulk and boundary dynamics in {BTZ} black holes,''
  \href{http://dx.doi.org/10.1103/PhysRevD.59.104001}{{\em Phys. Rev. D}
  {\bfseries 59} (Mar, 1999) 104001},
  \href{http://arxiv.org/abs/hep-th/9808037}{{\ttfamily hep-th/9808037}}.

\bibitem{Balasubramanian2013}
V.~Balasubramanian, A.~Bernamonti, B.~Craps, V.~Ker{\"a}nen, E.~Keski-Vakkuri,
  B.~M{\"u}ller, L.~Thorlacius, and J.~Vanhoof, ``Thermalization of the
  spectral function in strongly coupled two dimensional conformal field
  theories,'' \href{http://dx.doi.org/10.1007/JHEP04(2013)069}{{\em Journal of
  High Energy Physics} {\bfseries 2013} no.~4, (Apr, 2013) 69},
  \href{http://arxiv.org/abs/1212.6066}{{\ttfamily 1212.6066}}.

\bibitem{Horowitz1998}
G.~T. Horowitz and D.~Marolf, ``{A new approach to string cosmology},''
  \href{http://dx.doi.org/10.1088/1126-6708/1998/07/014}{{\em JHEP} {\bfseries
  07} (1998) 014},
\href{http://arxiv.org/abs/hep-th/9805207}{{\ttfamily arXiv:hep-th/9805207
  [hep-th]}}.

\bibitem{Banerjee2016}
P.~Banerjee, S.~Datta, and R.~Sinha, ``{Higher-point conformal blocks and
  entanglement entropy in heavy states},''
  \href{http://dx.doi.org/10.1007/JHEP05(2016)127}{{\em JHEP} {\bfseries 05}
  (2016) 127},
\href{http://arxiv.org/abs/1601.06794}{{\ttfamily arXiv:1601.06794 [hep-th]}}.

\bibitem{Anous2016}
T.~Anous, T.~Hartman, A.~Rovai, and J.~Sonner, ``{Black Hole Collapse in the
  1/c Expansion},'' \href{http://dx.doi.org/10.1007/JHEP07(2016)123}{{\em JHEP}
  {\bfseries 07} (2016) 123},
\href{http://arxiv.org/abs/1603.04856}{{\ttfamily arXiv:1603.04856 [hep-th]}}.

\bibitem{Anous2017}
T.~Anous, T.~Hartman, A.~Rovai, and J.~Sonner, ``{From Conformal Blocks to Path
  Integrals in the Vaidya Geometry},''
  \href{http://dx.doi.org/10.1007/JHEP09(2017)009}{{\em JHEP} {\bfseries 09}
  (2017) 009},
\href{http://arxiv.org/abs/1706.02668}{{\ttfamily arXiv:1706.02668 [hep-th]}}.

\bibitem{Nielsen2000}
M.~Nielsen and I.~Chuang, {\em {Quantum Information and Quantum Computation}}.
\newblock Cambridge University Press,
\newblock 2000.

\bibitem{Bennett1996}
C.~H. Bennett, D.~P. DiVincenzo, J.~A. Smolin, and W.~K. Wootters,
  ``Mixed-state entanglement and quantum error correction,''
  \href{http://dx.doi.org/10.1103/PhysRevA.54.3824}{{\em Phys. Rev. A}
  {\bfseries 54} (Nov, 1996) 3824--3851},
  \href{http://arxiv.org/abs/quant-ph/9604024}{{\ttfamily quant-ph/9604024}}.

\bibitem{Wootters1998}
W.~K. Wootters, ``Entanglement of formation of an arbitrary state of two
  qubits,'' \href{http://dx.doi.org/10.1103/PhysRevLett.80.2245}{{\em Phys.
  Rev. Lett.} {\bfseries 80} (Mar, 1998) 2245--2248},
  \href{http://arxiv.org/abs/quant-ph/9709029}{{\ttfamily quant-ph/9709029}}.

\bibitem{Renyi1961}
A.~R{\'e}nyi, ``On measures of information and entropy,''  {\em Proceedings of
  the fourth Berkeley Symposium on Mathematics, Statistics and Probability
  1960} (1961) 547--561.

\bibitem{Peres1996}
A.~Peres, ``Separability criterion for density matrices,''
  \href{http://dx.doi.org/10.1103/PhysRevLett.77.1413}{{\em Phys. Rev. Lett.}
  {\bfseries 77} (Aug, 1996) 1413--1415},
  \href{http://arxiv.org/abs/quant-ph/9604005}{{\ttfamily quant-ph/9604005}}.

\bibitem{Vollbrecht2002}
K.~G.~H. Vollbrecht and M.~M. Wolf, ``Conditional entropies and their relation
  to entanglement criteria,'' \href{http://dx.doi.org/10.1063/1.1498490}{{\em
  Journal of Mathematical Physics} {\bfseries 43} no.~9, (2002) 4299--4306},
  \href{http://arxiv.org/abs/quant-ph/0202058}{{\ttfamily quant-ph/0202058}}.

\bibitem{Horodecki1996}
M.~Horodecki, P.~Horodecki, and R.~Horodecki, ``Separability of mixed states:
  necessary and sufficient conditions,''
  \href{http://dx.doi.org/https://doi.org/10.1016/S0375-9601(96)00706-2}{{\em
  Physics Letters A} {\bfseries 223} no.~1, (1996) 1 -- 8},
  \href{http://arxiv.org/abs/quant-ph/9605038}{{\ttfamily quant-ph/9605038}}.

\bibitem{Zyczkowski1998}
K.~\ifmmode~\dot{Z}\else \.{Z}\fi{}yczkowski, P.~Horodecki, A.~Sanpera, and
  M.~Lewenstein, ``Volume of the set of separable states,''
  \href{http://dx.doi.org/10.1103/PhysRevA.58.883}{{\em Phys. Rev. A}
  {\bfseries 58} (Aug, 1998) 883--892},
  \href{http://arxiv.org/abs/quant-ph/9804024}{{\ttfamily quant-ph/9804024}}.

\bibitem{Vidal2001}
G.~Vidal and R.~F. Werner, ``Computable measure of entanglement,''
  \href{http://dx.doi.org/10.1103/PhysRevA.65.032314}{{\em Phys. Rev. A}
  {\bfseries 65} (Feb, 2002) 032314},
  \href{http://arxiv.org/abs/quant-ph/0102117}{{\ttfamily quant-ph/0102117}}.

\bibitem{Plenio2005a}
M.~B. Plenio, ``Logarithmic negativity: A full entanglement monotone that is
  not convex,'' \href{http://dx.doi.org/10.1103/PhysRevLett.95.090503}{{\em
  Phys. Rev. Lett.} {\bfseries 95} (Aug, 2005) 090503},
  \href{http://arxiv.org/abs/quant-ph/0505071}{{\ttfamily quant-ph/0505071}}.

\bibitem{Huang2014}
Y.~Huang, ``Computing quantum discord is np-complete,'' {\em New Journal of
  Physics} {\bfseries 16} no.~3, (2014) 033027,
  \href{http://arxiv.org/abs/1305.5941}{{\ttfamily 1305.5941}}.

\bibitem{Audenaert2003}
K.~Audenaert, M.~B. Plenio, and J.~Eisert, ``Entanglement cost under
  positive-partial-transpose-preserving operations,''
  \href{http://dx.doi.org/10.1103/PhysRevLett.90.027901}{{\em Phys. Rev. Lett.}
  {\bfseries 90} (Jan, 2003) 027901},
  \href{http://arxiv.org/abs/quant-ph/0207146}{{\ttfamily quant-ph/0207146}}.

\bibitem{Phoenix1988}
S.~Phoenix and P.~Knight, ``Fluctuations and entropy in models of quantum
  optical resonance,''
  \href{http://dx.doi.org/10.1016/0003-4916(88)90006-1}{{\em Annals of Physics}
  {\bfseries 186} no.~2, (09, 1988) 381--407}.

\bibitem{Gea-Banacloche1990}
J.~Gea-Banacloche, ``Collapse and revival of the state vector in the
  {J}aynes-{C}ummings model: An example of state preparation by a quantum
  apparatus,''  \href{http://dx.doi.org/10.1103/PhysRevLett.65.3385}{{\em Phys.
  Rev. Lett.} {\bfseries 65} (Dec, 1990) 3385--3388}.

\bibitem{Casini2009}
H.~Casini and M.~Huerta, ``Entanglement entropy in free quantum field theory,''
  {\em Journal of Physics A: Mathematical and Theoretical} {\bfseries 42}
  no.~50, (2009) 504007,  \href{http://arxiv.org/abs/0905.2562}{{\ttfamily
  0905.2562}}.

\bibitem{Amico2008}
L.~Amico, R.~Fazio, A.~Osterloh, and V.~Vedral, ``Entanglement in many-body
  systems,'' \href{http://dx.doi.org/10.1103/RevModPhys.80.517}{{\em Rev. Mod.
  Phys.} {\bfseries 80} (May, 2008) 517--576},
  \href{http://arxiv.org/abs/quant-ph/0703044}{{\ttfamily quant-ph/0703044}}.

\bibitem{Laflorencie2015}
N.~Laflorencie, ``Quantum entanglement in condensed matter systems,''
  \href{http://dx.doi.org/10.1016/j.physrep.2016.06.008}{{\em Physics Reports}
  {\bfseries 646} (2016) 1--59},
  \href{http://arxiv.org/abs/1512.03388v3}{{\ttfamily 1512.03388v3}}.

\bibitem{Eisert2006}
J.~Eisert and T.~J. Osborne, ``General entanglement scaling laws from time
  evolution,'' \href{http://dx.doi.org/10.1103/PhysRevLett.97.150404}{{\em
  Phys. Rev. Lett.} {\bfseries 97} (Oct, 2006) 150404},
  \href{http://arxiv.org/abs/quant-ph/0603114}{{\ttfamily quant-ph/0603114}}.

\bibitem{Bravyi2006}
S.~Bravyi, M.~B. Hastings, and F.~Verstraete, ``Lieb-robinson bounds and the
  generation of correlations and topological quantum order,''
  \href{http://dx.doi.org/10.1103/PhysRevLett.97.050401}{{\em Phys. Rev. Lett.}
  {\bfseries 97} (Jul, 2006) 050401},
  \href{http://arxiv.org/abs/quant-ph/0603121}{{\ttfamily quant-ph/0603121}}.

\bibitem{Calabrese2005}
P.~Calabrese and J.~Cardy, ``Evolution of entanglement entropy in
  one-dimensional systems,'' {\em Journal of Statistical Mechanics: Theory and
  Experiment} {\bfseries 2005} no.~04, (2005) P04010,
  \href{http://arxiv.org/abs/cond-mat/0503393}{{\ttfamily cond-mat/0503393}}.

\bibitem{Liu2014a}
H.~Liu and S.~J. Suh, ``Entanglement tsunami: Universal scaling in holographic
  thermalization,''
  \href{http://dx.doi.org/10.1103/PhysRevLett.112.011601}{{\em Phys. Rev.
  Lett.} {\bfseries 112} (Jan, 2014) 011601},
  \href{http://arxiv.org/abs/1305.7244}{{\ttfamily 1305.7244}}.

\bibitem{Hartman2013}
T.~Hartman and J.~Maldacena, ``Time evolution of entanglement entropy from
  black hole interiors,'' \href{http://dx.doi.org/10.1007/JHEP05(2013)014}{{\em
  Journal of High Energy Physics} {\bfseries 2013} no.~5, (May, 2013) 14},
  \href{http://arxiv.org/abs/1303.1080}{{\ttfamily 1303.1080}}.

\bibitem{Bianchi2018}
E.~Bianchi, L.~Hackl, and N.~Yokomizo, ``Linear growth of the entanglement
  entropy and the {Kolmogorov-Sinai} rate,''
  \href{http://dx.doi.org/10.1007/JHEP03(2018)025}{{\em Journal of High Energy
  Physics} {\bfseries 2018} no.~3, (Mar, 2018) 25},
  \href{http://arxiv.org/abs/1709.00427}{{\ttfamily 1709.00427}}.

\bibitem{Zurek2003}
W.~H. Zurek, ``Decoherence, einselection, and the quantum origins of the
  classical,'' \href{http://dx.doi.org/10.1103/RevModPhys.75.715}{{\em Rev.
  Mod. Phys.} {\bfseries 75} (May, 2003) 715--775},
  \href{http://arxiv.org/abs/quant-ph/0105127}{{\ttfamily quant-ph/0105127}}.

\bibitem{Zanardi2001}
P.~Zanardi, ``Virtual quantum subsystems,''
  \href{http://dx.doi.org/10.1103/PhysRevLett.87.077901}{{\em Phys. Rev. Lett.}
  {\bfseries 87} (Jul, 2001) 077901},
  \href{http://arxiv.org/abs/quant-ph/0103030}{{\ttfamily quant-ph/0103030}}.

\bibitem{Yang2017a}
I.-S. Yang, ``Entanglement timescale,''
  \href{http://dx.doi.org/10.1103/PhysRevD.97.066008}{{\em Phys. Rev. D}
  {\bfseries 97} (Mar, 2018) 066008},
  \href{http://arxiv.org/abs/1707.05792}{{\ttfamily 1707.05792}}.

\bibitem{Kim1996}
J.~I. Kim, M.~C. Nemes, A.~F.~R. de~Toledo~Piza, and H.~E. Borges,
  ``Perturbative expansion for coherence loss,''
  \href{http://dx.doi.org/10.1103/PhysRevLett.77.207}{{\em Phys. Rev. Lett.}
  {\bfseries 77} (Jul, 1996) 207--210}.

\bibitem{Li2008}
H.~Li and F.~D.~M. Haldane, ``Entanglement spectrum as a generalization of
  entanglement entropy: Identification of topological order in non-{A}belian
  fractional quantum {H}all effect states,''
  \href{http://dx.doi.org/10.1103/PhysRevLett.101.010504}{{\em Phys. Rev.
  Lett.} {\bfseries 101} (Jul, 2008) 010504},
  \href{http://arxiv.org/abs/0805.0332}{{\ttfamily 0805.0332}}.

\bibitem{Shore1993}
B.~W. Shore and P.~L. Knight, ``The {J}aynes-{C}ummings model,''
  \href{http://dx.doi.org/10.1080/09500349314551321}{{\em Journal of Modern
  Optics} {\bfseries 40} no.~7, (1993) 1195--1238}.

\bibitem{Daley2012}
A.~J. Daley, H.~Pichler, J.~Schachenmayer, and P.~Zoller, ``Measuring
  entanglement growth in quench dynamics of bosons in an optical lattice,''
  \href{http://dx.doi.org/10.1103/PhysRevLett.109.020505}{{\em Phys. Rev.
  Lett.} {\bfseries 109} (Jul, 2012) 020505},
  \href{http://arxiv.org/abs/1205.1521}{{\ttfamily 1205.1521}}.

\bibitem{Schachenmayer2013}
J.~Schachenmayer, B.~P. Lanyon, C.~F. Roos, and A.~J. Daley, ``Entanglement
  growth in quench dynamics with variable range interactions,''
  \href{http://dx.doi.org/10.1103/PhysRevX.3.031015}{{\em Phys. Rev. X}
  {\bfseries 3} (Sep, 2013) 031015},
  \href{http://arxiv.org/abs/1305.6880}{{\ttfamily 1305.6880}}.

\bibitem{Islam2015a}
R.~Islam, R.~Ma, P.~M. Preiss, M.~Eric~Tai, A.~Lukin, M.~Rispoli, and
  M.~Greiner, ``Measuring entanglement entropy in a quantum many-body system,''
  {\em Nature} {\bfseries 528} (Dec., 2015) 77,
  \href{http://arxiv.org/abs/1509.01160v1}{{\ttfamily 1509.01160v1}}.

\bibitem{Kaufman2016}
A.~M. Kaufman, M.~E. Tai, A.~Lukin, M.~Rispoli, R.~Schittko, P.~M. Preiss, and
  M.~Greiner, ``Quantum thermalization through entanglement in an isolated
  many-body system,'' \href{http://dx.doi.org/10.1126/science.aaf6725}{{\em
  Science} {\bfseries 353} no.~6301, (2016) 794--800},
  \href{http://arxiv.org/abs/1603.04409}{{\ttfamily 1603.04409}}.

\bibitem{Elben2018}
A.~Elben, B.~Vermersch, M.~Dalmonte, J.~I. Cirac, and P.~Zoller, ``R\'enyi
  entropies from random quenches in atomic {H}ubbard and spin models,''
  \href{http://dx.doi.org/10.1103/PhysRevLett.120.050406}{{\em Phys. Rev.
  Lett.} {\bfseries 120} (Feb, 2018) 050406},
  \href{http://arxiv.org/abs/1709.05060v2}{{\ttfamily 1709.05060v2}}.

\bibitem{Jaynes1963}
E.~T. Jaynes and F.~W. Cummings, ``Comparison of quantum and semiclassical
  radiation theories with application to the beam maser,''
  \href{http://dx.doi.org/10.1109/PROC.1963.1664}{{\em Proceedings of the IEEE}
  {\bfseries 51} no.~1, (Jan, 1963) 89--109}.

\bibitem{Quesada2013}
N.~Quesada and A.~Sanpera, ``Bound entanglement in the {J}aynes-{C}ummings
  model,'' {\em Journal of Physics B: Atomic, Molecular and Optical Physics}
  {\bfseries 46} no.~22, (2013) 224002,
  \href{http://arxiv.org/abs/1305.2604}{{\ttfamily 1305.2604}}.

\bibitem{Eberly1980}
J.~H. Eberly, N.~B. Narozhny, and J.~J. Sanchez-Mondragon, ``Periodic
  spontaneous collapse and revival in a simple quantum model,''
  \href{http://dx.doi.org/10.1103/PhysRevLett.44.1323}{{\em Phys. Rev. Lett.}
  {\bfseries 44} (May, 1980) 1323--1326}.

\bibitem{Pimenta2016}
H.~Pimenta and D.~F.~V. James, ``Characteristic-function approach to the
  {J}aynes-{C}ummings-model revivals,''
  \href{http://dx.doi.org/10.1103/PhysRevA.94.053803}{{\em Phys. Rev. A}
  {\bfseries 94} (Nov, 2016) 053803},
  \href{http://arxiv.org/abs/1608.06653}{{\ttfamily 1608.06653}}.

\bibitem{Gerry2005}
C.~Gerry and P.~Knight, {\em Introductory Quantum Optics}.
\newblock Cambridge University Pr.,
\newblock 2004.

\bibitem{Baker2017}
D.~Baker, D.~Kodwani, U.-L. Pen, and I.-S. Yang, ``A self-consistency check for
  unitary propagation of {H}awking quanta,''
  \href{http://dx.doi.org/10.1142/S0217751X17501986}{{\em International Journal
  of Modern Physics A} {\bfseries 32} no.~33, (2017) 1750198},
  \href{http://arxiv.org/abs/1701.04811v1}{{\ttfamily 1701.04811v1}}.

\bibitem{Yang2017}
I.-S. Yang, ``Secret loss of unitarity due to the classical background,''
  \href{http://dx.doi.org/10.1103/PhysRevD.96.025005}{{\em Phys. Rev. D}
  {\bfseries 96} (Jul, 2017) 025005},
  \href{http://arxiv.org/abs/1703.03466v2}{{\ttfamily 1703.03466v2}}.

\bibitem{Lashkari2014}
N.~Lashkari, ``Relative entropies in conformal field theory,''
  \href{http://dx.doi.org/10.1103/PhysRevLett.113.051602}{{\em Phys. Rev.
  Lett.} {\bfseries 113} (Jul, 2014) 051602},
  \href{http://arxiv.org/abs/1404.3216}{{\ttfamily 1404.3216}}.

\bibitem{Ruggiero2017}
P.~Ruggiero and P.~Calabrese, ``Relative entanglement entropies in 1 +
  1-dimensional conformal field theories,''
  \href{http://dx.doi.org/10.1007/JHEP02(2017)039}{{\em Journal of High Energy
  Physics} {\bfseries 2017} no.~2, (Feb, 2017) 39},
  \href{http://arxiv.org/abs/1612.00659}{{\ttfamily 1612.00659}}.

\bibitem{Hollands2017}
S.~Hollands and K.~Sanders, {\em Entanglement Measures and Their Properties in
  Quantum Field Theory}, vol.~34 of {\em SpringerBriefs in Mathematical
  Physics}.
\newblock Springer International Publishing, 2018.
\newblock
\newblock \href{http://arxiv.org/abs/1702.04924}{{\ttfamily 1702.04924}}.

\bibitem{Donnelly2014}
W.~Donnelly, ``Entanglement entropy and nonabelian gauge symmetry,''
  \href{http://dx.doi.org/10.1088/0264-9381/31/21/214003}{{\em Classical and
  Quantum Gravity} {\bfseries 31} no.~21, (2014) 214003},
  \href{http://arxiv.org/abs/1406.7304}{{\ttfamily 1406.7304}}.

\bibitem{Avery2014}
S.~G. Avery and M.~F. Paulos, ``Universal bounds on the time evolution of
  entanglement entropy,''
  \href{http://dx.doi.org/10.1103/PhysRevLett.113.231604}{{\em Phys. Rev.
  Lett.} {\bfseries 113} (Dec, 2014) 231604},
  \href{http://arxiv.org/abs/1407.0705}{{\ttfamily 1407.0705}}.

\bibitem{Liu2014}
H.~Liu and S.~J. Suh, ``Entanglement growth during thermalization in
  holographic systems,''
  \href{http://dx.doi.org/10.1103/PhysRevD.89.066012}{{\em Phys. Rev. D}
  {\bfseries 89} (Mar, 2014) 066012},
  \href{http://arxiv.org/abs/1311.1200}{{\ttfamily 1311.1200}}.

\bibitem{Belin2013}
A.~Belin, A.~Maloney, and S.~Matsuura, ``Holographic phases of {R}{\'e}nyi
  entropies,'' \href{http://dx.doi.org/10.1007/JHEP12(2013)050}{{\em Journal of
  High Energy Physics} {\bfseries 2013} no.~12, (Dec, 2013) 50},
  \href{http://arxiv.org/abs/1306.2640}{{\ttfamily 1306.2640}}.

\bibitem{Cresswell2015}
J.~C. Cresswell and D.~N. Vollick, ``Lorenz gauge quantization in conformally
  flat spacetimes,'' \href{http://dx.doi.org/10.1103/PhysRevD.91.084008}{{\em
  Phys. Rev. D} {\bfseries 91} (Apr, 2015) 084008},
  \href{http://arxiv.org/abs/1504.05914}{{\ttfamily 1504.05914}}.

\bibitem{Caputa2014}
P.~Caputa, M.~Nozaki, and T.~Takayanagi, ``Entanglement of local operators in
  large-{N} conformal field theories,''
  \href{http://dx.doi.org/10.1093/ptep/ptu122}{{\em Progress of Theoretical and
  Experimental Physics} {\bfseries 2014} no.~9, (2014) 093B06},
  \href{http://arxiv.org/abs/1405.5946}{{\ttfamily 1405.5946}}.

\bibitem{Horodecki2009}
R.~Horodecki, P.~Horodecki, M.~Horodecki, and K.~Horodecki, ``Quantum
  entanglement,'' \href{http://dx.doi.org/10.1103/RevModPhys.81.865}{{\em Rev.
  Mod. Phys.} {\bfseries 81} (Jun, 2009) 865--942},
  \href{http://arxiv.org/abs/quant-ph/0702225}{{\ttfamily quant-ph/0702225}}.

\bibitem{Bennett1993}
C.~H. Bennett, G.~Brassard, C.~Cr\'epeau, R.~Jozsa, A.~Peres, and W.~K.
  Wootters, ``Teleporting an unknown quantum state via dual classical and
  {E}instein-{P}odolsky-{R}osen channels,''
  \href{http://dx.doi.org/10.1103/PhysRevLett.70.1895}{{\em Phys. Rev. Lett.}
  {\bfseries 70} (Mar, 1993) 1895--1899}.

\bibitem{Bouwmeester1997}
D.~Bouwmeester, J.-W. Pan, K.~Mattle, M.~Eibl, H.~Weinfurter, and A.~Zeilinger,
  ``Experimental quantum teleportation,''  {\em Nature} {\bfseries 390} (Dec.,
  1997) 575.

\bibitem{Barrett2004}
M.~D. Barrett, J.~Chiaverini, T.~Schaetz, J.~Britton, W.~M. Itano, J.~D. Jost,
  E.~Knill, C.~Langer, D.~Leibfried, R.~Ozeri, and D.~J. Wineland,
  ``Deterministic quantum teleportation of atomic qubits,''  {\em Nature}
  {\bfseries 429} (June, 2004) 737.

\bibitem{Riebe2004}
M.~Riebe, H.~H\"affner, C.~F. Roos, W.~H\"ansel, J.~Benhelm, G.~P.~T.
  Lancaster, T.~W. K\:orber, C.~Becher, F.~Schmidt-Kaler, D.~F.~V. James, and
  R.~Blatt, ``Deterministic quantum teleportation with atoms,''  {\em Nature}
  {\bfseries 429} (June, 2004) 734.

\bibitem{Mitchell2004}
M.~W. Mitchell, J.~S. Lundeen, and A.~M. Steinberg, ``Super-resolving phase
  measurements with a multiphoton entangled state,'' {\em Nature} {\bfseries
  429} (May, 2004) 161,
  \href{http://arxiv.org/abs/quant-ph/0312186}{{\ttfamily quant-ph/0312186}}.

\bibitem{Pezze2009}
L.~Pezz\'e and A.~Smerzi, ``Entanglement, nonlinear dynamics, and the
  {H}eisenberg limit,''
  \href{http://dx.doi.org/10.1103/PhysRevLett.102.100401}{{\em Phys. Rev.
  Lett.} {\bfseries 102} (Mar, 2009) 100401},
  \href{http://arxiv.org/abs/0711.4840}{{\ttfamily 0711.4840}}.

\bibitem{giovannetti2011advances}
V.~Giovannetti, S.~Lloyd, and L.~Maccone, ``Advances in quantum metrology,''
  {\em Nature Photonics} {\bfseries 5} (Mar., 2011) 222,
  \href{http://arxiv.org/abs/1102.2318}{{\ttfamily 1102.2318}}.

\bibitem{gross2012nonlinear}
C.~Gross, T.~Zibold, E.~Nicklas, J.~Est\`eve, and M.~K. Oberthaler, ``Nonlinear
  atom interferometer surpasses classical precision limit,'' {\em Nature}
  {\bfseries 464} (Mar., 2010) 1165,
  \href{http://arxiv.org/abs/1009.2374}{{\ttfamily 1009.2374}}.

\bibitem{Ekert1991}
A.~K. Ekert, ``Quantum cryptography based on {B}ell's theorem,''
  \href{http://dx.doi.org/10.1103/PhysRevLett.67.661}{{\em Phys. Rev. Lett.}
  {\bfseries 67} (Aug, 1991) 661--663}.

\bibitem{Gisin2002}
N.~Gisin, G.~Ribordy, W.~Tittel, and H.~Zbinden, ``Quantum cryptography,''
  \href{http://dx.doi.org/10.1103/RevModPhys.74.145}{{\em Rev. Mod. Phys.}
  {\bfseries 74} (Mar, 2002) 145--195},
  \href{http://arxiv.org/abs/quant-ph/0101098}{{\ttfamily quant-ph/0101098}}.

\bibitem{Curty2004}
M.~Curty, M.~Lewenstein, and N.~L\"utkenhaus, ``Entanglement as a precondition
  for secure quantum key distribution,''
  \href{http://dx.doi.org/10.1103/PhysRevLett.92.217903}{{\em Phys. Rev. Lett.}
  {\bfseries 92} (May, 2004) 217903},
  \href{http://arxiv.org/abs/quant-ph/0307151}{{\ttfamily quant-ph/0307151}}.

\bibitem{Tang2014}
Z.~Tang, Z.~Liao, F.~Xu, B.~Qi, L.~Qian, and H.-K. Lo, ``Experimental
  demonstration of polarization encoding measurement-device-independent quantum
  key distribution,''
  \href{http://dx.doi.org/10.1103/PhysRevLett.112.190503}{{\em Phys. Rev.
  Lett.} {\bfseries 112} (May, 2014) 190503},
  \href{http://arxiv.org/abs/1306.6134}{{\ttfamily 1306.6134}}.

\bibitem{Audenaertetal2002}
K.~Audenaert, J.~Eisert, M.~B. Plenio, and R.~F. Werner, ``Entanglement
  properties of the harmonic chain,''
  \href{http://dx.doi.org/10.1103/PhysRevA.66.042327}{{\em Phys. Rev. A}
  {\bfseries 66} (Oct, 2002) 042327},
  \href{http://arxiv.org/abs/quant-ph/0205025}{{\ttfamily quant-ph/0205025}}.

\bibitem{Anders2008}
J.~Anders, ``Thermal state entanglement in harmonic lattices,''
  \href{http://dx.doi.org/10.1103/PhysRevA.77.062102}{{\em Phys. Rev. A}
  {\bfseries 77} (Jun, 2008) 062102},
  \href{http://arxiv.org/abs/0803.1102}{{\ttfamily 0803.1102}}.

\bibitem{EislerZimboras2014}
V.~Eisler and Z.~Zimbor{\'a}s, ``Entanglement negativity in the harmonic chain
  out of equilibrium,'' {\em New Journal of Physics} {\bfseries 16} no.~12,
  (2014) 123020,  \href{http://arxiv.org/abs/1406.5474}{{\ttfamily 1406.5474}}.

\bibitem{HelmesWessel2014}
J.~Helmes and S.~Wessel, ``Entanglement entropy scaling in the bilayer
  {H}eisenberg spin system,''
  \href{http://dx.doi.org/10.1103/PhysRevB.89.245120}{{\em Phys. Rev. B}
  {\bfseries 89} (Jun, 2014) 245120},
  \href{http://arxiv.org/abs/1403.7395}{{\ttfamily 1403.7395}}.

\bibitem{EislerZimboras2015}
V.~Eisler and Z.~Zimbor{\'a}s, ``On the partial transpose of fermionic
  {G}aussian states,'' {\em New Journal of Physics} {\bfseries 17} no.~5,
  (2015) 053048,  \href{http://arxiv.org/abs/1502.01369}{{\ttfamily
  1502.01369}}.

\bibitem{Shermanetal2016}
N.~E. Sherman, T.~Devakul, M.~B. Hastings, and R.~R.~P. Singh,
  ``Nonzero-temperature entanglement negativity of quantum spin models: {A}rea
  law, linked cluster expansions, and sudden death,''
  \href{http://dx.doi.org/10.1103/PhysRevE.93.022128}{{\em Phys. Rev. E}
  {\bfseries 93} (Feb, 2016) 022128},
  \href{http://arxiv.org/abs/1510.08005}{{\ttfamily 1510.08005}}.

\bibitem{Shimetal2018}
J.~Shim, H.-S. Sim, and S.-S.~B. Lee, ``Numerical renormalization group method
  for entanglement negativity at finite temperature,''
  \href{http://dx.doi.org/10.1103/PhysRevB.97.155123}{{\em Phys. Rev. B}
  {\bfseries 97} (Apr, 2018) 155123},
  \href{http://arxiv.org/abs/1808.08506}{{\ttfamily 1808.08506}}.

\bibitem{Calabreseetal2012}
P.~Calabrese, J.~Cardy, and E.~Tonni, ``Entanglement negativity in quantum
  field theory,'' \href{http://dx.doi.org/10.1103/PhysRevLett.109.130502}{{\em
  Phys. Rev. Lett.} {\bfseries 109} (Sep, 2012) 130502},
  \href{http://arxiv.org/abs/1206.3092}{{\ttfamily 1206.3092}}.

\bibitem{Coser2014}
A.~Coser, E.~Tonni, and P.~Calabrese, ``Entanglement negativity after a global
  quantum quench,'' {\em Journal of Statistical Mechanics: Theory and
  Experiment} {\bfseries 2014} no.~12, (2014) P12017,
  \href{http://arxiv.org/abs/1410.0900}{{\ttfamily 1410.0900}}.

\bibitem{Watson1992}
G.~Watson, ``Characterization of the subdifferential of some matrix norms,''
  \href{http://dx.doi.org/10.1016/0024-3795(92)90407-2}{{\em Linear Algebra and
  its Applications} {\bfseries 170} (1992) 33 -- 45}.

\bibitem{Hjorungnes2011}
A.~Hj{\o}rungnes, {\em Complex-Valued Matrix Derivatives With Applications in
  Signal Processing and Communications}.
\newblock Cambridge University Press,
\newblock 2011.

\bibitem{Magnus1985}
J.~R. Magnus and H.~Neudecker, ``Matrix differential calculus with applications
  to simple, {H}adamard, and {K}ronecker products,''
  \href{http://dx.doi.org/https://doi.org/10.1016/0022-2496(85)90006-9}{{\em
  Journal of Mathematical Psychology} {\bfseries 29} no.~4, (1985) 474 -- 492}.

\bibitem{Magnus2007}
J.~R. Magnus and H.~Neudecker, {\em Matrix Differential Calculus with
  Applications in Statistics and Econometrics}.
\newblock Wiley, 3rd~ed.,
\newblock 2007.

\bibitem{Havel2002}
T.~F. Havel, ``Robust procedures for converting among {L}indblad, {K}raus and
  matrix representations of quantum dynamical semigroups,''
  \href{http://dx.doi.org/10.1063/1.1518555}{{\em Journal of Mathematical
  Physics} {\bfseries 44} no.~2, (2003) 534--557},
  \href{http://arxiv.org/abs/quant-ph/0201127}{{\ttfamily quant-ph/0201127}}.

\bibitem{Zyczkowski2004}
K.~{\.{Z}}yczkowski and I.~Bengtsson, ``On duality between quantum maps and
  quantum states,''
  \href{http://dx.doi.org/10.1023/B:OPSY.0000024753.05661.c2}{{\em Open Systems
  {\&} Information Dynamics} {\bfseries 11} no.~1, (Mar, 2004) 3--42},
  \href{http://arxiv.org/abs/quant-ph/0401119}{{\ttfamily quant-ph/0401119}}.

\bibitem{Jaroslaw2011}
J.~A. Miszcak, ``Singular value decomposition and matrix reorderings in quantum
  information theory,'' \href{http://dx.doi.org/10.1142/S0129183111016683}{{\em
  International Journal of Modern Physics C} {\bfseries 22} no.~09, (2011)
  897--918},  \href{http://arxiv.org/abs/1011.1585}{{\ttfamily 1011.1585}}.

\bibitem{Hjorungnes2007}
A.~Hj{\o}rungnes and D.~Gesbert, ``Complex-valued matrix differentiation:
  Techniques and key results,''
  \href{http://dx.doi.org/10.1109/TSP.2007.893762}{{\em IEEE Transactions on
  Signal Processing} {\bfseries 55} no.~6, (June, 2007) 2740--2746}.

\bibitem{Higham2008}
N.~J. Higham, {\em Functions of Matrices: Theory and Computation}, vol.~104.
\newblock SIAM,
\newblock 2008.

\bibitem{Finketal2008}
J.~M. Fink, M.~G{\"o}ppl, M.~Baur, R.~Bianchetti, P.~J. Leek, A.~Blais, and
  A.~Wallraff, ``Climbing the {J}aynes-{C}ummings ladder and observing its
  nonlinearity in a cavity {QED} system,'' {\em Nature} {\bfseries 454} (July,
  2008) 315,  \href{http://arxiv.org/abs/0902.1827}{{\ttfamily 0902.1827}}.

\bibitem{Horodeckietal1998Bound}
M.~Horodecki, P.~Horodecki, and R.~Horodecki, ``Mixed-state entanglement and
  distillation: Is there a ``bound'' entanglement in nature?,''
  \href{http://dx.doi.org/10.1103/PhysRevLett.80.5239}{{\em Phys. Rev. Lett.}
  {\bfseries 80} (Jun, 1998) 5239--5242},
  \href{http://arxiv.org/abs/quant-ph/9801069}{{\ttfamily quant-ph/9801069}}.

\bibitem{Lavoieetal2010}
J.~Lavoie, R.~Kaltenbaek, M.~Piani, and K.~J. Resch, ``Experimental bound
  entanglement in a four-photon state,''
  \href{http://dx.doi.org/10.1103/PhysRevLett.105.130501}{{\em Phys. Rev.
  Lett.} {\bfseries 105} (Sep, 2010) 130501},
  \href{http://arxiv.org/abs/1005.1258}{{\ttfamily 1005.1258}}.

\bibitem{BenattiFloreaniniPiani2003}
F.~Benatti, R.~Floreanini, and M.~Piani, ``Environment induced entanglement in
  {M}arkovian dissipative dynamics,''
  \href{http://dx.doi.org/10.1103/PhysRevLett.91.070402}{{\em Phys. Rev. Lett.}
  {\bfseries 91} (Aug, 2003) 070402},
  \href{http://arxiv.org/abs/quant-ph/0307052}{{\ttfamily quant-ph/0307052}}.

\bibitem{Lopezetal2008}
C.~E. L\'opez, G.~Romero, F.~Lastra, E.~Solano, and J.~C. Retamal, ``Sudden
  birth versus sudden death of entanglement in multipartite systems,''
  \href{http://dx.doi.org/10.1103/PhysRevLett.101.080503}{{\em Phys. Rev.
  Lett.} {\bfseries 101} (Aug, 2008) 080503},
  \href{http://arxiv.org/abs/0802.1825}{{\ttfamily 0802.1825}}.

\bibitem{Wangetal2016}
L.-D. Wang, L.-T. Wang, M.~Yang, J.-Z. Xu, Z.~D. Wang, and Y.-K. Bai,
  ``Entanglement and measurement-induced nonlocality of mixed maximally
  entangled states in multipartite dynamics,''
  \href{http://dx.doi.org/10.1103/PhysRevA.93.062309}{{\em Phys. Rev. A}
  {\bfseries 93} (Jun, 2016) 062309},
  \href{http://arxiv.org/abs/1506.06878}{{\ttfamily 1506.06878}}.

\bibitem{Almeidaetal2007}
M.~P. Almeida, F.~de~Melo, M.~Hor-Meyll, A.~Salles, S.~P. Walborn, P.~H.~S.
  Ribeiro, and L.~Davidovich, ``Environment-induced sudden death of
  entanglement,''  \href{http://dx.doi.org/10.1126/science.1139892}{{\em
  Science} {\bfseries 316} no.~5824, (2007) 579--582}.

\bibitem{AlQasimiJames2008}
A.~Al-Qasimi and D.~F.~V. James, ``Sudden death of entanglement at finite
  temperature,'' \href{http://dx.doi.org/10.1103/PhysRevA.77.012117}{{\em Phys.
  Rev. A} {\bfseries 77} (Jan, 2008) 012117},
  \href{http://arxiv.org/abs/0707.2611}{{\ttfamily 0707.2611}}.

\bibitem{Breueretal2009}
H.-P. Breuer, E.-M. Laine, and J.~Piilo, ``Measure for the degree of
  non-{M}arkovian behavior of quantum processes in open systems,''
  \href{http://dx.doi.org/10.1103/PhysRevLett.103.210401}{{\em Phys. Rev.
  Lett.} {\bfseries 103} (Nov, 2009) 210401},
  \href{http://arxiv.org/abs/0908.0238}{{\ttfamily 0908.0238}}.

\bibitem{Rivasetal2010}
A.~Rivas, S.~F. Huelga, and M.~B. Plenio, ``Entanglement and non-{M}arkovianity
  of quantum evolutions,''
  \href{http://dx.doi.org/10.1103/PhysRevLett.105.050403}{{\em Phys. Rev.
  Lett.} {\bfseries 105} (Jul, 2010) 050403},
  \href{http://arxiv.org/abs/0911.4270}{{\ttfamily 0911.4270}}.

\bibitem{Aaronsonetal2013}
B.~Aaronson, R.~L. Franco, G.~Compagno, and G.~Adesso, ``Hierarchy and dynamics
  of trace distance correlations,'' {\em New Journal of Physics} {\bfseries 15}
  no.~9, (2013) 093022,  \href{http://arxiv.org/abs/1307.3953}{{\ttfamily
  1307.3953}}.

\bibitem{Deffner2013}
S.~Deffner and E.~Lutz, ``Quantum speed limit for non-{M}arkovian dynamics,''
  \href{http://dx.doi.org/10.1103/PhysRevLett.111.010402}{{\em Phys. Rev.
  Lett.} {\bfseries 111} (Jul, 2013) 010402},
  \href{http://arxiv.org/abs/1302.5069}{{\ttfamily 1302.5069}}.

\bibitem{Deffner2017}
S.~Deffner and S.~Campbell, ``Quantum speed limits: from {H}eisenberg's
  uncertainty principle to optimal quantum control,'' {\em Journal of Physics
  A: Mathematical and Theoretical} {\bfseries 50} no.~45, (2017) 453001,
  \href{http://arxiv.org/abs/1705.08023}{{\ttfamily 1705.08023}}.

\bibitem{Deffner2017a}
S.~Deffner, ``Geometric quantum speed limits: a case for {Wigner} phase
  space,'' \href{http://dx.doi.org/10.1088/1367-2630/aa83dc}{{\em New Journal
  of Physics} {\bfseries 19} no.~10, (Oct, 2017) 103018},
  \href{http://arxiv.org/abs/1704.03357}{{\ttfamily 1704.03357}}.

\bibitem{Tracy1988}
D.~S. Tracy and K.~G. Jinadasa, ``Patterned matrix derivatives,''  {\em The
  Canadian Journal of Statistics / La Revue Canadienne de Statistique}
  {\bfseries 16} no.~4, (1988) 411--418.

\bibitem{Hjorungnes2008}
A.~Hj{\o}rungnes and D.~P. Palomar,
  \href{http://dx.doi.org/10.1109/SAM.2008.4606875}{``Patterned complex-valued
  matrix derivatives,''} in {\em 2008 5th IEEE Sensor Array and Multichannel
  Signal Processing Workshop}, pp.~293--297.
\newblock
\newblock July, 2008.

\bibitem{Hjorungnes2008a}
A.~Hj{\o}rungnes and D.~P. Palomar,
  \href{http://dx.doi.org/10.1109/ISABEL.2008.4712619}{``Finding patterned
  complex-valued matrix derivatives by using manifolds,''} in {\em 2008 First
  International Symposium on Applied Sciences on Biomedical and Communication
  Technologies}, pp.~1--5.
\newblock
\newblock Oct, 2008.

\bibitem{Johnston2018}
N.~Johnston and E.~Patterson, ``The inverse eigenvalue problem for entanglement
  witnesses,'' \href{http://dx.doi.org/10.1016/j.laa.2018.03.043}{{\em Linear
  Algebra and its Applications} {\bfseries 550} (2018) 1 -- 27},
  \href{http://arxiv.org/abs/1708.05901}{{\ttfamily 1708.05901}}.

\bibitem{Lu2018}
T.-C. Lu and T.~Grover, ``Singularity in entanglement negativity across
  finite-temperature phase transitions,''
  \href{http://dx.doi.org/10.1103/PhysRevB.99.075157}{{\em Phys. Rev. B}
  {\bfseries 99} (Feb, 2019) 075157},
  \href{http://arxiv.org/abs/1808.04381}{{\ttfamily 1808.04381}}.

\bibitem{Fedorovetal2018}
K.~G. Fedorov, S.~Pogorzalek, U.~Las~Heras, M.~Sanz, P.~Yard, P.~Eder,
  M.~Fischer, J.~Goetz, E.~Xie, K.~Inomata, Y.~Nakamura, R.~Di~Candia,
  E.~Solano, A.~Marx, F.~Deppe, and R.~Gross, ``Finite-time quantum
  entanglement in propagating squeezed microwaves,'' {\em Scientific Reports}
  {\bfseries 8} no.~1, (Apr., 2018) 6416,
  \href{http://arxiv.org/abs/1703.05138}{{\ttfamily 1703.05138}}.

\bibitem{Layden2012}
D.~Layden, M.~F.~G. Wood, and I.~A. Vitkin, ``Optimum selection of input
  polarization states in determining the sample {M}ueller matrix: a dual
  photoelastic polarimeter approach,''
  \href{http://dx.doi.org/10.1364/OE.20.020466}{{\em Opt. Express} {\bfseries
  20} no.~18, (Aug, 2012) 20466--20481}.

\bibitem{Simon2010}
B.~N. Simon, S.~Simon, F.~Gori, M.~Santarsiero, R.~Borghi, N.~Mukunda, and
  R.~Simon, ``Nonquantum entanglement resolves a basic issue in polarization
  optics,'' \href{http://dx.doi.org/10.1103/PhysRevLett.104.023901}{{\em Phys.
  Rev. Lett.} {\bfseries 104} (Jan, 2010) 023901},
  \href{http://arxiv.org/abs/0906.2467}{{\ttfamily 0906.2467}}.

\bibitem{Balasubramanian2018}
V.~Balasubramanian, B.~Craps, T.~De~Jonckheere, and G.~S{\'a}rosi,
  ``Entanglement versus entwinement in symmetric product orbifolds,''
  \href{http://dx.doi.org/10.1007/JHEP01(2019)190}{{\em Journal of High Energy
  Physics} {\bfseries 2019} no.~1, (Jan, 2019) 190},
  \href{http://arxiv.org/abs/1806.02871}{{\ttfamily 1806.02871}}.

\bibitem{Rubin2002}
B.~Rubin, ``Radon, cosine and sine transforms on real hyperbolic space,''
  \href{http://dx.doi.org/https://doi.org/10.1006/aima.2002.2074}{{\em Advances
  in Mathematics} {\bfseries 170} no.~2, (2002) 206 -- 223}.

\bibitem{Espindola2018}
R.~Esp{\'i}ndola, A.~G{\"u}ijosa, and J.~F. Pedraza, ``Entanglement wedge
  reconstruction and entanglement of purification,''
  \href{http://dx.doi.org/10.1140/epjc/s10052-018-6140-2}{{\em The European
  Physical Journal C} {\bfseries 78} no.~8, (Aug, 2018) 646},
  \href{http://arxiv.org/abs/1804.05855}{{\ttfamily 1804.05855}}.

\bibitem{Balasubramanian2018a}
V.~Balasubramanian and C.~Rabideau, ``The dual of non-extremal area:
  differential entropy in higher dimensions,''
  \href{http://arxiv.org/abs/1812.06985}{{\ttfamily 1812.06985}}.

\bibitem{Pippenger2003}
N.~{Pippenger}, ``The inequalities of quantum information theory,''
  \href{http://dx.doi.org/10.1109/TIT.2003.809569}{{\em IEEE Transactions on
  Information Theory} {\bfseries 49} no.~4, (April, 2003) 773--789}.

\bibitem{Hernandez2019}
S.~Hern\'andez~Cuenca, ``Holographic entropy cone for five regions,''
  \href{http://dx.doi.org/10.1103/PhysRevD.100.026004}{{\em Phys. Rev. D}
  {\bfseries 100} (Jul, 2019) 026004},
  \href{http://arxiv.org/abs/1903.09148}{{\ttfamily 1903.09148}}.

\bibitem{Rangamani2016}
M.~Rangamani and T.~Takayanagi, {\em {Holographic Entanglement Entropy}}.
\newblock Springer, 2017.
\newblock
\newblock \href{http://arxiv.org/abs/1609.01287}{{\ttfamily 1609.01287}}.

\bibitem{Ruskai2002}
M.~B. Ruskai, ``Inequalities for quantum entropy: {A} review with conditions
  for equality,'' \href{http://dx.doi.org/10.1063/1.1497701}{{\em Journal of
  Mathematical Physics} {\bfseries 43} no.~9, (Apr., 2002) 4358--4375},
  \href{http://arxiv.org/abs/quant-ph/0205064}{{\ttfamily quant-ph/0205064}}.

\bibitem{Hayden2004}
P.~Hayden, R.~Jozsa, D.~Petz, and A.~Winter, ``Structure of states which
  satisfy strong subadditivity of quantum entropy with equality,''
  \href{http://dx.doi.org/10.1007/s00220-004-1049-z}{{\em Communications in
  Mathematical Physics} {\bfseries 246} no.~2, (Apr, 2004) 359--374},
  \href{http://arxiv.org/abs/quant-ph/0304007}{{\ttfamily quant-ph/0304007}}.

\bibitem{Casini2008}
H.~Casini and M.~Huerta, ``Remarks on the entanglement entropy for disconnected
  regions,'' \href{http://dx.doi.org/10.1088/1126-6708/2009/03/048}{{\em
  Journal of High Energy Physics} {\bfseries 2009} no.~03, (Mar, 2009)
  048--048},  \href{http://arxiv.org/abs/0812.1773}{{\ttfamily 0812.1773}}.

\bibitem{Rota2016}
M.~Rota, ``Tripartite information of highly entangled states,''
  \href{http://dx.doi.org/10.1007/JHEP04(2016)075}{{\em Journal of High Energy
  Physics} {\bfseries 2016} no.~4, (Apr, 2016) 75},
  \href{http://arxiv.org/abs/1512.03751}{{\ttfamily 1512.03751}}.

\bibitem{Casini2005}
H.~Casini, C.~D. Fosco, and M.~Huerta, ``Entanglement and alpha entropies for a
  massive {D}irac field in two dimensions,''
  \href{http://dx.doi.org/10.1088/1742-5468/2005/07/p07007}{{\em Journal of
  Statistical Mechanics: Theory and Experiment} {\bfseries 2005} no.~07, (Jul,
  2005) P07007--P07007},
  \href{http://arxiv.org/abs/cond-mat/0505563}{{\ttfamily cond-mat/0505563}}.

\bibitem{Pearl1988}
J.~Pearl, {\em Probabilistic Reasoning in Intelligent Systems: Networks of
  Plausible Inference}.
\newblock Morgan Kaufman, revised 2nd~ed.,
\newblock 1988.

\bibitem{Rodriguez2007}
G.~Rodriguez, ``Lecture notes on generalized linear models: Log-linear models
  for contingency tables.'' http://data.princeton.edu/wws509/notes/, 2007.

\bibitem{Koller2009}
D.~Koller and N.~Friedman, {\em Probabilistic Graphical Models}.
\newblock MIT Press,
\newblock 2009.

\end{thebibliography}\endgroup


\end{document}